\documentclass[11pt,dvipsnames]{article}

\usepackage{jheppub}
\usepackage[utf8]{inputenc}
\usepackage{amsmath,amsthm,amsfonts,amssymb,mathtools,bbm,mathdots,tensor}
\usepackage[sans]{dsfont}
\usepackage[mathscr]{eucal}
\usepackage[normalem]{ulem}
\usepackage{enumitem}
\usepackage{slashed,extarrows,bigints,cancel}
\usepackage{graphicx}
\usepackage{float,caption,multirow,makecell}
\usepackage{bm}
\usepackage[]{xcolor}
\usepackage[hyperpageref]{backref}
\usepackage[]{hyperref}
\hypersetup{bookmarks=true, colorlinks = true, urlcolor=MidnightBlue, citecolor=red, linkcolor=ao!90!black}
\definecolor{ao}{rgb}{0.0, 0.5, 0.0}

\usepackage{multirow,booktabs} 
\usepackage{longtable}

\usepackage{tikz}
\usetikzlibrary{decorations.markings,intersections, decorations.pathreplacing,decorations.pathmorphing, angles,quotes, shapes.geometric, arrows,matrix,positioning, patterns}
\tikzset{snake it/.style={decorate, decoration=snake}}
\usepackage{tikz-cd}

\newcommand{\mscr}[1]{\mathscr{#1}}

\newcommand{\mfk}[1]{\mathfrak{#1}}
\newcommand{\mbs}[1]{\boldsymbol{#1}}
\newcommand{\mbb}[1]{\mathbb{#1}}

\newcommand{\mbf}[1]{\mathbf{#1}}
\newcommand{\tsf}[1]{\textsf{#1}}
\newcommand{\mcal}[1]{\mathcal{#1}}

\newcommand{\wh}[1]{\widehat{#1}}
\newcommand{\wt}[1]{\widetilde{#1}}

\newcommand{\br}{\bar{r}}
\newcommand{\bg}{\bar{g}}

\newcommand{\iso}{\msf{iso}}

\newcommand{\tenofo}[1]{\text{\normalfont #1}}
\newcommand{\pa}[1]{\partial_{#1}}
\newcommand{\rd}{\msf{d}}
\newcommand{\bd}{\mbs{d}}

\newcommand{\phasespace}{\mbs{\Gamma}}

\newcommand{\cJ}{{\cal J}}
\newcommand{\cN}{{\cal N}}
\newcommand{\scri}{\mathcal I}

\newcommand{\cP}{{\mathcal P}}
\newcommand{\cM}{{\mathcal M}}
\newcommand{\tcM}{\widetilde{\mathcal M}}

\allowdisplaybreaks
\numberwithin{equation}{section}

\newcommand{\sds}{\,\tikz[baseline=1]{
		\draw[line width=.6pt] (0,.13) circle (.8ex);
		\draw[line width=.6pt] (-.05ex,.27) -- (-.05ex,0);
		\draw[line width=.6pt] (0,.13) -- (.8ex,.13);}\,}

\newcommand{\vsds}{\tikz[baseline=1]{
		\draw[line width=.6pt] (0,.13) circle (.8ex);
		\draw[line width=.6pt] (0,.23) -- (0,+.03);
		\draw[line width=.6pt] (0,.13) -- (.8ex,.13);}\,}

\newenvironment{eqaligned}
{%
\begin{equation}
    \begin{aligned}
    } 
{%
\end{aligned}
\end{equation}
\ignorespacesafterend}

\newenvironment{eqaligned*}
{%
\begin{equation*}
    \begin{aligned}
    } 
{%
\end{aligned}
\end{equation*}
\ignorespacesafterend}

\newenvironment{eqgathered}
{%
\begin{equation}
    \begin{gathered}
    } 
{%
\end{gathered}
\end{equation}
\ignorespacesafterend}

\newenvironment{eqgathered*}
{%
\begin{equation*}
    \begin{gathered}
    } 
{%
\end{gathered}
\end{equation*}
\ignorespacesafterend}

\newcommand{\f}{\frac}

\def\p{\partial}

\def\d{\delta}

\def\x{\xi}

\def\s{\sigma}

\def\D{\Delta}

\def\rd{\mathsf{d}}
\def\pa{\partial }

\def\bq{\gamma}
\def\d{\delta }

\newcommand{\bit}{\begin{itemize}}
	\newcommand{\eit}{\end{itemize}}

\def\be#1\ee{\begin{align}#1\end{align}}

\newcommand{\bc}{\begin{center}}
	\newcommand{\ec}{\end{center}}

\newcommand{\va}{\scriptscriptstyle}

\newcommand{\C}{{\mathbb C}}

\newcommand{\Z}{{\mathbb Z}}

\newcommand{\cI}{{\mathcal I}}

\newcommand{\cL}{{\mathcal L}}

\newcommand{\cD}{{\mathcal D}}
\newcommand{\cC}{{\mathcal C}}

\newcommand{\SL}{\mathrm{SL}}
\newcommand{\SO}{\mathrm{SO}}

\newcommand{\so}{{\msf{so}}}

\newcommand{\bea}{\begin{eqnarray}}
	\newcommand{\eea}{\end{eqnarray}}
\newcommand{\bs}{\begin{subequations}}
	\newcommand{\es}{\end{subequations}}

\newcommand{\la}{\label}

\newcommand{\bms}{\textsf{bms}}
\newcommand{\Poinc}{\textsf{Poinc}}
\newcommand{\ebms}{\textsf{ebms}}
\newcommand{\gbms}{\textsf{gbms}}
\newcommand{\g}{\textsf{g}}

\newcommand{\diff}{\textsf{diff}}

\newcommand{\msf}[1]{\mathsf{#1}}

\newcommand{\bfe}{{ \mbs \epsilon}}

\newcommand{\bfs}{{ \mbs \rho}}

\theoremstyle{definition}
\newtheorem{thr}{Theorem}[section]

\theoremstyle{definition}

\newtheorem{rmk}{Remark}[section]

\title{\centering On the definition of the spin charge \\ in asymptotically-flat spacetimes}

\author[a]{Laurent Freidel,}
\author[b,c]{Seyed Faroogh Moosavian}%
\author[d,e]{and Daniele Pranzetti}

\affiliation[a]{Perimeter Institute for Theoretical Physics,\\ 31 Caroline Street North, Waterloo, Ontario N2L 2Y5,  Canada}
\affiliation[b]{Department of Physics, McGill University, Ernest Rutherford Physics Building, \\ 3600 Rue University, Montr\'eal, QC H3A 2T8, Canada}
\affiliation[c]{Mathematical Institute, University of Oxford, Woodstock Road, Oxford, OX2 6GG, UK}
\affiliation[d]{Universit\`a degli Studi di Udine,
via Palladio 8,  I-33100 Udine, Italy}
\affiliation[e]{Institute for Fundamental Physics of the Universe (IFPU),
Via Beirut 2, 34151 Trieste, Italy}

\emailAdd{lfreidel@perimeterinstitute.ca}
\emailAdd{sfmoosavian@gmail.com}
\emailAdd{daniele.pranzetti@uniud.it}

    \abstract{We propose a solution to a classic problem in gravitational physics consisting of defining the spin associated with asymptotically-flat spacetimes. We advocate that the correct asymptotic symmetry algebra to approach this problem is the generalized--BMS algebra $\gbms$ instead of the BMS algebra used hitherto in the literature for which a notion of spin is generically unavailable.
     We approach the problem of defining the spin charges from 
    the perspective of coadjoint orbits of $\gbms$ and construct the complete set of Casimir invariants that determine $\gbms$ coadjoint orbits, using the notion of vorticity for $\gbms$. This allows us to introduce spin charges for $\gbms$ as the generators of area-preserving diffeomorphisms forming its isotropy subalgebra.
    To elucidate the parallelism between our analysis and the Poincar\'e case, we clarify several features of the Poincar\'e embedding in $\gbms$  and reveal the presence of condensate fields associated with the symmetry breaking from $\gbms$ to Poincar\'e. We also introduce the notion of a rest frame available only for this extended algebra. This allows us to construct, from the spin generator, the gravitational analog of the Pauli--Luba\'nski pseudo-vector. Finally, we obtain the $\gbms$ moment map, which we use to construct the gravitational spin charges and gravitational Casimirs from their dual algebra counterparts.}

\begin{document} 

\maketitle

\section{Introduction}

\label{sec:introduction}

The power of symmetry, encoded in a Lie algebra $\g$ (or its corresponding group), in the study of dynamical physical systems has been known for quite some time. In ideal-enough cases, it may happen that the severe constraints of symmetries can be leveraged to completely unveil the classical dynamics of the system which leads to the notion of integrability \cite{Liouville1853}. This power came to prominence with the work of Emmy Noether on the relation between symmetries of a system and conservation laws \cite{Noether1918}.
On the other hand, in the Hamiltonian formulation of dynamical system, the phase space $\phasespace$, the space of all possible classical configurations of the system, is a symplectic manifold, and $\g$
naturally acts on this space \cite{Hamilton1833,Arnold1989}. The physical (or reduced) phase space is the quotient space with respect to this action, and under some mild requirements, the quotient space has a canonical symplectic structure. The procedure to construct this quotient space, called symplectic reduction, is the classical counterpart of the study of representation theory of the group \cite{MarsdenWeinstein1974,SjamaarLerman1991}. An essential tool in this construction is the notion of moment map $\mu_\g:\phasespace\to \g^* $ for the Hamiltonian $\g$-action whose key property is its commutativity with the $\g$-action or $\g$-equivariance \cite{Souriau1970,Souriau1997}. This basically means that the action of $\g$ on $\phasespace$ commutes with the coadjoint action of $\g$ on $\g^*$. In particular, 
\begin{equation}
        \{F,G\}_{\g^*}\circ\mu_{\g}=\{F\circ \mu_{\g},G\circ \mu_{\g} \}_{\phasespace}, \qquad \forall F,G\in 
\mcal{F}(\g^*),
    \end{equation}
where $\{\cdot,\cdot\}_{\g^*}$ is the linear Poisson structure on $\g^*$, $\{\cdot,\cdot\}_{\phasespace}$ is the natural symplectic structure on $\phasespace$, and $\mcal{F}(\g^*)$ is the space of functions on $\g^*$. Therefore, moment map is the tool to translate algebraic statements about $\g^*
$ to statements about the phase space $\phasespace$. It simply represents the mathematical essence of Noether's theorems \cite{Noether1918}. Furthermore,
$\g$-invariant objects on $\g^*$, the Casimir functions, lead to $\g$-invariant quantities on the phase space $\phasespace$. Casimirs are important because they lead to  conserved quantities and therefore play a key role in understanding the dynamical evolution of the system under study.

 \smallskip Symmetries become even more important within quantum mechanics. As shown by Wigner \cite{Wigner1931}, the generators of symmetries can be represented as linear operators acting on the Hilbert space of the quantum system, as the space of all possible quantum states. Hence, it can be realized as certain representation spaces of symmetries.\footnote{One needs to distinguish between the notion of global symmetries and gauge invariance. The physical Hilbert space carries a nontrivial action of the first while is inert to the action of the second.} In the simplest case, the translation symmetry leads to the conservation of linear momentum. More interestingly is the angular momentum of the system which is conserved under the rotational symmetry and its quantization leads to many well-known theorems about quantum-mechanical dynamics of angular momentum \cite{Schwinger195201}. Furthermore, the quantization amounts to replace $\{\cdot,\cdot\}_{\g^*}$ with the Lie algebra bracket $-\mfk{i}[\cdot,\cdot]_{\g}$, where the latter is the Lie bracket of $\g$. In particular, under this quantization map, Casimir functions on coadjoint orbits turn to Casimir elements of the algebra. On the other hand, there are systematic methods to construct the representation spaces of symmetry algebras among which are the method of induced representation of George Mackey \cite{Mackey195201,Mackey195309,Mackey1978} and the method of coadjoint orbits of Alexandre Kirillov
\cite{Kirillov196202,Kirillov1976,Kirillov199908,Kirillov200407}. In the latter method, the existence of the symmetry action on the phase space can be used construct unitary representations. Indeed each coadjoint orbits of the action would lead to such a representation. Therefore, the study of coadjoint orbit is an important tool in the phase space quantization. 

\smallskip Other than the construction of conserved quantities, by the pioneering work of Bargmann and Wigner, elementary particles are defined to be unitary irreducible representations of the isometry of the spacetime on which the theory lives \cite{BargmannWigner194805}. In the absence of gravity, it is well-known that the isometry algebra is simply the Poincar\'e algebra, a case that has been extensively studied in the past including the classical treatments in  \cite{Wigner193901,Dirac194502,HarishChandra194705,GelfandNaimark194711,Naimark1964,KimWigner198705}. 

\smallskip The inclusion of gravity poses certain puzzles, the most important of which is the identification of a proper notion of symmetry of a given gravitational system that replaces the notion of isometry of a spacetime. In fact, there is no fixed spacetime in quantum gravity. Hence studying the isometry of a single spacetime and then, based on Bargmann--Wigner philosophy \cite{BargmannWigner194805}, defining elementary particles as its unitary irreducible representations do not make much sense. Furthermore, there is not a sensible notion of symmetry associated with compact spacetimes without boundaries. However, in the presence of boundaries, there is a way out as follows: One can fix an asymptotic structure\footnote{The so-called universal structure in the language of \cite{Ashtekar201409}.} $\mcal{S}_\infty$ and then define the asymptotic symmetry group as those elements of the bulk diffeomorphisms\footnote{As the diffeomorphism group depends on the choice of a differentiable structure (up to equivalences), we also need to fix this structure first. Therefore, we are indeed considering all spacetimes with a fixed differentiable structure, fixed topology, and fixed asymptotic structure $\mcal{S}_\infty$. One then needs to include all possible topologies and differentiable structure to obtain quantum gravity with a fixed asymptotic structure.} of a physical spacetime $M$ that fix $\mcal{S}_\infty$. More precisely,\footnote{Here, it is understood that $M$ is the conformal completion of a physical space-time. We refer to \cite{Compere:2018aar} for the general definition of asymptotic symmetries in gravity.}
\begin{equation}
    \text{Asymptotic symmetries}:=\frac{\text{Diff}(M,\mcal{S}_\infty)}{\text{Diff}(M,1_\infty)},
\end{equation}
where the group $\text{Diff}(M,\mcal{S}_\infty)$ consists of diffeomorphisms of bulk physical spacetime preserving $\mcal{S}_\infty$ and $\text{Diff}(M,1_\infty)$ is the group of those diffeomorphisms that goes off to identity asymptotically. In general, this asymptotic structure {\it is not} the isometry group of the boundary of the (conformal completion) of a spacetime, denoted as $\mcal{I}$ in the literature, and hence is not a property of a spacetime. Instead, one fixes a certain structure depending on $\mcal{I}$ being spacelike (asymptotically-dS), null (asymptotically-flat), or timelike (asymptotically-AdS).\footnote{The question of which structure needs to be fixed is an important question. Depending on the type of structure that is being fixed, one obtains different asymptotic symmetries.} For the case of asymptotically-flat spacetimes, $\mcal{I}\simeq\mbb{R}\times S$, where $S$ is the celestial sphere and the isomorphism is global. 

\smallskip The first example of asymptotic symmetries in gravity appeared in the foundational works of Bondi–Van der Burg–Metzner \cite{BondivanderBurgMetzner196208} and Sachs \cite{Sachs196212}, the so--called BMS group, whose Lie algebra is denoted as $\bms$. It is an asymptotic symmetry group of four-dimensional asymptotically-flat, Lorentzian spacetimes whose properties and representations have been extensively studied in the past including the classical works of McCarthy \cite{McCarthy197211,McCarthy197305,McCarthy197311,CrampinMcCarthy197408,McCarthy197505,CrampinMcCarthy197610,McCarthy197801}. The analogous analysis for three-dimensional BMS group is performed in \cite{Barnich:2014kra, Barnich:2015uva, Barnich:2021dta}. The study of asymptotic symmetries has been revived in recent years mainly due to the unraveling of the so-called infrared triangle, the three-way relationship between asymptotic symmetries, soft theorems and memory effects \cite{Strominger201812}. These studies have uncovered larger asymptotic symmetry groups of four--dimensional asymptotically-flat, Lorentzian spacetimes such as Extended--BMS group \cite{Barnich:2009se,BarnichTroessaert201001,BarnichTroessaert201102}, Generalized--BMS group \cite{Campiglia:2014yka,CampigliaLaddha201502}, and Weyl--BMS group \cite{Freidel:2021fxf}. 

\smallskip As asymptotic symmetries provide a proper notion of symmetry in the gravitational context, in the sense that they act non-trivially on the phase space of the system, one can apply the above-mentioned procedure and ask the following two questions:

\begin{itemize}
    \item[--] Does it make sense to define the notion of ``elementary particles" in the presence of gravity as representations of an asymptotic symmetry group? 

    \item[--] Can asymptotic symmetries be used to construct conserved quantities in the presence of gravity? 
\end{itemize}

Regarding the first question, despite the fact that asymptotic symmetries have nothing to do with isometry of a spacetime, the view of defining ``elementary particles" in the presence of gravity as unitary irreducible representations of an asymptotic symmetry group has been advocated in \cite{Sachs196212,Komar:1965zz,NEWMAN1965} and especially by McCarthy as the main motivation for his detailed study of the representation theory of the BMS group \cite{McCarthy197211,McCarthy197305,McCarthy197311,CrampinMcCarthy197408,McCarthy197505,CrampinMcCarthy197610,McCarthy197801} (see in particular \cite{McCarthy197209} and also the related work \cite{Arens197112}). This perspective has also been recalled in \cite{Barnich:2014kra} as a motivation to study the representations of three-dimensional BMS group. Hence, one can naturally apply the same philosophy to a larger asymptotic symmetry group, in particular the generalized BMS group and its algebra $\gbms$, which would be the main focus of this work. 

\smallskip Regarding the second question, the possibility of constructing conserved quantities, such as energy--momentum and spin, in asymptotically-flat Lorentzian spacetimes has important implications not only for the quantum theory, as argued above. For instance, an unambiguous definition of spin would provide a key tool in the study of compact binary coalescences emitting gravitational waves, as it would allow for a more rigorous treatment  of the asymptotic frame at $\scri$ in
matching the quasi-normal mode  modeling with the  numerical relativity  strain waveforms
\cite{Mitman:2021xkq,MaganaZertuche:2021syq,Mitman:2022kwt}.

The idea of using $\bms$ as the ``right" asymptotic symmetry algebra has led to many attempts in the past  \cite{Winicour196806,AshtekarHansen197808,GerochWinicour198104,Bramson197501,Penrose198205,AshtekarWinicour198212,DrayStreubel198401,Moreschi1986,ChenWangYau201312,chrusciel2003hamiltonian,Compere:2019gft,Elhashash:2021iev,ChenWangWangYau202103,ChenWangWangYau202107,Compere:2023qoa,ChenParaizoWaldWangWangYau202207,Javadinezhad:2022hhl,Wang202303,FuentealbaHenneauxTroessaert202305} (see \cite{Szabados2009} for a thorough review prior to recent developments). 
 Let us briefly clarify the status of these attempts. The $\bms$ algebra has a unique translation subalgebra that enables one to define the so-called Bondi--Sachs energy-momentum tensor on $\mcal{I}$ \cite{Bondi:1960jsa,BondivanderBurgMetzner196208, Sachs196212}. However, there are the following undesired features of this algebra that prevent one from defining angular momentum and spin for asymptotically-flat spacetimes.

    \begin{enumerate}
    \item [(1)] {\it Supertranslation ambiguity}: The fact that $\bms$ does not have a canonical Lorentz subalgebra \cite{Sachs196212,Penrose:1964ge,Geroch1977}.

    \item [(2)] {\it Center-of-mass ambiguity}: The fact that the frames where the shear vanishes at early and late times are generically different in the presence of radiation \cite{AshtekarDeLorenzoKhera201910}.

    \item  [(3)] {\it Rest frame ambiguity}: There is no canonical notion of a rest frame, in which the total momentum is zero, associated with $\bms$, as the condition is cut-dependent.  

   \item [(4)]  {\it Undefined angular momentum aspect}: 
   The fact that the knowledge of the angular momentum generator does not define uniquely the  angular momentum aspect.
\item[(5)] {\it Absence of spin generator}: It turns out that if the mass aspect is generic, one cannot associate a spin generator to the $\bms$ algebra for the simple reason that the little group is $\mbb{Z}_2$, which is not a Lie group and hence there is no corresponding Lie algebra. This fact has been pointed out as early as the pioneering work of Sachs \cite{Sachs196212}.

    \end{enumerate}

We will further elaborate on these issues in \S\ref{sec:issues with the bms algebra}. The main proposal of this work is that these issues will be resolved once the algebra of asymptotic symmetries is enriched to be the $\gbms$ algebra, which is defined as \cite{Campiglia:2014yka,CampigliaLaddha201502}
\begin{equation}\label{eq:gbms algebra}
    \gbms=\diff(S)\sds\mbb{R}^S_{-1},
\end{equation}
where $\diff(S)$ is the diffeomorphism algebra of the two-sphere $S$ generated by a generic smooth vector field on $S$ (or super-Lorentz transformations), and $\mbb{R}^S_{-1}$ is the algebra of supertranslations.\footnote{The subscript $-1$ refers to the conformal weight  of functions on the celestial sphere parametrizing supertranslations. See \S\ref{sec:issues with the bms algebra} for details.} The dual $\gbms^*$ is labeled by a pair $(j_A,m)$, where $j_A$ and $m$ are the angular momentum and the mass aspect, respectively. Note that $\mbs{j}=j_A\rd\sigma^A$ is a one-form on sphere equipped with coordinates $\sigma^A$ and $m=m(\sigma)$ is simply a function on sphere.

\smallskip Let us now briefly summarize the main results and formulas in this work.

\subsection{Contributions of the paper}

In this paper, we shed light on various aspects of the algebra $\gbms$, including some of the issues that might be known to the experts but are obscure in the literature. The main results and formulas can be summarized as follows.

\paragraph{Poincar\'e embeddings (\S\ref{sec:gbms and poincare embeddings}).} In \S\ref{sec:gbms goldstone modes} we  revisit in our context  previous approaches  to the angular momentum
problem  within
the standard $\bms$ framework. More precisely, we introduce the notion of a supertranslation invariant {\it intrinsic angular momentum}, which requires the introduction of
an electric supertranslation Goldstone mode. We show that this intrinsic angular momentum generator satisfies the Lorentz Lie algebra $\SL(2,\C)$, when restricting to conformal Killing vector fields.  The need to extend the Poincar\'e algebra with the addition of a Goldstone mode introduces a cut dependence which leads to the difficulties reviewed  in \S\ref{sec:issues with the bms algebra}; Moreover, differently from the Pauli--Luba\'nski pseudo-vector, which is constructed purely from the generators of the Poincar\'e algebra itself, the intrinsic angular momentum does not satisfy an algebra isomorphic to the little group algebra. This is what motivates our analysis carried out in the following sections.

\smallskip Before moving on to this investigation, in \S\ref {sec:poincare embedding} we clarify how the Poincar\'e embedding in $\gbms$ follows a symmetry-breaking pattern, which leads us to the   notion of  {\it condensate modes}, canonically conjugated to the Goldstone modes. These supertranslation and super-Lorentz condensates define an equivalence relation for the mass and angular momentum aspects,  thus representing respectively the geometry of the quotient spaces $\bms/ \Poinc,  \gbms/ \bms$. We then show how to explicitly construct a Poincar\'e charge
algebra inside $\gbms$ for a general non-round
sphere metric.

\paragraph{Algebraic aspects of $\gbms$ (\S \ref{sec:algebraic aspects of gbms} and \S\ref{sec:reference frames and goldstone modes}).} We then focus on purely algebraic properties of $\gbms$, with no phase space involved. These consist of (1) the coadjoint orbits and their Casimir invariants and (2) the generators of the isotropy subalgebra. The motivation to study the first aspect has been explained above: the quantization of coadjoint orbits provides the representation spaces of $\gbms$. Furthermore, a proper quantization of $\gbms$ involves the notion of spin, and as such, we construct $\gbms$ spin generator. Finally, we introduce the notion of rest frame for $\gbms$, with no counterpart in $\bms$ algebra.

\smallskip In \S\ref{sec:algebraic aspects of gbms}, we first consider the $\gbms$ coadjoint action on a point of its dual labelled by the mass aspect and the angular momentum aspect 1-form $(m, \mbs{j})$  (see \eqref{eq:coadjoint action on gbms*} for the explicit action). Similar to the analysis of coadjoint orbits of corner symmetry algebra \cite{DonnellyFreidelMoosavianSperanza202012}, we can classify the coadjoint orbits through an analogy with fluid mechanics. A crucial role is played by the notion of density $\rho = m^{\frac23}$ whose $\gbms$ transformations are given in \eqref{eq:rho transformation under gbms}) and by the  vorticity. For the algebra $\gbms$, the vorticity 2-form is given by 
\begin{equation}
    \mbs{w}=m^{-\frac53 }\left(m\rd \mbs{j} - \frac23  \rd m  \wedge \mbs{j}\right)\,.
\end{equation}
The two crucial properties of this quantity are as follows: (1) Its invariance under supertranslation, and (2) the fact that it transforms as a 2-form (see \eqref{wT}). It then follows that the Casimir invariants of coadjoint orbits are given by ($\rho:=m^{\frac{2}{3}}$ whose $\gbms$ transformations are given in \eqref{eq:rho transformation under gbms})
\begin{equation}
    \mathsf{C}_n(\gbms) := \bigintsss_S    w^n \rho\,\bfe\,, \qquad n=0,1,2,\ldots,
\end{equation}
where $w$ is the vorticity scalar associated to the two-form $\mbs{w}= w \rho \bfe$, and $\bfe$ defines an area form on $S$ which is preserved by the $\gbms$ action (see \eqref{dq0}). 

\smallskip We next move to the construction of spin generators for $\gbms$. What we will call a spin generator has the following two defining properties: (1) it is a generator of the isotropy subalgebra of $\gbms$. By definition, it is the subalgebra of $\gbms$ that preserves a given mass aspect $m$; (2) it is invariant under any supertranslation. It turns out that the isotropy subalgebra of $\gbms$ is given by
\begin{equation}\label{eq:gbms isotropy subalgebra, introduction}
    \mbf{iso}(\gbms)=\tsf{sdiff}_{\bfs}(S)\sds\mbb{R}^S_{-1},
\end{equation}
where $\tsf{sdiff}_{\bfs}(S)$ is the algebra of area-preserving diffeomorphisms on $S$ for the area element $\mbs{\rho}:=\rho\bfe=m^{\frac{2}{3}}\bfe$. Having the isotropy subalgebra at hand, we identify its (smeared) generators as
\begin{equation}\label{eq:gbms spin generator, introduction}
   \msf{S}_\chi =\bigintsss_S \chi\,\mbs{w}, \qquad \chi\in C^\infty(S),
\end{equation}
where $C^\infty(S)\simeq \mbb{R}^C_0$ is the space of functions on sphere. From the $\gbms$ transformations of $\rho$ and $w$, it then turns out that $\msf{S}_\chi$ is supertranslation-invariant. As \eqref{eq:gbms spin generator, introduction} has all the right properties, we define it to be the spin generator of $\gbms$. 

\smallskip An overview of these results  paralleled by the comparison with the familiar Poincar\'e algebra is summarized in Table \ref{tab:comparison between poincare and gbms}.
\begin{table}[h]
    \centering
    \renewcommand{\arraystretch}{1.4}
    \begin{tabular}{c c c}
         & Poincar\'e &  GBMS
         \\ \hline 
    Lie Group & $\tenofo{SO}(3,1)^\uparrow\ltimes\mbb{R}^4$  &   $\tenofo{Diff}(S)\ltimes\mbb{R}^S_{-1}$
    \\ 
    Lie Algebra  & $\msf{so}(3,1)\sds\mbb{R}^4$ &  $\msf{diff}(S)\sds\mbb{R}^S_{-1}$
    \\
    \makecell{Lie Coalgebra \\ Elements} & $(j_{\mu\nu},p_\mu)$ & $(j_A,m)$
    \\
    \makecell{Type of Orbits} & Massive & Massive
    \\ 
    \makecell{Label of Orbits} & $-p^2>0$  &   $m>0$
    \\
    \makecell{Isotropy Subalgebra} & $\msf{so}(3)\sds\mbb{R}^4$ & $\msf{sdiff}_{\mbs{\rho}}(S)\sds\mbb{R}^S_{-1}$
    \\
    Spin Generators & $w_\mu$ & $\msf{S}_\chi$ $$%
    \\
    Casimirs &  $(-p^2,w^2)$  &  $\msf{C}_n,\, n=0,1,\ldots$
    \\ \hline 
    \end{tabular}
    \caption{The comparison between the Poincar\'e algebra  in four dimensions $\msf{iso}(3,1)$ and $\gbms$. In this table, $w_\mu$ denotes the  components of the (classical) Pauli--Luba\'nski pseudo-vector.}
    \label{tab:comparison between poincare and gbms}
\end{table}

\smallskip As $\bms$ suffers from the ambiguity of defining a notion of rest frame (unless we are in the particular case of a stationary
 spacetime), we investigate the possibility of defining rest frame for $\gbms$ in \S\ref{sec:reference frames and goldstone modes}. The impossibility  of reaching such a frame is one of the main reasons why the definition of spin is ambiguous within the BMS framework.
At the same time, it turns out that due to the access to the full diffeomorphism group, it is possible to define such a notion in GBMS. This allows us to understand the spin generator \eqref{eq:gbms spin generator, introduction} as the generator of
superrotation in the generalised rest frame, in complete analogy to the Poincar\'e case.

\smallskip Recalling that the Bondi frame is defined by the condition of a constant curvature $R$ of $S$ and non-constancy of the mass aspect $m$, we define the rest mass reference frame for $\gbms$ as the frame that satisfies
\begin{equation}
    \partial_A m=0, \qquad \partial_A R\ne 0,
\end{equation}
where $A=1,2$ denotes the coordinates on $S$. One can then derive an explicit $\gbms$ transformation that maps the Bondi frame to the rest frame and vice versa (see \eqref{eq:from rest to the bondi frame}).  

\smallskip As a concrete example of $\gbms$ reference frames, we consider the simple example of stationary spacetimes. It turns out that for such spacetimes, one can go to a frame which is both Bondi and at rest. We derive the form of mass and angular moment aspects of stationary spacetimes in the rest frame in \eqref{jRCM}, as well as in a general boosted Lorentz frame in \eqref{jAv}. 

\smallskip The last question that we touch on in \S\ref{sec:Multi} is whether the mass aspect of a configuration of massive particles can be obtained by applying $\bms$ transformations to a constant mass aspect. The motivation to study this question is obvious. If the answer is positive, then the necessity of enlarging the symmetry algebra to $\gbms$ would become dubious. However, we give a definite negative answer to this question using the harmonic decomposition of mass aspect, which we develop in \S\ref{sec:Gold-mass}. Furthermore, we obtain the harmonic decomposition of angular momentum aspect in \S\ref{sec:Gold-angular}.  These decompositions reveal the following intriguing feature. While the $\ell=0,1$ global modes reproduce the standard 4-momentum and Lorentz piece of the angular momentum of a particle, 
the $\ell\geq 2 $ condensate contributions of the mass and angular momentum aspects reproduce exactly the soft factors in the
leading and subleading soft graviton theorems respectively. 

\paragraph{Pauli--Luba\'nski generators (\S\ref{sec:pauli-lubanski generators}).} We next consider the question of constructing Pauli--Luba\'nski generators for $\bms$ algebra. We construct it for the case of a mass aspect on the coadjoint
orbit of the constant mass (see \eqref{rhov}) in \eqref{PL2} and show that it reproduces the Poincar\'e analog expression in \eqref{W-comp}. 

\paragraph{Applications (\S\ref{sec:applications}).} After developing the necessary tools, we define the $\gbms$ moment map and study several applications as follows.

\paragraph{Moment map for the $\gbms$ action (\S \ref{sec:the gbms moment map}).} Thinking of $\gbms$ as the symmetry algebra which acts on a phase space, the two natural questions are the following
\begin{itemize}
    \item [$-$] What is the phase space which carries a Hamiltonian action of $\gbms$?

    \item [$-$] What is the moment map for the Hamiltonian action? 
\end{itemize}
The answer to the first question is provided by certain simplifying conditions on the radiative phase space studied in \cite{CampigliaPeraza202002,Freidel:2021qpz}. These conditions are
\begin{equation}
    \mcal{I}^A=\mcal{N}^{AB}=\wt{\mcal{M}}=0,
\end{equation}
where $\mcal{I}^A$, $\mcal{N}^{AB}$, and $\wt{\mcal{M}}$ are the covariant energy current, radiative news, and covariant dual mass, respectively. We will call this phase space electric and  non-radiative, and denote it as $\phasespace_{\va \text{ENR}}$. The moment map $\mu_{\gbms}:\phasespace_{\va \text{ENR}}\to \gbms^*$ is then given by 
\begin{equation}
    \mu_{\gbms}(\mcal{M})=m,\qquad \mu_\gbms\left(\frac{1}{2}\left(\mcal{J}_A-u\mcal{D}_A\mcal{M}\right)\right)=j_A,
\end{equation}
where $\mcal{M}$ and $\mcal{J}_A$  the covariant mass and the covariant angular momentum, respectively. 

\paragraph{Gravitational Casimirs and the gravitational spin charge (\S\ref{sec:phase space quantities and their evolution}).} Equipped with the moment map, we can translate the algebraic properties of $\gbms^*$. In particular, the gravitational Casimirs are given by
\begin{equation}\label{eq:gravitational gbms casimirs, introduction}
    \mscr{C}_n(\phasespace_{\va \text{ENR}}):=\mu^*_{\gbms}\msf{C}_n(\gbms)=\bigintsss_S\,\mcal{M}^{\frac{2}{3}}\textbf{w}^n\,\bfe,\qquad n=0,1,2,\ldots. 
\end{equation}
with the gravitational vorticity $\mbf{w}$  defined as
\begin{equation}
    \mbf{w}:=\mu_{\gbms}^*w=\frac{1}{2}\mcal{M}^{-\frac{2}{3}}\epsilon^{AB}\partial_A\left(\mcal{M}^{-\frac{2}{3}}\mcal{J}_B\right). 
\end{equation}
Furthermore, the gravitational spin charge, which generates the action of \eqref{eq:gbms isotropy subalgebra, introduction} on $\phasespace_{\va \text{ENR}}$, is
\begin{equation}\label{eq:gbms gravitational spin charge}
    \mscr{S}_\chi:=\mu^*_\gbms \msf{S}_\chi=\frac{1}{2}\bigintsss_S \chi\,\epsilon^{AB}\partial_A\left(\mcal{M}^{-\frac{2}{3}}\mcal{J}_B\right)\bfe.
\end{equation}
A particular feature of \eqref{eq:gravitational gbms casimirs, introduction} and \eqref{eq:gbms gravitational spin charge} is their cut-independence, i.e. they do not depend on the Bondi coordinate $u$ along $\mbb{R}\subset\mcal{I}$.

\paragraph{Gravitational Casimirs for the Kerr spacetime (\S\ref{sec:gravitational Casimirs for the kerr metric}).} Note that the expression for Casimirs involve the covariant mass $\mcal{M}$ and the covariant angular momentum $\mcal{J}_A$. These quantities can be read-off from the asymptotic expansion of the metric in the Bondi-Sachs coordinates. We give an explicit example of the computation of gravitational Casimirs for the Kerr metric, where they are given by
\begin{equation}
    \mscr{C}_n(\phasespace^{\va \text{Kerr}}_{\va \text{ENR}})=
    \left\{
    \begin{aligned}
    \frac{(-3a)^n}{n+1}&M^{\frac{2-n}{3}}, &\qquad n&=0,2,4,\ldots,
    \\
    &0, &\qquad n&=1,3,5,\ldots,
    \end{aligned}
    \right.
\end{equation}
where $M$ and $a$ are the mass and the reduced angular momentum (defined in \eqref{eq:reduced angular momentum}), respectively. 

\smallskip Technical reviews and proofs of various results are collected in  Appendices \ref{sec:moment maps and coadjoint orbits}, \ref{sec:pauli-lubanski pseudo-vector as a constant of motion}, \ref{App:comp}, \ref{App:conf}, \ref{App:decom}, \ref{sec:details on the kerr metric and casimirs}.

\section{Background material and review of the issues}
This section is devoted to providing certain background material and motivation for the rest of the paper. In \S\ref{sec:motivation and background}, we review general motivations and background on the study of coadjoint invariants, their relation to conserved quantities, and also asymptotic symmetries in gravity. We then turn to the issues with the $\bms$ algebra in \S\ref{sec:issues with the bms algebra}. These issues motivate (and indeed demand) to consider an enlarged asymptotic symmetry algebras in the context of asymptotically-flat spacetimes. Finally, as a precursor and warm-up to the study of symmetry algebras in gravity, we review in some detail some relevant aspects of Poincar\'e algebra (such as massive representations, Casimir elements, and coadjoint orbits) in \S\ref{sec:Poincare algebra}.

\subsection{Motivation and background}\label{sec:motivation and background}

By the pioneering work of Bargmann and Wigner, elementary particles are {\it defined} to be the unitary irreducible  representations of the isometry group of a spacetime \cite{BargmannWigner194805} (see also \cite{Arens197112,McCarthy197209}).  Given the isometry (Lie) algebra $\g$ of a spacetime, 
 Casimir elements, namely elements that are
invariant under the whole algebra $\g$, provide a convenient way to distinguish its irreducible
representations, which are further constrained by the unitarity requirement. Casimir elements for (the universal enveloping algebra of) $\g$ are identified with a $\g$-conserved (i.e. invariant under the $\g$-action) quantity $\wh{\mscr{O}}$. At the quantum level and through the Noether theorem, we know that the group action is given by the adjoint action 
\begin{equation}\label{eq:variation defined by Lie bracket}
    \delta_{\text{J}} \wh{\mscr{O}}:= [\text{J},\wh{\mscr{O}}], \qquad \text{J}=\sum_{a=1}^{\dim\g} C_a \text{J}^a\,,
\end{equation}
where $C_a$s are constants   and $\text{J}^a$ denote a basis of $\g$ generators. A Casimir is, by definition, a $\g$-conserved operator  $\wh{\mscr{C}}$ satisfying
    \begin{equation}\label{eq:definition of g-variation, quantum-mechanical}
        \delta_{\text{J}} \wh{\mscr{C}}=0, \qquad \forall\text{J}\in\g. 
    \end{equation}
    The knowledge of the value of a complete set of Casimir elements associated with 
an isometry algebra $\g$ labels the representation $\mcal{R}$ of $\g$. 

\smallskip The Lie algebras we are interested in this work (Poincar\'e, $\bms$ or $\gbms$) are in the form of a semi-direct sum
\begin{equation}
    \g=\msf{k}\sds\msf{n},
\end{equation}
where $\msf{k}$ is  the quotient algebra, $\msf{n}$ is the Abelian normal ideal, and $\sds$ means that $\msf{k}$ acts on $\msf{n}$ by the Lie bracket (i.e. $[\msf{k},\msf{n}]\in \msf{n}$).
The corresponding group is denoted $\mathrm{G}=\mathrm{K}  \ltimes \mathsf{n}$ where the Abelian subgroup is simply equal to its Lie algebra when it is represented additively.

\paragraph{Induced representations and coadjoint orbits.} An established method for studying the representations of such Lie algebras is the so-called induced representation \cite{Wigner193901,Mackey195201,Mackey195309,Mackey1978}. In this method, one constructs representations of $\g$ from representations of the so-called Wigner's little (or isotropy) subgroup, first introduced by Wigner in his study of representations of in-homogeneous Lorentz algebra \cite{Wigner193901}.
To construct induced representations, one first starts by selecting an element $p \in \msf{n}^*$ of the dual normal Lie algebra. Such an element represents a character
$\chi_p(x)= e^{i\langle p|x\rangle}$, with $x\in \msf{n}$ and $\langle\cdot|\cdot\rangle$ being a pairing between $\msf{n}$ and $\msf{n}^*$, for the Abelian group represented additively. The adjoint action of $\tenofo{K}$ on $\msf{n}$ leads to a \emph{coadjoint action}\footnote{This action is such that $\langle \mathsf{ad}^*_up| v\rangle = -\langle p|[u,v]\rangle$, for $p\in \msf{n}^*$ and $u,v \in \msf{k}$.} $\mathsf{ad}^*$ on $\msf{n}^*$. The corresponding group action is denoted $\tenofo{Ad}^*_U$ for $U\in \text{K}$. The sets of elements $\tenofo{K}p$ that can be reached from $p$ by the coadjoint action of $\tenofo{K}$ is called the orbit of $p$. Given an element $p\in \msf{n}^*$ one defines its little or isotropy subgroup to be given by  
\begin{equation}\label{eq:generic definition of isotropy subalgebra for adjoint action}
    \tenofo{Iso}({p}):=\left\{U\in\tenofo{K}\,\big|\,\tenofo{Ad}^*_U\,p=p\right\}.
\end{equation}
The corresponding Lie algebra is $\mbf{iso}({p}):=\left\{u\in\msf{k}\,\big|\,\mathsf{ad}^*_u\, p =0\right\}$. The isotropy groups of different points of $\tenofo{K}p$ are isomorphic, so we can refer to the orbit isotropy group as designing the equivalence class.

\smallskip Induced representations are therefore characterized by the choice of coadjoint orbits of $\mathsf{n}^*$ and by the choice of representation of the isotropy subgroup. Roughly speaking, once the orbit is fixed then the question of classification of all possible unitary irreducible representations boils down to studying the same question for its isotropy subalgebra, which has a much simpler structure \cite{Rawnsley197502,Baguis1998}. The subtle and often difficult question that has to be addressed is whether all representations of interests come from this construction. For example, in the case of the Poincar\'e algebra, the choice of orbit is encoded into the mass. For a positive mass, the spin corresponds to a choice of a representation of the little group's orbit. For positive mass, the little group is $\SO(3)$ and we recover the usual definition of spin as a little group representation label. It then turns out that all unitary irreducible representations of the Poincar\'e group come from this construction \cite{Wigner193901}.

\smallskip Coming back to our initial question then, it is crucial to understand whether the BMS group is big enough to construct (quasi-local) conserved quantities in the presence of gravity. This idea has led to many attempts in the past to construct conserved quantities in asymptotically-flat spacetimes  \cite{Winicour196806,AshtekarHansen197808,GerochWinicour198104,Bramson197501,Penrose198205,AshtekarWinicour198212,DrayStreubel198401,Moreschi1986,ChenWangYau201312,chrusciel2003hamiltonian,Compere:2019gft,Elhashash:2021iev,ChenWangWangYau202103,ChenWangWangYau202107,Compere:2023qoa,ChenParaizoWaldWangWangYau202207,Javadinezhad:2022hhl,Wang202303,FuentealbaHenneauxTroessaert202305} (see \cite{Szabados2009} for a thorough review prior to recent developments). 
However, as advocated above,
in order to construct well-defined physical conserved quantities in an asymptotically-flat spacetime with a fixed asymptotic structure $\mcal{S}_\infty$, one needs first to identify
the Casimir elements of
the algebra of asymptotic symmetries. This viewpoint is a generalization of the familiar example of the Poincar\'e algebra where the values of the Casimir elements in a unitary irreducible  representation determine the (bare) mass and the spin of an elementary particle, as we will briefly review in \S \ref{sec:Poincare algebra}.

\subsection{Issues with the $\bms$ algebra}\label{sec:issues with the bms algebra}

In this section, we introduce the first example of asymptotic symmetry groups in gravity, the BMS group discovered in \cite{BondivanderBurgMetzner196208,Sachs196212}, and furthermore elaborate on various issues in defining conserved quantities associated with this group.

\paragraph{The BMS group.} Recalling that for asymptotically-flat spacetimes $\mcal{I}\simeq\mbb{R}\times S$, where $S$ denotes the two-sphere, the BMS group is
\begin{equation}\label{eq:bms group}
    \text{BMS}=\mathrm{SO}(3,1)^{\uparrow}\ltimes \mbb{R}_{-1}^S,
\end{equation}
where  $\rm{SO}(3,1)^{\uparrow}$ denotes the proper orthochronous subgroup of the Lorentz group and
$\mbb{R}^S_{-1}$ denotes the infinite-dimensional Abelian ideal of supertranslations. Its elements are given by functions on $S$ which transform under $\rm{SO}(3,1)^{\uparrow}$ as densities of weight\footnote{In the following we denote by $\mathbb{R}^S_\Delta$ the space of functions on the sphere that transform under $\rm{SO}(3,1)^{\uparrow}$ as densities of weight $\Delta/2$. The unit sphere measure $\bfe$ transforms under Lorentz transformation $g$ as $\bfe \to g^*\bfe = \omega_g^{-2} \bfe$, where $\omega_g$ is given in \eqref{eq:bms action on supertranslations}.} $-1/2$.
  More explicitly, if one choose 
 complex coordinates $(z,\bar{z})$ on the unit round sphere with metric $\rd s^2 = \frac{4 \rd z\rd \bar{z}}{(1+|z|^2)^2} $, Lorentz transformations are represented by $2\times 2$ complex matrices 
 \be g= \left(\begin{matrix}
a & c \\
b & d 
\end{matrix}\right)\in \tenofo{SL}(2,\mathbb{C}). \label{mat}
\ee 
The group $\widetilde{\text{BMS}}= \tenofo{SL}(2,\mathbb{C})\ltimes \mbb{R}_{-1}^S$ is the universal cover of $\text{BMS}$ \cite{McCarthy197211}.
Its elements act on $\mathbb{R}_{\Delta}^S$ as fractional transformation 
\begin{eqaligned}\label{eq:bms action on supertranslations}
    (g\cdot f)(z,\bar{z})&:= \omega_g^{-\Delta} f\left(\frac{az+b}{cz+d}, \frac{\bar{a}\bar{z}+\bar{b}}{\bar{c}\bar{z}+\bar{d}}\right),
 \\
 \omega_g &:= \frac{|az+b|^2+|cz+d|^2}{1+|z|^2}.
\end{eqaligned}
 In the following, we refer to $\Delta$ as the conformal weight or simply the weight of $f$.
 One has to remember that $\Delta$ is twice the density weight.\footnote{A density $\rho$ on the sphere is such that under a map $(z,\bar{z})\to (F(z),\bar{F}(\bar{z}))$, it transforms as
 \begin{equation*}
    \rho \to \rho_F= |\pa F|^2  (\rho \circ F),
 \end{equation*}
 and therefore $\rho_F/\rho = \omega_g^{-2} $ when $\rho =2/(1+|z|^2)^2$ is the round sphere measure and $F=g$ is a conformal transformation. Thus, a density of weight $w$ is of conformal weight $\Delta =2w$.\label{ftn:transformation of density under diffeomorphism}}

\paragraph{The BMS algebra.} The algebra of BMS group \eqref{eq:bms group} is given by
\begin{equation}\la{bms-alg}
    \msf{bms}=\msf{so}(3,1)\sds\mbb{R}^S_{-1},
\end{equation}
where $\msf{so}(3,1)$ is the Lorentz algebra and $\mbb{R}^S_{-1}$ denotes the algebra of supertranslations, and the symbol $\sds$ denotes the action of Lorentz part on supertranslations given by the $\msf{bms}$ Lie bracket. In the following, we also need to consider the dual $\msf{bms}^*$ of this Lie algebra, which is labeled by a pair $(j_A,m)$, where $j_A$ and $m$ are the $\msf{bms}$ angular momentum and mass aspects, respectively.

\smallskip The pairing between elements of $\mbb{R}^S_\Delta$ and its dual is given through an integral over the sphere. In order for this integral to be invariant under diffeomorphism it needs to be of conformal weight $2$. This means that the duality operation maps $\mbb{R}_{\Delta}^S$ into 
 $\mbb{R}_{2-\Delta}^S$, i.e. 
  \be
  (\mbb{R}_{\Delta}^S)^*= \mbb{R}_{2-\Delta}^S.
  \ee 
  This means that the space of \emph{mass aspects} which contains elements dual to supertranslations is given by $\mbb{R}_3^S$, i.~e. scalar operators of conformal weight $3$. This fact was first established by Bondi \cite{BondivanderBurgMetzner196208}.

\paragraph{Issues in constructing BMS-conserved quantities.} As our main aim in this work is to construct the spin generator in asymptotically-flat spacetimes, we need to explain first why the $\bms$ algebra, as a symmetry of asymptotically-flat spacetimes, is {\it not} suited for this purpose. To do this, let us elaborate on each of the issues   (\hyperlink{bms-issue1}{1}--\hyperlink{bms-issue5}{5}) listed in the introduction.

\begin{itemize}

\item [(1)]\hypertarget{bms-issue1}{}\emph{Supertranslation ambiguity.}  
It is well-known that the set of 
conformal Killing vectors (CKV) $ Y^A\pa_A\in TS $ preserving, up to scale, the round sphere metric ${q}$ form a closed algebra under Lie bracket isomorphic to the Lorentz algebra. 
In order to  embed this representation of the Lorentz algebra into  BMS, i.~e. as a BMS subalgebra, one  needs to choose a cut $\cC$ of $\scri$. Such a cut is located at $u=G(\sigma)$ and  $G$ is a given function on the sphere that determines the supertranslation frame.\footnote{Here $G$ stands for `Goldstone', as it will become clear shortly below.}
The  Lorentz generators associated with the cut $\cC$ are then represented  by vectors on $S$ of the form
\be 
\xi^C_Y := Y^A \pa_A
+ Y[G] \pa_u+ \frac12 D_AY^A (u-G)\pa_u, 
\ee
where we denoted $Y[G]:= Y^A\pa_A G$. 
This vector is  the only element of BMS which is  tangent to the cut $u=G(\sigma)$ and acts as the CKV on the cut.
From this definition, it is clear that two different cuts lead to two different notions of Lorentz transformations that are related to each other by a supertranslation 
$(Y[\Delta G] -\frac12 D_AY^A \Delta G)\pa_u$ with  parameter $\Delta G= G-G'$ given by the difference between the two cuts. This  is simply a reflection of the fact that the supertranslation group plays the role that translations have in special relativity.
In special relativity a translation changes the orbital component of the  total angular momentum. In general relativity a supertranslation also change the angular momentum.\footnote{In special relativity, this is reflected in the fact that the commutator between generators of translations and angular momentum is non-zero (see \eqref{eq:poincare Lie algebra}). Considering BMS as the asymptotic symmetry group, the generator of supertranslations do not commute with angular momentum (generators  of Lorentz rotations). In both cases, the semi-direct product structure of the associated group (or equivalently, the semi-direct sum structure of the corresponding Lie algebra) leads to this conclusion.}

\item[(2)] \hypertarget{bms-issue2}{}{\it Center-of-mass ambiguity.} This is a related puzzle to the above due to the memory effect:
In non-radiative regions the news tensor vanishes by definition. 
The shear is therefore time independent and can be decomposed as a sum of an electric and a magnetic component $C_{AB}= -2(D_{\langle A} D_{B\rangle} G +
D_{\langle A} \widetilde{D}_{B\rangle} \widetilde{G})$ where $G$ and  $\widetilde{G}$ are respectively  the electric and magnetic supertranslation Goldstones, defined as the variables conjugated to the soft charges.
The angle bracket denotes the symmetric trace-free projection and 
$\widetilde{D}_A=\epsilon_A{}^B D_B$ denotes the magnetic derivative. In stationary regions such as the asymptotic regions $u\to \pm \infty$, it is further assumed that the magnetic component of $C_{AB}$ vanishes \cite{Bieri:2018asm}. Therefore in stationary regions, we can choose the cuts to be  such that $C_{AB}=0$, hence $G=\widetilde{G}=0$ after fixing the supertranslation ambiguity. This choice is called the \emph{center-of-mass frame}. The problem is that the center-of-mass frames of early and late time are generically different.
Indeed the supertranslation memory effect  \cite{Christodoulou:1991cr, thorne1992gravitational} implies that in the presence of asymptotic radiation $G^+\neq G^-$ where $C^\pm_{AB}=\lim_{u\to \pm \infty} C_{AB}$. Therefore the different Poincar\'e subgroup chosen by the good cut condition are different at the initial and final time.

\smallskip If we are in a non-radiative region which is not stationary then there are even further issues since we can no longer impose that $C_{AB}=0$. In this case we have to decide how one can fix the center-of-mass frame. There are two, not equivalent, viable  proposals in literature to do so by fixing the supertranslations. The first one due to Newman--Penrose, is the \emph{electric good cut condition} which imposes that $G=0$ even when $\widetilde{G}\neq0$ \cite{Newman:1966ub, rizzi1998angular}. The other one due to Moreschi is the \emph{nice cut condition} \cite{Moreschi:1988pc}  which demands that the higher multipole moments
\begin{equation}
    \msf{P}^{\va \mathrm{Mor}}_{\ell m}:= \bigintsss_S Y_{\ell m} m_{\va \mathrm{Mor}},
\end{equation}
of the Moreschi mass aspect all vanish for $\ell\geq 1$. The Moreschi mass aspect is given in terms of the Bondi mass aspect by
\begin{equation}
    m_{\va \mathrm{Mor}} := m_{\text{B}} - \frac14 D_AD_B C^{AB}\,,
\end{equation}
where $m_{\text{B}}$ is the Bondi mass aspect.

\item [(3)] \hypertarget{bms-issue3}{}{\it Rest frame  ambiguity.}
Given a supertranslation generator $T(\sigma)\pa_u$ and a Lorentz generator $Y$ associated to a reference cut $u=0$, we can construct the corresponding Noether charges $\msf{P}_T$ that design the supermomentum charges associated with supertranslation parameter $T$, while $\msf{J}_Y$  design the angular momentum charges associated with the vector field $Y=Y^A\pa_A$. These charges are obtained as integrals over the sphere of a corresponding mass and angular momentum aspects. The point is that, since $T$ is an arbitrary function on the sphere, it can be expanded in spherical harmonics. This means that we have in general an infinity of supertranslation charges $\msf{P}_{\ell m}$, with $\ell\in \mathbb{N}$ and $|m|\leq \ell$, given by the charge evaluation for $T=Y_{\ell m}$. $\msf{P}_{00}$ represents the global energy and $\msf{P}_{1i}$ the global momentum.
 We can still perform a Lorentz transformation that imposes $\msf{P}_{1i}=0$ which means that we are in a naive rest frame. Generically, in this rest frame, we will have that $\msf{P}^+_{\ell m}\neq 0$ for $\ell\geq 2$.\footnote{The only exception is if we are in a stationary black hole spacetime (see \S\ref{sec:statST}).} Therefore, this means that the initial and final rest frames are completely different frames. Note that since $\msf{P}_{T}$ is a supertranslation generator, it is invariant under supertranslations.\footnote{Note that the Moreschi supermomentum $\msf{P}_T^{\va \mathrm{Mor}}$ discussed earlier is not supertranslation invariant in non-radiative spacetime and therefore does not qualify as a supertranslation charge in the sense of Noether while the Bondi supermomentum does. } Therefore we see that the rest frame ambiguity is different from the previous ambiguity, that is, it cannot be cured by performing a supertranslation.

\item[(4)] \hypertarget{bms-issue4}{}{\it Undefined angular momentum aspect.}
    Given the round sphere metric $q_{AB}$ on $S$, 
the Lorentz part of the BMS algebra is a subalgebra of Diff$(S)$ generated by conformal Killing vectors $Y_{\mathrm{CKV}}$ of $(S,q_{AB})$. It is important to appreciate that, 
while the knowledge of 
\begin{equation}
    \msf{J}_Y= \bigintsss_S Y^A j_A\,\bfe,  
\end{equation}
for $Y\in \msf{diff}(S)$ determines the angular momentum aspect $j_A(\sigma)$, the knowledge of the BMS charges $J_{Y_{\mathrm{CKV}}}$ only determines  an equivalence class $[j_A]$ of angular momentum aspects, where $j_A$ is equivalent to $j_A'$ if there exists a symmetric traceless tensor $\tau_{AB}$ such that 
\be 
j'_A= j_A + D^B \tau_{BA}.
\ee
This clearly shows that for BMS it is hopeless to try to identify what is the angular momentum aspect from BMS symmetry. BMS only determine the \emph{equivalence class} $[j_A]$ of angular momentum aspects  under the transformations $j_A\to j'_A$ (see \S \ref{sec:poincare embedding} for further implications on this important point). A similar conclusion applies to extended BMS (EBMS) group \cite{Barnich:2009se, Barnich:2011mi}, where in the corresponding algebra, the Lorentz part in \eqref{bms-alg} is enlarged to the Lie algebra of the local conformal
diffeomorphisms of $S$. 
For instance, this ambiguity can help resolve the discussion about the  
quadratic ambiguity  in the angular momentum aspect discussed in \cite{CompereNichols202103, ChenParaizoWaldWangWangYau202207} and given by  $\bigintsss_S Y^A \left(4C_{AB}D_C C^{BC}+ D_A(C_{BC}C^{BC})\right)\bfe$. Elhashash and Nichols \cite{Elhashash:2021iev} have shown that this quadratic ambiguity vanishes when $Y$ is a conformal Killing vector and when $C_{AB}$ is purely electric.

    \item[(5)]\hypertarget{bms-issue5}{} {\it Absence of  spin generator.}
    It is well known that the spin generator for Poincar\'e algebra, also called Pauli--Luba\'nski pseudo-vector, is defined as the canonical generator of transformations that preserves the momentum generator. This vector is invariant under translation. Analogously, in the case of $\bms$, the isotropy algebra is defined as the subalgebra preserving the mass aspect $m$, and the spin aspect should be the generator of the isotropy subalgebra. However, 
    it turns out that, {unless the mass aspect is in the orbit of the constant mass aspect,} one cannot associate a spin aspect to the $\bms$ algebra. This fact, which has been pointed out as early as the pioneering work of Sachs \cite{Sachs196212}, can be seen as follows.
    Given a mass aspect $m(z,\bar{z})$, the BMS  isotropy subgroup consists of elements 
    $g\in \SL(2,\C)$ such that 
    \be 
  m\left(\frac{az+b}{cz+d}, \frac{\bar{a}\bar{z}+\bar{b}}{\bar{c}\bar{z}+\bar{d}}\right) = \omega_g^{3} m(z,\bar{z}),
    \ee 
    where $\omega_g$ is given in \eqref{eq:bms action on supertranslations}. The classification of orbits was first done by McCarthy in    \cite{McCarthy197211}, where it was shown that the isotropy subgroups are given by 
    \begin{table}[H]
        \centering
        \begin{tabular}{c c}
        \multirow{2}{2.5cm}{isotropy group} & \multirow{2}{2.5cm}{Fixed Points}   
        \\\\
        \midrule
        $\mbb{Z}_2$ &  $m(z,\bar{z})$
        \\
        $\tenofo{U}(1)$ & $m(|z|)$
        \\
        $\tenofo{SU}(2)$ &  $m_0$
        \end{tabular}
        \caption{Possible isotropy groups of the BMS group.}
        \label{tab:isotropy subalgebras of BMS}
    \end{table}
    In this table  $m(z,\bar{z})$ denotes a generic function on the sphere representing the  mass aspect that cannot be written as a function of $|z|$,  and  $m_0$ is a constant.
    Therefore, a generic mass aspect, described by a function $m(z,\bar{z})$ that cannot necessarily be put into an axisymmetric form $m(|z|)$ or a constant $m_0$ does not have a Lie algebra. Hence, the isotropy algebra is not well-defined and as such there is no generator (i.e. spin) associated to it. 
    This means that in $\bms$ there is no way to define the spin aspect for a generic mass aspect. 
    
    One  exception to this result is for the most degenerate orbit where $m$ is constant. In this case we recover  as isotropy group the  SU$(2)$ group and, hence, we expect that there exists for this particular orbit a definition of the spin generator. This spin generator has been constructed explicitly by Compere et al. in   \cite{Compere:2023qoa}. Assuming that the mass aspect is in the Lorentz orbit of a constant mass, they produced an explicit expression for a spin generator invariant under supertranslation that reduces to the spatial components of the Pauli--Luba\'nski pseudo-vector. We will discuss this result further in \S \ref{sec:PL-BMS constant mass} and show how it connects to our result. However, what is important to realize is that in a multi-particle scattering process the mass aspect is usually generic, as we will show in   \S \ref{sec:Multi}, hence one cannot restrict to the constant mass aspect.

     Another exception is the axisymmetric case where the mass aspect is only a function of $|z|$. No explicit construction of the U$(1)$ generator has been achieved yet and we won't consider it here.
    
\end{itemize}

In order to formulate a theory that can be quantized using the gravitational symmetry group one needs to have a non-trivial spin aspect that would carry at the quantum level a nontrivial notion of spin.
Such a notion cannot be generally defined using the algebra $\bms$.

\smallskip It is important to stress that the puzzles \hyperlink{rest frame ambiguity}{3}, \hyperlink{undefined angular momentum}{4} and \hyperlink{absence of spin aspect}{5} are present even in the {\it non-radiative} regions. This important fact seems to have been missed by most studies on angular momentum and it is the main point we would like to address.  As we are going to see in \S\ref{sec:spin generator of gbms}, these undesirable features are due to the fact that the symmetry algebra ($\bms$ in this case) is taken to be {\it too small} and can be dealt with once we consider the Lie algebra of generalized BMS group  \cite{Campiglia:2014yka}, which we denote  by $\gbms$, as the symmetry algebra for asymptotically-flat spacetimes. A similar remark was already made by Barnich--Oblak in the asymptotically-flat  3D case, where the Virasoro extension (including central charges) of the 3D  Lorentz group  played a pivotal role to define an intrinsic angular momentum  free from supertranslation ambiguities \cite{Barnich:2015uva, Barnich:2014kra}.

\smallskip 
We will argue that the quest of trying to define the proper notion of  spin is only meaningful in the context of $\gbms$. In fact,
the main point of our paper is to show that we can resolve  the issues (\hyperlink{bms-issue2}{2}), (\hyperlink{bms-issue3}{3}), (\hyperlink{bms-issue3}{4}) and (\hyperlink{bms-issue3}{5})  simply by enlarging the symmetry algebra from $\bms$ to $\gbms$. In $\gbms$ the quotient subgroup is the group of all diffeomorphisms on the sphere. This simple fact allows us to circumvent the ambiguity in the angular momentum aspect and  to provide a well-defined notion of rest frame and spin charge  for non-radiative regions. This definition does not depend on a specific choice of cut, is thus free of the center-of-mass ambiguity. Therefore, the cut-independent nature of the spin charge vs. the angular momentum charge makes the supertranslation ambiguity (\hyperlink{bms-issue1}{1}) irrelevant in our construction, as no specific rotational vector field needs to be fixed in order to select  a preferred   Poincar\'e subgroup.

\smallskip Alternatively in the literature, the first two puzzles (\hyperlink{bms-issue1}{1}) and (\hyperlink{bms-issue2}{2}) are often resolved in the non-radiative regions by resorting to the use of a supertranslation Goldstone $G$ in order to define a spin generator through the notion of {\it intrinsic angular momentum}  (see \S \ref{sec:Poincare algebra} for a review of this construction in special relativity). The supertranslation Goldstone transforms under supertranslation as $\delta_T G =T$ and it  is not invariant under boost as it transforms as a weight $-1$ field
\be\label{Ctrans}
(g\cdot G)(z,\bar{z})= \omega_g(z,\bar{z})  G\left(\frac{az+b}{cz+d}, \frac{\bar{a}\bar{z}+\bar{b}}{\bar{c}\bar{z}+\bar{d}}\right).
\ee 
If one has access to such a Goldstone mode, one can define a notion of intrinsic angular momentum which is supertranslation invariant. Using the description of 
Comp\`ere--Oliveri--Seraj \cite{Compere:2019gft}
the intrinsic angular momentum can be written as 
\be \la{SYbms}
\msf{I}_Y:= \msf{J}_Y - \msf{P}_{Y[G]}{+}\tfrac12 \msf{P}_{{G} D_AY^A}.
\ee 
The last two terms that depend on a product of mass aspect and Goldstone mode represent a notion of super orbital momentum which is subtracted from the total angular momentum.
This is a $\bms$ analog of the angular momentum subtraction \eqref{PoincS} for the Poincar\'e algebra.
 As shown in Appendix \ref{sec:proofs in section 3}, this operator is invariant under supertranslation by construction and  satisfies
 \be\la{S-algebra}
\{\msf{I}_{Y}, \msf{I}_{Y'}\}_{\bms^*}&=\msf{I}_{\left[
Y,Y'
  \right]_{S}}\,,
\ee
where, $\{\cdot,\cdot\}_{\bms^*}$ denotes the linear Poisson structure on $\bms^*$\footnote{This is defined by \eqref{eq:gbms charge algebra} for $\gbms$, with the charges as in \eqref{eq:gbms charges}, by restricting $Y$ to be a conformal Killing vector (instead of a general vector field generating $\msf{diff}(S)$).} and  $[Y,Y']_{S}$ denotes the Lie bracket of vector fields on $S$.
The first one to propose such a definition for an intrinsic angular momentum was Moreschi \cite{Moreschi1986, Moreschi:2002ii}, since then other groups have developed similar formulation from different motivations, such as Comp\`ere--Oliveri--Seraj \cite{Compere:2019gft}, Chen--Wang--Yau \cite{ChenWangWangYau202107,  ChenWangWangYau202103},  Javadinezhad--Kol--Porrati \cite{Javadinezhad:2022hhl}.

\smallskip It is important to appreciate though that the intrinsic angular momentum $\msf{S}_Y$ can only be defined once we extend the Poincar\'e algebra with the addition of a Goldstone mode dual to supertranslation and that it forms a representation of the full Lorentz algebra $\SL(2,\C)$. 
This is very different from the construction of a spin operator such as the Pauli--Luba\'nski pseudo-vector, which is constructed purely from the generator of the Poincar\'e algebra itself, without the introduction of any Goldstone mode operator.  In particular, $\msf{S}_Y$ does not provide a definition of the rotational component of the $\bms$ algebra and it does not satisfy an algebra isomorphic to the little group algebra. 

\smallskip
Moreover, the challenge of defining the super  orbital angular momentum (i.e. the two terms on the RHS of \eqref{SYbms} depending on $G$) consists of the possibility to define properly the Goldstone operator $G$, as reviewed above. In particular, whether one uses the good cut, electric good cut or nice cut conditions, the Goldstone mode is always defined from a condition on some of the shear components. 
The shear only determines the $\ell\geq 2$ components of $G$, so defining a Goldstone mode requires more information than simply the shear to be well defined. It requires the specification of the angular component $\ell=0$ and $\ell=1$. These components did not originally appear in the definition of the intrinsic angular momentum in \cite{Compere:2019gft, ChenWangWangYau202107,  ChenWangWangYau202103, Javadinezhad:2022hhl}, as only invariance under proper supertranslations (those of mode $\ell\geq 2$) was demanded. They were later included in \cite{JavadinezhadPorrati202211,Riva:2023xxm} in order to have a non-vanishing angular momentum flux in a perturbative expansion
in powers of the Newton constant $G_N$ starting at $\mcal{O}(G_N^2)$, as obtained in \cite{Damour:2020tta, Manohar:2022dea, DiVecchia:2022owy, DiVecchia:2022piu, Veneziano:2022zwh}. However, fixing the  $\ell=0,1$ components amounts to a choice of origin at future null infinity, namely a choice of a particular section of it---which is the equivalent of the choice of origin for the coordinate system used to define
the intrinsic angular momentum in special relativity---.   Moreover, these components  depend on the choice of Lorentz frame as well, as seen from  \eqref{Ctrans}; in particular, a boost transformation mixes the higher harmonics of $G$ with the
$\ell = 0, 1$ ones.
Therefore, neither of these definitions are  indeed supertranslation invariant nor Lorentz covariant. A proposal for determining the first two harmonics of the Goldstone operator $G$ in a Lorentz covariant manner was recently introduced in \cite{Javadinezhad:2023mtp}.

\subsection{A warm-up example: The Poincar\'e algebra}\label{sec:Poincare algebra}

To clarify the method we employ in this work, we provide in this section  a  brief review of the familiar example of the Poincar\'e algebra.
We explain the Poincar\'e algebra in some detail, considering its massive representation, and the crucial role of the Pauli--Luba\'nski pseudo-vector in the definition of intrinsic spin of a Poincar\'e particle (an irreducible representation of the Poincar\'e algebra);
we then review its coadjoint action and coadjoint orbits. We refer to classic and standard literature for a complete discussion \cite{Wigner193901,KimWigner198705,BargmannWigner194805,Dirac194502,HarishChandra194705,GelfandNaimark194711,Naimark1964,Knapp198612,BogolubovLogunovOksakTodorov1990,Onishchik200402,Hall2015,OhnukiKitakadoSugiyama198804}.

\paragraph{The Poincar\'e algebra and the Lie bracket.}

The isometry group of the four-dimensional\footnote{There is nothing special about four dimensions and the discussion can be easily generalized to any dimensions.} Minkowski spacetime, which is relevant to construct four-dimensional special-relativistic (classical or quantum) field theories, is the Poincar\'e algebra, denoted as $\msf{iso}(3,1)$ 
\begin{equation}\label{eq:the poincare algebra}
    \msf{iso}(3,1)=\tsf{so}(3,1)\sds\mbb{R}^4,
\end{equation}
where $\tsf{so}(3,1)$ denotes the Lie algebra of $\tenofo{SO}(3,1)^\uparrow$.\footnote{Technically-speaking, one has to distinguish between the in-homogeneous Lorentz group $\tenofo{SO}(3,1)\ltimes \mbb{R}^4$ and its in-homogeneous proper orthochronous subalgebra $\tenofo{SO}(3,1)^{\uparrow}\ltimes\mbb{R}^4$. It is the latter one that is identified as the Poincar\'e group and \eqref{eq:the poincare algebra} is its algebra.} It is generated by the momentum $\mbs{P}_\mu$, generating translations, and angular momentum $\mbs{J}_{\mu\nu}$, generating Lorentz transformations. They satisfy the following Lie bracket
\begin{eqgathered}\label{eq:poincare Lie algebra}
    [\mbs{P}_\mu,\mbs{P}_\nu]=0, \qquad [\mbs{J}_{\mu\nu},\mbs{P}_\rho]=\mfk{i}(\eta_{\mu\rho}\mbs{P}_\nu-\eta_{\nu\rho}\mbs{P}_\mu),
    \\
    [\mbs{J}_{\mu\nu},\mbs{J}_{\rho\sigma}]=\mfk{i}(\eta_{\mu\rho}\mbs{J}_{\nu\sigma}+\eta_{\nu\sigma}\mbs{J}_{\mu\rho}-\eta_{\mu\sigma}\mbs{J}_{\nu\rho}-\eta_{\nu\rho}\mbs{J}_{\mu\sigma}),
\end{eqgathered}
where $\eta=[\eta_{\mu\nu}]=\tenofo{diag}(-1,+1,+1,+1)$ is the Minkowski metric. 

\paragraph{Construction of Casimir elements.} As it is well-known, to study irreducible representation of Poincar\'e algebra, one needs to construct the Casimir elements for the Poincar\'e Lie algebra \eqref{eq:the poincare algebra}. Let us explain the construction briefly. We first need to define the change in a Poincar\'e-covariant object $\mscr{O}$ under the transformations generated by the Poincar\'e algebra
\begin{equation}\label{eq:the change of an object under the poincare-algebra action}
    \delta_{X}\mscr{O}=[X,\mscr{O}], \qquad \forall X\in\msf{iso}(3,1).
\end{equation}
From the general structure of \eqref{eq:the poincare algebra}, it is clear that on the orbit of a given covariant object\footnote{By the orbit of an object $\mscr{O}$ under an algebra $\g$, we mean
\begin{equation*}
    \tenofo{Orb}_{\g}(\mscr{O}):=\left\{\mscr{O}'\,\big|\,\mscr{O}'=X\rhd\mscr{O},\;\forall X\in\g\right\},
\end{equation*}
where $X\rhd\mscr{O}$ denotes the action of $X$ on $\mscr{O}$. In the case of Lorentz algebra $\tsf{so}(3,1)$, and say a vector $V^\mu$, this is given by 
\begin{equation}
    \tenofo{Orb}_{\tsf{so}(3,1)}(V^\mu)=\left\{V'^\mu\,\big|\,{V'}^\mu=\mbs{\Lambda}\indices{^\mu_\nu}V^\nu,\;\forall \mbs{\Lambda}\in \tsf{so}(3,1)\right\}.
\end{equation}} (like a vector, tensor, etc) under the Lorentz algebra, a Lorentz-invariant object built out of generators of the normal subalgebra $\mbb{R}^4$ is fixed. As $\mbb{R}^4$ is generated by $\mbs{P}_\mu$, one thus has to look for the simplest Lorentz-invariant object built out of $\mbs{P}_\mu$. This is given by $-\mbs{P}^2=-\mbs{P}_{\mu}\mbs{P}^{\mu}$, where we have included an extra minus sign for later convenience.\footnote{For $\mbs{P}^2\le 0$, one also need to consider the sign of $\mbs{P}_0$. We ignore this extra subtlety.} \eqref{eq:the change of an object under the poincare-algebra action} can be written as $\delta_{\mbs{J}_{\mu\nu}}\mbs{P}^2=[\mbs{J}_{\mu\nu},\mbs{P}^2]=0$. $\mbs{P}^2$ is already invariant under the normal subalgebra $\mbb{R}^4$, i.e. $\delta_{\mbs{P}_\mu}\mbs{P}^2=[\mbs{P}_\mu,\mbs{P}^2]=0$. As any element $\mbs{X}$ of the Poincare algebra can be written as a pair $\mbs{X}=(x^\mu\mbs{P}_\mu,\frac{1}{2}\psi^{\mu\nu}\mbs{J}_{\mu\nu})$ for constant coefficients $x^\mu$ and $\psi^{\mu\nu}$, we conclude that $\delta_{\mbs{X}} \mbs{P}^2=[\mbs{X},\mbs{P}^2]=0$ for any $\mbs{X} \in\msf{iso}(3,1)$. Hence, we see that the (quadratic) Casimir element of the Poincar\'e algebra $\msf{iso}(3,1)$ is
\begin{equation}\label{eq:quadratic casimir of poincare algebra}
    \wh{\mscr{C}}_2(\msf{iso}(3,1))=-\mbs{P}^2.
\end{equation}
It turns out that the construction of the next Casimir element is more tricky \cite{Wigner193901}. First notice that by inspection,  no cubic Casimir element exists since no Poincar\'e-invariant combination of $(\mbs{J}_{\mu\nu},\mbs{P}_\mu)$ can be constructed. However, there is a (quartic) Casimir element of the Poincar\'e algebra given by
\begin{equation}\label{eq:quartic Casimir of the Poincare algebra}
    \wh{\mscr{C}}_4(\msf{iso}(3,1))=\mbs{W}^2=\mbs{W}_\mu \mbs{W}^\mu\,,
\end{equation}
where $\mbs{W}_\mu$ is the so-called Pauli--Luba\'nski element\footnote{Note that $\mbs{W}_\mu$ is a polynomial in the $\iso(3,1)$ generators and naturally belongs to the universal enveloping algebra.} \cite{Lubanski194203a,Lubanski194203b,Wigner193901}
\begin{equation}\label{eq:definition of PL pseudo-vector}
    \mbs{W}_\mu:= \frac{1}{2}\varepsilon_{\mu\nu\rho\sigma}\mbs{P}^\nu \mbs{J}^{\rho\sigma}.
\end{equation}
The key properties of this vector are as follows 

\begin{enumerate}
    \item [(1)] Under Lorentz transformations, it transforms as
    \begin{equation}\label{eq:pauli-lubanski under lorentz rotation}
        [\mbs{J}_{\mu\nu},\mbs{W}_\rho]=\mfk{i}(\eta_{\nu\rho}\mbs{W}_\mu-\eta_{\mu\rho}\mbs{W}_\nu),
    \end{equation}
    and most importantly, it is invariant under translation
\begin{eqgathered}\label{eq:algebra relations for the Pauli--Lubanski vector}
    [\mbs{P}_\mu,\mbs{W}_\nu]=0,
\end{eqgathered}
Moreover, it satisfies the algebra
\begin{equation}\label{eq:Lie bracket of components of Pauli--Lubanski pseudo-vector}
[\mbs{W}_\mu,\mbs{W}_\nu]=\mfk{i}\varepsilon_{\mu\nu\rho\sigma}\mbs{P}^\rho \mbs{W}^\sigma.
\end{equation}
    \item[(2)]\hypertarget{pro:pauli-lubanski as generator of little group in fixed-momentum subspace}{$\mbs{W}_\mu$ is orthogonal to the 4-momentum operator $\mbs{P}_\mu \mbs{W}^\mu=0$. Moreover, if we diagonalize $\mbs{P}_\mu$ and evaluate  $\mbs{W}_\mu$ when $\mbs{P}_\mu=k_\mu$, we find that they satisfy the isotropy subalgebra of Lorentz transformation preserving the vector $k$.}\footnote{Consider states satisfying $\mbs{P}_\mu|p\rangle=k_\mu|p\rangle$. On such states, $\mbs{W}_\mu$ generates Lorentz transformations that leave $k^\mu$ invariant. This can be seen by noting that $\mbs{P}_\mu \mbs{W}_\nu|p\rangle=\mbs{W}_\nu \mbs{P}_\mu|p\rangle=k_\mu \mbs{W}_\nu|p\rangle$, which implies that $\mbs{W}_\nu|p\rangle\propto |p\rangle$, hence $\mbs{W}_\mu$ rotates the state $|p\rangle$ in the constant-momentum subspace determined by $k_\mu$. However, due to $\mbs{P}^\mu \mbs{W}_\mu=0$, there are less than four independent components. Indeed on such states, \eqref{eq:Lie bracket of components of Pauli--Lubanski pseudo-vector} is the defining relation of the isotropy subalgebra leaving $k_\mu$ fixed. For $k_\mu$ being time-like or light-like, the isotropy subalgebra is $\so(3)$ or $\iso(2)$, respectively.\label{ftn:pauli-lubanski generates isotropy subalgebra}}
    
    \item[(3)] Under a parity transformation, $x^0\to x^0$ and $x^i\to -x^i$, \eqref{eq:adjoint action of poincare group} implies that

    \begin{eqaligned}\label{eq:transformation of poincare generators under parity}
        \mbs{P}_0&\mapsto +\mbs{P}_0, &\quad \mbs{P}_i&\mapsto -\mbs{P}_i,
        \\
        \mbs{J}_{0i}&\mapsto - \mbs{J}_{0i}, &\qquad  \mbs{J}_{ij}&\mapsto + \mbs{J}_{ij},
    \end{eqaligned}
    which in turn leads to
    \begin{equation}
         \mbs{W}_0 \mapsto -\mbs{W}_0, \qquad \mbs{W}_i\mapsto +\mbs{W}_i.
    \end{equation}
    This is the reverse of the transformation of a four-vector under the parity. Hence, by definition, $\mbs{W}_\mu$ is a pseudo-vector.\footnote{This can also be seen by noting that for a general coordinate transformation $\mbs{\Lambda}$ of Minkowski spacetime, we have $\mbs{W}_{\mu'}=\det(\mbs{\Lambda})\mbs{\Lambda}\indices{_{\mu'}^\mu}\mbs{W}_\mu$, which shows the axial nature of $\mbs{W}_\mu$.}
\end{enumerate}

As \eqref{eq:quartic Casimir of the Poincare algebra} is manifestly Lorentz-invariant, we have $\delta_{\mbs{J}_{\mu\nu}}\mbs{W}^2=0$. Furthermore, \eqref{eq:algebra relations for the Pauli--Lubanski vector} implies $\delta_{\mbs{P}_\mu}\mbs{W}^2=[\mbs{P}_\mu,\mbs{W}^2]=0$. Therefore, $\delta_{\mbs{X}}\mbs{W}^2=0$ for any $\mbs{X}\in \msf{iso}(3,1)$, and hence $\mbs{W}^2$ is indeed a Casimir element of $\msf{iso}(3,1)$. It turns out that these are all the Casimir elements for the Poincar\'e algebra \cite{Wigner193901}.

\paragraph{The physical interpretation of Casimir elements.}

The above discussion implies that a general irreducible representation of the Poincar\'e algebra (i.e. an elementary particle in a Poincar\'e-invariant world {\it \`a la} Bargmann--Wigner philosophy) is labeled by the values of $-\mbs{P}^2$ and $\mbs{W}^2$. We denote this representation as $(m,s)$ and a state in this representation as $|m,s\rangle$, where $m$ and $s$ are the values of $-\mbs{P}^2$ and $\mbs{W}^2$, respectively. We thus need to understand the physical significance of these Casimir elements. 

\smallskip The physical meaning of $-\mbs{P}^2$ is clear: it determines the (bare) physical mass $m$ of the corresponding particle via $-\mbs{P}^2|m,s\rangle=m^2|m,s\rangle$. Depending on the value of $-\mbs{P}^2$ (positive, zero, or negative), one has different unitary irreducible representations (massive, massless, and tachyonic, respectively). 

\smallskip The physical meaning of $\mbs{W}^2$ is more subtle. Using $\mbs{P}_\mu=(E,\mbs{P}_i)$, $\mbs{J}$ (the rotation generators) with $\mbs{J}_i:= \varepsilon\indices{_i^{jk}}\mbs{J}_{jk}$, and $\mbs{K}$ (the boost generators) with $\mbs{K}_i:= \mbs{J}_{0i}$ for $i,j=1,2,3$, satisfying
\begin{eqaligned}
    \label{eq:lorentz algebra in terms of rotation and boost generators}
    [\mbs{J}_i,\mbs{J}_j]&=\mfk{i}\varepsilon\indices{_{ij}^k}\mbs{J}_k, 
    \\
    [\mbs{J}_i,\mbs{K}_j]&=\mfk{i}\varepsilon\indices{_{ij}^k}\mbs{K}_k,
    \\
    [\mbs{K}_i,\mbs{K}_j]&=-\mfk{i}\varepsilon\indices{_{ij}^k}\mbs{J}_k,
\end{eqaligned}
we can write the Pauli--Luba\'nski pseudo-vector in the component form
\begin{eqgathered}\label{eq:components of the Pauli--Lubanski vector}
    \mbs{W}_0=\mbs{J}^i\mbs{P}_i, \qquad \mbs{W}_i=E\mbs{J}_i-(\mbs{P}\times\mbs{K})_i.
\end{eqgathered}
This form is suitable to extract some general lessons about representations of Poincar\'e algebra. In the following, we only consider massive representations. 

\smallskip   Given a momentum $\mbs{P}_\mu=(E,\mbs{P}_i)$ one can always find a pure boost transformation $\mbs{\Lambda}_{\mbs{P}}$ which maps the rest frame momentum $\mbs{P}^{\mathsf{rest}}_\mu=(m,\mbs{0})$ onto $\mbs{P}_\mu$, i.e. $(\mbs{P}^{\mathsf{rest}} \cdot \mbs{\Lambda}_{\mbs{P}})_\mu= \mbs{P}_\mu$ and we use the notation
$(\mbs{P} \cdot  \mbs{\Lambda})_\nu:=\mbs{P}_\mu \mbs{\Lambda}^\mu{}_\nu$.
Such a boost transformation is given explicitly by 
    \begin{eqgathered}\label{eq:lorentz transformation from the rest frame to a general frame}
        (\mbs{\Lambda}_{\mbs{P}})\indices{^0_0}=\frac{E}{m}, \qquad  (\mbs{\Lambda}_{\mbs{P}})\indices{^0_i}=\frac{\mbs{P}_i}{m}, \qquad  (\mbs{\Lambda}_{\mbs{P}})\indices{^i_0}=\frac{\mbs{P}^i}{m},
        \\
        (\mbs{\Lambda}_{\mbs{P}})\indices{^i_j}=\delta\indices{^i_j}+\frac{\mbs{P}^i\mbs{P}_j}{m(m+E)}.
    \end{eqgathered}
Its inverse is given by $\mbs{\Lambda}_{\mbs{P}}^{-1}= \mbs{\Lambda}_{\wh{\mbs{P}}}$ where $\wh{\mbs{P}}_\mu=(E,-\mbs{P}_i)$ is the parity-reversed momentum.
It is important to appreciate that in general, given a Lorentz transformation $\mbs{\Lambda}$ we have that 
$\mbs{\Lambda}_{\mbs{P}} \mbs{\Lambda}$ differs from $\mbs{\Lambda}_{\mbs{P}\cdot \mbs{\Lambda}} $ by a rotation
\be\label{precession}
\mbs{\Lambda}_{\mbs{P}} \mbs{\Lambda} = \mbs{R}(\mbs{\Lambda},\mbs{P}) \mbs{\Lambda}_{\mbs{P}\cdot \mbs{\Lambda}},
\ee
where $\mbs{P}^{\mathsf{rest}} \mbs{R}(\mbs{\Lambda},\mbs{P})  = \mbs{P}^{\mathsf{rest}}$.

\smallskip It turns out that the physical interpretation of the quartic Casmir \eqref{eq:quartic Casimir of the Poincare algebra} can be given in terms of the notion of spin. By definition, the spin generator of an algebra is the generator of its isotropy subalgebra. In the case of Poincar\'e algebra, the spin generator  $\mbs{S}_i$ is defined as the spatial component of  the Pauli--Luba\'nski pseudo-vector in the rest frame. It is obtained by boosting the Pauli--Luba\'nski vector into the rest frame  \cite[\S 7.2]{BogolubovLogunovOksakTodorov1990}
\be\la{SW}
m \mbs{S}_i :=   ( \mbs{W}\cdot \mbs{\Lambda}_{\mbs{P}}^{-1} )_i.
\ee 
We  find that
\be 
 \label{eq:spin 3-vector in general Lorentz frame}
        \mbs{S}_i=\frac{1}{m}\left(\mbs{W}_i-\frac{\mbs{W}_0}{m+E}\mbs{P}_i\right).
\ee

One can also verify that the spin vector corresponds to the rotation vector in the rest frame, namely
\be\la{Jrest}
\varepsilon\indices{_i^{jk}}\mbs{J}^{\mathsf{rest}}_{jk}
=\varepsilon\indices{_i^{jk}}
  ( \mbs{J}\cdot \mbs{\Lambda}_{\mbs{P}}^{-1}\cdot \mbs{\Lambda}_{\mbs{P}}^{-1} )_{jk}=\mbs{S}_i\,.
\ee

From its definition or through a direct evaluation one can show that the spin vector commutes with $\mbs{P}_\mu$ and satisfy the $\msf{so}(3)$ algebra as expected\footnote{Since it is the isotropy subalgebra of Poincar\'e algebra for a generic massive momentum.}
\be
[\mbs{P}_\mu, \mbs{S}_i]=0, \qquad [\mbs{S}_i, \mbs{S}_j]=i \epsilon_{ij}{}^k \mbs{S}_k. 
\ee
It can be also be checked that $\mbs{S}_i$ is preserved by rotations, i.~e. $[\mbs{J}_i, \mbs{S}_j]=i\epsilon_{ij}{}^k\mbs{S}_k$ and that it is a pseudo-vector.\footnote{Note that orbital angular momentum ${L}_i=\varepsilon_{i}^{\enspace jk}x_jp_k$ is an axial vector under space reflection (taking into account that the Levi--Civita symbol is a tensor density of weight $-1$).}
It is shown in  \cite[\S 7.2]{BogolubovLogunovOksakTodorov1990} that $\mbs{S}$ is uniquely determined by these conditions. On the other hand, the spin vector is not preserved under boost. One can check directly that $[\mbs{K}_i, \mbs{S}_j]\neq 0$.
In particular, this means that $(0,\mbs{S}_i)$ does not transform as a vector under Lorentz transformations   as expected from the nontrivial precession \eqref{precession}. The spin vector is such that  $\mbs{W}^2 =m^2\mbs{S}_i \mbs{S}^i$. Since $\mbs{S}^2=\mbs{S}_i \mbs{S}^i$ is the Casimir element for $\tsf{so}(3)$, its eigenvalues on an irreducible spin-$s$ representation of $\tsf{so}(3)$\footnote{Here, we use the fact that irreducible representations of $\tsf{so}(3)$ are spin-$s$ representations of dimension $2s+1$.} are $s(s+1)$. Hence, we have
    \begin{equation}
        \mbs{W}^2|m,s\rangle=m^2s(s+1)|m,s\rangle.
    \end{equation}

    \smallskip The upshot of the above discussion is that $\mbs{W}^2$ as one of the Casimir elements of the Poincar\'e algebra provides an unambiguous definition of the spin of a particle in any special-relativistic system. Therefore, each elementary particle in such a system is distinguished by its mass and spin. 
 
\paragraph{Intrinsic angular momentum.}
The Poincar\'e algebra only contains momentum and angular momentum operators, and the spin generator is constructed purely in terms of them. 
If we extend the Poincar\'e algebra with the addition of 
a position operator $x^\mu$ conjugated to $\mbs{P}_\mu$, one can easily define now a translation invariant ``intrinsic angular momentum''. 
The position operator $x^\mu$ can be understood as a  translation Goldstone mode analogous to 
 $C$ described in \S \ref{sec:issues with the bms algebra}, we can define the intrinsic angular momentum operator given by the difference between the total and orbital angular momentum
\be \label{PoincS}
S_{\mu\nu} = \mbs{J}_{\mu\nu} - x_\mu \mbs{P}_\nu -\mbs{P}_\mu x_\nu.
\ee 

This object is obviously translation invariant but it doesn't have an algebraic interpretation since it depends on the position $x$.
A canonical choice of position called Newton--Wigner (NW) position \cite{Pryce194811,Newton:1949cq, Choi:2014nea,Freidel:2020ayo}
is achieved if we demand that the spin-angular momentum is related to the spin-vector \eqref{eq:spin 3-vector in general Lorentz frame}
\be \la{spinspin}
\frac12 \epsilon_i{}^{jk} S_{jk}= \mbs{S}_i.
\ee 
 
 The NW position is uniquely defined by this requirement up to time translation $x_\mu \to x_\mu -\tau \mbs{P}_\mu$.
In the Newton--Wigner frame, we have that  
\be 
\mbs{K}_i = x_i\mbs{P}_0-\mbs{P}_ix_0 + \frac{(\mbs{P}\times \mbs{S})_i}{m+\mbs{P}_0}.
\ee 
What distinguishes the Newton--Wigner coordinates from any other choice of coordinates is that they are entirely determined by the momentum and angular momentum.\footnote{They are explicitly given by \cite{Newton:1949cq, Freidel:2020ayo}
\begin{equation*}
    x_\mu^{\mathrm{NW}} = \frac{\mbs{J}_{\mu \nu} (\mbs{P}^{\nu} + \mbs{P}_{\mathrm{rest}}^{\nu})}{\mbs{P}\cdot (\mbs{P} + \mbs{P}_{\mathrm{rest}})}.
\end{equation*}} 
However the Newton-Wigner coordinates do not transform covariantly under boost.

\paragraph{Coadjoint orbits of Poincar\'e group.} Up to now, we have explained the well-known representation theory of Poincar\'e algebra. However, beginning in \S \ref{sec:gbms coadjoint action and charge algebra}, we work with coadjoint orbits of gravitational symmetry algebras. Therefore, as a warm-up to that study, we explain the coadjoint orbits of the Poincar\'e group. Here, we just collect the basic formulas and relegate all the details to Appendix \ref{sec:pauli-lubanski pseudo-vector as a constant of motion}. For a discussion of coadjoint orbits of Poincar\'e group from a more general and abstract perspective see \cite[\S 2.3]{DonnellyFreidelMoosavianSperanza202012}. 

\smallskip We denote a basis for $\iso(3,1)$ and $\iso(3,1)^*$ as $\{\mbs{P}_\mu,\mbs{J}_{\mu\nu}\}$ and $\{\mbs{X}^\mu,\mbs{\Psi}^{\mu\nu}\}$, respectively. Generic elements $ \mbs{X}\in\iso(3,1)$ and  $ \mbs{X}^*\in\iso(3,1)^*$ have the following expansions 
\begin{equation}
    \mbs{X}=x^\mu\mbs{P}_\mu+\frac{1}{2}\psi^{\mu\nu}\mbs{J}_{\mu\nu},\qquad 
    \mbs{X}^*= p_\mu\mbs{X}^\mu+\frac{1}{2}j_{\mu\nu}\mbs{\Psi}^{\mu\nu}\,,
\end{equation}
where $(x,\psi)$ are Lie algebra coordinates, while 
$(p,j)$ are dual Lie algebra coordinates and the pairing is simply
\be 
\langle \mbs{X}| \mbs{X}^*\rangle= x^\mu p_\mu +\psi^{\mu\nu}j_{\mu\nu}.
\ee 
The coadjoint actions of an element $g=(\mbs{\Lambda},\mbs{a})\in\tenofo{Iso}(3,1)$ on the dual coordinates $(p_\mu,j_{\mu\nu})$ are \cite{Rawnsley197502}
\begin{eqaligned}
    \label{eq:coadjoint action of the poincare group}
    \tenofo{Ad}^*_g p_\mu&=( p\cdot \mbs{\Lambda})_\mu,
    \\
    \tenofo{Ad}^*_g j_{\mu\nu}&=\left(\mbs{\Lambda}\cdot j \cdot\mbs{\Lambda}^{-1}\right)_{\mu\nu}-\left[(p \cdot  \mbs{\Lambda})_\mu \mbs{a}_\nu-\mbs{a}_\mu (p\cdot \mbs{\Lambda} )_\nu\right].
\end{eqaligned}
As $(p_\mu,j_{\mu\nu})$ can be thought of as a basis for linear functions on $\iso(3,1)^*$, we can compute the linear Poisson structure on $\iso(3,1)^*$ as
\begin{eqgathered}\label{eq:linear poisson structure on iso(3,1)*}
    \{p_\mu,p_\nu\}_{\iso(3,1)^*}=0,\qquad \{p_\mu,j_{\nu\rho}\}_{\iso(3,1)^*}=\eta_{\mu\nu}p_\rho-\eta_{\mu\rho}p_\nu,
    \\
    \{j_{\mu\nu},j_{\rho\sigma}\}_{\iso(3,1)^*}=\eta_{\mu\sigma}j_{\nu\rho}-\eta_{\mu\rho}j_{\nu\sigma}+\eta_{\nu\rho}j_{\mu\sigma}-\eta_{\nu\sigma}j_{\mu\rho}.
\end{eqgathered}
This familiar form is the classical analog of the Poincar\'e algebra \eqref{eq:the poincare algebra}.

\smallskip Analogous to irreducible representations, which are distinguished by a complete set of Casimir elements, coadjoint orbits can be distinguished by the complete set of Casimir functions. The classical analogs of two Casimir elements $\wh{\mscr{C}}_2(\iso(3,1))$ and $\wh{\mscr{C}}_4(\iso(3,1))$ of $\iso(3,1)$, defined in \eqref{eq:quadratic casimir of poincare algebra} and \eqref{eq:quartic Casimir of the Poincare algebra} respectively, are the two Casimir functions on coadjoint orbits of $\tenofo{Iso}(3,1)$, which are given by\footnote{Note that the Lie bracket of $\iso(3,1)$ encodes the adjoint action, the Lie Poisson structure on $\iso(3,1)^*$ encodes the coadjoint action. Since \eqref{eq:linear poisson structure on iso(3,1)*} is the classical version of \eqref{eq:poincare Lie algebra}, the proof of the invariance of these Casimir functions is the same as the construction of Casimir elements $\wh{\mscr{C}}_2(\iso(3,1))$ and $\wh{\mscr{C}}_4(\iso(3,1))$ by using Poisson brackets \eqref{eq:linear poisson structure on iso(3,1)*} instead of Lie brackets \eqref{eq:poincare Lie algebra}.}
\begin{eqgathered}\label{eq:casimir functions on coadjoint orbits of poincare}
    \mscr{C}_2(\iso(3,1)):=-\eta^{\mu\nu}p_\mu p_\nu,\qquad \mscr{C}_4(\iso(3,1)):=\eta^{\mu\nu} w_\mu w_\nu,
\end{eqgathered}
where 
\begin{equation}\label{eq:dual PL vector}
    w_\mu:=\frac{1}{2}\varepsilon_{\mu}\!^{\nu\rho\sigma}p_\nu j_{\rho\sigma},
\end{equation}
is the dual\footnote{Here, dual means an element of the coalgebra $\iso(3,1)^*$.} Pauli--Luba\'nski pseudo-vector \eqref{eq:definition of PL pseudo-vector}, are invariant under the coadjoint action of $\tenofo{Iso}(3,1)$. \eqref{eq:linear poisson structure on iso(3,1)*} implies the classical analogs of \eqref{eq:algebra relations for the Pauli--Lubanski vector}, \eqref{eq:pauli-lubanski under lorentz rotation}, 
\begin{eqgathered}\label{eq:poisson bracket of dual pauli-lubanski pseudo-vector}
    \{p_\mu,w_\nu\}_{\iso(3,1)^*}=0, \qquad \{j_{\mu\nu},w_\rho\}_{\iso(3,1)^*}=\eta_{\mu\rho}w_\nu-\eta_{\nu\rho}w_\mu,
\end{eqgathered}
and \eqref{eq:Lie bracket of components of Pauli--Lubanski pseudo-vector}
\begin{equation}\label{eq:poisson bracket of dual pauli-lubanski pseudo-vector with itself}
    \{w_\mu,w_\nu\}_{\iso(3,1)^*}=\varepsilon_{\mu\nu}\,^{\rho\sigma}p_\rho w_\sigma.
\end{equation}
\eqref{eq:linear poisson structure on iso(3,1)*} and \eqref{eq:poisson bracket of dual pauli-lubanski pseudo-vector} imply that $\mscr{C}_2(\iso(3,1))$ and $\mscr{C}_4(\iso(3,1))$ are invariant under the coadjoint action \eqref{eq:coadjoint action of the poincare group}, hence, generic coadjoint orbits of $\tenofo{Iso}(3,1)$, which are eight-dimensional, are labeled by the values of these functions.

\paragraph{$w_\mu$ as a constant of motion.} We would like to derive the evolution of $w_\mu$. The motivation is to compare the result with the evolution of the analogous quantity for $\gbms$, which we study in \S\ref{sec:evolution equations}. This can be achieved by defining a set of coordinates on a coadjoint orbit of $\tenofo{Iso}(3,1)$. One can then show that $w_\mu$ is a constant of motion 
\begin{equation}
    \frac{\rd w_\mu}{\rd \tau}:=\{w_\mu,H\}_{\iso(3,1)^*}=0\,,
\end{equation}
where the evolution in $\tau$ is determined through the Hamiltonian $H$. All the details can be found in Appendix \ref{sec:evolution of dual pauli-lubanski pseudo-vector}.

\smallskip After a rather long de tour of the Poincar\'e algebra, we now move on to the main symmetry algebra of our interest, i.e. the generalized BMS algebra.

\section{$\gbms$ algebra and Poincar\'e embeddings}\label{sec:gbms and poincare embeddings}

In this section, we first describe in detail in \S \ref{sec:gbms algebra and bracket} the algebraic aspects of the $\gbms$ algebra. 
 To avoid functional analysis and representation theoretic questions, we present the analysis at the semi-classical level, i.e. at the level of coadjoint orbits, and study the $\gbms$ coadjoint action in \S \ref{sec:gbms coadjoint action and charge algebra}. 
 We also introduce in \S \ref{sec:gbms goldstone modes} the notion of intrinsic angular momentum charge, which has appeared in previous literature as a proposal to overcome some of the ambiguities reviewed above, and we derive its algebra.
 Finally, we  explain in \S \ref{sec:poincare embedding} how to embed the Poincar\'e algebra in $\gbms$, which allows us to show the relationship of the $\gbms$ Casimirs with the usual notion of mass and spin associated with a Poincar\'e subalgebra. 
 This analysis provides a precursory step in the program of asymptotic quantization  of gravity in spacetimes with $\gbms$ as their asymptotic-symmetry group. For the sake of brevity, we sometime denote $\gbms$ as $\g$ and its dual $\gbms^*$ as $\g^*$. 
 
\subsection{$\gbms$: Algebra and  Lie bracket}\la{sec:gbms algebra and bracket}

Let us start by reviewing the $\gbms$ algebra and its Lie bracket. This algebra was first proposed by Campiglia and Laddha to account for the subleading soft theorem from asymptotic symmetries and includes  super-Lorentz transformations on the  celestial sphere \cite{Campiglia:2014yka}. Since its introduction, it has been extensively studied \cite{CampigliaLaddha201502,Compere:2018ylh, CampigliaPeraza202002, CampigliaLaddha202106, Freidel:2021fxf} and shown to be the relevant symmetry algebra for the study of  memory effects \cite{Pasterski:2015tva, Nichols:2017rqr, Nichols:2018qac}. A first generalization of this algebra, allowing for local conformal rescaling of the 2D sphere metric, was introduced in \cite{Freidel:2021fxf}. This Weyl extension of the BMS algebra provides a useful tool, that we will exploit below, to classify the primary fields of the asymptotic symmetry group in terms of their conformal dimension and spin. Moreover, it allows us to recast the leading (in a large-$r$ expansion) asymptotic dynamics in a compact and elegant form \cite{Freidel:2021qpz}.

Afterwards, the $\gbms$ algebra was  extended by a spin-$2$ generator  realizing the sub-subleading symmetry \cite{Freidel:2021dfs}, and then further generalized to include an entire tower of higher spin symmetry generators \cite{Freidel:2021ytz}. We now define this algebra and its Lie bracket.

\smallskip The asymptotic ends of an asymptotically-flat spacetime  includes spacelike infinity $\iota_0$, timelike infinity $\iota_\pm$ and 
null infinity  $\scri_\pm$. Null infinity has the topology of a trivial fiber bundle  $\scri_\pm = \mathbb{R} \times S$ over the celestial sphere $S$.
Let us  introduce Bondi coordinates $x^\mu=(u,r,\sigma^A)$  on spacetime where $y^a=(u,\sigma^A)$ denotes coordinates on $\scri_+$.  $u$ is the retarded time  labeling null outgoing geodesic congruences which intersect $2d$ spheres, $r$ measures  the  sphere's radius along these geodesics and 
 $\sigma^A$ denote coordinates  on the celestial sphere.\footnote{Similarly on $\scri_-$ we chose advanced time coordinates $(v,\sigma')$.} 
It is essential  to note  that $\gbms$ depends on the choice of a normalized reference area form on $\scri$
\begin{equation}\label{eq:area-form preserved by gbms}
   \bfe := \epsilon\, \rd^2 \sigma = \frac12 \epsilon_{AB} \rd \s^A\wedge \rd \s^B, \qquad  \bigintsss_S \bfe =1.
\end{equation}
Here $\bfe$ denotes the two-form, $\epsilon$ denotes the density and $\epsilon_{AB}$ denotes the anti-symmetric two-form components. Importantly, $\epsilon$ is taken to be strictly positive.\footnote{As we will see \S\ref{sec:gbms coadjoint orbits}, the isotropy subalgebra preserves a rescaled area form. The assumption of strict positivity of the area form enters the study of coadjoint orbits. This is analogous to the role of mass in the case of Poincar\'e algebra reviewed in \S\ref{sec:Poincare algebra}, where mass can be positive (massive orbits), zero (massless orbits), and negative (tachyonic orbits). It would be interesting to work out the implications of zero and negative area forms. For a related discussion, see \cite{DonnellyFreidelMoosavianSperanza202012}.} The reference area form $\bfe$ defines a choice of conformal frame at asymptotic infinity. It enters the definition of algebra as a structure constant and we denote $\gbms_{\bfe}$ the corresponding generalized BMS algebra.  
Note that by Moser Theorem \cite{Moser196502}, any two such area measure are related by a diffeomorphism.\footnote{More precisely if $\bfe$ and $\bfe'$ are two such measure then there exist a sphere diffeomorphism $F: S\to S$ such that $\bfe'=F^*\bfe$.}
This means  that, while $\gbms_{\bfe}$ and $\gbms_{\bfe'}$ represent two \emph{different} subalgebras of $\gbms$, they  are canonically isomorphic. 

\smallskip The algebra $\gbms_{\bfe}$ depends on a pair $(Y,T)$, where $T(\sigma)$ is  a function on $S$ representing supertranslations and $Y=Y^A(\sigma)\pa_A$ is a vector field on $S$ representing infinitesimal super-Lorentz transformations (infinitesimal diffeomorphisms in $\msf{diff}(S)$). The data $(Y,T)$ determine a spacetime vector field given by 
\be\label{vectorfield}
\xi_{(Y,T)} = T(\sigma) \pa_u  + Y^A(\sigma)\pa_A + W_Y(\sigma) (u\pa_u -r\pa_r),
\ee
where $W_Y$, called the Weyl factor \cite{Freidel:2021fxf}, is restricted to be proportional to the divergence of $Y$ with respect to the given volume form
\begin{equation}\label{Wres}
	W_Y:= \f12 \mathrm{div}_{\epsilon}(Y)= \f12 \epsilon^{-1} \pa_A (\epsilon Y^A)\,.
\end{equation}
This expression for $W_Y$ shows that it depends only on $\epsilon$.
From \eqref{vectorfield} it is direct to see that $\gbms$ has the following Lie bracket
\be\la{gbms}
	[\x_{(Y_1,T_1)},\x_{(Y_2,T_2)}]_{\gbms}=\x_{(Y_{12},T_{12})},
\ee
with
\begin{equation}\label{gbmsvec}
	\begin{gathered}
	    Y_{12}=[Y_1,Y_2]_{S},
	    \\
	    T_{12}=(Y_1[T_2]- T_2 W_{Y_1})  -(Y_2[T_1]-T_1 W_{Y_2}).
	\end{gathered}
\end{equation}
Here, $[Y_1,Y_2]_{S}^A= (Y_1^B\pa_BY_2^A-Y_2^B\pa_BY_1^A)$ denotes the Lie bracket of vector fields on $S$ and $Y[T]:=Y^A\pa_A T$ denotes the action of a vector field on the function $T$. This means that the $\gbms$ Lie algebra can be written as a semi-direct sum
\be
        \gbms=\diff(S) \sds \, \mbb{R}_{-1}^S\,.
\ee
As in \S\ref{sec:issues with the bms algebra}, $  \mbb{R}_{\Delta}^S
$ denotes the space of functions on the celestial sphere of conformal weight $\Delta$  and  $\sds $ denotes  the action
\be
\delta_Y \phi_\Delta = Y[\phi_\Delta(\sigma)] + \Delta  W_Y \phi_\Delta(\sigma),
\ee
of   $\diff(S)$   on $  \mbb{R}_{\Delta}^S$.  This action can be extended to the space of spin-$s$ tensor fields\footnote{By definition a  spin $s$ tensor is a  symmetric and  traceless covariant tensor $\tau_{A_1\cdots A_s}$, while a spin $-s$ tensor is  a  symmetric and traceless contravariant tensor
$\tau^{A_1\cdots A_s}$.} of weight $\Delta$, denoted as $V_{(\Delta,s)}$, with \cite{Freidel:2021dfs}
\be
\delta_Y \mcal{O}_{(\Delta,s)} = \cL_Y[\mcal{O}_{(\Delta,s)} ] + (\Delta-s)W_Y \mcal{O}_{(\Delta,s)}, \qquad \forall~\mcal{O}_{(\Delta,s)}\in V_{(\Delta,s)},
\ee
where $\cL_Y$ denotes the Lie derivative. With this notation, we have $T\in V_{(-1,0)}$ and $Y\in V_{(-1,-1)}.$ 

\smallskip We assume that the sphere is equipped with a 2D metric $\gamma_{AB}$ compatible with $\bfe$, which means that the metric area form  coincides with the given area form
\be
\epsilon=\sqrt{\gamma}\,,
\ee
with $\gamma:=\mathrm{det}(\gamma)$.
The $\gbms$ algebra acts on the metric by conformal diffeomorphism.
\be\la{dgamma}
\delta_{(Y,T)} \gamma_{AB} = \cL_Y \gamma_{AB} - 2 W_Y \gamma_{AB}
= 2\cD_{\langle A} Y_{B\rangle},
\ee
where the angle bracket means that we take the symmetric traceless components
and $\cD_A$ is the covariant derivative that preserves $\gamma_{AB}$, in terms of which the Weyl factor can simply be written as
\begin{equation}\label{Wres2}
	W_Y=\frac1{2} \epsilon^{-1}\pa_A (\epsilon Y^A)= \f12 \cD_A Y^A .
\end{equation}
In particular, the transformation \eqref{dgamma} shows that a supertranslation leaves the metric invariant. Moreover, it also makes  explicit the fact that  $\gbms$ preserves the area form
\begin{equation}\label{dq0}
	\d_{(Y,T)}\sqrt{\gamma}=0=\d_{(Y,T)}\bfe\,.
\end{equation}
This condition means that $\sqrt{\gamma}=\epsilon$ is a constant for the symmetry algebra, and $\gbms_{\bfe}$ is an algebra, not an algebroid.

\smallskip We have seen that, by Moser Theorem, all $\gbms_{\bfe}$ algebras are canonically isomorphic. Therefore, we do not lose any information by fixing the reference area form.
From now on we will work with $\bfe=\mathring{\bfe}$ being fixed to be the round sphere area form 
\be \label{eq:round-sphere area form}
\mathring{\bfe} = \frac1{4\pi} \sin \theta \rd \theta \wedge \rd \phi  = \frac1{2i\pi} \frac{\rd z  \wedge \rd \bar{z} }{(1+|z|^2)^2}\,,
\ee 
where, as in \S\ref{sec:issues with the bms algebra}, $(z,\bar{z})$ denotes the local complex coordinate on the sphere endowed with the round-sphere metric. We denote $\gbms:= \gbms_{\mathring\bfe}$ for $\mathring\bfe$ given in \eqref{eq:round-sphere area form}.

\paragraph{The relation between $\bms$ and $\gbms$.} { To connect with the previous section,  we recall that $\bms_{q}$ is a subalgebra of $\gbms_\epsilon$ which is characterized by the choice of a round sphere metric $q_{AB}$ such that $R(q)=2$ and such that $\sqrt{q} = \epsilon$. $\bms_q$ is obtained  from the elements of $\gbms$ that preserve $q$, i.~e. $(T,Y)\in \bms_q$ if $\delta_{(T,Y)}q_{AB}=0$. 
} Explicitly, this means restricting the vector fields $Y$ to be global conformal Killing vectors of the round sphere metric $q_{AB}$, namely 
\be \label{confinv}
D_A Y_B + D_B Y_A = D_C Y^C {q}_{AB}\,,
\ee 
where $D_A$ is the covariant derivative that preserves $q_{AB}$.
The subalgebra of these vector fields forms a finite dimensional algebra isomorphic to $\msf{sl}(2,\C)$.
In \cite{BarnichTroessaert201001, BarnichTroessaert201601}, Barnich and Troessaert proposed to promote this algebra to an infinite dimensional symmetry algebra called extended $\bms$ and denoted $\ebms$ by keeping the condition of local conformal  Killing vectors \eqref{confinv}, but allowing the vector fields to admit poles on $S$. 
We have that 
\be\label{eq:the embedding of bms, ebms, gbms}
\bms_q \subset \ebms_q \subset \gbms.
\ee 
{
Given two different round sphere metrics $q$ and $q'$, there exists a sphere diffeomorphism $F\in \mathrm{Diff}(S)$ such that $q'= F^*(q)$. Therefore, the algebras $\bms_q$
and $\bms_{q'}$ are isomorphic, but they correspond to different subalgebras of $\gbms$.
In particular one can easily see that $\bms_q$ and $\bms_{q'}$ \emph{do not commute} when $q\neq q'$. This situation is analogous to the choice of a rotation subgroup inside Lorentz. On the one hand, the rotation group SO$(3)$ is well-defined. On the other hand, there are infinite ways to view it as a subgroup of the Lorentz group labelled by the choice of a unit timelike vector $t$. As subgroups of the Lorentz group, SO$_t(3)$ and SO$_{t'}(3)$ are isomorphic but distinct and they do not commute.
}

 In the following, when  the round sphere metric representative is chosen to be the usual metric in spherical coordinates
\be \label{q0}
\mathring{q}_{AB}\rd \sigma^A \rd \sigma^B = \rd\theta^2 + \sin\theta^2 \rd \varphi^2\,,
\ee 
we use the label $\mathring{q}$.

\subsection{$\gbms$: Coadjoint action and  charge algebra}\label{sec:gbms coadjoint action and charge algebra}

Having collected the required background, we now study coadjoint actions and coadjoint orbits for $\gbms$. As is well established \cite{Kirillov200407,Kirillov196202,Kirillov1976,Kirillov199908,Kostant1965,Kostant1970,Souriau1970,Souriau1997}, one way to construct the representations of any Lie algebra (subject to topological subtleties) is to first understand the coadjoint action of $\gbms$ on its dual $ \gbms^*$ and to describe its orbits. This is the subject of this section and \S\ref{sec:gbms coadjoint orbits}. Let us emphasize again that the main reason we focus on $\gbms$ coadjoint orbits is that the orbits of $\bms$ are too degenerate to admit an interesting notion of spin \cite{McCarthy197211}. 

\paragraph{The coadjoint action.} We denote elements of $\gbms^*$ by a pair of  coordinates $(m, j)$, where $m$ is a scalar dual to $T$ and $j=j_A\rd \sigma^A$ is a one form  dual to $Y=Y^A\partial_{A}$. $m$ and $j$ are called the \emph{mass aspect}  and the \emph{angular momentum aspect}, respectively. Since $\gbms$ is a local group on the sphere, the canonical charges are realized as the integral of the charge aspects on the sphere 
\be\label{eq:gbms charges}
\mathsf{P}_T:=\bigintsss_S T m \,\bfe, \qquad \mathsf{J}_{Y} :=\bigintsss_S Y^Aj_A \, \bfe, 
\ee
where $\bfe$ is defined in \eqref{eq:area-form preserved by gbms}. Note that these charges  generate the coadjoint action of $\gbms$ on $\gbms^*$, which we make more precise below (see around \eqref{eq:gbms charge action}). The total charge is given by the sum of these charges. The canonical pairing   $\langle\cdot|\cdot\rangle:\gbms\times\gbms^*\to\mbb{R}$ is given by
\begin{equation}
    \langle j,m | Y,T \rangle %
    =\mathsf{P}_T +  \mathsf{J}_Y\,.
\end{equation}
As we will see in \S \ref{sec:the gbms moment map}, the coadjoint-orbit angular momentum defined here through the canonical pairing is \emph{half} the gravitational angular momentum.\footnote{The normalization of the angular momentum is conventional and has varied over the years in the gravity literature. In recent years, the convention has settled down \cite{Flanagan:2015pxa, Compere:2019gft, ChenWangYau201312, Freidel:2021dfs}. If we require  the coadjoint action to match the action on the gravitational angular momentum, one needs to choose $m=\f1{4\pi G} m^{\mathrm{Grav}}$  and $j_A =\f1{8\pi G}  j_A^{\mathrm{Grav}}$. The numerical factors in the coadjoint action \eqref{eq:coadjoint action on gbms*} match the ones obtained in \cite{Barnich:2021dta}.
} 
 Using this pairing,  we can define the infinitesimal coadjoint action of $(Y,T)\in \gbms$ on $(j,m ) \in \gbms^*$ from
\begin{equation}\la{coadj}
		\langle \delta_{( Y_1, T_1)} ( j,m)| Y_2,T_2 \rangle =-\langle j,m | Y_{12},T_{12} \rangle.
\end{equation}
Given this definition, one obtains that the coadjoint action reads  (see Appendix \ref{sec:proofs in section 3} for the calculation)
\begin{eqaligned}\label{eq:coadjoint action on gbms*}
    \delta_{(Y,T)}  m&=	Y^A \pa_A m+  3 W_Y m,
    \\
    \delta_{(Y,T)} j_A&=\mcal{L}_Yj_A + 2W_Y j_A + \f32 m\pa_AT+\f{T}{2}\pa_Am,
\end{eqaligned}
where $\mcal{L}_Yj_A=Y^B\cD_Bj_A+j_B\cD_AY^B$ is the Lie derivative of components of the one-form $j$. 
	These transformation laws reveal that $m$ and $j_A$ are both of conformal weight $3$
	\be
m\in V_{(3,0)}\,,\qquad	j_A\in V_{(3,1)} \,. 
	\ee
This result, which is in agreement with \cite{Freidel:2021qpz, Freidel:2021dfs},  ensures that the integrand $(Tm+Y^Aj_A)$ of the pairing belongs to $V_{(2,0)}$, as $T\in V_{(-1,0)}$ and $Y\in V_{(-1,-1)}$. In other words, it is a scalar density that can be integrated over the sphere and it leads to a  quantity invariant under diffeomorphism.

\paragraph{The $\gbms$ charge algebra.} It is well-known that the dual of a Lie algebra is endowed with a canonical Poisson bracket \cite{Berezin196704,Kirillov197608,BayenFlatoFronsdalLichnerowiczSternheimer197803,Weinstein198301} (see Appendix \ref{sec:moment maps and coadjoint orbits}). Due to the existence of this Lie--Poisson structure on $\gbms^*$, we can define the change of an object on $\gbms^*$ using the charges \eqref{eq:gbms charges}. This provides the (classical) analog of \eqref{eq:variation defined by Lie bracket}. For any function $\mscr{O}\in C^\infty(\gbms^*)$, the coadjoint action of $\gbms$ on $\gbms^*$ is given by
\begin{equation} \label{eq:gbms charge action}
    \delta_{(Y,T)} \mscr{O} := \{\mscr{O} , \mathsf{P}_T + \mathsf{J}_{Y}\}_{\g^*}.
\end{equation}
In particular, applying this formula to linear functions of the charges, we can read-off the $\gbms$ charge algebra readily from \eqref{eq:coadjoint action on gbms*}
\begin{eqaligned}\label{eq:gbms charge algebra}
        \{\mathsf{J}_{Y_1}, \mathsf{J}_{Y_2}\}_{\g^*}(j,m)&= \mathsf{J}_{ [Y_1, Y_2]_S},
        \\
        \{\mathsf{J}_{Y}, \mathsf{P}_{T}\}_{\g^*}(j,m)&= \mathsf{P}_{(Y[T]- T W_{Y})},
    \\
    \{\mathsf{P}_{T_1}, \mathsf{P}_{T_2}\}_{\g^*}(j,m)&=0.
\end{eqaligned}
Note that the $\gbms$ algebra is analogous to the Poincar\'e algebra \eqref{eq:poincare Lie algebra} where  $\mathsf{J}_Y$ and $\mathsf{P}_T$ are the analogs of $\mbs{J}_{\mu\nu}$ and $\mbs{P}_\mu$, respectively.
One sees that the supertranslation generators $\msf{P}_T$ commute, while the action of a super-angular momentum $\msf{J}_Y$ on a supermomentum is a supermomentum with parameter $\delta_YT= Y[T]-TW_Y$. This is the standard transformation for a weight $-1$ scalar that we already encountered.
The same commutation relation can also be interpreted as the fact that the super-angular momentum is not invariant under supertranslation. This fact arises due to the semi-direct sum structure of the $\gbms$ algebra.

\smallskip As a particular instance of \eqref{eq:gbms charge algebra}, we can consider the linear Poisson brackets of the coordinates on $\gbms^*$. They are given by (see Appendix \ref{sec:proofs in section 3})
\begin{eqgathered}\label{eq:poisson brackets of coordinates on gbms*}
    \begin{aligned}
    \{j_A(\sigma),j_B(\sigma')\}_{\g^*}&=j_A(\sigma')\pa_B\delta^{(2)}(\sigma-\sigma')-j_B(\sigma)\partial'_A\delta^{(2)}(\sigma-\sigma'),
    \\
    \{j_A(\sigma),m(\sigma')\}_{\g^*}&=\frac{m(\sigma')}{2}\pa_A\delta^{(2)}(\sigma-\sigma')-m(\sigma)\pa'_A\delta^{(2)}(\sigma-\sigma').
    \end{aligned}
    \\ 
    \{m(\sigma'),m(\sigma')\}_{\g^*}=0.
\end{eqgathered}

\subsection{$\gbms$: Supertranslation  Goldstone mode}\label{sec:gbms goldstone modes}

Before defining the spin generator for $\gbms$ in \S \ref{sec:algebraic aspects of gbms} by introducing the $\gbms$ isotropy algebra, let us revisit in our context the previous approaches in \cite{Compere:2019gft, ChenWangWangYau202107,  ChenWangWangYau202103, Javadinezhad:2022hhl} to the angular momentum problem in asymptotically-flat spacetimes based on the original proposal of  \cite{Moreschi1986, Moreschi:2002ii} and within the standard $\bms$ framework.

\smallskip As mentioned in \S \ref{sec:issues with the bms algebra}, the idea pursued in previous references to introduce the notion of a supertranslation invariant intrinsic angular momentum
\eqref{SYbms} requires the introduction of an electric supertranslation Goldstone, whose transformation properties are given by
\be 
\d_TG=T, \qquad \d_Y G=(\cL_Y-W_Y)G\,.
\ee
Let us consider that case where $(Y,T)$ are vector fields generating $\gbms$ transformations. By definition, the shear associated to this Goldstone is purely electric and given by 
\be 
C_{ AB}=-2\cD_{\langle A} \cD_{B\rangle} G.
\ee 
Given a Goldstone $G$, we can introduce the orbital angular momentum generator \cite{Compere:2019gft}
\be 
\mathsf{L}_Y:= \mathsf{P}_{Y[G]-W_Y  G} 
= \frac12 \bigintsss_S Y^A ( 3 m \p_A G + G \p_Am)\,,
\ee 
and the intrinsic angular momentum is then defined as the difference between the total angular momentum and the orbital one as 
\be\la{SY}
\mathsf{I}_Y := \mathsf{J}_Y - \mathsf{L}_Y\,.
\ee
It is direct to check that  this operator is translation invariant
\be 
\delta_T\mathsf{I}_Y=
\{ \mathsf{I}_Y , \mathsf{P}_T \} = 0.
\ee
In order to verify that this object has the desired properties, we need to check it satisfies the right algebra. 
 In Appendix \ref{sec:proofs in section 3}, we verify first that the orbital momentum generator indeed transforms as a vector  and it satisfies the $\diff(S)$ algebra for  $\gbms$ vector fields $Y, Y'$, namely
\be\la{LLJ}
\{\mathsf{L}_{Y},\mathsf{J}_{Y'}\}_{\g^*}&= \mathsf{L}_{[Y,Y']_S}\,,
\cr
\{\mathsf{L}_{Y},\mathsf{L}_{Y'}\}_{\g^*}&= \mathsf{L}_{[Y,Y']_S}\,.
\ee
Given the first bracket in \eqref{eq:gbms charge algebra}, this implies that also the intrinsic angular momentum generator
represents the  $\diff(S)$ algebra through the $\gbms^*$ linear Poisson structure (see Appendix \ref{sec:proofs in section 3})
\be\la{SSS}
\{\msf{I}_{Y},\msf{I}_{Y'}\}_{\g^*}&= \msf{I}_{[Y,Y']_S}\,.
\ee
As we show in \S \ref{sec:poincare embedding}, restricting to the BMS case as in \cite{Compere:2019gft}, where the vector fields $Y, Y'$ correspond to conformal Killing vectors, the  Lie bracket of vector fields on $S$   reduces to the Lorentz Lie algebra $\msf{sl}(2,\C)$.
The validity of the algebra \eqref{LLJ} and \eqref{SSS} form a particularly interesting result which, up to our knowledge, was not derived before.

\subsection{$\gbms$: Poincar\'e embeddings}\label{sec:poincare embedding}

As Poincar\'e algebra is embedded in $\gbms$, a natural task is to identify the charges generating the action of this algebra on $\gbms^*$. The aim of this section is to construct the Poincar\'e charge algebra by first identifying the generators of this subalgebra, constructing the associated charges, and finally computing their brackets. 

\paragraph{Symmetry-breaking pattern.} We would like to show that the Poincar\'e embedding follows a symmetry-breaking pattern. More precisely, given  a choice of round sphere metric\footnote{We recall that $q$ is compatible with the given volume form that defines $\gbms=\gbms_\epsilon$} $q_{AB}$ and angular momentum aspect $j_A$, we can uniquely identify a $\bms_{(j,q)}$
 subalgebra of $\gbms$. As the expressions \eqref{eq:the charge for lorentz generator in gbms} and \eqref{eq:analog of four-momentum generator for gbms} demonstrate, the choice of subalgebra generators depends on $(j_A,q_{AB})$. Similarly, the data $(m,j_A,q_{AB})$  determine a choice of Poincar\'e subalgebra inside $\gbms$. In summary, we have the embeddings
\be 
\mathsf{Poinc}_{(m,j,q)} \subset \bms_{(j,q)} \subset \gbms.
\ee 
The embedding of the $\bms$ and Poincar\'e subalgebras determines an equivalence class. Understanding  the equivalence relation requires introducing  the notion of \emph{condensate}\footnote{In condensed matter the condensate operators are all the operations that commute with the Goldstone operator \cite{Colangelo:2000dp, Burnell:2017otf}.}  modes 
$(C_0,C_1^{AB})$, where $C_{0}$ is a scalar of dimension $-1$ while $C_{1}^{AB}$ is symmetric traceless tensor of dimension $0$. Given $C_0$ we denote
\begin{equation}
    C_0^{AB} :=-2 D^{\langle A} D^{B\rangle}C_0,
\end{equation}
its symmetric traceless image  of dimension $+1$ --- we use the notation  $\langle AB\cdots \rangle$ to denote the symmetric traceless components.
The equivalence relation is then described as follows. We say that  
$(m,j)\sim (m',j')$ if there exists $(C_0,C_1^{AB})$ such that 
\be 
m'= m + D_{\langle A }D_{B \rangle } C_0^{AB}\,,
\qquad 
j_A' = j_A + D_{\langle A } D_B D_{C \rangle  } C_1^{BC}\,.
\ee
As we will see in  \S\ref{sec:reference frames and goldstone modes},
$C_0$ represents the  supertranslation   condensate while $C_1^{AB}$ represents a super-Lorentz condensate.
If $(m,j)\sim (m',j')$ then the Poincar\'e subalgebra is unchanged
\be 
\mathsf{Poinc}_{(m',j',q)}=\mathsf{Poinc}_{(m,j,q)}, \qquad
\bms_{(j',q)}=\bms_{(j,q)}.
\ee
An important point about the condensate is that it can be understood as a physical entity dual (or canonically conjugated) to the Goldstone modes. 
Let us illustrate this for supertranslations where the Goldstone operator is denoted by $G$ and defined in \S\ref{sec:gbms goldstone modes}. 
The action of the Golstone operator on $m$ modifies the condensate. The integrated Goldstone operator associated with the label $c_0$, is defined as
\be 
G_{c_0}:= \bigintsss_S c_0^{AB} D_{\langle A }D_{B \rangle } G\,.
\ee 
The canonical action of this operator changes the  supertranslation condensates
\be 
\{ G_{c_0}, m\}_{\g^*} = D_{\langle A }D_{B \rangle } c_0^{AB}. 
\ee 
This corresponds to the shift $C_0\to C_0+c_0$.
At the quantum level this means that quantum states $|p\rangle$ and $ e^{i\wh{G}_{c_0}} |p\rangle$ carry the same value of the broken symmetry charges $\wh{p}^\mu$, but correspond to different condensates shifted by $c_0$.

\smallskip The analysis presented here shows that the condensate  modes $C_0$  and $C_1^{AB}$ represent respectively  the geometry of the quotient spaces
\be 
 \bms/ \Poinc, \qquad \gbms/ \bms.
\ee
The total quotient group $\gbms/ \mathsf{Poinc}$  labels the different Poincar\'e vacua inside $\gbms$. The perspective connecting the condensate operators $(C_0,C_1^{AB})$ to a  symmetry breaking mechanism has been developed by Kapec et al. in \cite{Kapec:2017tkm, Kapec:2021eug, Kapec:2022axw,Kapec:2022hih}. As we will see in \S\ref{sec:reference frames and goldstone modes}, there is a deep connection between the condensates and the soft factors appearing in the soft theorems.

\subsubsection{Poincar\'e embedding inside $\gbms$} \la{sec:Poimc-em}

With this conceptual characterization of the embedding pattern in mind, let us analyze its realization in more detail. In this section, we provide an explicit construction of the Poincar\'e generators. 
To achieve this goal, a first natural question is whether the restriction to a particular round metric $\mathring{q}$ is necessary, and what happens when we change from the given round sphere metric to another one $q=F^*(\mathring{q})$. Moreover we also investigate whether choosing a round sphere metric is necessary at all to define Poincar\'e generators.  We find that any choice of metric is admissible.
This question arises since, let us recall, in order to define the GBMS group, we only introduced an area form, but no metric was needed (see \S\ref{sec:gbms algebra and bracket}).

\smallskip More precisely, we  can define  the Poincar\'e subalgebra of $\gbms$ associated to any given non-round sphere metric $\gamma$, and this corresponds to the same Poincar\'e subalgebra one can associate to a general round sphere metric $q=F^*(\mathring{q})$,
with $F:S\to S$ an orientation preserving diffeomorphism of the sphere, satisfying $R(q)=2$ but  not associated with a specific choice of spherical coordinates. In fact, the condition $R(q)=2$ is invariant under diff$(S)$, so we can extend the standard BMS symmetry group to include a more general choice of sphere coordinates. The diffeomorphism $F$ labels different embeddings of $\bms$ into $\gbms$.
While this last fact might be well known by the experts in the field, we have not seen it explained in some detail anywhere; that is why, in our attempt to be as pedagogical as possible, we are detailing in the following a clear explanation of this point.\footnote{{The relevance of this discussion was also revealed to us in discussion with A. Rignon-Bret and S. Speziale  (see also \cite{pirsa_PIRSA:23110062}).}}

\smallskip Therefore, we will proceed in two steps. We are first going to show that $\bms_\gamma$ depends only on the conformal equivalence class of $\gamma$ and is therefore given by $\bms_q$,
 for $q$ a round sphere metric in that class. This ensures that the Lorentz sectors associated with the two metrics match; we then introduce a definition of translation generators invariant under conformal rescalings, which guarantees that $\Poinc_\gamma=\Poinc_{q}$. In a second step, we explicitly construct the generators of $\Poinc_{q}$, by parametrizing a general round sphere metric $q=F^*(\mathring{q})$ in terms of null vectors defining an embedding of the 2-sphere into Minkowski space.

\paragraph{$\gbms$  and  sphere metrics.}
Let us start with some preliminaries on non-round sphere metrics.  Given a metric $\gamma_{AB}$ on the sphere we denote $\gamma  =\det(\gamma_{AB})$ its area form. We also denote $\cD_A$ its covariant derivative, $R(\gamma)$ its curvature tensor and $T_{AB}(\gamma)$ its stress tensor. $T_{AB}(\gamma)$ is defined as the unique symmetric traceless tensor that satisfies
\be 
\cD_B T_{A}{}^B(\gamma) + \frac12 \pa_A R(\gamma)=0
\ee 
This tensor appears as the traceless component of the Geroch tensor \cite{Geroch1977, Compere:2018ylh, Freidel:2021qpz}.\footnote{ The Geroch tensor is $\rho_{AB} = \frac{q_{AB}}{2} R - T_{AB}$.}
The $\gbms$ condition \eqref{dq0} imposes  that 
\be
\sqrt\gamma= {\mathring{\epsilon}}
\ee
where $\mathring\epsilon$ is the  given area form that defines $\gbms=\gbms_{\mathring\epsilon}$.
Given such $\gamma_{AB}$,
there exists a round sphere metric $q_{AB}$ and a conformal factor $\phi$ such  that 
\be \label{confrescaling}
\gamma_{AB} = e^{-2\phi} q_{AB}, \qquad e^{2\phi} =\frac{ \sqrt{\det q}}{\mathring{\epsilon} }, \qquad R(q)=2.
\ee 
The second equality shows that $\phi=\phi_q$ is uniquely characterized by the choice of round sphere metric $q$.
Since the stress tensor of a round sphere metric vanishes, the conformal   rescaling for the curvature and stress tensor implies that\footnote{
If $q$ is an arbitrary metric we have more generally that (see App. \ref{App:conf})
\be 
T_{AB}( e^{-2\phi} q) = T_{AB}(q) - 2 e^{-\phi} D_{\langle A} D_{B\rangle} e^\phi, 
\qquad
R(e^{-2\phi} q) = e^{2\phi}( R(q) + 2 \Delta \phi).
\ee
}
\be 
T_{AB}( e^{-2\phi} q) = - 2 e^{-\phi} D_{\langle A} D_{B\rangle} e^\phi, 
\qquad
R(e^{-2\phi} q) = 2 e^{2\phi}( 1 + \Delta \phi).
\ee 
The action \eqref{dgamma} of Diff$(S)$ on $\gamma_{AB}$ is extended to the pair $(\phi, q_{AB})$ as 
\be \label{diffactionq}
\delta_Y \phi = Y[\phi] - W_Y, \qquad 
\delta_Y q_{AB} = \cL_Y  q_{AB}.
\ee
This implies the following transformations for the stress tensor
\be 
\delta_Y T_{AB}(\gamma) = \cL_Y T_{AB} + \cD_{\langle A} \cD_{B\rangle} W_Y. 
\ee

Note that the condition \eqref{confrescaling}  does not uniquely determine the metric $q_{AB}$. A rescaling $q_{AB}\to e^{-2\varphi} q_{AB}$ where $\varphi $ is solution of the equations
\be\label{eqcond}
D_{\langle A} D_{B\rangle} e^\varphi=0, \quad 
  (1+ \Delta \varphi)=e^{-2\varphi},
\ee  
with $D_A$  the round sphere covariant derivative,
is admissible.
A way to fix this ambiguity and achieve \eqref{diffactionq} is to choose a reference round sphere metric and parametrise $\gamma $ in terms of $F \in \mathrm{Diff}(S)$ by demanding that $ \gamma_{AB} = e^{{-2} \phi} F^*(\mathring{q}_{AB})$. This amounts to choosing $\varphi_{\mathring{q}}=0$. 

\paragraph{What is $\mathsf{Poinc}_\gamma$?}
In order to answer this question, we first provide the general definition of  $\mathsf{Poinc}_\gamma$ as 
the set of translations and Lorentz transformations generated by the element $T\in V_{(-1,0)}$ and $Y\in V_{(-1,-1)}$ solutions of  
\be\la{Poin-g-def}
\cD_{ A}(\gamma_{B C} Y^C)  
+ \cD_{ B}(\gamma_{A  C} Y^C)
= \gamma_{AB} (\cD_C Y^C),
\qquad
\left(\cD_{\langle A} \cD_{B\rangle} + \frac12 T_{AB}(\gamma)\right) T
= 0\,.
\ee
As we are going to see in more detail below, this space is 4-dimensional \cite{Geroch1977} and it forms a Poincar\'e subalgebra of $\gbms$. In particular, given
$(T_1,Y_1) ,(T_2,Y_2) \in \mathsf{Poinc}_\gamma$ then $[(T_1,Y_1) ,(T_2,Y_2)]_{\g}\in \mathsf{Poinc}_\gamma$.

We now want to use this definition \eqref{Poin-g-def} to relate $\Poinc_\gamma$ to $\Poinc_{q}$.
First of all, it is easy to see that if $Y$ is a conformal Killing vector (CKV) of $q_{AB}$, solution of $ \cL_Y q_{AB} =  (D_C Y^C) q_{AB}$, it is a CKV of $\gamma_{AB}$ with a shifted Weyl factor. More precisely, from the  definition of the divergence 
we have
\be\la{cDY}
\cD_C Y^C&=\f1{\sqrt{\gamma}}\pa_C( e^{-2\phi }\sqrt{q} Y^C)=D_CY^C-2Y^C \p_C \phi \,.
\ee
From this, we obtain
\be
\cL_Y \gamma_{AB}
= \cL_Y( e^{-2\phi} q_{AB})
= e^{-2\phi} q_{AB}( D_C Y^C - 2Y[\phi])
= \gamma_{AB} \cD_C Y^C\,.
\ee

Hence, while the action of $\gbms$ maps the reference round sphere metric $\mathring{q}$ onto  $ \gamma_{AB} = e^{{-2} \phi} F^*(\mathring{q}_{AB})$, we have that $\bms_\gamma=\bms_{F^*\mathring{q}}$. This way, we can label the $\bms$ subgroups of $\gbms$ by the space of round sphere metrics $q=F^*(\mathring{q})$.
This means that, for a given round sphere metric $q_{AB}$, the  generators of the Lorentz subalgebra of  $\bms_q$ identify as well the Lorentz embedding in $\gbms$.

\smallskip Next, in terms of the metric $q=F^*(\mathring{q})$, the second condition in \eqref{Poin-g-def} is given by
\be \la{DDT}
\left(\cD_{\langle A} \cD_{B\rangle} + \frac12 T_{AB}(\gamma)\right) T
=e^{-\phi} D_{\langle A} D_{B\rangle} (e^\phi T)=0\,,
\ee
The first equality is shown in App. \ref{App:conf} where we used $T_{AB}(q)=0$.
Therefore, the   translation sectors associated to the two metrics coincide as well  and thus 
\be
\Poinc_\gamma=\Poinc_{q}= 
F(\Poinc_{\mathring{q}})\ee
under the map $(Y,T)\to (Y, e^\phi T)$.
From this we see that the space of solution is 4-dimensional: A general solution is labelled by four contants $X_\mu$ and  reads 
\be 
T = e^{-\phi} X_\mu (\mathring{n}^\mu\circ F)
\ee where $\mathring{n}^0=1$ and $\mathring{n}^i $ are the component of the unit  vector 
 \be\la{nring}
 \mathring{n}^i:=(\sin{\theta}\cos{\varphi},  \sin{\theta}\sin{\varphi}, \cos{\theta})\,.
 \ee

\paragraph{Spherical metrics and vectors.}

Having identified the Poincar\'e algebra embedded in $\gbms$ with $\Poinc_{q}$, we want to provide next an explicit construction of its generators. To do so it will prove convenient to parametrized the round metric metric $q$ 
in terms of its embedding 
  in 3D Euclidean space given, for a set of two intrinsic coordinates $\sigma^A$ on the sphere, by the map
\be \label{Fmap}
\iota: S \to \mathbb{R}^3,\quad  \sigma^A \to n^i(\sigma), \qquad n^i n_i =1\,.
\ee
Such an embedding can be related to the canonical embedding $\mathring{\iota}$ with image $\mathring{n}^i$ by a diffeomorphism $F= \mathring{\iota}^{-1}\circ \iota$. This implies
 that 
 \be
 n^i = R^i{}_j (\mathring{n}^j\circ F).
\ee
where $R$ is a rotation matrix.

We can then extend the embedding to Minkowski space introducing a pair of null vectors $n, \bar n$ such that
\be\la{nn}
n\cdot n =0, \qquad n \cdot \bar{n}=-2,\qquad \bar{n}\cdot\bar{n}=0\,;
\ee 
these can be written as   
$
n^\mu := \alpha (1,n^i)$ and  
$\bar{n}^\mu:= \alpha^{-1}(1,-n^i)$.
Without loss of generality, we can take $\alpha=1$ so that 
\be\la{pnA}
\pa_A( n+ \bar{n})=0\,.
\ee
 A general embedding is then defined in terms of the two space-like embedding frame fields \cite{Riva:2023xxm} $e_A{}^\mu$ and a 2-sphere metric given by 
\be\la{eAmu}
e^\mu_A:=\p_A n^\mu \,,\qquad 
q_{AB} := e^\mu_A e^\nu_B \eta_{\mu\nu}\,.
\ee
The two frames $e_A$ form,  together with the two null vectors,  a basis of Minkowski space. In particular, 
$
n_\mu e^\mu_A=0=\bar n_\mu e^\mu_A$,
and the inverse 2-sphere  metric is mapped to the rank two symmetric tensor
\be\la{q-ind}
q^{\mu\nu}
= e_A^\mu e_B^\nu q^{AB}
= \eta^{\mu\nu}+\f12(n^\mu \bar n^\nu+\bar n^\mu  n^\nu)\,.
\ee 
From the determinant definition, we obtain the identity
\be\la{epsnn}
\epsilon_q^{AB} e^\mu_A e^\nu_B =
\f12\varepsilon^{\mu\nu}{}_{\rho \sigma} \bar n^\rho n^\sigma\,,
\ee
where $\epsilon_q^{AB}=q^{-\f12} \varepsilon^{AB}$ is the Levi--Civita tensor.

\smallskip 
In Appendix \ref{sec:proofs in section 3} we show that the frame field satisfies the following key  compatibility condition 
\be\la{cDe}
D_B e^\mu_A=\f12(\bar n^\mu - n^\mu) q_{AB}\,,
\ee
from which we extract that 
\be\la{DDn}
D^A D_A n^\mu=\bar n^\mu - n^\mu\,,\quad 
D_{\langle A}D_{B\rangle} n^\mu =0\,.
\ee
Furthermore, \eqref{epsnn} together with \eqref{cDe} imply the identity
\be\la{ndn}
n^\mu \p_A n^\nu -n^\nu \p_A n^\mu
=\varepsilon^{\mu\nu}{}_{\rho \sigma}  n^\rho  \epsilon_A{}^B \p_B n^\sigma\,.
\ee
We can think of these relations as a connection between intrinsic coordinates $\sigma^A$ and embedding coordinates $n^\mu$ for a general {round} sphere metric. We also see that $n^\mu$ is by construction, a basis of solution of \eqref{DDT} for the metric $q$ in \eqref{eAmu}.

One can directly check that  the 2-sphere metric \eqref{eAmu} has scalar curvature $R(q)=2$. To do so, we can apply the general identity for a 2D metric
\be
[D_A, D_B] V^A=\f R2 V_B
\ee
to the 2D vectors $q^{AC} e^\mu_C$; this yields
\be
\f {R(q)}2 e^\mu_B&=q^{AC}[D_A, D_B] e^\mu_C
=\f{q^{AC}}2\left[D_A (\bar n^\mu - n^\mu)q_{BC}-D_B (\bar n^\mu - n^\mu)q_{AC} \right]
\cr
&=-\f12\p_B (\bar n^\mu - n^\mu) =e^\mu_B\,,
\ee
from which we get that 
$
R(q)=2$.

\subsubsection{Lorentz and Poincar\'e charge algebras}\la{sec:Poi-charges}

We now write down the charges which generate the coadjoint action of a Lorentz and Poincar\'e   subalgebras of $\gbms$ on $\gbms^*$. 

\paragraph{The Lorentz charge algebra.} To define the generator of a Lorentz subalgebra $\so(3,1)$ of $\gbms$, we use the null vectors 
\be 
n^\mu = (1,n^i), \qquad 
\bar{n}^\mu=(1,-n^i)\,.
\ee 
These satisfy the scalar products \eqref{nn}.
The identities \eqref{DDn} for the unit vectors become the differential qualities
\be \label{keydiff}
\Delta n^i =-2 n^i\,,\qquad
D_{\langle A}D_{B\rangle} n^i =0\,,
\ee 
where $\Delta=D_AD^A$ denotes the sphere Laplacian associated with the round metric $q$.
These express the fact that $n^i$ spans the $\ell=1$ spherical harmonics.
Similarly, the identities \eqref{q-ind}, \eqref{epsnn},  \eqref{ndn} for the null vector spatial components yield respectively
\begin{eqgathered}\la{nid}
q^{AB}\pa_A n^i \pa_B n^j
=\eta^{ij}-n^i n^j\,,\qquad 
\epsilon_q^{AB}\pa_A n^i \pa_Bn^j=\varepsilon^{ij}\!_k n^k,
\\
n^i\pa_A n^j- n^j\pa_A n^i = \varepsilon^{ij}\!_k \epsilon_A{}^{B}\pa_B n^k.
\end{eqgathered}
The vector field associated with a Lorentz transformation $\Lambda$ is  given by \cite{Flanagan:2015pxa,Riva:2023xxm} 
\be 
Y^A_\Lambda &=-\Lambda^{\mu \nu}  {n}_\mu \pa^A \bar n_\nu 
= ( \Lambda^{k} q^{AB}
+ \widetilde{\Lambda}^k \epsilon_q^{AB}) {\pa}_B n_k
\ee  where $\Lambda^k:= \Lambda^{0k}$ and $\widetilde\Lambda^k := \frac12  \Lambda^{ij} \varepsilon_{ij}{}^k$ denote the boost and rotation components respectively. This means that the vector fields  generating the rotation and boost  are given by \cite{Barnich:2011mi,Flanagan:2015pxa, Compere:2019gft}
\begin{eqaligned}\label{eq:conformal killing vectors on sphere}
    \widetilde{Y}^A_{i}&:=\epsilon_q^{AB}\pa_B n_i, \qquad 
    Y^A_{i}&:=  q^{AB}\pa_B n_i.
\end{eqaligned}
These vectors satisfy the conformal Killing property \eqref{confinv}.
Their divergences are
\be \la{ndiv}
 D_A \widetilde{Y}^A_{i}=0, \qquad 
 D_A Y^A_{i}= - 2 n_i\,.
\ee
Hence, by substituting\footnote{The definition given here is  consistent with the definition $\mbs{K}_i:= \mbs{J}_{0i}$}  $Y^A_{\mu\nu} =- {n}_{[\mu} \pa^A \bar n_{\nu]}$  into the charge $\mathsf{J}_Y$ in \eqref{eq:gbms charges}, we define the charge generating the coadjoint action of a Lorentz subalgebra as
\be\label{eq:the charge for lorentz generator in gbms}
 J_{\mu\nu}:= \bigintsss_S Y_{\mu\nu}^A\, j_A\,\bfe\,.
\ee
The  rotation $J_i$ and boost $K_i$ generators are therefore  given by substituting \eqref{eq:conformal killing vectors on sphere} in 
\eqref{eq:the charge for lorentz generator in gbms}
\be \la{Lor-gen}
J_i:= \bigintsss_S j_A \epsilon_q^{AB} \pa_B n_i \, \bfe,
\qquad
K_i:= \bigintsss_S j_A q^{AB} \pa_B n_i \,  \bfe.
\ee
These generators satisfy the Lorentz charge algebra (see Appendix \ref{app:Lor} for the derivation)
\begin{eqaligned}\la{Lor-alg}
    \{J_i, J_j\}_{\g^*}&=-\varepsilon_{ij}\!^k J_k,
    \\
    \{J_i,K_j\}_{\g^*}&=-\varepsilon_{ij}\!^k K_k,
    \\
    \{K_i,K_j\}_{\g^*}&=+\varepsilon_{ij}\!^k J_k.
\end{eqaligned}
The meaning of these relations is that $(J_i,K_i)$ are the charges generating the coadjoint action of a Lorentz subalgebra of $\gbms$.\footnote{The coadjoint action of the full $\gbms$ charge algebra is \eqref{eq:gbms charge algebra} for the charges \eqref{eq:gbms charges}. \eqref{Lor-gen} is the specialization of those charges for particular choices, given in \eqref{eq:conformal killing vectors on sphere}, of the vector field $Y^A$ in \eqref{eq:gbms charges}
\begin{equation*}
    J_i=\mathsf{J}_{\widetilde{Y}_{i}}, \qquad K_i=\mathsf{J}_{Y_{i}},\qquad J_{\mu\nu} = \mathsf{J}_{Y_{\mu\nu}}.
\end{equation*}
Since the vector fields in \eqref{eq:conformal killing vectors on sphere} generate the Lorentz algebra $\msf{so}(3,1)$, \eqref{Lor-gen} are generating the coadjoint action of $\msf{so}(3,1)$ on $\gbms^*$.} 

\paragraph{The Poincar\'e charge algebra.}
The analog of four-momentum $P_\mu:=(E,{P}_i)$ is given by
\begin{equation}\label{eq:analog of four-momentum generator for gbms}
    P_\mu:=\bigintsss_S n_\mu\,m\,\bfe,
\end{equation}
satisfying (which is a trivial consequence of $\d_T m=0$ \eqref{eq:coadjoint action on gbms*} or equivalently the last bracket in \eqref{eq:poisson brackets of coordinates on gbms*})
\begin{equation}\label{eq:poisson bracket of momentum, gbms}
    \{P_\mu,P_\nu\}_{\g^*}=0.
\end{equation} 
Furthermore, we have (see Appendix \ref{app:Lor})
\begin{equation}\label{eq:poisson bracket of momentum and angular momentum, gbms}
    \{P_\mu,J_{\nu\rho}\}_{\g^*}=\eta_{\mu\nu}P_\rho-\eta_{\mu\rho}P_\nu\,.
\end{equation}
Taking into account \eqref{eq:poisson bracket of momentum, gbms} and \eqref{eq:poisson bracket of momentum and angular momentum, gbms} enhances the Lorentz charge algebra \eqref{Lor-alg} to that of Poincar\'e algebra $\iso(3,1)$. These are the charges that generate the coadjoint action of $\iso(3,1)$ on $\gbms^*$. It should be clear by now that the charges associated with different round sphere metrics corresponds to different Poincar\'e generators inside GBMS labelled by the vector $n^\mu =\mathring{n}^\mu \circ F$.

\smallskip Having obtained the Poincar\'e charges and their algebra for a Poincar\'e embedding inside $\gbms$, the natural next step is to define the notion of Pauli--Luba\'nski generator (hence the spin) for such an embedding. We take up this task in \S\ref{sec:bms pauli-lubanski} after providing the ground in \S\ref{sec:spin generator of gbms} and \S\ref{sec:(m,j) in general lorentz frame for stationary spacetimes}. %

\section{$\gbms$ coadjoint orbits and  spin generator}\la{sec:algebraic aspects of gbms}

In \S\ref{sec:gbms coadjoint action and charge algebra}, we derived the coadjoint action of $\gbms$ (see \eqref{eq:coadjoint action on gbms*}). The natural next step  is studying the orbits of this action, which we do in \S\ref{sec:gbms coadjoint orbits} and it involves the construction of the invariants of a typical coadjoint orbit. A crucial role in the construction of these invariants is played by the notion of vorticity. Using this notion, and under the hypothesis that the mass aspect is positive, we then construct the generator of the isotropy subalgebra of $\gbms$ for a generic $m\in(\mbb{R}_{-1})^*$, which furthermore turns out to be supertranslation-invariant. Hence, in \S\ref{sec:spin generator of gbms} we naturally identify it as the spin generator for $\gbms$.

\subsection{$\gbms$: Coadjoint orbits}\label{sec:gbms coadjoint orbits}

Having the coadjoint action \eqref{eq:coadjoint action on gbms*} of $\gbms$ at hand, we can study its coadjoint orbits. The coadjoint orbits are labelled by    $\gbms^*$ Casimirs. These are the functionals of $(j,m)$ which are invariant under the coadjoint action \eqref{eq:coadjoint action on gbms*} and therefore  represent the orbit invariants. When quantized, they provide labels for the representations of the quantum algebra.

\smallskip The first remark is that the structure of the algebra $\gbms$ is similar to the symmetry algebra of 2-dimensional barotropic fluids \cite{MarsdenRatiuWeinstein198401,KhesinChekanov198901,ArnoldKhesin1999,Morrison199802}
\begin{equation}\label{eq:hydrodynamical algebra}
    \tsf{h}:= \diff(S) \sds  \mbb{R}_{0}^S\,.
\end{equation}
The key difference is that the conformal dimension of the normal factor $\mbb{R}^S_0$ is $0$ (unlike the case of $\gbms$, which is $-1$). The variables parametrizing the dual Lie algebra $\tsf{h}^*$ are the fluid  density $\rho\in V_{(2,0)}$ and the fluid densitized momentum
$p_A\in V_{(1,1)}$.
The Casimirs for this algebra have been constructed explicitly \cite{DonnellyFreidelMoosavianSperanza202012}. They are given by the {\it enstrophies} which are moments of the vorticity, defined by the two-form $\mbs{w}_{\va \tenofo{Fluid}}:=\rd p$, where $p=p_A\rd\sigma^A$.

\paragraph{The vorticity for $\gbms$.} To exploit these results, the strategy is  to construct quantities in $\gbms^*$ that behave as $\rho$ and $p_A$. We notice that (1) $m$ has conformal dimension $+3$ and hence $m^{2/3}$ (like $\rho$) is in  $V_{(2,0)}$; and (2) $j_A$ has conformal dimension $+3$ and hence $ j_A/ m^{2/3}$ (like $p_A$) belongs to $V_{(1,1)}$. Hence, we identify the fluid-type variables $(\rho,p_A)$ for $\gbms$ as
\be\label{eq:relation between fluid and gravity coadjoint elements}
\rho:= m^{\frac23}, \qquad p_A := \rho^{-1} j_A. 
\ee
This map can be inverted
\begin{equation}\label{eq:inverse relatin between fluid and gravity coadjoint elements}
    m=\rho^{3/2}, \qquad j_A=\rho\,p_A.
\end{equation}
For this to hold, we have to assume that $m>0$,\footnote{Since we are taking a square root and $\rho$ is always positive. It would be interesting to investigate the consequences of zero and negative mass aspects for our analysis.} i.e. the mass aspect $m=m(\sigma)$ is a strictly positive function on $S$. From now on, we make this assumption.

\smallskip Using the procedure developed in \cite{DonnellyFreidelMoosavianSperanza202012}, we need to construct the analogue of vorticity in the case of $\gbms$. This can be achieved as follows. From the transformations \eqref{eq:coadjoint action on gbms*}, we can easily see that $\rho$ is (1) invariant under supertranslations and (2) is a density under $\diff(S)$ 
\be\label{eq:rho transformation under gbms}
\delta_T \rho=0\,, \qquad 
\delta_Y\rho =  \cD_A(\rho  Y^A)\,,
\ee 
where $\delta_T:=\delta_{(0,T)} $ and $\delta_{Y} := \delta_{(Y,0)}$.
On the other hand, $p_A$ transforms as a one form under diffeomorphism and in a simple manner under supertranslation (see Appendix \ref{sec:proofs in section 4})
\be\label{eq:p transformation under gbms}
\delta_T p_A=\frac32 \pa_A ( \sqrt{\rho} T), \qquad 
\delta_Yp_A  = \cL_Y p_A. 
\ee
Note that $\sqrt{\rho} T \in V_{(0,0)}$ is a dimensionless scalar. To define the vorticity, let us recall that $\epsilon_{AB}$ is the tensor that defines the sphere measure
	\begin{equation}\label{eq:the relation between measure and volume form in GBMS}
	    \bfe= \sqrt{\gamma}\rd^2\sigma = \frac12 \epsilon_{AB} \rd \s^A\wedge \rd \s^B,
	\end{equation}
	 where the right-hand side denotes the 2-form on $S$ corresponding to the normalized density $\bfe$ \eqref{eq:area-form preserved by gbms}.\footnote{Since $S$ is an orientable manifold, $1$-densities, as sections of the density bundle on $S$, can be canonically identified with non-zero $2$-forms, as sections of the second exterior power of $T^*S$. 
We assume that this identification between densities and two-forms is understood in the rest of the paper.} The inverse of $\epsilon_{AB}$  is denoted $\epsilon^{AB}$. It satisfies $\epsilon^{AC} \epsilon_{BC} =\delta^A{}_B$ and it can be obtained by raising indices  $\epsilon^{AB}=\gamma^{AA'}\gamma^{BB'} \epsilon_{A'B'}$ with respect to the metric $\gamma$, which is compatible with $\bfe$ in \eqref{eq:the relation between measure and volume form in GBMS}. We can now define the vorticity as
\be \label{vortdef}
	w:= \rho^{-1}\epsilon^{AB} \pa_A p_B 
	= m^{-\frac73 }\left( m\epsilon^{AB} \pa_A j_B -\frac23  \epsilon^{AB} \pa_A m  \, j_B\right) \,.
 \ee
From this definition, together with \eqref{eq:rho transformation under gbms} and \eqref{eq:p transformation under gbms}, one obtains (see Appendix \ref{sec:proofs in section 4})
	\be\label{wT}
\delta_T w = 0 \,, \qquad 
\delta_Y w = Y[w]\,. 
\ee
The first identity expresses the fact that the vorticity $w$ is invariant under supertranslation.    The second equality expresses the fact that $w \in V_{(0,0)}$ transforms as a scalar. Indeed terms as $m \epsilon^{AB}\pa_A j_B$ are scalars of dimension $7$.\footnote{The spatial derivative $\p_A$ raises the  conformal dimension by +1.} This is offset by the factor $m^{-\frac{7}{3}}$. This key result is the fundamental property that we were looking for and allows us to define the spin aspect for $\gbms$.

\paragraph{Orbit invariants for $\gbms$.} We can now easily construct the Casimir functionals for $\gbms$. They are given by all the moments of the vorticity
\be\label{eq:gbms casimir functions}
\mathsf{C}_n(\gbms) := \bigintsss_S    w^n \rho  \bfe\,, \qquad n=0,1,\ldots. 
\ee
The fact that they are invariant under $\gbms$ can be easily seen (see Appendix \ref{sec:proofs in section 4})
\begin{equation}\label{eq:invariance of casimirs under gbms}
\delta_{(Y,T)} \mathsf{C}_n(\gbms)=
\bigintsss_S \pa_A(w^n\rho Y^A\bfe)= 0.
\end{equation}
Therefore, these generic real values of these invariants provide the labels for coadjoint orbits of $\gbms$. This completes our construction of Casimir functionals on coadjoint orbits of $\gbms$.

\paragraph{Supertranslation orbits.} One distinguished feature of the angular momentum aspect $j_A$ is its non-trivial transformation under a supertranslation, as \eqref{eq:coadjoint action on gbms*} implies. Under a supertranslation with parameter $T$,  $j_A$ shift to $j_A'$ where 
\begin{equation}\label{eq:new angular momentum aspect from a supertranslation}
    j_A' -j_A= \frac{3}{2}m\partial_A T+\frac{T}{2}\partial_Am.
\end{equation}
These equations can be written in a simpler form by multiplying the both sides with $\frac23 m^{-\frac{2}{3}}$ and rearranging the equality. The result is
\begin{equation}\label{eq:differential of supertranlation parameter}
\frac{2}{3}m^{-\frac{2}{3}}\left(\bm{j}'-\bm{j}\right) = \rd(m^{\frac13} T),    
\end{equation} 
where we denote $\bm{j}= j_A\rd\sigma^A$ and we recall that the parameter  $m^{-\frac13} T \in V_{(0,0)}$ is a scalar. In this relation, $\rd:=\rd\sigma^A\partial_A$ is the de-Rham differential on $S$, satisfying $\rd^2=0$. This means that the translation orbits are labelled by the vorticity  2-form 
\begin{equation}
    \bm{w}_{\bm{j}}:= \rd ( m^{-\frac23} \bm{j}) = m^{\frac23} w \bfe.
\end{equation}
This means that two elements $\bm{j},$ and $\bm{j}'$ in the same translation orbit, if $\bm{w}_{\bm{j}}=\bm{w}_{\bm{j}'}$.
Equivalently this means that we have  the consistency relation
\begin{equation}\label{eq:consistency of de-Rham differential}
\rd\left(m^{-\frac{2}{3}}\left(\bm{j}'-\bm{j}\right)\right)=0,
\end{equation}
which follows from applying $\rd$ to both sides of \eqref{eq:consistency of de-Rham differential}. The supertranslation parameter $T$ relating $\bm{j}$ and $\bm{j}'$ can be determined through \eqref{eq:differential of supertranlation parameter} if and only if \eqref{eq:consistency of de-Rham differential} is satisfied. This, in particular, means that not any two angular momentum aspects $\bm{j}$ and $\bm{j}'$ can be related through a supertranslation. 

\smallskip Having \eqref{eq:consistency of de-Rham differential} in mind, \eqref{eq:differential of supertranlation parameter} can be integrated to determine $T$. It is given up to an overall constant as
 \begin{equation}\label{eq:the coordinate-dependent formula for supertranslation between j and j'}
    T(\sigma)=m(\sigma)^{-\frac{1}{3}}\left((m^{\frac{1}{3}}T)(\sigma_{0})+\frac{2}{3}\bigintsss_{\sigma_{0}}^\sigma m^{-\frac{2}{3}}(\bm{j}'-\bm{j})\right),
\end{equation}
where $\sigma_0$ denotes a reference point on the sphere and  the integration in the second term can be done over an {\it arbitrary curve}\footnote{For those $\mbs{j}$, $\mbs{j}'$, and $m$s satisfying \eqref{eq:consistency of de-Rham differential}, Stokes Theorem guarantees that 
\begin{equation*}
    \bigointssss_\Gamma \left(m^{-\frac{2}{3}}\left(\bm{j}'-\bm{j}\right)\right)=0,
\end{equation*}
where $\Gamma$ is any arbitrary possibly non-simple closed curve on $S$, which is due to $S$ being simply-connected. It then follows that the integral in \eqref{eq:the coordinate-dependent formula for supertranslation between j and j'} is independent of the curve between initial and final points. It of course depends on the endpoints, as the notation in \eqref{eq:the coordinate-dependent formula for supertranslation between j and j'} indicates.} between $\sigma_0$ and a generic point labelled by $\sigma$.

\subsection{$\gbms$: The spin generator}\label{sec:spin generator of gbms}

As we have seen in our brief recap of the construction of Pauli--Luba\'{n}ski pseudo-vector, which gives the definition of spin angular momentum of the Poincar\'e algebra in \eqref{eq:spin 3-vector in general Lorentz frame}, a prominent role is played by the isotropy algebra of the Poincar\'e algebra. It was defined as the subalgebra of the Lorentz algebra that preserves a fixed three-momentum. As we have seen, for a massive particle,\footnote{The analog of mass in our situation is $m$ and the fact that we are working with $m>0$ is the analog of massiveness for the Poincar\'e algebra.} this is the subalgebra $\msf{so}(3)$ generated by the  spin element, which is invariant under translations and it corresponds to the  Pauli--Luba\'{n}ski pseudo-vector in the rest frame. Our aim in this section is to repeat this exercise for $\gbms$: we first determine the isotropy subalgebra of $\gbms$. The knowledge of isotropy subalgebra would help us to construct its generators, the spin generator. We deffer the task of  constructing an analog of the Pauli--Luba\'{n}ski pseudo-vector for $\gbms$ to \S\ref{sec:bms pauli-lubanski}. The outcome of the construction will be an object whose phase space counterpart is the spin charge on the asymptotic phase space of asymptotically-flat spacetimes with $\gbms$ as their asymptotic symmetry group. We construct this phase space quantity in \S\ref{sec:gravitational casimirs and spin generator}. 

\paragraph{The isotropy subalgebra for $\gbms$.} To determine the isotropy (or little) algebra of $\gbms$, we choose a generic\footnote{By generic, we mean a generic function $m=m(\sigma)=m(z,\bar{z})$. There are more specialized choices such as constant $m$ or $m(|z|)$, which would affect the corresponding isotropy algebra.} $m\in (\mbb{R}_{-1})^*$, and study those $\gbms$ transformations that preserves $m$, i.e. $\delta_{(Y,T)}m=0$. Hence, by definition, the isotropy subalgebra of $\gbms$ associated with a given mass aspect $m$ is given by a pair $(Y,T)$ such that 
\begin{equation}\la{dYTm}
    \delta_{(Y,T)}  m =	Y^A \pa_A m+  3 W_Y m=0,
\end{equation}
which can be written as 
\begin{eqaligned}
    Y^A \pa_A m+{ \f32 m \cD_AY^A}  
		&= \f{3m^{\f13}}2\,  \cD_A \left(m^{\f23} Y^A \right)
		= \f{3\sqrt{\rho}}2\, \mathrm{div}_{\bfs}(Y)=0,
\end{eqaligned}
where we have defined a rescaled measure 
\begin{equation}
    \bfs :=\rho\bfe=m^{\frac{2}{3}}\bfe\,,
\end{equation}
and $\mathrm{div}_{\bfs}(Y) : = \f{1}{\rho} \pa_A \left(\rho Y^A \right)$ is the divergence of $Y$ with respect to the measure $\bfs$. Note that the condition \eqref{dYTm} is comparable to the one adopted in 3D gravity by  \cite{Barnich:2015uva} to define the intrinsic angular momentum.
Since we have assumed that $m>0$, this relation shows that the isotropy subalgebra of $\gbms$ is the one that preserves $\bfs$, namely (compare with the isotropy subalgebra of the Poincar\'e algebra $\msf{so}(3)\sds\mbb{R}^4$)
\begin{equation}\label{eq:isotropy subalgebra of gbms}
\mbf{iso}(\gbms)=\msf{sdiff}_{\bfs}(S)\sds\mbb{R}^S_{-1}\,,
\end{equation}
where $\msf{sdiff}_{\bfs}(S)$ is the subalgebra of area-preserving diffeomorphisms that preserves the measure $\bfs$ (rather than the round-sphere area form $\bfe$). These diffeomorphisms are  globally defined at the celestial sphere \cite{ArnoldKhesin1999}. 
This subalgebra is generated by vector fields of the form
\begin{equation}\label{eq:vector fields generating isotropy subalgebra of gbms}
    Y_\chi :=-\rho^{-1} \epsilon^{AB}\pa_A \chi \pa_B,\qquad \epsilon^{AB}=\frac{\varepsilon^{AB}}{\sqrt{\gamma}},\qquad  \chi\in C^\infty(S).
\end{equation}
The Lie bracket of these vector fields is (see Appendix \ref{sec:proofs in section 4})
\begin{equation}\label{eq:algebra of area-preserving vector fields for rescaled area form}
 [Y_\chi,Y_\psi]_S=Y_{\{\chi,\psi\}_\rho},
\end{equation}
where
\begin{equation}\la{rho-PB}
    \{\chi,\psi\}_\rho:= \rho^{-1}\epsilon^{AB}\partial_A\chi\partial_B\psi
\end{equation}
is the Poisson bracket on $S$ given by the inverse of the  rescaled symplectic structure
\begin{equation}
    \rho_{AB}:=\rho\epsilon_{AB}.
\end{equation}
\eqref{eq:algebra of area-preserving vector fields for rescaled area form} is the Lie bracket of $\msf{sdiff}_{\bfs}(S)$. Note that the area-preserving diffeomorphisms form a subalgebra and not an algebroid, as the measure $\rho$
is preserved by these transformations; this is 	analogous to the full $\gbms$ case due to the condition \eqref{dq0}.
Finally, note that the $\mbb{R}^S_{-1}$ factor of $\msf{iso}(\gbms)$ does not act on $m$ at all, and hence in the following we only focus on $\msf{sdiff}_{\bfs}(S)$.\footnote{This is analog to the situation of the Poincar\'e algebra where a chosen momentum is invariant under translations. Hence, the isotropy subalgebra is enlarged from the isotropy subalgebra of the Lorentz part, which is $\so(3)$, to the isotropy subalgebra of the Poincar\'e algebra, which is $\msf{so}(3)\,\vsds\mbb{R}^4$.}

\paragraph{The spin generator of $\gbms$.}

As we have seen in \S \ref{sec:Poincare algebra}, for a massive particle, the spin angular momentum is defined to be the value of the angular momentum of a particle in its rest frame, which in turn can be written in terms of the spatial components of the Pauli--Luba\'nski pseudo-vector (see \eqref{eq:components of the Pauli--Lubanski vector}). Our goal is to  provide an analogous construction for $\gbms$ and provide the generator of spin for $\gbms$.

\smallskip We propose that the $\gbms$ spin charge is the smeared version of the vorticity $w$. The charge is labelled by a function on $S$, $\chi \in \mathbb{R}_0^S=C^\infty(S)$, and is defined as 
\begin{equation}\label{eq:gbms spin generator}
   \msf{S}_\chi :=\bigintsss_S \chi \rho w\bfe=\bigintsss_S \chi\,\epsilon^{AB} \pa_A {p}_B \,\bfe, \qquad \chi\in C^\infty(S).
\end{equation}
From the transformation properties of $w$ under $\gbms$, given in \eqref{wT}, we can see that (see Appendix \ref{sec:proofs in section 4})
\begin{equation}\label{eq:coadjoint transformation of vorticity}
    \delta_T \msf{S}_\chi=0, \qquad \delta_Y \msf{S}_\chi=-\msf{S}_{Y[\chi]}.
\end{equation}
In particular, this implies
\be\la{mw}
\big\{\mathsf{P}_T, \msf{S}_\chi \big\}_{\g^*}=0.
\ee
Furthermore, we can then show that the bracket of the spin charge forms a closed subalgebra (see Appendix \ref{sec:proofs in section 4})
\begin{equation}\label{eq:poisson lie bracket of smeared vorticities}
    \big\{\msf{S}_\chi,\msf{S}_\psi\big\}_{\g^*}=-\msf{S}_{\{\chi,\psi\}_\rho}\,.
\end{equation}
This relation implies that $\msf{S}_\chi$ implements the action of $\msf{sdiff}_{\mbs{\rho}}(S)$ on $\gbms^*$. Therefore, both properties above are satisfied, namely
\begin{enumerate}
    \item [(1)] $ \mathsf{S}_\chi$ is invariant under supertranslations, i.e. $\delta_T\msf{S}_\chi=0$ ;

    \item [(2)] $ \mathsf{S}_\chi$ generates the isotropy algebra $\msf{sdiff}(S)_\rho$ of $\gbms$ \eqref{eq:poisson lie bracket of smeared vorticities}.
\end{enumerate}
Hence, $\mathsf{S}_\chi$ represents the spin generator for the $\gbms$ algebra.
The first property is the analog of the fact that the (spatial components of) Pauli--Luba\'nski pseudo-vector commutes with the generators of translations, as encoded in \eqref{eq:algebra relations for the Pauli--Lubanski vector}. The second property is the analog of the fact that the spin vector generates in Poincar\'e  the isotropy algebra $\msf{so}(3)$.

\section{$\gbms$ reference frames and Goldstone modes}\label{sec:reference frames and goldstone modes}

In this section, we study natural reference frames associated with $\gbms$, in order to make more transparent the contextualization of the $\gbms$ spin operator we just constructed within the familiar Poincar\'e framework reviewed in \S\ref{sec:Poincare algebra}.

Recall that in the Poincar\'e setting, we define two important frames of reference: (1) the rest frame to be the frame where the velocity of the particle vanishes and (2) the center-of-mass frame 
to be the frame where
where the center-of-mass remains at the origin.
The rest frame is reached by performing a boost, while the center-of-mass is achieved by performing a translation. 

 In the gravitational context, most of the discussion about a choice of $
\bms$ frame has been, so far, around a choice of center-of-mass (supertranslation) frame and about the definition of the $\bms$ orbital angular momentum. However, in order to understand the physical nature of the spin operator we have constructed, we need  to find the gravitational analog of the rest frame condition. In fact,  since the spin operator is by construction invariant under supertranslations, it doesn't depend on which center-of-mass frame we are in. At the same time though, 
here we show that the spin generator can be understood as the generator of superrotation in the generalised rest frame defined by the condition $\pa_A m=0$.
Indeed, when this rest frame condition is satisfied, we have that 
\be
	w= m^{-\frac43 }\epsilon^{AB} \pa_A j_B  \,,
 \ee
 which is the generator of superrotation. This is the analog of \eqref{Jrest} for $\gbms$.
To reach such a rest frame, given a general mass aspect, is not possible with the BMS action, but it requires the action of GBMS.

In \S\ref{sec:gbms rest vs bondi}, we define the notion of rest frame for $\gbms$ and clarify its distinction with the usual notion of Bondi frame. In \S \ref{sec:statST} we specialize to a stationary spacetime and derive the expressions for the mass and angular momentum aspects in the center-of-mass and rest frame. Furthermore, we compute these quantities in a generic boosted frame. In \S \ref{sec:Gold-mass} we provide an explicit expansion of the boosted mass into spherical harmonics, which we use in \S \ref{sec:Multi} to show that the multi-particle mass aspect does not belong to the BMS coadjoint orbit with constant total mass representative.  
Finally, in \S \ref{sec:Gold-angular} we perform the spherical harmonic decomposition for the boosted angular momentum aspect.

\subsection{$\gbms$ reference frames: Rest vs Bondi frames}\label{sec:gbms rest vs bondi}

In this section, we elaborate on two reference frames associated with $\gbms$. Let us consider an orientation preserving diffeomorphism of the sphere $F:S\to S$. For  concreteness we denote $\sigma^A$ the original coordinates and $\widetilde{\sigma}^a= F^a(\sigma)$ the coordinates on the image sphere; also,  the corresponding derivatives are respectively denoted as $\pa_A$ and $\pa_a$. We also denote the metric on the original sphere as $\gamma$. Given the area form $\bfe=\frac12 \epsilon_{AB}(\sigma) \rd \sigma^A \wedge \rd \sigma^B$ with $\epsilon_{AB}=\sqrt{\gamma}\varepsilon_{AB}$, we can construct a density $\rho_F$ such that 
$F^*\bfe = \rho_F \bfe$. One finds that 
\be \la{rhoF}
\rho_F  = \frac12 \epsilon^{AB}\pa_A F^a \pa_BF^b \epsilon_{ab}\circ F 
=\sqrt{\f{\gamma\circ F}{\gamma}} \mathrm{\det}\left(\pa_A F^a\right),
\ee 
where $\epsilon_{ab}\circ F$ is the area tensor in the new coordinates.
The inversion formula implies that 
$\rho_{F^{-1}}\circ F = 1/ \rho_F$.
Under such a diffeomorphism we have that the mass aspect and angular momentum aspect transform as 
\be \label{mjtrans}
m\to m^F := \rho_F^{\frac32} \,(m\circ F), 
\qquad 
j_A \to j_A^F := \rho_F\, (j_a\circ F) \pa_AF^a.
\ee 
Note that these come from the exponentiation of the coadjoint action \eqref{eq:coadjoint action on gbms*} (see also \cite{Flanagan:2015pxa,Freidel:2021fxf}). 
Furthermore, the metric and its curvature tensor (of the celestial sphere $S$) transform\footnote{We use that $ 
R(e^{2\phi} \gamma)= e^{-2\phi} (R(\gamma)-2 \Delta \phi)
$.} as
\begin{eqgathered}
    \gamma_{AB}\to \gamma_{AB}^F =  \frac{(\gamma\circ F)_{ab}\,\pa_A F^a \pa_BF^b}{\rho_F}, 
\\
R\to R^F:= \rho_F( R\circ F + \Delta \ln\rho_F)\,,
\end{eqgathered}
which again follow from the exponentiation of the $\gbms$ infinitesimal transformation \eqref{dgamma}. Let us emphasize that this Diff$(S)$ action of the GBMS group, called super-Lorentz transformations,  on the metric differs from the   naive Diff$(S)$ action; the latter is recovered for the  diffeomorphisms such that $\rho_F=1$. The subgroup of such diffeomorphisms is denoted SDiff$(S)$. It is composed of diffeomorphism preserving the area form. Infinitesimally this means that $W_Y$ in \eqref{Wres2} vanishes, and these transformations are called superrotations. While superrotations $F\in \mathrm{SDiff}(S)$ preserve the Bondi gauge condition $R(\gamma)=2$, they generically do not preserve the metric: $F^*\gamma_{AB}\neq \gamma_{AB}$ unless $F$ belongs to the isotropy group of $\gamma$.

From this transformation, we can define two distinguished $\gbms$ frames by putting certain constraints on the doublet $(m,R)$. 

\begin{itemize}

    \item[$-$] {\bf\small Rest frame.} In the rest frame, the pair $(m^{\va \msf{R}},R^{\va \msf{R}})$ satisfy
    \begin{equation}\label{eq:gbms rest frame}
        \pa_a m^{\va  \mathsf{R}}=0, \qquad \pa_a R^{ \va \mathsf{R}} \neq 0,
    \end{equation}
    i.e. the mass aspect is constant.

    \item[$-$] {\bf\small Bondi frame.} In the Bondi frame, the pair $(m^{\va \msf{B}},R^{\va \msf{B}})$ instead satisfies the following relations
    \begin{equation}\label{eq:gbms bondi frame}
        \partial_A m^{\va \msf{B}}\ne 0\,, \qquad \partial_A R^{\va \msf{B}}=0\,.
    \end{equation}
     Note that the \emph{Bondi frame} is such that the curvature is constant. This frame is used in the $\bms$ literature since the condition $\partial_A R^{\va \msf{B}}=0$ basically means that the celestial sphere necessarily has the round-sphere metric (or a constant rescaled version thereof).
\end{itemize}

Finally, we see that the two frames are related by a diffeomorphism $F: S^{ \mathsf{B}}\to S^{ \mathsf{R}}$, which we take to go from the rest frame to the Bondi frame for convenience. We have 
\be \label{eq:from rest to the bondi frame}
m^{\va \mathsf{B}} = m^{\va  \mathsf{R}}\rho_{F}^{\frac32}, 
\qquad
R^{ \va \mathsf{R}}= \rho_{F^{-1}}(R^{\va \mathsf{B}} + \Delta \ln\rho_{F^{-1}} ).
\ee

{ As shown later, the only spacetimes which can be, at the same time, in the rest and Bondi frame  are stationary ones. }

\smallskip An important comment is in order. Having the possibility to reach the rest frame is essential in our ability to define the spin as the angular momentum aspect in the rest frame, namely the gravitational analog of  \eqref{Jrest}.
 In other words, in the case of Poincar\'e, one needs to be able to reach to the rest frame in order  to define an intrinsic notion of spin, where the boost component of the angular momentum is zero. Once this intrinsic spin is defined, one can boost it to an arbitrary Lorentz frame. Similarly, in the case of $\gbms$, we are able to go to the $\gbms$ rest frame, where $m$ is constant, and this  provides an intrinsic notion of spin. One can then  ``boost" this spin to an arbitrary $\gbms$ frame by performing a diffeomorphism (as supertranslations will not change the mass aspect). This important fact should be compared and contrasted with the fact that (in general) there is no   notion of rest frame for $\bms$. 
This is the source of many puzzles and ambiguities. The root of these puzzles is that for $\bms$, one fixes a Bondi frame such that $\pa_A R^{\va \mathsf{B}}=0$, in which $\pa_A m^{\va \mathsf{B}}\neq 0$ unless we are in the particular case of a stationary black hole spacetime (see \S\ref{sec:Multi}). \eqref{eq:from rest to the bondi frame} shows that the extension to $\gbms$ takes care of this fundamental issue, since we  can always, by 
 a choice of diffeomorphism, reach the rest frame which is such that $\pa_A m^{\va \mathsf{R}}=0$. In this frame we  have that $\pa_A R^{\va \mathsf{R}}\neq 0$.

\subsection{Stationary  spacetimes}\la{sec:statST}

We now study the stationary spacetime condition and show, following \cite{Flanagan:2015pxa}, that in the rest and center-of-mass frame,
the angular momentum aspect $j_A$ is time-independent, and its
 electric parity component, as well as the $\ell\geq 2$ spherical harmonics of the magnetic component, can be set to zero. Let us see this explicitly.

\subsubsection{Mass and angular momentum aspects for stationary spacetimes}

The stationarity condition $\pa_u m=\pa_u j_A=\pa_u C_{AB}=0$, is much stronger than the non-radiative condition
$\pa_u C_{AB} = 0$.\footnote{The more general non-radiative condition reads (see \S\ref{sec:Norad})
\be\la{stat-con}
\pa_u^2 C_{AB}=0,\qquad
 \cD_B \pa_u C_{A}{}^B={-} \frac12 \pa_A R,\qquad \cD_{[A} C_{B]C} =0\,.
\ee 
}
Stationarity  demands that spacetime is non-radiative, i.~e. that the news vanish, but also that the mass, angular momentum and higher-spin charges (see \cite{Freidel:2021qpz} and \S \ref{sec:Norad}) are conserved in time.
 This means that
\bea
\cD_A m + \widetilde{\cD}_A\widetilde{m} &=&0, \label{m} \\
\cD_{\langle A} j_{B \rangle} &=& -\frac34 (C_{AB} m +\widetilde{C}_{AB} \widetilde{m}), \label{j}
\eea
where $\wt{m}= \cD_A \cD_B \wt{C}^{AB}$ denotes the dual mass, $\wt{\cD}_A=\epsilon_A{}^B\cD_B$ is the dual derivative, and $\wt{C}_{AB}=\epsilon_{A}{}^C C_{CB}$ is the dual of the shear tensor. The equation \eqref{m} implies that $\cD_A m=0=\cD_A\widetilde{m}$. This means that $m=m^{\va \msf{R}}$ is constant and that $\widetilde{m}=0$.\footnote{Since  $\widetilde{m} =\cD_A \cD_B \widetilde{C}^{AB}$  only contains $\ell\geq 2$ terms, it vanishes if constant.} The constancy of the mass means that we are in the rest frame, while the vanishing of the dual mass means that the shear can be written entirely in terms of a Goldstone field $G \in \mathbb{R}^S_{-1}$ which characterises the supertranslation frame
\be 
C_{AB}= -2  \cD_{\langle A} \cD_{B\rangle} G.
\ee 
$G$ is uniquely determined by $C_{AB}$ if we assume that $G_{\ell=0}=G_{\ell=1}=0$.
\eqref{j} can then be written as 
\be 
\cD_{\langle A} (j_{B \rangle}- \tfrac32 m^{\mathsf{R}}\cD_{B\rangle} G)= 0,
\ee
which means that $j_{B}- \frac32 m^{\va  \mathsf{R}}\cD_{B} G$ is a constant $\ell=1$ spherical harmonic.
In other words, we find that for a stationary spacetime, we have 
\be\label{eq:mass and angular momentum of stationary spacetimes in the rest frame}
m= m^{\va  \mathsf{R}}, \qquad 
j_A(n)= \frac12 J_{\mu\nu} n^{[\mu} \pa_A n^{\nu]} +
\tfrac32 m^{\va  \mathsf{R}} \pa_{A } G(n)\,,
\ee 
where $J_{\mu\nu}$ is a constant angular momentum. Denoting $(J_i,K_i)$ the rotational and boost components of $J_{\mu\nu} $ we can rewrite this as ($\wt{\partial}_A=\epsilon_A{}^B\partial_B$) 
\be \la{jR}
j_A(n)= \widetilde{\pa}_A n^i J_i + \pa_A\left(\frac32 m^{\va  \mathsf{R}}G(n) + K_i n^i\right).
\ee
The second term corresponds to a supertranslation with parameter $T= G + \frac{ 2 K_in^i}{3m^{\va \mathsf{R}}}$.
We can therefore eliminate this term by going to the rest and center-of-mass frame (RCM frame).
The expressions for mass and angular momentum aspects in the RCM frame can be read-off from \eqref{eq:mass and angular momentum of stationary spacetimes in the rest frame} as\footnote{Recall that a supertranslation does not change the mass aspect, as can be seen in \eqref{eq:coadjoint action on gbms*}.}
\be \label{jRCM}
m^{\va \msf{RCM}}=m, \qquad j_A^{\va  \mathsf{RCM}}(n) =  \widetilde{\pa}_A n^k J_k = (n^i\pa_An^j) \epsilon_{ijk}J^k.
\ee 
So we find that up to a supertranslation and a rotation, we can always choose the stationary angular momentum to be purely magnetic.
We can still perform a rotation that rotates $J_i= J\delta_i^3$ along the 3rd axis say, where $J$ is the black hole spin.

\paragraph{The case of Kerr spacetime.} As we will show in \S\ref{sec:gravitational Casimirs for the kerr metric} and Appendix \ref{sec:details on the kerr metric and casimirs}, for the Kerr metric, 
\begin{equation}
    j_A^{\va \tenofo{Kerr}} \rd \sigma^A={\f12} 3aM(\cos\theta\,\rd\theta-\sin^2\theta\,\rd\varphi).
\end{equation}
As $\widetilde\p_{\varphi} n^3= -\sin\theta \pa_\theta n^3= \sin^2{\theta}$  for the round sphere metric,
comparing with \eqref{jR} we find that 
the Kerr metric angular momentum  corresponds to the rest frame expression \eqref{jR} with
\be
J^{\va \tenofo{Kerr}}=-\f{3aM}2\,,\quad 
G^{\va \tenofo{Kerr}}= a \left(\sin\theta - \frac{\pi}{4}\right)
\,,\quad 
K_i^{\va \tenofo{Kerr}}=0\,,
\ee
consistently with \cite{chrusciel2003hamiltonian, Flanagan:2015pxa, Compere:2019gft}. 
The substraction by $\pi/4$ ensure that  $G$ only contains $ \ell\ge 2$ spherical harmonics as $ \int_S \epsilon \sin\theta =\frac{\pi}{4}$.\footnote{
One can write $\sin \theta=\f{2\sqrt{z\bar z}}{1+z\bar z}$ so the spherical harmonic expansion of $\sin \theta$ contains all $\ell=2 k, k\in\Z^{0+}$ and $m=0$ components.}

\subsubsection{Mass and angular momentum aspects in a generic Lorentz frame}\label{sec:(m,j) in general lorentz frame for stationary spacetimes}

Next, we  would like to derive expressions for mass and angular momentum aspects of stationary spacetimes in a generic boosted frame. For this, we can use \eqref{mjtrans}. However, to use that equation, we first need to determine the conformal factor that corresponds to a Lorentz boost, which we now turn to. 

\paragraph{Conformal rescaling as a Lorentz boost.}  The boost transformation simply results from a conformal map of the celestial sphere.
 To do so let  us recall that we work with the null vector 
$n^\mu = \tau^\mu + \widehat{n}^\mu$ where $\tau^\mu=(1,0,0,0)$  is the rest frame vector and $\widehat{n}^\mu=(0, n^i)$ is spacelike. Let $\Lambda_v$ denote the boost that maps $\tau$ onto a unit vector in the hyperboloid of velocity $v\in \mathbb{R}^3$, with $v^\mu=(0,v^i)$, and let us denote the corresponding unit-norm momentum as
\be\label{eq:4-momentum in a generic lorentz frame}
 p_v^\mu:= 
 (\Lambda_v \tau)^\mu 
 = \gamma_v(\tau^\mu + v^\mu) = \gamma_{v}(1,v^i),
\ee
where
\begin{equation}
    \gamma_v:=\frac{1}{\sqrt{1-v^2}}.
\end{equation}
Going to such a boosted frame corresponds to applying an inverse boost   $n^\mu\to (\Lambda_{-v}  n )^\mu:= (\Lambda_{-v})^\mu{}_\nu n^\nu $ to $n$. 
Under a boost the null vector stays null,
hence we have that 
\be\la{Ln}
\Lambda_{-v} n = \omega_{v}(n) n_v,\ee
where $\omega_v(n)$\footnote{Since this is just a factor coming from a boost, we have used the same notation as the factor $\omega_g$, defined in \eqref{eq:bms action on supertranslations}, as  will see in \eqref{eq:omega_v as omega_g}.} is a rescaling factor given by 
\be \la{On}
\omega_v(n) = -   [\Lambda_{-v}n] \cdot \tau =
 n\cdot (\Lambda_{v}\tau) = -(n\cdot {p_v})
 = \gamma_{v}(1-v\cdot \widehat{n}).
\ee
Therefore, using \eqref{eq:lorentz transformation from the rest frame to a general frame}, we get that  a boost  yields a transformation of $n \to n_v$  given  by 
\be\label{eq:boosted n}
n_v^i= \omega_v^{-1}(n)\left(n^i +  \left[\frac{\gamma_{v} {v\cdot \widehat{n}}}{\gamma_{v}+1}-1\right] \gamma_{v} v^i\right).
\ee

\smallskip To relate the rescaling factor $\omega_v$ to the factor $\rho_{\Lambda_{-v}}$ defined in \eqref{rhoF} for $F=\Lambda_{-v}$, we use the fact that $\rho_{F_v}$ appears as the conformal rescaling of the metric, namely (see Footnote \ref{ftn:transformation of density under diffeomorphism})
\bea
\rd s_v^2 &=& (\rd \widehat{n}_v)^2
= (\rd n_v)^2
= \frac{(\rd [\omega_v n_v])^2}{\omega_v^2}=
\frac{(\rd [\Lambda_{-v} n])^2}{\omega_v^2}=
\frac{(\rd n)^2}{\omega_v^2}
= \frac{(\rd \widehat{n})^2}{\omega_v^2}.
\eea  
This shows that the metric is simply rescaled under boost and satisfy the transformation $\Lambda^*_{-v} q = \rho_{\Lambda_{-v}} q $ with 
\be 
\rho_{\Lambda_{-v}} = \omega_v^{-2}.
\ee 
This results are valid when $n^i= \mathring{n}^i\circ F$ for any $F\in \mathrm{Diff}(S)$.
We can illustrate them using 
 complex coordinates where  
\be \la{nz2}
\mathring{n}^i(z,\bar{z}) = \left(
\frac{z+\bar{z}}{1+|z|^2}, \frac{-i(z-\bar{z})}{1+|z|^2},\frac{1-|z|^2}{1+|z|^2} \right).
\ee 
Then
$\mathring{n}_v^i(z,\bar{z})= \mathring n^i(z', \bar{z}')$ with 
$z'=\frac{az+b}{cz+d}$ and the coefficients $a,b,c,d$ can explicitly be expressed in terms of the velocity, see for instance \cite[Eqs. (A.7)--(A.9)]{Compere:2019gft}.

We can then express the RHS of \eqref{rhoF} in complex coordinates as 
\be\la{rhoF-complex}
\rho_F (z, \bar z)=|\p_z F|^2\f{(1+|z|^2)^2}{(1+|F|^2)^2}\,,
\ee
for any conformal transformation.
Taking $F=\Lambda$ to be an $\SL(2, \C)$ transformation $F(z)=\frac{az+b}{cz+d}$ we get   
\be\label{eq:omega_v as omega_g}
\omega_v=  \frac{|cz+d|^2 + |az+b|^2}{1+|z|^2}= \gamma_{v}(1-v\cdot \widehat{\mathring n})\,.
\ee

\paragraph{Boosted mass and angular momentum aspects.}

Now that we have given the expressions for the mass and angular momentum aspects \eqref{jRCM} of a stationary spacetime in the center of mass frame, we can construct their general expressions for a boosted stationary spacetime.   For a particle of mass $m$ and spin $J$ with velocity $v$ these can be explicitly computed from the general transformation formulas  \eqref{mjtrans}  for a boost ($F=\Lambda_{-v}$); they are given by
\be 
m_{v}(n) = \frac{m}{[\gamma_{v}(1-v\cdot\widehat{n})]^3}\,,
\qquad
j_A^v(n)= \frac{n^\mu \pa_A n^\nu J^{v}_{\mu\nu}}{[\gamma_{v}(1-v\cdot\widehat{n})]^4}\,,
\la{jAv}
\ee 
where, in terms of the boost transformations $\Lambda_v$ defined in \eqref{eq:lorentz transformation from the rest frame to a general frame} with $\mbs{P}_i=\gamma_v m v_i$,
\be
J_v^{\mu \nu} :=  
\Lambda_{v}^\mu{}_j
 \Lambda_{v }^\nu{}_k \varepsilon^{ijk} J_i
\ee 
is a boosted spin, constant on the celestial sphere and  such that ${p_{v\mu}} J_{v}^{ \mu\nu} =0.$ 
For the mass aspect, the derivation is direct and the expression was already given by Bondi et al. in \cite{BondivanderBurgMetzner196208}. For the angular momentum aspect this expression does not seem to have appeared in the gravity literature  (see the statement in \cite{Compere:2019gft} and the discussion around Eq. (5.10)). One notable exception is the work of Campiglia which derives a similar expression from the subleading soft theorem \cite{Campiglia:2015lxa}.  
To prove the statement we use the expression \eqref{jRCM} for the rest and center-of-mass frame angular momentum   to rewrite the transformation rule \eqref{mjtrans} in order to express the angular momentum aspect in a boosted frame. We start with 
$j_A^{\mathsf{RCM}} = j_A^{\mu\nu} J^0_{\mu\nu}$, where 
$J^0_{0i}=0$ and $J^{ 0}_{ij}=\varepsilon_{ij}{}^k J_k$ is the rest {and center-of-mass} frame angular momentum and after applying \eqref{mjtrans}, we get
\be 
j^{[\mu\nu]}_A = \frac{ n_v^{[\mu}\pa_A F^a \pa_a n_v^{\nu]}}{\omega_v^2(n)}\,.
\ee 
Now we use that 
\be 
 n_v^{[\mu}\rd n_v^{\nu]}  =
 \frac{\omega_v n_v^{[\mu}\rd (\omega_v n_v^{\nu]}) }{\omega_v(n)^2}= \Lambda_{-v}^\mu{}_\alpha
 \Lambda_{-v}^\nu{}_\beta
 \frac{ n^{[\alpha}\rd n^{\beta]}}{\omega_v(n)^2}
= \Lambda_{v}^{\alpha\mu}
 \Lambda_{v}^{\beta\nu}
 \frac{ n_{[\alpha}\rd n_{\beta]}}{\omega_v(n)^2}
\ee 
and use chain rule  $ \p_A= \p_A F^a \p_a $ to establish the final identity for the {boosted} angular momentum aspect in \eqref{jAv}.

\subsection{Condensate fields for  mass and angular momentum aspects}\label{sec:Condensate}

One of the motivations to study the algebra $\gbms$ rather than $\bms$ is to see whether a general multi-particle configuration belongs to the orbit of $\bms$ with a constant-mass representative. It turns out that the study of this question requires us to understand the decomposition of the mass in spherical harmonics. This section is devoted to the study of this question. We first determine the harmonic decomposition of mass in \S\ref{sec:Gold-mass}. Equipped with this result, in \S\ref{sec:Multi}, we show that the mass aspect of a generic multi-particle configuration {\it does not} belong to the $\bms$ orbit with constant mass-aspect representative. Finally,  we explain the harmonic decomposition of the angular momentum aspect in \S\ref{sec:Gold-angular} and give the explicit construction of the respective condensate fields.

\subsubsection{Supertranslation condensate} \la{sec:Gold-mass}

Let us start with the mass aspect and treat the angular momentum in \S \ref{sec:Gold-angular}. 
One defines the global momentum as $P^\mu = \mathsf{P}_{n^\mu}= Mp_{v}^\mu$,  given by the integral
\begin{equation}\label{momenta}
    {P}^\mu=\bigintsss_S mn^\mu\,,
\end{equation}
which represents the  $\ell=0,1$ modes of the mass aspect.
The  higher $\ell\geq 2$ modes are determined by a supertranslation condensate $C_0=C_{m}$ (see \S \ref{sec:poincare embedding}) where the  label $m$ emphasize that the spin $0$ condensate depends on the mass aspect. 
By definition, the supertranslation condensate $C_{m}$ is such that the Bondi mass admits the decomposition  
\begin{eqaligned}\label{mPn}
m(n) &= 
 m(n)|_{\ell=0,1}
 - \frac12 \Delta (\Delta +2 ) C_{m}(n) 
 \\
& =   m(n)|_{\ell=0,1}
 +  D_A D_B C^{AB}_{m}(n)\,,
\end{eqaligned}
where $\Delta$, as before, is the Laplacian on the sphere, and 
\be\label{eq:l=0,1 component of mP(n)}
m_P(n)|_{\ell=0}= - \tau \cdot P \,,
\qquad
m_P(n)|_{\ell=1}=  3 \widehat{n} \cdot P \,,
\ee
where $\tau^\mu=(1, \vec{0})$ and $\widehat{n}^\mu = (0, \vec{n})$.  Since $-\Delta (\Delta +2 )C_{m}|_{\ell} = (\ell-1)\ell(\ell+1) (\ell+2)C_{m}|_{\ell} $,
only the components $C_{m}|_{\ell\geq 2}$  of the condensate  are determined by this equation.
We know that the mass aspect is invariant under supertranslation, i.e $\delta_T m=0$. Therefore the condensate $C_{m}$ being a function of $m$ is also supertranslation invariant
\be 
\delta_T C_{m}=0.
\ee 
This important property means that we can use the condensate to fix the supertranslation frame, i.e. to chose a relationship between the supertranslation Goldstone and the supertranslation condensate.

This is exactly how Moreschi proposed   to fix the supertranslation frame by demanding $G= 2 C_{m}$.
To see this, lets recall that  the  Moreschi mass aspect \cite{ Moreschi:1988pc} (see also \cite{Moreschi1986, Moreschi:2002ii}) is given by 
\be 
m_{\mathrm{Mor}} = m- \frac14 D_AD_B C^{AB}\,,
\ee
and it is such that it is strictly decreasing over time
\be 
\pa_u m_{\mathrm{Mor}} 
= -\frac18 N_{AB}N^{AB}.
\ee
Moreover, its transformation under supertranslation is 
\be\label{transmor}
\d_{T} m_{\va \mathrm{Mor}} &= T \pa_u m_{\va \mathrm{Mor}}
+ \frac14( \D+R(\gamma)) \D T + \f12 \p^A R(\gamma)\p_A T.
\ee 
It is valid even when we are in the radiative phase space, and follows from the following  supertranslation transformations (see e.g. \cite{Freidel:2021qpz})
\be 
\d_{T} ( D\!\cdot\! C)^B &= T\pa_u (D\!\cdot\! C)^B
+ N^{BA}\pa_A T
-(R(\bq)\pa^BT +\pa^B\D T)\,,
\la{dDC}
\ee
which yield
\be
\d_{T} (D_AD_B C^{AB}) &= 
 D_A(\d_{T} D_B C^{AB})
 \cr
&=T\pa_u (D_AD_B C^{AB})
+ 2D_BN^{BA}\pa_A T \cr
& 
-D_B(R(\bq)\pa^B T +\pa^B\D T) + N^{AB} D_AD_B T  \,,
\la{dDDC}
\ee
and 
\be
\d_{T} M&=T \pa_u M
+ \left( \f12 D_B N^{AB} +\frac{\pa^A R(\gamma)}{4} \right)\pa_A T +
\f14  N^{AB} D_A\pa_B T  \,.
\la{dM}
\ee

The transformation \eqref{transmor} simplifies when $R(\gamma)=2$, which we now assume. 
Therefore, this means that, given a supertranslation Goldstone $G$  (see \S\ref{sec:gbms goldstone modes}) which determines a cut $u=G$ of $\scri$, we have that 
\be 
m_{\va \mathrm{Mor}} - \frac14 \Delta (\Delta+2) G 
\ee
is supertranslation invariant. Choosing $G=2 C_{m}$ amounts to chose a supertranslation frame where the Moreschi mass aspect only contains  global modes $\ell=0,1$  \cite{ Moreschi:1988pc}. This is equivalent to the Bondi mass  decomposition  \eqref{mPn}.

\smallskip For a single black hole spacetime the global momentum  \eqref{momenta} determines the mass aspect 
\begin{eqaligned}
    m_{\va P}(n)&= \frac{P^4}{({-}P\cdot n)^3}=\frac{M}{\gamma_v^3(1-v\cdot n)^3},
\end{eqaligned}
where we have used \eqref{eq:4-momentum in a generic lorentz frame}. We can decompose this expression into  spherical modes $P_{\ell m}(P):= \mathsf{P}_{Y_{\ell m}}$ and evaluate the supertranslation condensate.
In Appendix \ref{App:Gold}, we show by an explicit calculation that the single black hole condensate can be taken to be  \cite{JavadinezhadPorrati202211,Veneziano:2022zwh,Riva:2023xxm}  
\be \la{Cp}
C_{\va P}(n) =  (n\cdot P) \ln  \left(\f{-n\cdot P}{M}\right)\,,
\ee 
where we denote $C_{\va P}:= C_{m_P}$.
An interesting aspect of this expression is that it contains $\ell=0$ and $\ell=1$ components for $C$ which are usually left undetermined (see \cite{Javadinezhad:2023mtp} for a discussion on  how to fix the global modes of $C_{\va P}$.).
The corresponding shear is therefore given by 
\be \la{CAB}
C_{\va P}^{AB}(n) =-2D^{\langle A} D^{B \rangle } C_{\va P}(n)= 4\frac{ (D^{\langle A} n \cdot P)( D^{B \rangle } n \cdot P)}{ (n\cdot P)}\,.
\ee 
Quite remarkably, this expression for the shear reproduces exactly the soft factor in the leading soft graviton theorem \cite{Weinberg:1965nx}, with $n^\mu$ representing the soft graviton 4-momentum and $D_A n^\mu$ its polarization tensor.

Under a boost transformation, by means of \eqref{Ln} and \eqref{On}, we have that 
\be
C_{\va P}(n_\Lambda) &= C_{\va P}\left( \frac{ \Lambda^{-1} n }{(- n\cdot \Lambda p)}\right)  =  \f12  \frac{(n\cdot \Lambda P)}{\omega_\Lambda(n)} \ln  \left(\f{-n\cdot \Lambda p}{M \omega_\Lambda(n)}\right)\cr
&= 
\frac { C_{\va \Lambda p}(n) }{\omega_\Lambda(n)} 
- \frac12 (n\cdot \Lambda P) \frac{ \ln(\omega_\Lambda(n)) }{\omega_\Lambda(n)}\,,\la{eq:S'}
\ee  
where we denote $\omega_\Lambda(n)= -(n \cdot \Lambda \tau)$ and $p=P/M$. The first term in \eqref{eq:S'} corresponds to the expected transformation under boost of a weight $-1$ scalar. The second term  implies that $C_{\va P}$ transforms anomalously under boost transformation.
This last term vanishes  after we sum the contribution from in and out particles due to momentum conservation.

\subsubsection{Multi-particle mass aspect}\label{sec:Multi}

In this section, we show that multi-particle states belong to the same $\gbms$ orbit but different $\bms$ orbits. This is one of the main motivation for studying $\gbms$ instead of $\bms$.\footnote{We would like to thank G. Compere for a discussion that led to this section.} Consider a scattering process consisting of $N$ particles of momenta $P_I=M_I (\gamma_I , \gamma_I v_I^i)$  for $I=1,\ldots,N$. From \eqref{jAv}, the general mass aspect of this collection of $N$ particles is given by 
\be \la{totm}
m_{\va N}(n):=\sum_{I=1}^N m_{ \va P_{ I}}(n)
= \sum_{I=1}^N \frac{ M_{\va  I}}{\left[\gamma_{\va I} \left(1- v_{\va I}^in_i \right)\right]^3 }\,.
\ee
The important fact we want to establish is that such a multi-particle state \emph{does not} belong to a BMS coadjoint orbit with
constant total  mass $M$ representative,
namely $m_{\va N} \neq m_{P_{\mathrm{tot}}}$ where $P_{\mathrm{tot}}=\sum_i P_i$.   
The main point is to establish that while the two aspects have the same global charge, they differ significantly in the value of their supertranslation condensates.

\smallskip Let us consider the simple case of two particles. We have
\be\la{2p-bad}
m_2(n)
&:= m_{\va P_1}(n)+m_{\va P_2}(n)=\ \frac{ M_1}{\left[\gamma_1 \left(1- v_1^i  n_i\right) \right]^3 }
+\ \frac{ M_2}{\left[\gamma_2 \left(1- v_2^i  n_i \right)\right]^3 }\,.
\ee
Energy and momentum 
conservation ${ P_{\mathrm{tot}}}=P_1+P_2$  imply that 
\be
m_{\va P_{\mathrm{tot}}}(n)
=
\frac{\left(
 M_1^2+ M_2^2+2\gamma_1 \gamma_2M_1M_2(1-v_1^iv_{2i})
\right)^{2}}{\left[
\gamma_1 M_1(1-v_1^i   n_i)+\gamma_2 M_2(1-v_2^i   n_i)
\right]^3}\,.
\la{2p-good}
\ee

 Information about whether the multi-particle state belongs to the BMS orbit with constant mass representative given by \eqref{2p-good} can be obtained by comparing the spherical harmonic components of \eqref{2p-bad} and \eqref{2p-good}. As it is clear from the expression \eqref{mPn}, the $\ell=0,1$ components of the mass aspects agree by construction
 $m_{\va P_{\mathrm{tot}}}(n)|_{\ell=0,1}=m_2(n)|_{\ell=0,1}$.
 On the other hand,  the $\ell\geq2$ components of the mass aspects \eqref{2p-bad} and \eqref{2p-good} are different. This follows from the fact that  they carry different condensates (see Appendix \ref{App:Gold})
 \be
 D_A D_B (C_{\va P_1}^{AB} + C_{\va P_2}^{AB} - C_{\va  P_{\mathrm{tot}}}^{AB})\neq 0 \,,
\ee 
where $C_{\va P}$ and $C_{\va P}^{AB}$ are given in  \eqref{CAB} and \eqref{Cp}.
The combination $C_{\va P_1} + C_{\va P_2} - C_{ \va P_{\mathrm{tot}}}$ measures how much the initial state condensate differs from a single black-hole condensate.  
 
\smallskip This result  means that in a general scattering process, the mass aspect of the initial and final states does not belong to the $\bms$ orbit with a constant-mass representative.\footnote{If the entire initial state ends up in a black hole, this means that while the initial state does not belong to the constant mass orbit, the final state does. } 
 In other words, we cannot start with an arbitrary configuration consisting of many particles in generic Lorentz frames, and reach a constant mass representative by applying a $\bms$ transformation.\footnote{Recall that only the Lorentz part of $\bms$ changes the mass aspect, while a supertranslation preserves it (see \eqref{eq:coadjoint action on gbms*} with $Y$ restricted to be a conformal Killing vector).}. On the other hand,
using the bigger group $\gbms$ we can map the multi-particle condensate onto the single particle one. Indeed as shown in \S\ref{sec:gbms rest vs bondi}, given $m_2$  there exists a diffeomorphism $F_2\in \mathrm{Diff}(S)$ such that $m_2^{F_2}$ is constant (see  \eqref{mjtrans}). This diffeomorphism maps the condensate $C_{\va P_1}+ C_{\va P_2}$ onto the trivial condensate.
Similarly, for $m_{\va P}$ one can find a diffeomorphism $F_{\va P}$ that maps the condensate $C_{\va P} \to 0$. Therefore we find that $m_P$ and $m_2$ are in the same $\gbms$ orbit.

\smallskip We thus need to be able to have access to a larger set of transformations than the Lorentz transformations of $\bms$ to understand the symmetries of a scattering process. This is an important motivation to consider $\gbms$, in which Lorentz transformations are replaced with $\msf{diff}(S)$.

 \subsubsection{Angular momentum condensate}\la{sec:Gold-angular}

In the previous section we have given the decomposition of a general mass aspect in terms of the sum of total momentum aspect plus the condensate contribution. 
We give, here, a similar  decomposition for the angular momentum aspect. 

Given a mass and angular momentum aspect $(m,j_A)$, we have  that  the total angular momentum is given by 
\be
 J_{\mu\nu}:= \bigintsss_S j_A\, ({n}_{[\nu} \pa^A \bar n_{\mu]})\, \bfe\,.
\ee

Inverting this relation means that we can write the angular momentum aspect in terms of a global angular momentum aspect $J_A(n):= j_A(n)|_{\ell=1}$ plus a piece the contains the $\ell\geq 2$ modes. That second piece is encoded into the super-boost condensate denoted 
$C_{\va (P,j)}$. In practice this means that we have 
\be 
j^{\va (P,J)}_A(n) = J_A(n) + D_{\langle A} D_B D_{C\rangle} C_{\va (P,J)}^{BC}\,.
\ee

In fact, if we start with the  
  the boosted angular momentum aspect in \eqref{jAv} in the covariant form 
\be \la{JAcov}
j_A^{\va (P,J)}(n) = P^4 \frac{(n^\mu D_An^\nu)}{(-n\cdot P)^4} J_{\mu\nu}\,,
\ee  
we show in  Appendix \ref{App:AMGold}, in exact analogy to the mass aspect, 
that  this aspect can be decomposed into a contribution of $\ell=1$ harmonics  (corresponding to the standard Lorentz piece) 
\be
J_A(n) =(n^\mu D_An^\nu) J_{\mu\nu}\,,
\ee
plus a $\ell\geq 2$ contribution
 given by the subleading soft factor, namely
\be
C^{BC}_{\va (P,J)}(n)=\frac23 \f{(D^{\langle B} n \cdot P)( n^\mu D^{C \rangle }  n^\nu J_{\mu\nu})}{(n\cdot P)}\,.
\ee
It is quite remarkable that, as for the mass aspect, also in this case the identification between the condensate and the corresponding soft factor continues to hold \cite{Kapec:2014opa, Kapec:2016jld}.

\section{Pauli--Luba\'nski generator}\label{sec:pauli-lubanski generators}

We have verified above that the Poisson bracket of charges \eqref{eq:the charge for lorentz generator in gbms} with vector fields \eqref{eq:conformal killing vectors on sphere} and \eqref{eq:analog of four-momentum generator for gbms} generate a Poincar\'e subalgebra of $\gbms$.
Since the Pauli--Luba\'nski pseudo-vector \eqref{eq:definition of PL pseudo-vector} has a crucial role in constructing irreducible representations of the Poincar\'e algebra \eqref{eq:the poincare algebra}, a natural issue is the construction of analogous quantity for the Poincar\'e subalgebra of $\bms$. More precisely, for $\bms$ the spin generator is non-trivial only for constant mass aspects, in which case the isotropy subalgebra is isomorphic to the algebra satisfied by Pauli--Luba\'nski generator. In  \S\ref{sec:PL-BMS constant mass} we show that, when the mass aspect is constant,  our construction agrees with \cite{Compere:2023qoa}. This means, namely,  that the $\gbms$ spin generator we construct agrees with the $\bms$ spin generator. In other words,  when  the mass aspect is constant the $\gbms$ spin generator is the Pauli--Luba\'nski generator.

\smallskip In fact, as we have explained in \S\ref{sec:issues with the bms algebra}, a generic mass aspect does not have an isotropy subalgebra in $\bms$. On the other hand, from the work of  McCarthy \cite{McCarthy197211} summarized in Table  \ref{tab:isotropy subalgebras of BMS}, we know that for a constant mass aspect $m(z,\bar{z})=\text{constant}$, there is a non-trivial isotropy Lie subalgebra of $\bms$. Hence, one can ask whether there is an analog of Pauli--Luba\'nski generator, when restricting to coadjoint
orbits with constant mass representatives, as the generator of this isotropy subalgebra. In this section, we answer this  question: We construct the Pauli--Luba\'nski generator for a Poincar\'e embedding inside $\bms$. We will do this in two steps.  First, in \S\ref{sec:bms pauli-lubanski}, we use the spin charge constructed in \S\ref{sec:spin generator of gbms} to define an object that transforms covariantly under Poincar\'e transformations in analogy to \eqref{eq:poisson bracket of dual pauli-lubanski pseudo-vector}, as this is the first requisite  for the  Pauli--Luba\'nski generator. This can be achieved without any restriction on the mass aspect. 
Secondly, in order reproduce the algebra   \eqref{eq:poisson bracket of dual pauli-lubanski pseudo-vector with itself} representing the second defining property, we will need to restrict in \S\ref{sec:PL-BMS constant mass} to the constant mass aspect orbit of $\bms$ in order to define the   Pauli--Luba\'nski generator from the object previously introduced; we also  verify  that its spatial components can be written as the gravitational analog of the Poincar\'e expression
 \eqref{eq:components of the Pauli--Lubanski vector}.

\subsection{Lorentz covariance}\label{sec:bms pauli-lubanski}

Our aim is to  define 
the analog of the Pauli--Luba\'nski pseudo-vector for a Poincar\'e subalgebra of $\bms$ associated with a metric $q$ and corresponding vector $n_\mu$. We recall the definition of the spin charge\footnote{We use the bracket notation $\msf{S}[\chi]$ instead of the index notation $\msf{S}_\chi$ for readability.}  \eqref{eq:gbms spin generator}
\begin{equation}
    \msf{S}[\chi]:=\bigintsss_S\,\chi\rho w\bfe, \qquad \chi\in C^\infty(S).
\end{equation}
We then define the following quantity 
\begin{equation}\label{eq:4-vector W for gbms}
    S_\mu := \msf{S}[\sqrt{\rho} n_\mu]\,,
\end{equation}
and show that it satisfies the analog of \eqref{eq:poisson bracket of dual pauli-lubanski pseudo-vector}, which is the first property ascribed to the Pauli--Luba\'nski generator.

To begin, we immediately have
\be\la{PW}
\{P_\mu,S_\nu\}_{\g^*}&=-\d_{T=n_\mu} S_\nu=0
\ee
due to \eqref{eq:rho transformation under gbms} and \eqref{wT}. Furthermore, in Appendix  \ref{app:YW} we show that  the action of diffeomorphism on $S_\mu$ is given by 
\be\la{YW}
\d_Y S_\mu =  \f12 \msf{S}[\sqrt{\rho}  D_AY^A n_\mu] 
- \msf{S}[\sqrt{\rho} Y[n_\mu]].
\ee
In Appendix \ref{app:Wcov} we show that when $Y$ is a conformal vector field this expression, quite remarkably, reduces to 
\be\label{eq:gbms bracket of W and angular momentum generators}
\{J_{\mu\nu},S_\rho\}_{\g^*}=(\eta_{\nu\rho}S_\mu-\eta_{\mu\rho}S_\nu).
\ee
\smallskip If we denote by $\d_{\mu}$ the transformations generated by  \eqref{eq:4-vector W for gbms}, by anti-symmetry of the canonical action \eqref{PW} and \eqref{YW}, we derive the transformations 
\be\la{dn}
\d_{\mu} m=0\,,\qquad
\d_{\mu} j_A=
 \f12  n_\mu \p_A \left(\sqrt{\rho}  \,\epsilon^{BC}\partial_B p_C\right)
 +\f32 \sqrt{\rho} \,\p_A n_\mu \,\epsilon^{BC}\partial_Bp_C.
\ee
Using these transformations, one can verify that the vector \eqref{eq:4-vector W for gbms} satisfies the algebra
\be
&-\{S_\mu,S_\nu\}_{\g^*}=\d_{\mu}S_\nu
=
\msf{S}[\{\rho^{\f12}n_\mu,\rho^{\f12}n_\nu\}_\rho],
\ee
as expected from the bracket \eqref{eq:poisson lie bracket of smeared vorticities} and the fact that the $\bms^*$ element $\rho$ entering the smearing function in \eqref{eq:4-vector W for gbms} commutes with the vorticity \eqref{mw}.

\smallskip Therefore $S_\mu$ commutes with the momentum and transforms covariantly under Lorentz, as expected. As such it is a good candidate for a Pauli--Luba\'nski generator. However, as we haven't put any restriction on  the mass aspect entering the definition \eqref{eq:4-vector W for gbms}, its algebra doesn't close generally.  It turns out though  that its algebra reproduces the Pauli--Luba\'nski algebra  \eqref{eq:poisson bracket of dual pauli-lubanski pseudo-vector with itself} if one restricts our construction to the  mass  aspect of a boosted black hole. That is what we do next.

\subsection{Pauli--Luba\'nski embedding inside $\bms$}\la{sec:PL-BMS constant mass}

In order to complete our construction of the Pauli--Luba\'nski generator, we can  now use    \eqref{eq:4-vector W for gbms} and restrict to the constant mass aspect orbit of $\bms$ to show that the second defining property 
 \eqref{eq:poisson bracket of dual pauli-lubanski pseudo-vector with itself} can also be satisfied.
In other words, we now restrict  our attention to the case of a boosted stationary spacetime. In this case, we have that the density  aspect is given by  
\be
\rho=\rho_{v}=\frac{\rho^{\va \msf{R}}}{\gamma^2_v(1-v\cdot \widehat{n})^2},\label{rhov}
\ee where $\rho^{\va\msf{R}}:= M^{\f23}$ see \eqref{jAv}. 
In this case, we can evaluate the expression for the  spatial components of \eqref{eq:4-vector W for gbms} as
\be\la{W-comp}
M^\f13 S_{i}
&=-\sqrt{\rho^{\va \msf{R}}}  
\bigintsss_S \epsilon^{AB} \pa_A (\sqrt{\rho_{v}} n_i)\, \rho_{v}^{-1} j_B \,\bfe
\cr
&=-\gamma_{v} \bigintsss_S  (1-v^jn_j) \epsilon^{AB} \pa_A n_i\,  j_B \,\bfe
- 
\gamma_{v}\bigintsss_S v^j n_i \epsilon^{AB} \pa_An_j\, j_B \,\bfe
\cr
&=\gamma_{v} J_i
- 
\gamma_{v}\bigintsss_S v^j\epsilon^{AB}( n_i  \pa_An_j- n_j  \pa_An_i)\, j_B \,\bfe
\cr
&=\gamma_{v} J_i
- 
\varepsilon_i{}^{jk} \gamma_{v}v^j\bigintsss_S \epsilon^{AB}\epsilon_A{}^C \p_C n_k\, j_B \,\bfe
\cr
&=\gamma_{v} ( J_i
- 
\varepsilon_{ijk} v^j K^k)\cr
&= M^{-1}\left( E J_i
- 
\varepsilon_{ijk} P^j K^k \right) = M^{-1} W_j
\ee
where $P^\mu=(E, P_i)= M\gamma_{v}(1, v^i)$.
We have used the last identity in \eqref{nid}, and the rotation and boost generator definitions \eqref{Lor-gen}. We thus recover the expression \eqref{eq:components of the Pauli--Lubanski vector} for the spatial components of the Pauli--Luba\'nski pseudo-vector in Poincar\'e. 

\smallskip It is important to note that this expression is invariant under supertranslation.
To understand how this is possible we 
need to recall that McCarthy showed that the orbits associated with the 
the constant  mass aspects of $\bms$ algebra  have non-trivial isotropy subalgebras.
 From Table \ref{tab:isotropy subalgebras of BMS}, the isotropy group is isomorphic to SU(2). In this restrictive case, we therefore expect a formula for the spin in terms of the $\bms$ generators. This is the formula we just unraveled in \eqref{W-comp}, that was first given  in \cite[Eq. 287]{Compere:2023qoa} and where it was shown to be supertranslation invariant.
 
\smallskip For completeness,  we show in Appendix \ref{app:PLalg} that the  Pauli--Luba\'nski pseudo-vector algebra \eqref{eq:Lie bracket of components of Pauli--Lubanski pseudo-vector} is reproduced when we perform the constant mass orbit restriction \eqref{rhov}.  Namely, we have 
\be\la{PLalg}
 \msf{S}[\{\rho_v^{\f12}n_\mu,\rho_v^{\f12}n_\nu\}_\rho]=- {M}^{-\f43} \varepsilon_{\mu\nu \rho \sigma} P^\rho S^\sigma\,.
\ee
 The main steps of the proof involve expanding 
\be
\msf{S}[\{\rho_{v}^{\f12}n_\mu,\rho_{v}^{\f12}n_\nu\}_\rho] &=\varepsilon_{\mu\nu}^{\hphantom{\mu\nu}\rho 0}
\bigintsss_S 
n_\rho
 \,\epsilon^{CD}\partial_Cp_D\,
\bfe
\cr
&
+\f{1}2 \bigintsss_S 
\epsilon^{AB} (n_\mu\p_B n_\nu- n_\nu\p_B n_\mu) \rho_{v}^{-1}\, \partial_A  \rho_{v}
 \,\epsilon^{CD}\partial_Cp_D\,
\bfe\,,
\ee
and  using the relations 
\be
&\f12 \rho_{v}^{-1}\, \partial_A  \rho_{v}=\f{v^\ell\p_A n_\ell}{(1-v\cdot \widehat{n})}\,,\cr
&\epsilon^{AB} (n_i\p_B n_j- n_j\p_B n_i)
=-\varepsilon_{ij}\!^k\p^A n_k\,,
\cr
&
\epsilon^{AB} (n_i\p_B n_0- n_0\p_B n_i)
=-\epsilon^{AB}\p_B n_i\,.
\ee
In light of these results, we see that if we introduce the {\it Pauli--Luba\'nski generator}  for a Poincar\'e embedding inside $\bms$
\be\label{PL2}
W_{\mu} :=M^{\f43} \msf{S}[\sqrt{\rho_{v}} n_\mu]\,,
\qquad
P^{\mu} :=  \msf{P}[ n_\mu].
\ee
which is:
\begin{itemize}
\item[$-$] supertranslation invariant, as encoded in \eqref{PW},
\item[$-$] covariant under Lorentz transformations, as \eqref{eq:gbms bracket of W and angular momentum generators}  implies, and
\item[$-$] 
satisfying the algebra 
\be\label{eq:poisson bracket of pauli-lubanski with itself, gbms}
\{W_{\mu},W_{\nu}\}_{\g^*}=\varepsilon_{\mu\nu \rho \sigma} P^\rho W^\sigma\,.
\ee
\end{itemize}

\section{Applications}\label{sec:applications}

In this section, we first introduce in  \S\ref{sec:the gbms moment map} a phase space on which $\gbms$ acts by Hamiltonian transformations and construct the moment map for this action. 
We then consider two applications for the formalism developed so far: (1)  In \S \ref{sec:phase space quantities and their evolution}, we use the moment map to construct the gravitational spin charge and gravitational Casimirs in asymptotically-flat spacetimes with $\gbms$ as their asymptotic symmetries;   (2)in \S \ref{sec:gravitational Casimirs for the kerr metric}, we write down explicitly the gravitational Casimir invariants for the Kerr metric.

\subsection{$\gbms$: The moment map}\label{sec:the gbms moment map}
 In \S\ref{sec:gbms and poincare embeddings} and \S\ref{sec:algebraic aspects of gbms}, we studied algebraic aspects of the $\gbms$ algebra including the study of its coadjoint orbits, the invariants of these orbits, its isotropy subalgebra and defined its generator as the spin generator of $\gbms$.  We now want to translate these results into statements about gravitational physics. As usual, this is achieved by noting that there is a gravitational phase space $\phasespace$ that carries the Hamiltonian action of the $\gbms$ algebra \cite{CampigliaPeraza202002,Freidel:2021qpz,Sudhakar:2023uan},
 and hence there is a moment map $\mu_{\gbms}:\phasespace\to\gbms^*$. By definition, $\mu_{\gbms}$ is a smooth map, and as such can be used to construct the phase space quantities from those of $\gbms^*$, which we have constructed in the previous sections, by the pull-back operation. Therefore, the aim of this section is two-fold: (1) we construct the gravitational phase space that carries a Hamiltonian action of $\gbms$ in \S \ref{sec:Norad}, and then (2) we build the moment-map $\mu_{\gbms}$ for the $\gbms$ action on this phase space in \S \ref{sec:gravitational moment map}. The construction of $\mu_{\gbms}$ will be used for the formulation of the gravitational spin charge and gravitational Casimirs in asymptotically-flat spacetimes, which is the subject of \S\ref{sec:phase space quantities and their evolution}.

\subsubsection{Radiative phase space at null-infinity}\la{sec:PS-null-infinity}

In  this section, we review some of the main results established in \cite{Freidel:2021qpz}, as these will be pivotal to the construction of the moment map in the following sections.
 In the Bondi--Sachs coordinates, the metric is given by \cite{BondivanderBurgMetzner196208, Sachs196212}
\be \label{eq:BondiMetric}
\rd s^2 = -2e^{2\beta} \rd u \left( \rd r  + \Phi \rd u\right) + r^2 \Gamma_{AB} \left(\rd \sigma^A -\f{\Upsilon^A}{r^2}\rd u\right)\left(\rd \sigma^B -\f{\Upsilon^B}{r^2}\rd u\right).
\ee
This metric satisfies the Bondi gauge conditions given by  
$
g_{rr}=0, g_{rA}=0, \pa_r \sqrt{\Gamma}=0$.  In addition to the gauge conditions we 
impose \emph{extended}\footnote{Note that we do not require Ricci scalar of the 2D sphere metric $R(\gamma)=2$,  which is why we call our boundary conditions \emph{extended}.} Bondi asymptotic boundary conditions 
\cite{Barnich:2009se,BarnichTroessaert201001,BarnichTroessaert201601}, which means that the metric components $(\Phi, \beta, \Gamma_{AB}, \Upsilon^A)$ admit the following asymptotic expansion 
\begin{subequations}\la{eq:FallOff}
\begin{align}
\Phi&= \frac{R({\gamma})}4 - \frac{  M}{r}+\mcal{O}\left(\f1{r^2}\right)\,,\\
\beta&=-\frac{1}{32 } \frac{  C_{AB} C^{AB}}{r^2}+\mcal{O}\left(\f1{r^3}\right)\,,\\
\Upsilon^A&=-\frac12 {\cD}_B C^{BA}- 
\frac{1}{  r } 
\left(\f23 \cJ^A- \frac12 C^{AB} {\cD}^C C_{CB} - \frac{1}{16 } \pa^A  \left(C_{BC} C^{BC}\right) \right) +\mcal{O}\left(\f1{r^2}\right)\,,\label{PA}\\
\Gamma_{AB}&=  {\gamma}_{AB}  + \frac{1}{r}C_{AB} + \frac{1}{ 4r^2} \gamma_{AB}\left(C_{CD}C^{CD} \right)
+  \mcal{O}\left(\f1{r^3}\right)\,,
\la{gamma}
\end{align}
\end{subequations}
where $M$ is the Bondi mass, $C_{AB}$ the asymptotic shear, while ${\cal J}_A$ is   the covariant angular momentum  
\cite{Freidel:2021qpz}. 
When evaluating  asymptotic quantities, we use the metric ${\gamma}_{AB}$  to lower and raise the indices $\{A, B, \dots\}$ on the 2-sphere. 
		
\smallskip The expansions of the different coefficients are  needed to obtain the 
expansion of the metric $ \rd s^2 $ 
to order\footnote{Since $\rd r$ is of order $\mcal{O}(r)$, $g_{ur}$ needs to be expanded to order $\mcal{O}(r^{-2})$, since $g_{AB}= r^2 \Gamma_{AB}$,  $\Gamma_{AB}$ needs to be expanded to order $\mcal{O}(r^{-3})$ and since $g_{uA} = \Gamma_{AB} \Upsilon^A$,  $\Upsilon^A$ needs to be expanded to order $\mcal{O}(r^{-1})$,  to achieve  $\mcal{O}(r^{-1})$ for the expansion of  the metric $g_{ab} \rd x^a \rd x^b$.} $\mcal{O}(r^{-1})$. The demand that the metric is asymptotically-flat and asymptotic Einstein’s equations (EEs) {are satisfied would} impose that the asymptotic sphere metric is time independent $\pa_u \gamma_{AB}=0$.
In the following, we denote by ${\cD_A}$  the covariant derivative associated with $\gamma_{AB}$.
All the other coefficients are time-dependent functions on the sphere, i.e. $M= M(u,\s^A)$.
Because of the Bondi determinant gauge condition $r^4\sin^2\theta=\det(\gamma_{AB})$, the symmetric tensor $C_{AB}$ is traceless  when contracted with the inverse asymptotic metric $\gamma^{AB}$.
The  $\mcal{O}(r^{-2})$ factor in the metric expansion is uniquely  determined by the Bondi gauge condition and the demand that logarithmic anomalies vanish  \cite{Winicour16} (see e.g. \cite{Compere:2018ylh,Freidel:2021fxf, Freidel:2024tpl} for a relaxation of this gauge condition).

\smallskip We can now summarize the covariance properties  revealed in \cite{Freidel:2021qpz} and  the nested structure that organizes them.
To study the dynamics of asymptotic gravity, it is important to split the observables associated into radiative observables and corner observables.
To do so let us introduce some notation where $N^{AB}$ is the covariant news and $\cN^{AB}$ is its time derivative
\be
 N^{AB} := \dot{C}^{AB} - T^{AB},\qquad \cN^{AB} := \dot N^{AB}=\ddot C_{AB}\,,
\ee
where  $T^{AB}$ is the traceless component of the Geroch tensor \cite{Geroch1977} introduced above. Their  transformations under the GBMS action are
\begin{eqaligned}
    \la{BMSW-trans}
\d_{(\tau,Y)} q^{AB} &=\left[  \cL_Y + 2\dot{\tau} \right]  q^{AB}, \cr
\d_{(\tau,Y)} C^{AB} &=\left[  \cL_Y + 3\dot{\tau} \right]  C^{AB}
 - \left(2 \cD^{\langle A}\cD^{B\rangle} - \dot{C}^{AB}\right)\tau, \cr
 \d_{(\tau,Y)}T_{AB}&=\cL_Y T_{AB}-2\cD_{\langle A} \cD_{B\rangle}\dot \tau\,,
 \cr
 \d_{(\tau,Y)} N^{AB} &=\left[\tau\pa_u+  \cL_Y + 4\dot{\tau} \right]  N^{AB}
 \,, \cr
  \d_{(\tau,Y)} \cN^{AB} &=\left[\tau\pa_u+  \cL_Y + 5\dot{\tau} \right]  \cN^{AB},
\end{eqaligned}  
where we have introduced the parameter  $\tau=\tau(T,Y)$ given by
\begin{equation}\label{eq:tau and its time derivatives}
    \tau :=T+ \frac{u}{2} \cD_AY^A\,,\qquad
\dot{\tau} = \frac12 \cD_AY^A\,,
\qquad \ddot{\tau}=0.
\end{equation}

One also introduces the energy current\footnote{In order to conform to the standard convention  for the angular momentum, we have implemented the following change of notation w.r.t. \cite{Freidel:2021qpz}: $\cJ_{\rm here}^A= \cP^A_{\rm there}, \cI^A_{\rm here}=\cJ^A_{\rm there}$.} 
\be\la{cI}
\cI^A :=   \f12  \cD_B \dot{C}^{AB} + \f14 \pa^A R=  \f12 \cD_B  N^{AB}\,,
\ee
 as well as  the covariant mass and the covariant dual mass \cite{Godazgar:2018qpq,Godazgar:2018dvh,Godazgar:2019dkh, Freidel:2021qpz}
\begin{eqaligned}
{\cM}&:= M + \f18C_{AB} (N^{AB} +T^{AB})\,,
\cr
\tcM&:= \frac14 \cD_{A} \cD_{B}\widetilde{C}^{AB} - \f18 \widetilde C_{AB} (N^{AB} +T^{AB})\,,\la{tcM}
\end{eqaligned}
where we used the complex structure to define the duality operation
	\be\la{dualC}
	\widetilde{C}^{AB}:= \epsilon^{A}{}_C C^{C B}\,,\qquad {\widetilde \cD}_A:=\epsilon_{A}{}^B \cD_B\,.
	\ee
The equations of motions are given by \cite{Freidel:2021qpz}
\begin{eqaligned}\la{ceom-intro}
\dot \cI^A & = \f12  \cD_B \cN^{AB}\,,\\ 
\dot \cM &= \f12  \cD_A \cI^A  +  \f18  C_{AB} {\cN}^{AB}\,,\\
\dot{ \tcM} &= \f12  \cD_A \widetilde \cI^A  +  \f18  C_{AB} {\widetilde \cN}^{AB}\,,\\
\dot\cJ_A &=  \cD_A \cM + \widetilde{\cD}_A\tcM  + C_{AB} \cI^B\,.
\end{eqaligned}

\subsubsection{Non-radiative phase space}\la{sec:Norad}
As shown in \cite{Freidel:2021qpz}, the no-radiation condition is obtained by demanding that
\be\la{NR}
\cN^{AB}=0\,, \qquad \cI^A=0\,.
\ee
This condition is preserved by the symmetry transformation and implies the presence  of conserved corner charges constructed in  \cite{Freidel:2021qpz}. This means that $\cN^{AB}, \cI^A$  represent radiative data.  	We denote the  non-radiative phase space  implementing  \eqref{NR} with $\phasespace_{\va \mathrm{NR}}$. The ultimate goal of our study is to understand whether $\phasespace_{\va \mathrm{NR}}$ can be understood  as a gravitational phase space isomorphic to coadjoint orbits of some extended symmetry group.\footnote{A more precise statement is made in Appendix \ref{sec:moment maps and coadjoint orbits}.}

\smallskip In this paper, we tackle  a simpler, but still highly non-trivial endeavour, which is to understand a more restricted phase space as the union of coadjoint orbits of the GBMS group. The simplification we  take is to restrict the set of gravitational configurations which are  non-radiative and    also assume  that  the covariant dual mass \eqref{tcM} vanishes. 
This last phase space is the one universally studied in  the asymptotic symmetry literature and exclusively studied in the celestial holography literature. It imposes 
\be 
\cN^{AB}=0\,, \qquad \cI^A=0\,,\qquad {\tcM}=0\,.
\ee
We call this phase space electric\footnote{This refers to the fact that the mass and dual mass are the gravitational analog of the electric and  magnetic charge aspects.}  and   non-radiative (ENR) and denote it as $\phasespace_{\va \rm ENR}$. Obviously, we have the inclusion
$
\phasespace_{\va \mathrm{ENR}}\subset \phasespace_{\va \mathrm{NR}}$. 
More precisely, we have the following relation between these phase spaces 
\be
\phasespace_{\va \mathrm{ENR}}= \phasespace_{\va \mathrm{NR}}/\{\tcM =0\}. 
\ee
The important point is that the condition  $\tcM=0$ is also preserved by the $\gbms$ symmetry transformations \cite{Freidel:2021qpz}.

\paragraph{Corner symmetry transformations.}

Within the electric  and  strongly non-radiative phase space,
the  set of observables under consideration are the covariant corner observables   $(\cM,\cJ_A)$. 
These variables correspond to the leading asymptotic contributions of the Weyl scalars ${\rm Re}(\Psi_2),\Psi_1$ respectively ($\cN^{AB}$ corresponding to the asymptotic value of $\Psi_4$ with $\cI^A$ corresponding to the asymptotic value of $\Psi_3$ and $\widetilde{\cM}$ corresponding to the asymptotic value of $\mathrm{Im}(\Psi_2)$).
	Within $\phasespace_{\va \mathrm{ENR}}$, the corner symmetry transformations of these variables are given by\footnote{Analog transformations were found  in \cite{Barnich:2011ty, BarnichTroessaert201601, Barnich:2019vzx}   in the Penrose--Newman formalism for the extended BMS group \cite{Barnich:2009se, Barnich:2011mi}, where the diffeomorphisms are restricted to be local Killing vector fields. The derivation in \cite{Freidel:2021qpz} relaxed this restriction by including the full group of sphere diffeomorphism.}  \cite{Freidel:2021qpz}
	
		\begin{subequations}\la{dc}
		\bea
		\d_{(\tau,Y)} {\cal M} &=&
		\left[\cL_Y + 3 \dot{\tau} \right]{\cal M}\,,\label{eq:transformation of mass aspect} \\
		\d_{(\tau,Y)} \cJ_A &=& \left[ \cL_{Y} + 2 \dot\tau\right] \cJ_A + 
		3 {\cal M} \pa_A \tau+ \tau\p_A \cM \,,\label{eq:transformation of angular momentum aspect} 
		\eea
	\end{subequations}
	where we have used the asymptotic equations of motion (in $\phasespace_{\va \mathrm{ENR}}$, following from \eqref{ceom-intro})
		\begin{subequations}\la{ceom}
		\be
		\dot \cM = 0\,,\qquad 
		\dot\cJ_A =  \cD_A \cM \,.\la{ceom-P}
		\ee
	\end{subequations}

\paragraph{Conserved charges.}%
In the following we will treat the metric variable $\gamma^{AB}$, $C^{AB}, N^{AB}$ as background structure and refer to the phase space purely as functionals of the conserved charge aspects $(\msf{m}, \msf{j_A})$, namely
		\be\label{eq:shear as functional of m and j}
		\phasespace_{\va \mathrm{ENR}} = C[\msf{m},\msf{j}_A]\,, 
		\ee
		keeping the metric and shear  dependence as implicit.
	From the evolution equations \eqref{ceom}, it is immediate to see that these $u$-independent aspects on the sphere  are given by\footnote{The shift of the angular momentum aspect by the derivative of the mass appeared in \cite{Hawking:2016sgy}.}
\be\label{eq:charge aspects in terms of phase space quantities}
\msf{m}&=\cM\,,\\
\msf{j}_A&=\f12\left[\cJ_A- u \cD_A\cM \right]\,.\la{pA} 
 \ee
These conserved charge aspects on
the asymptotic sphere parameterize the electric and strongly non-radiative phase space, as \eqref{eq:shear as functional of m and j} entails. 
The associated   charges which are conserved under the time evolution in $\phasespace_{\va \mathrm{ENR}}$ are given by
\be\label{eq:general expression of phase space conserved charges}
	Q_{(Y,T)}
	&=  \bigintsss_S \left( T\msf{m} +  Y^A \msf{j}_A \right)\bfe\,.
	\ee
  These are the phase space analogs of the sum of the $\gbms$ charges \eqref{eq:gbms charges}.
	
\smallskip From  \eqref{dc}, the $\gbms$ action on the conserved charge aspects parametrizing $\phasespace_{\va \rm ENR}$ is given by 
	\begin{subequations}\la{ddc3}
		\be
		\d_{(Y,T)} \msf{m} &=
		\left[\cL_Y + 3 W_Y \right]\msf{m}\, , \la{A1}\\
		\d_{(Y,T)} \msf{j}_A &= \left[ \cL_{Y} + 2 W_Y \right] \msf{j}_A + 
		\f32 \msf{m} \pa_A T
		+ \f{T}{2} \p_A \msf{m} \,.\la{A2} 
		\ee
	\end{subequations}
Having these transformations at hand, we are now ready to construct the moment map for the $\gbms$ action on the gravitational phase $\phasespace_{\va \text{ENR}}$, which we turn to.

\subsubsection{Gravitational moment map}
\label{sec:gravitational moment map}

From the comparison of the first two of the symmetry transformations \eqref{ddc3} 
  with the infinitesimal coadjoint action  \eqref{eq:coadjoint action on gbms*}, the moment-map $\mu_\gbms : \phasespace_{\va \rm ENR} \to \gbms^{*}$ for the $\gbms$ action on 
$
\phasespace_{\va \rm ENR}=   C[\,\msf{j}_A,\msf{m}]$ can be straightforwardly found. It is simply given by  
	\be\label{eq:gbms moment map}
	\mu_{\gbms}(\msf{m})=m,\qquad \mu_\gbms(\msf{j}_A)=j_A.
	\ee
	In other words, $\msf{m}$ is the charge aspect for the supertranslations and $\msf{j}_A$ is the charge aspect for the diffeomorphisms.
	An analog result for the BMS group was first shown by Barnich and Ruzziconi \cite{Barnich:2021dta} who recognized, 
	using the Newman--Penrose formalism, that under the conditions that $\Psi_4=\Psi_3=\mathrm{Im}(\Psi_2)=0$, the asymptotic Weyl scalars $\mathrm{Re}(\Psi_2) $ and $\Psi_1$ could be understood as coadjoint orbit labels for the BMS group.
	We recover the same results from \eqref{eq:gbms moment map} in a different and somewhat more direct fashion.

\subsection{Phase space quantities and their evolution}
\label{sec:phase space quantities and their evolution}

In \S\ref{sec:spin generator of gbms} and \S\ref{sec:bms pauli-lubanski}, we have defined three quantities belonging to $\gbms^*$: the Casimirs $\msf{C}_n(\gbms)$ in \eqref{eq:gbms casimir functions}, the spin generator $\msf{S}_\chi$ in \eqref{eq:gbms spin generator}, and the Pauli--Luba\'nski-like element $S_{\mu}$ in \eqref{eq:4-vector W for gbms}. As we have the moment map, the natural next step is to find the quantities defined on $\phasespace_{\va \text{ENR}}$ corresponding to these quantities. This can be achieved by pulling-back $\msf{C}_n(\gbms)$, $\mathsf{S}_\chi$, and $W_{\mu}$ by $\mu_{\gbms}$ (as a smooth map) to $\phasespace_{\va \text{ENR}}$. It is the aim of this section to determine these phase space quantities, and compute their time evolution.

\subsubsection{Gravitational Casimirs and the spin generator}\label{sec:gravitational casimirs and spin generator}

By construction, $\mu_\gbms$ is an equivariant momentum map, hence is a Poisson map (see Theorem \ref{thr:equivariant moment maps are poisson}). From \eqref{eq:charge aspects in terms of phase space quantities} and \eqref{eq:gbms moment map}, we see that the momentum map for $\gbms$ is given explicitly as
\begin{equation}\label{eq:moment map for gbms}
    \mu_{\gbms}({\mcal{M}})=m, \qquad \mu_\gbms(2^{-1}(\mcal{J}_A-u\cD_A\mcal{M}))=j_A.
\end{equation}
It then follows from \eqref{vortdef} that
\begin{eqaligned}\label{eq:pulled-back vorticity}
  \textbf{w}:=  \mu^{*}_\gbms w
    =\frac{1}{2}\mcal{M}^{-\frac{2}{3}}\epsilon^{AB}\partial_A\left(\mcal{M}^{-\frac{2}{3}}\mcal{J}_B\right)\,,
\end{eqaligned}
which shows that the phase space quantity associated with the vorticity only depends on the corner data on $S$. 

\paragraph{Gravitational Casimir functionals.} 
We can now define the phase space quantities corresponding to Casimir functionals \eqref{eq:gbms casimir functions} as
\begin{eqaligned}\label{eq:pulled-back casimirs}
    \mscr{C}_{n}(\phasespace_{\va \text{ENR}}):=\mu^*_\gbms\msf{C}_n(\gbms)=\bigintsss_S\,\mcal{M}^{\frac{2}{3}}\textbf{w}^n\,\bfe,
\end{eqaligned}
where
\begin{equation}
    \textbf{w}^n=\frac{1}{2^n}\mcal{M}^{-\frac{2}{3}n}\epsilon^{A_1B_1}\ldots \epsilon^{A_nB_n}\partial_{A_1}(\mcal{M}^{-\frac{2}{3}}\mcal{J}_{B_1})\ldots\partial_{A_n}(\mcal{M}^{-\frac{2}{3}}\mcal{J}_{B_n}).
\end{equation}
We call $\mscr{C}_{n}(\phasespace_{\va \text{ENR}})$ the {\it gravitational Casimir functionals}. By construction, these quantities are invariant under $\gbms$ action on $\phasespace_{\va \text{ENR}}$, namely $\delta_{(Y,T)}\mscr{C}_n(\phasespace_{\va \text{ENR}})=0$. This fact can be checked by a direct computation analog to the proof of \eqref{eq:invariance of casimirs under gbms}.

\paragraph{Gravitational spin charge for $\gbms$.} In \S\ref{sec:spin generator of gbms}, we have identified the smeared vorticity \eqref{eq:gbms spin generator} as the spin charge of $\gbms$. We can define the associated phase space quantity by pulling it back with $\mu_\gbms$.
Hence, we define the {\it gravitational spin charge} in asymptotically-flat spacetimes with $\gbms$ as their asymptotic symmetry group as
\begin{equation}\label{eq:the gravitational spin charge}
   \mscr{S}_\chi:=\mu^*_\gbms \msf{S}_\chi=\frac{1}{2}\bigintsss_S \chi\,\epsilon^{AB}\partial_A\left(\mcal{M}^{-\frac{2}{3}}\mcal{J}_B\right)\bfe,
\end{equation}
which, by construction, is the generator for the action of the isotropy algebra $\diff_{\mbs{\rho}}(S)$ of $\gbms$ on $\phasespace_{\va \text{ENR}}$, i.e.
\begin{equation}
    \{\mscr{S}_\chi,\mscr{S}_\psi\}_{\phasespace_{\va \tenofo{ENR}}}=-\mscr{S}_{\{\chi,\psi\}_\rho}.
\end{equation}
This relation follows from \eqref{eq:poisson lie bracket of smeared vorticities} and \eqref{eq:property of poisson morphisms}.

\paragraph{Gravitational Pauli--Luba\'nski generator.}  The next important quantity that we have constructed for $\gbms$ is the Pauli--Luba\'nski generator for a Poincar\'e embedding. Its gravitational counterpart can be defined by considering first  the pullback of the quantity $S_\mu$ in \eqref{eq:4-vector W for gbms} 
\begin{eqaligned}
    \mscr{S}_\mu&:=\mu^*_{\gbms}{S}_\mu
    \\
    &=\frac{1}{2}\bigintsss_S\,\mcal{M}^{\frac{1}{3}}\epsilon^{AB}\partial_A\left(\mcal{M}^{-\frac{2}{3}}\mcal{J}_B\right)\,n_\mu\,\bfe\,,
\end{eqaligned}
 From \eqref{PW}, \eqref{eq:gbms bracket of W and angular momentum generators}, and \eqref{eq:property of poisson morphisms}, it follows that these quantities satisfy  
\begin{equation}
    \{\mscr{P}_\mu,\mscr{S}_\nu\}_{\phasespace_{\va \text{ENR}}}=0, \qquad \{\mscr{J}_{\mu\nu},\mscr{S}_\rho\}_{\phasespace_{\va \text{ENR}}}=\eta_{\nu\rho}\mscr{P}_\mu-\eta_{\mu\rho}\mscr{P}_\nu,
\end{equation}
 where
\begin{eqaligned}
    \mscr{P}_\mu&:=\mu^*_\gbms P_\mu=\bigintsss_S\,n_\mu\mcal{M}{\bfe}\,,
    \\
    \mscr{J}_{\mu\nu}&:=\mu^*_\gbms J_{\mu\nu}=\frac{1}{2}\bigintsss_S\,Y^A_{\mu\nu}\mcal{J}_A{\bfe}\,,
\end{eqaligned}
 are charges that generate the action of four-translations and Lorentz transformations on $\phasespace_{\va \text{ENR}}$, respectively. 

\smallskip Finally, using \eqref{PL2}, we can then define the gravitational Pauli--Luba\'nski generator  as
\begin{eqaligned}
    \mscr {W}_{\mu}&:=\mu^*_\gbms {W}_{\mu}
    \\
    &=\frac{1}{2}\mcal{M}_{\msf{R}}^{\frac{4}{3}}\bigintsss_S\,\left(\frac{\epsilon^{AB}\partial_A\left(\mcal{M}^{-\frac{2}{3}}\mcal{J}_B\right)}{\gamma_v(1-v\cdot \widehat{n})}\right)n_\mu{\bfe}\,.
\end{eqaligned}
where,  $\mcal{M}_{\msf{R}}$, denotes the total rest mass.
Note that \eqref{eq:poisson bracket of pauli-lubanski with itself, gbms} and \eqref{eq:property of poisson morphisms} imply the algebra of gravitational Pauli--Luba\'nski generator
\begin{equation}
    \{\mscr{W}_{\mu},\mscr{W}_{\nu}\}_{\phasespace_{\va \text{ENR}}}=\varepsilon_{\mu\nu\rho\sigma}\mscr{P}^\rho\mscr{W}^\sigma. 
\end{equation}

\subsubsection{Evolution equations}\label{sec:evolution equations}

In the presence of radiation (while keeping the condition ${\tcM}=0$), the charge aspects \eqref{pA} are no longer conserved, but satisfy  the equations of motion 
\be
\dot \cM &= \f12  \cD_A \cI^A  +  \f18  C_{AB} {\cN}^{AB}\,,\la{ceom-m}\\
 \dot \cJ_A &=\p_A \cM +C_{AB} \cI^B
 \,.
\ee
Accordingly, the gravitational spin generator evolves as
\be
\dot {\mscr{S}}_\chi &=\frac{1}{2}\bigintsss_S \chi\,\epsilon^{AB}\partial_A\left(\dot \cM^{-\frac{2}{3}}\mcal{J}_B
+\mcal{M}^{-\frac{2}{3}}\dot \cJ_B\right)\bfe
\cr
&=
\bigintsss_S  \cM^{-\f23} \epsilon^{AB}\pa_A \chi \left (
 \f13 \cM^{-1}\dot \cM \cJ_B
-\f12 C_{BC} \cI^C
\right) \bfe\,.
\ee
This charges are conserved when the non-radiative conditions $\cN^{AB}=\mcal{I}^A=0$ are imposed.

\subsection{Gravitational Casimirs for the Kerr metric}\label{sec:gravitational Casimirs for the kerr metric}

In the previous section, we have constructed the phase space quantities associated with the Casimir invariants of $\gbms$. In this section, we construct these quantities concretely for the particular example of the Kerr metric. Note that the asymptotic symmetries (in our case, the $\gbms$ algebra) belong to the boundary of (conformal completion) of an asymptotically-flat spacetime. As we are going to construct the conserved quantities, i.e. Casimir functionals, for the $\gbms$ algebra, we need to write the large-distance form of the metric of a given spacetime in a suitable coordinate system. As usual, it is the Bondi--Sachs coordinates that will be the most convenient for our purpose \cite{BondivanderBurgMetzner196208,Sachs196212,TamburinoWinicour196610}. Therefore, we first write the Kerr metric in the Bondi--Sachs coordinates, which we then use to construct mass and angular momentum aspects, $\mcal{M}$ and $\mcal{J}$, respectively, for the Kerr metric. We then use these quantities to construct gravitational Casimirs for the Kerr metric. Finally, we comment on the implications of the No-Hair Theorem for our gravitational Casimir invariants.

\subsubsection{The Kerr metric in the Bondi--Sachs coordinates}

In this section, we write the large-distance behavior of the Kerr metric in the Bondi--Sachs coordinates (regarding this expansion, see also \cite{Barnich:2011mi,Compere:2019gft}). The procedure is to first consider the Kerr metric in the generalized Bondi--Sachs  coordinates (see Eq. \eqref{eq:definition of gbs coordinates} for the definition), and then do a change of coordinate to the Bondi--Sachs coordinates, fixed by the so-called Bondi--Sachs gauge. In the following, we only mention the final results, and relegate the details of derivation following the procedure explained in \cite{FletcherLun200309} (which is basically a change of coordinates at large radius) to Appendix \ref{sec:writing the kerr metric in the bondi-sachs coordinates}. 

\smallskip We denote the Bondi--Sachs coordinates by $\{u,r,\theta,\varphi\}$. The components of the metric, which we denote as $g_{\mu\nu}$, are given as follows (see Appendix \ref{sec:writing the kerr metric in the bondi-sachs coordinates}). The components along $u$ are
\begin{eqaligned}\la{guu}
    g_{uu}&=-1+\frac{2M}{r}+\mcal{O}(r^{-2}),
    \\
    g_{ur}&=-1+\frac{a^2}{r^2}\left(\frac{1}{2}-\cos^2\theta+\frac{1}{8}\left[\frac{1+4\sin^2\theta-8\sin^4\theta}{\sin^2\theta}\right]\right)+\mcal{O}(r^{-4}),
    \\
    g_{u\theta}&=\frac{a\cos\theta}{2\sin^2\theta}+\frac{a\cos\theta}{4r}\left(8M+\frac{a}{\sin^3\theta}\right)+\mcal{O}(r^{-2}),
    \\
    g_{u\varphi}&=-\frac{2aM\sin^2\theta}{r}+\mcal{O}(r^{-2}),
\end{eqaligned}
and the other components are 
\begin{eqaligned}\la{gtp}
    g_{\theta\theta}&=r^2+\frac{a}{\sin\theta}r+\frac{a^2}{2\sin^2\theta}+\mcal{O}(r^{-1}),
    \\
    g_{\theta\varphi}&=\frac{2a^2M\cos\theta}{r}+\mcal{O}(\br^{-2}),
    \\
    g_{\varphi\varphi}&=r^2\sin^2\theta-ar\sin\theta+\frac{a^2}{2}+\mcal{O}(r^{-1}).
\end{eqaligned}
In these expression, $M$, $J$, and $a$ are the mass, angular momentum, and the reduced angular momentum, defined in \eqref{eq:reduced angular momentum}. Equipped with this large-$r$ form of the Kerr metric, we can now construct the gravitational Casimir functionals, to which we now turn to.

\subsubsection{Constructing Casimirs for the Kerr Metric}\la{sec:Kerr-Casimirs}

The Casimirs $\mscr{C}_n$ have been defined in terms of the scalar vorticity $w$, which has been constructed in terms of $m$ and $j$ (see \eqref{vortdef}). The corresponding phase space quantity was obtained in \S \ref{sec:the gbms moment map} by means of the $\gbms$ moment map. This is given by the expression \eqref{eq:pulled-back vorticity}. 
Hence, we need to construct the covariant mass $\mcal{M}$ and covariant angular momentum $\cJ$ for the Kerr metric. They are given by (see Appendix \ref{sec:gravitational casimir functionals, appendix})
\begin{equation}\label{eq:kerr mass and momentum aspects in the phase space}
    \mcal{M}=M, \qquad \cJ=3aM(\cos\theta\,\rd\theta-\sin^2\theta\,\rd\varphi).
\end{equation}

We can thus construct the  Kerr gravitational vorticity $\mbf{w}^{\va \tenofo{Kerr}}$ as
\begin{equation}
    \mbf{w}^{\va \tenofo{Kerr}}=-3aM^{-\frac{1}{3}}\cos\theta.
\end{equation}
Finally, the gravitational Casimir functionals \eqref{eq:pulled-back casimirs}  for the Kerr metric are given by (see Appendix \ref{sec:gravitational casimir functionals, appendix})
\begin{equation}\label{eq:gravitational Casimir functionals for the Kerr metric, main text}
    \mscr{C}_n(\phasespace^{\va \text{Kerr}}_{\va \text{ENR}})=
    \left\{
    \begin{aligned}
    \frac{(-3a)^n}{n+1}&M^{\frac{2-n}{3}}, &\qquad n&=0,2,4,\ldots,
    \\
    &0, &\qquad n&=1,3,5,\ldots.
    \end{aligned}
    \right.
\end{equation}
In particular, all the odd gravitational Casimir functionals vanish.

\subsubsection{Casimirs, Kerr parameters, and the No-Hair Theorem}

Having the explicit expression of the gravitational Casimirs for the Kerr metric, we can now compute them explicitly for any value of $n$. The first few of them are
\begin{eqgathered}
\mscr{C}_0(\phasespace^{\va \text{Kerr}}_{\va \text{ENR}})=M^{\frac{2}{3}}, \qquad 
    \mscr{C}_1(\phasespace^{\va \text{Kerr}}_{\va \text{ENR}})=0, 
    \\
    \mscr{C}_2^{\va \tenofo{Kerr}}=3a^2=\frac{3J^2}{M^2c^2}\,.
\end{eqgathered}
 Therefore, we can express the Kerr parameters $M$ and $J$ in terms of $\msf{C}_0^{\va \tenofo{Kerr}}$ and $\msf{C}_2^{\va \tenofo{Kerr}}$ as
\begin{eqaligned}\label{eq:mass and angular momentum of kerr in terms of C0 and C2}
    M=\left(\mscr{C}_0(\phasespace^{\va \text{Kerr}}_{\va \text{ENR}})\right)^{\frac{3}{2}},\qquad J=\left(\frac{c^2}{3}\cdot \mscr{C}_0(\phasespace^{\va \text{Kerr}}_{\va \text{ENR}})^3\mscr{C}_2(\phasespace^{\va \text{Kerr}}_{\va \text{ENR}})\right)^{\frac{1}{2}}.
\end{eqaligned}
On the other hand, from \eqref{eq:gravitational Casimir functionals for the Kerr metric, main text}, $\mscr{C}_{2n}(\phasespace^{\va \text{Kerr}}_{\va \text{ENR}})$ is given by
\begin{eqaligned}\label{eq:higher casimirs in terms of C0 and C2}
    \mscr{C}_{2n}(\phasespace^{\va \text{Kerr}}_{\va \text{ENR}})&=\frac{3^{2n}}{2n+1}\cdot\frac{J^{2n}M^{\frac{2(1-2n)}{3}}}{c^{2n}}
    \\
    &=\frac{3^n}{2n+1}\mscr{C}_{0}(\phasespace^{\va \text{Kerr}}_{\va \text{ENR}})^{n+1}\mscr{C}_2(\phasespace^{\va \text{Kerr}}_{\va \text{ENR}})^n\,.
\end{eqaligned}
Therefore, all the non-zero higher Casimirs $\mscr{C}_{2n}(\phasespace^{\va \text{Kerr}}_{\va \text{ENR}})$ for $n=2,3,4,\ldots$ are determined by $\mscr{C}_{0}(\phasespace^{\va \text{Kerr}}_{\va \text{ENR}})$ and $\mscr{C}_{2}(\phasespace^{\va \text{Kerr}}_{\va \text{ENR}})$. 

\smallskip Recall that according to the No-Hair Theorem \cite{Israel196712,Carter197102,Robinson197504}, mass-type $M_n$ and spin-type $S_n$ multipole moments of the Kerr metric are determined as a function of $M$ and $J$ in a unique fashion by the relation $M_n+\mfk{i}S_n=M(\mfk{i}a)^n$ for $n\ge 0$ \cite{Geroch197008,Hansen197401}. This is compatible with our result in the sense that
all non-zero higher Casimirs are completely determined in terms of the first two through the relation \eqref{eq:higher casimirs in terms of C0 and C2}. It is expected that all Casimirs are generic functions of $J$ and $M$. This does not a priori determine what the precise dependence on $(M,J)$ of the higher Casimirs is. As we have shown we  can solve this infinite set of relations and express $M$ and $J$ in terms of just two of the Casimirs. This is the direct consequence of the No-Hair Theorem for our gravitational Casimirs.

\section{Conclusions}

In this work, we proposed a new definition of  spin in asymptotically-flat spacetimes using the notion of asymptotic symmetries. After reviewing the issues of defining conserved quantities using the $\bms$ algebra, we argued that the issues can be resolved once the symmetry algebra is enhanced to $\gbms$. 

\smallskip After elaborating on the $\gbms$ coadjoint action and charge algebra in \S\ref{sec:gbms coadjoint action and charge algebra}, we introduced a Goldstone field associated with $\gbms$ supertranslations using which the intrinsic angular momentum was defined in \S\ref{sec:gbms goldstone modes}. This quantity was shown to satisfy the right algebra. We then elaborated on the Poincar\'e embeddings inside $\gbms$ for a general metric on the celestial sphere in \S\ref{sec:poincare embedding} and revealed the role of condensate fields characterising the symmetry breaking. The explicit Poincar\'e charges and algebra for a given embedding were derived in \S\ref{sec:Poi-charges}.

\smallskip In \S\ref{sec:algebraic aspects of gbms}, we studied the algebraic aspect of $\gbms$. We showed that $\gbms$ has a very similar structure to the hydrodynamical algebra \eqref{eq:hydrodynamical algebra}, whose coadjoint orbits have been studied and classified previously in \cite{DonnellyFreidelMoosavianSperanza202012}. Similar to the hydrodynamical algebra, an important role in the construction of invariants of coadjoint orbits is played by the notion of vorticity, which we identified in \eqref{vortdef}. Equipped with this definition, we then constructed an infinite set of Casimir functionals \eqref{eq:gbms casimir functions} whose level sets label coadjoint orbits of $\gbms$. This motivated the definition of the spin charge for $\gbms$ as  the smeared version of the vorticity, which was shown to be  supertranslation-invariant \eqref{eq:coadjoint transformation of vorticity}, and  the generator of the isotropy subalgebra of $\gbms$ \eqref{eq:poisson lie bracket of smeared vorticities}. 

\smallskip \S\ref{sec:reference frames and goldstone modes} was devoted to the topic of reference frames for spacetimes with $\gbms$ as their asymptotic-symmetry group. We introduced the notion of rest frame for $\gbms$ and the transformation between the rest and the usual Bondi frame in \S\ref{sec:gbms rest vs bondi}. Using these reference frames, we then case-studied the stationary spacetimes and obtained their mass and angular momentum aspects in the rest and general Lorentz frames in \S\ref{sec:statST}. Finally, we obtained the harmonic decomposition of the mass and angular momentum aspects in terms of their condensate fields. Using the former one, we showed that a generic multi-particle configuration does not belong to a $\bms$ coadjoint orbit with a constant-mass representative. 

\smallskip \S\ref{sec:pauli-lubanski generators} was devoted to the construction of the Pauli--Luba\'nski generator associated with a Poincar\'e embedding inside $\bms$. We used the spin generator in \eqref{eq:4-vector W for gbms} to construct an object
which is supertranslation invariant and transforms as a vector under Lorentz symmetry.
This vector  satisfies a closed algebra only when we restrict the mass aspect to be an element of the constant mass orbit, which is the orbit associated with staionary spacetimes.
In this case our definition of spin recovers the usual 
Pauli--Luba\'nski generator
as shown in \S\ref{sec:PL-BMS constant mass}.

\smallskip Finally, we explored several applications of our analysis to the gravitational phase space in \S\ref{sec:applications}. In particular, we derived the moment map for  $\gbms$ in \S\ref{sec:gravitational moment map} and   the evolution of the gravitational spin charge in \S\ref{sec:evolution equations}. As a concrete example, we constructed the gravitational Casimirs for the particular example of the Kerr metric in \S\ref{sec:gravitational Casimirs for the kerr metric}.

\smallskip Our work opens the way towards three main different future explorations. First, one would like to understand more deeply the symmetry-breaking mechanism and the role of the corresponding condensate fields described in \S\ref{sec:poincare embedding} and \S\ref{sec:Condensate}.
Second, we would like to understand whether the Casimirs we have constructed are relevant for the numerical study of gravitational dynamics. In fluid mechanics, the Casimirs are the slow variables that lead to long-term predictive behaviour in chaotic dynamics. We still have to see whether a similar interpretation can be given in gravity.
Finally, our analysis is classical; it would be essential to develop the quantum representation theory  to study   S-matrix scattering amplitudes from the point of view of the unbroken GBMS symmetry group,  instead of using the broken symmetry representation of Poincar\'e, as one usually does.

\bigskip
\paragraph{Acknowledgement} 
We thank Glenn Barnich,  Geoffrey Comp\`ere, Lionel Mason, and Christopher Wright for helpful discussions. Research at Perimeter Institute is supported in part by the Government of Canada through the Department of Innovation, Science and Economic Development Canada and by the Province of Ontario through the Ministry of Colleges and Universities.
The work of S.F.M. is funded by the Natural Sciences and Engineering Research Council of Canada (NSERC), funding reference number SAPIN/00047-2020, and in part by the Alfred P. Sloan Foundation, grant FG-2020-13768, and also in part by ERC Consolidator Grant \#864828 “Algebraic Foundations of Supersymmetric Quantum Field Theory” (SCFTAlg). D.P. has received funding from the European Union's Horizon 2020 research and innovation programme under the Marie Sklodowska-Curie grant agreement No. 841923.
This work was supported by the Simons Collaboration on Celestial Holography.

\appendix

\section{Moment maps and coadjoint orbits}\label{sec:moment maps and coadjoint orbits}

In this appendix, we  remind several useful notions such as moment map, coadjoint action, and coadjoint orbits \cite{Audin1991,Kirillov200407}.

\paragraph{Moment map.} 

Let $G$ be a real Lie group whose Lie algebra is $\g$. As a vector space, $\g$ has a dual $\g^*$. Furthermore, let $(M,\omega)$ be a symplectic manifold  on which the group $G$ acts by symplectomorphism. This notion can be defined by considering the $G$-action as a map $G\to \tenofo{Diff}(M)$ which induces the map $\varphi:M\to M$ such that $\varphi^*\omega=\omega$, i.e. the symplectic form is preserved under a symplectomorphism. The action of $G$ on $M$ is Hamiltonian if there is a $G$-equivariant map $\mu:M\to\g^*$, called the {\it moment-map} satisfying certain properties as follows: Let $X^{\#}$ be the vector field on $M$ generated by the flow $\exp(tX)\cdot p$ of $X$, where $\cdot$ denotes the action of $G$ on $M$. This is the vector field generated by the one-parameter subgroup $\{\exp(tX)\,|\,X\in\g,t\in\mbb{R}\}\subset G$. Note that this map associates to any element $X\in\g$, a vector field $X^{\#}$ on $M$. More explicitly, this vector field is defined as
\begin{equation}\label{eq:the vector field defining the moment-map}
    X^{\#}(p):= \left.\frac{\rd}{\rd t}\varphi(\exp(tX))\right|_{t=0}(p), \qquad \forall p\in M. 
\end{equation}
This vector field satisfies 
\begin{equation}\label{eq:the defining equation of the moment-map}
    \rd\mu_X=\iota_{X^{\#}}\omega=\omega(X^{\#},\cdot),
\end{equation}
where $\mbs\mu_X(p):= \mu(p)(X)$ for any $p\in M$ denotes the component of $\mu$ along $X$. This relation basically means that $X^{\#}$ is the Hamiltonian vector field associated with the Hamiltonian function $\mu_X$. On the other hand, $G$-equivariance means that the moment-map commutes with the $G$-action, i.e.
\begin{equation}
    \mu(\varphi(p))=\tenofo{Ad}^*_g(\mu(p)), \qquad p\in M,
\end{equation}
where $\varphi:M\to M$ is the map we introduced above, and $\tenofo{Ad}^*_g$ denotes the coadjoint action of $G$ on $\g^*$, which we will define momentumrily. 

\hypertarget{Lie-Poisson on dual}{\paragraph{The Lie--Poisson structure on $\g^*$.}}
It is a well-known fact that for any Lie algebra $\g$, its dual $\g^*$ is equipped with a Poisson bracket, the so-called Lie--Poisson structure \cite{Berezin196704,Kirillov197608,BayenFlatoFronsdalLichnerowiczSternheimer197803,Weinstein198301}. For $F,G\in C^\infty(\g^*)$, we have
    \begin{equation}\label{eq:lie-poisson structure on dual lie algebra}
        \{F,G\}_{\g^*}(\mbs{p}):=\langle \mbs{p},[\bd_{\mbs{p}}F,\bd_{\mbs{p}}G]_{\g}\rangle,
    \end{equation}
    where $\bd_{\mbs{p}}F:T\g^*\to\mbb{R}$, and the differential should be thought of as maps into $\g$ (rather than $\g^{**}$). Note also that $T_{\mbs{p}}\g^*\simeq\g^*$, as a linear space.

\begin{rmk}{\bf (Equivariant moment maps as Poisson morphisms).}\label{rmk:equivariant moment maps as poisson morphisms}
    Let us make a comment about the equivariance of moment maps which we have assumed in its definition. Note that the action of a group on a symplectic (or Poisson) manifold does not need to have an equivariant moment map. The general scenario for the emergence of such non-equivariant moment maps has been discussed in \cite{MarsdenMisioekPerlmutterRatiu199808}. In general, having a smooth map between two Poisson manifolds, like a moment map, there is no natural way to pull-back the Poisson structure \cite{Meinrenken201612}. However, the elementary but important feature of equivariant moment maps is the following result \cite{Lie1890,GuilleminSternberg198006,GuilleminSternberg1990,HolmesMarsden198302,MarsdenRatiu1999}
\begin{thr}[Equivariant Moment Maps are Poisson]\label{thr:equivariant moment maps are poisson}
    Let $(M,\{\cdot,\cdot\}_\rho)$ be a Poisson manifold carrying a $G$-action for a Lie group $G$ with an equivariant moment map $\mu:M\to\g^*$, where $\g=\tenofo{Lie}(G)$. Then, $\mu$ is a Poisson map (or Poisson morphism) preserving the Poisson brackets in the following sense 
    \begin{equation}\label{eq:property of poisson morphisms}
        \mu^*\{F,G\}_{\g^*}=\{\mu^*F,\mu^* G\}_M, \qquad \forall F,G\in 
\mcal{F}(\g^*),
    \end{equation}
    where $\{\cdot,\cdot\}_{\g^*}$ is the Lie--Poisson structure on $\g^*$.
\end{thr}
The importance of this theorem for us is that we are mainly concerned with the action of a symmetry algebra $\g$ on (asymptotic) gravitational phase space, which is naturally equipped with a Poisson structure, induced by its symplectic structure. Theorem \ref{thr:equivariant moment maps are poisson} implies that this Poisson structure is induced from the Lie--Poisson structure on $\g^*$ through the corresponding equivariant moment map. Furthermore, in \S \ref{sec:the gbms moment map}, we explicitly construct the moment map for the action of $\gbms$ on a certain phase space (which we have dubbed as electric and  non-radiative) of an asymptotically-flat spacetime.
\end{rmk}

\paragraph{Coadjoint actions.} 

The coadjoint action of $G$ on $\g^*$ is defined as follows
\begin{equation}
    \langle \tenofo{Ad}^*_g \mbs{p},Y\rangle:= \langle \mbs{p},\tenofo{Ad}_{g^{-1}}Y\rangle, \qquad \forall g\in G,\; Y\in\g,\;\mbs{p}\in\g^*.
\end{equation}
$\tenofo{Ad}$ and $\tenofo{Ad}^*$ denote the adjoint and coadjoint actions of $G$, respectively. It can be shown that the vector fields \eqref{eq:the vector field defining the moment-map} corresponding to the adjoint and coadjoint actions are given by
\begin{equation}\label{eq:the definition of tangent vectors to adjoint and coadjoint orbits}
        \begin{aligned}
            \wt{X}_{Y}&:= \left.\frac{\rd}{\rd t}\tenofo{Ad}_{\tenofo{exp}(tX)}Y\right|_{t=0}\in \mfk{X}(\g),
            \\
            \wh{X}_{\mbs{p}}&:= \left.\frac{\rd}{\rd t}\tenofo{Ad}^*_{\tenofo{exp}(tX)}\mbs{p}\right|_{t=0}\in\mfk{X}(\g^*),
        \end{aligned}
\end{equation}
where $\mfk{X}(\g)$ and $\mfk{X}(\g^*)$ denote the space of vector fields on $\g$ and $\g^*$, respectively. A straightforward computation shows that
\begin{equation}
    \wt{X}_Y=[X,Y], \qquad \langle \wh{X}_{\mbs{p}},Y\rangle=-\langle\mbs{p},[X,Y]\rangle, \qquad \forall\, Y\in\g. 
\end{equation}
Similarly, one can define the coadjoint action of $\g$ on $\g^*$ by
\begin{equation}\label{eq:the definition of ad* action}
    \langle \tenofo{ad}^*_X \mbs{p},Y\rangle:= -\langle \mbs{p},\tenofo{ad}_{X}Y\rangle=-\langle \mbs{p},[X,Y]\rangle, \qquad \forall X,Y\in\g,\;\mbs{p}\in\g^*,
\end{equation}
where $\tenofo{ad}$ and $\tenofo{ad}^*$ denote the adjoint and coadjoint actions of $\g$. 

\paragraph{Coadjoint orbits and its properties.} The coadjoint orbit $\mcal{O}_{\mbs{p}}$ of the coadjoint action  with the initial point $\mbs{p}$ is defined by
\begin{equation}
    \mcal{O}_{\mbs{p}}:=\left\{\mbs{p}'\in\g^*\,\big|\,\mbs{p}'=\tenofo{Ad}_{g}^*\mbs{p},\;\;\forall g\in G\right\}.
\end{equation}
Coadjoint orbits have some nice properties. In the following, we will mention some of the relevant ones.

\begin{itemize}
    \item[$-$]{\it Coadjoint orbits as a quotient.} A coadjoint orbit $\mcal{O}_{\mbs{p}}$ is diffeomorphic to $G/G_{\mbs{p}}$, where $G_{\mbs{p}}$ is the isotropy subgroup of the point $\mbs{p}$ defined by
\begin{equation}\label{eq:definition of isotropy subgroup of coadjoint orbits}
    G_{\mbs{p}}:=\left\{g\in G\,\big|\,\tenofo{Ad}^*_{g}\mbs{p}=\mbs{p}\right\}. 
\end{equation}

    \item[$-$] {\it Tangent space of coadjoint orbits.} The tangent space at a point $\mbs{p}'\in \mcal{O}_{\mbs{p}}$ is generated by $\{\wh{X}_{\mbs{p}'}\,|\,X\in\g\}$ with $\wh{X}_{\mbs{p}'}$ is given by \eqref{eq:the definition of tangent vectors to adjoint and coadjoint orbits}. This is almost tautological since the definition given for $\wh{X}_{\mbs{p}'}$ in \eqref{eq:the definition of tangent vectors to adjoint and coadjoint orbits} is precisely the definition of tangent space at a point of coadjoint orbit. This and \eqref{eq:definition of isotropy subgroup of coadjoint orbits} imply 
\begin{equation}
    T_{\mbs{p}'}\mcal{O}_{\mbs{p}}\simeq\g/\g_{\mbs{p}'},
\end{equation}
where $\g_{\mbs{p}'}\subset\g$ is the isotropy subalgebra of $\mbs{p}'$, the Lie algebra of $G_{\mbs{p}'}$ defined in \eqref{eq:definition of isotropy subgroup of coadjoint orbits}.

    \item[$-$] {\it Canonical symplectic structure on coadjoint orbits.} Coadjoint orbits are endowed with the so-called Kirillov-Kostant-Souriau symplectic form \cite{Kirillov196202,Kirillov1976,Kirillov199908,Kostant1965,Kostant1970,Souriau1970,Souriau1997}
\begin{equation}
    \omega_{\mbs{p}'}(\wh{X},\wh{Y})=-\mbs{p}'([X,Y]), \qquad \forall X,Y\in\g,\quad \mbs{p}'\in\mcal{O}_{\mbs{p}},
\end{equation}
where $\wh{X}$ and $\wh{Y}$ are two vector fields on $\g^*$ generated by the $\g$-actions of $X$ and $Y$, respectively. More explicitly, we have (note that the tangent space of the orbit at $\mbs{p}'$ is generated by $\tenofo{ad}^*_X\mbs{p}'$ for all $X\in\g$, as we explained above)
\begin{equation}\label{eq:the explicit form of symplectic structure}
    \omega_{\mbs{p}'}(\tenofo{ad}^*_X\mbs{p}',\tenofo{ad}^*_Y\mbs{p}')=-\mbs{p}'([X,Y]), \qquad \forall X,Y\in\g, \qquad \mbs{p}'\in\mcal{O}_{\mbs{p}}.
\end{equation}
This is a closed $2$-form and furthermore, is invariant under the coadjoint action of $G$. This can be seen by choosing any pairing and noting that
\begin{equation}
  \mbs{p}([X,Y])=\langle\mbs{p},[X,Y]\rangle\quad\longmapsto\quad\langle\tenofo{Ad}^*_g\mbs{p},\tenofo{Ad}_g[X,Y]\rangle=\langle\mbs{p},[X,Y]\rangle.
\end{equation}
This means that $G$-action on $\g^*$ is a symplectomorphism and therefore, $\omega_{\mbs{p}}$ defines a canonical symplectic structure on $\mcal{O}_{\mbs{p}}$ which does not depend on the choice of an inner product on $\g$. As a resulting of being symplectic, coadjoint orbits are always even-dimensional (even in a suitable sense, in infinite dimensions).

    \item[$-$] {\it Coadjoint orbits as level sets of Casimir functions.} Another important property of coadjoint orbits is that they can be labeled by some functions $\mscr{C}\in C^\infty(\mcal{O}_{\mbs{p}})$ of the coadjoint-orbit coordinates called the Casimir functionals satisfying
\begin{equation}\label{eq:the invariance of Casimir functionals under coadjoint action}
    \tenofo{ad}^*_X\mscr{C}(\mbs{p})=\sum_a\left.\left(\frac{\partial \mscr{C}}{\partial p^a}\tenofo{ad}^*_X{p^a}\right)\right|_{\mbs{p}}=\rd\mscr{C}(\wh{X})=\omega_{\mbs{p}}(\mcal{V}_{\mscr{C}},\wh{X})=0,
\end{equation}
where $\{p^a\}$ is a set of local coordinates in the neighbourhood of $\mbs{p}$, $\rd{\mscr{C}}=\omega_{\mbs{p}}(\mcal{V}_{\mscr{C}},\cdot)$ for the Hamiltonian vector field $\mcal{V}_{\mscr{C}}$ associated to $\mscr{C}$, and $\wh{X}=\tenofo{ad}^*_{X}f^a\partial_{f^a}$ is a vector field on $\g^*$ corresponding to the action of $X\in\g$ on $\mbs{p}\in\g^*$. Equation \eqref{eq:the invariance of Casimir functionals under coadjoint action} just tells that Casimir functions are invariant under the coadjoint action. Thus, the level sets of a complete set of Casimir functions\footnote{By a complete set of Casimir functions, we mean a set which contains a maximal number of functions with two properties: 1) $\{\mscr{C},\mscr{C}'\}(\mbs{p})=0$ for any two functions $\mscr{C}$ and $\mscr{C}'$ belonging to the set, and 2) $\{\mscr{C},F\}(\mbs{p})=0$ for all functions $F\in\mcal{F}(\g^*)$, where $\mcal{F}(\g^*)$ being the space of functions on $\g^*$.} label coadjoint orbits, i.e. for any level set $\{\mscr{C}_n=\text{C}_n,\,n=1,2,\ldots\}$ for constants $\text{C}_n$, there is a coadjoint orbit. It then follows that any $F\in\mcal{F}(\g^*,\mbb{R})$ Poisson-commutes with all $\mscr{C}$s
\begin{equation}\label{eq:the Poisson-commutativity of any function on coadjoint orbit with Casimir functions}
    \{\mscr{C},F\}_{\g^*}(\mbs{p})=\omega_{\mbs{p}}(\mcal{V}_{\mscr{C}},\mcal{V}_F)=0, \qquad F\in\mcal{F}(\g^*),
\end{equation}
where $\rd F=\omega_{\mbs{p}}(\mcal{V}_{F},\cdot)$. Equation \eqref{eq:the Poisson-commutativity of any function on coadjoint orbit with Casimir functions} follows from \eqref{eq:the invariance of Casimir functionals under coadjoint action} by replacing $\wh{X}$ with $\mcal{V}_F$. As coadjoint orbits can be realized as level sets of Casimir functions, we have the following decomposition of $\g^*$
\begin{equation}
    \g^*=\bigsqcup_{\mbf{C}}\mcal{O}_\mbf{C},
\end{equation}
where $\mbf{C}=\{\text{C}_1,\text{C}_2,\ldots\}$ is a set of numbers determining the level sets of Casimir functions and $\mcal{O}_{\mbf{C}}$ is the corresponding coadjoint orbit.

 \item[$-$] {\it Momentum map for the coadjoint action.} The moment-map for the coadjoint action of $G$ of $\g^*$ is nothing but the inclusion map $\mbs{i}:\mcal{O}_{\mbs{p}}\hookrightarrow\g^*$. We can show this by proving that the inclusion map is a) an equivariant map and b) it satisfies \eqref{eq:the defining equation of the moment-map} with $\mu_X\to \mbs{i}_X$. The first property follows immediately by noting that the required action is the restriction of coadjoint action, which is already equivariant. The second property can be proven by noting that $\mbs{i}_X(\mbs{p}')=\langle \mbs{p}',X\rangle$  and hence $[\rd\mbs{i}_X(\wh{Y})](\mbs{p}')=\langle\mbs{p}',[X,Y]\rangle$, where to prove this we use the fact that $\mbs{i}$ is the inclusion map. On the other hand, $\iota_{\wh{X}}\omega_{\mbs{p}'}(\wh{Y},\cdot)=\langle\mbs{p}',[X,Y]\rangle$. Comparing these results, it follows that the inclusion map satisfies \eqref{eq:the defining equation of the moment-map}, i.e. $\rd\mbs{i}_X=\iota_{\wh{X}}\omega_{\mbs{p}}$, which implies that it is the moment-map for the coadjoint action. 

    \item[$-$] {\it Compatibility of the Lie--Poisson structure on {\normalfont$\g^*$} and the symplectic structure on its coadjoint orbits.} For $F,G\in\mcal{F}(\g^*)$, the Lie--Poisson structure $\{F,G\}_{\g^*}$ on $\g^*$ and the symplectic structure $\mbs{\omega}_\mcal{O}$ on the coadjoint orbit $\mcal{O}$ are compatible in the following sense
\begin{equation}\label{eq:the compatibility of lie-poisson struture and symplectic structure on a coadjoint orbit}
    \{F,G\}_{\g^*}(\mbs{p})\Big|_\mcal{O}=\{F|_\mcal{O},G|_\mcal{O}\}_{\g^*}(\mbs{p})=\mbs{\omega}_{\mcal{O}}(X_F,X_G)(\mbs{p}),
\end{equation}
where $\cdot|_\mcal{O}$ denotes the restriction to the coadjoint orbit $\mcal{O}$, and $X_F$ and $X_G$ are Hamiltonian vector fields generated by $F$ and $G$.
\end{itemize}

The relevance of these concepts for us is as follows. What we actually have in mind is to realize a gravitational phase space as a union of inverse images of coadjoint orbits of some extended symmetry group under the moment map. This can be made a bit more precise as follows although still many many functional analysis questions are swept under the carpet. Consider an extended symmetry algebra $\g$ acting by Hamiltonian transformations on a gravitational phase space $\phasespace$. It is well-known that 
\begin{equation*}
    \g^*:=\bigsqcup_{\mbf{C}} \mcal{O}_{\mbf{C}},
\end{equation*}
where $\mbf{C}$ labels the level set of Casimir functions of $\g$. Due to the existence of the moment map $\mu_\g:\phasespace\to\g^*$ for the $\g$-action on $\phasespace$, the latter can be partitioned into symplectic leaves of the pull-back of the Lie--Poisson structure \eqref{eq:lie-poisson structure on dual lie algebra} to the phase space $\phasespace$. Hence,
\begin{equation*}
    \phasespace=\bigsqcup_{\mbf{C}} \phasespace_{\mbf C}, \qquad \phasespace_{\mbf{C}}:=\mu_\g^{-1}(\mcal{O}_{\mbf{C}}),
\end{equation*} 
where $\mu_\g^{-1}(\mcal{O}_{\mbf{C}})$ is the inverse image of $\mcal{O}_{\mbf{C}}$ under the moment-map $\mu_\g$.

\paragraph{Coadjoint orbits of semi-direct products.}
In the main body of the paper, we are dealing with coadjoint orbits of semi-direct product groups (or algebras in the form of a semi-direct sum). Such groups take the form $G=G'\ltimes N$, where $G'$ is called the quotient factor, and $N$, called the normal factor, has the structure of an Abelian group.\footnote{The corner symmetry group, discovered in \cite{DonnellyFreidel201601}, is an example of a semi-direct product group with a non-Abelian normal factor. Using the reduction of this normal factor to an Abelian one, the classification of coadjoint orbits of this group is done in \cite{DonnellyFreidelMoosavianSperanza202012}.} The study of coadjoint orbits of such groups is well-known in the literature \cite{Rawnsley197502,Baguis1998,DonnellyFreidelMoosavianSperanza202012}. In brief, one first studies the isotropy subalgebra (the Lie algebra version of \eqref{eq:generic definition of isotropy subalgebra for adjoint action}), which by construction involves the normal part $N$. The invariants of coadjoint orbits for isotropy subalgebra would provide the analogs of spin for $G$. Furthermore, these invariants can then be lifted to the whole group by a judicious modification of these invariants. For an explicit example of this procedure relevant to gravity, see \cite{DonnellyFreidelMoosavianSperanza202012}.

\section{Dual Pauli--Luba\'nski pseudo-vector as a constant of motion}\label{sec:pauli-lubanski pseudo-vector as a constant of motion}

In this appendix, we study the time evolution of dual Pauli--Luba\'nski pseudo-vector \eqref{eq:dual PL vector} on a coadjoint orbit of Poincar\'e group. For this purpose, we first find a set of canonical coordinates on a coadjoint orbit (as a symplectic manifold) and then define the time evolution. We finally show that the time evolution of $w_\mu$, defined in \eqref{eq:dual PL vector}, vanishes. Hence, it is a constant of motion.

\subsection{Coadjoint orbits of Poincar\'e group}\label{sec:coadjoint orbit of poincare group}

The coadjoint orbits of Poincar\'e group is a very well-known subject. For a detailed construction and some applications see \cite{CushmanvanderKallen200305,CushmanvanderKallen200605,DonnellyFreidelMoosavianSperanza202012,Andrzejewski:2020qxt}. 

\paragraph{Coadjoint actions of $\iso(3,1)$ and $\tenofo{Iso}(3,1)$.}
The adjoint action of the Poincar\'e algebra $\msf{iso}(3,1)$ on itself is implemented by the brackets \eqref{eq:poincare Lie algebra}. Instead of coadjoint orbits, one can study the adjoint orbits since one can show that there is a one-to-one correspondence between adjoint and coadjoint orbits of Poincar\'e algebra \cite[\S 3]{Arathoon201806}. 

\smallskip Let us begin by studying the coadjoint orbits of the Poincar\'e group. As $\{\mbs{P}_\mu,\mbs{J}_{\mu\nu}\}$ is a basis for $\iso(3,1)$, a generic element of $\mbs{X}\in\iso(3,1)$ is expanded as
\begin{equation}
    \mbs{X}=x^\mu \mbs{P}_\mu+\frac{1}{2}\psi^{\mu\nu}\mbs{J}_{\mu\nu},
\end{equation}
with $\psi^{\nu\mu}=-\psi^{\mu\nu}$. Similarly, a generic element $\mbs{X}^*\in\iso(3,1)^*$ has the following expansion
\begin{equation}\label{eq:expansion of a generic element of iso(3,1)*}
    \mbs{X}^*=p_\mu\mbs{X}^\mu+\frac{1}{2}j_{\mu\nu}\mbs{\Psi}^{\mu\nu},\qquad \mbs{\Psi}^{\nu\mu}=-\mbs{\Psi}^{\mu\nu},
\end{equation}
where $\{\mbs{X}^\mu,\mbs{\Psi}^{\mu\nu}\}$ is a basis for $\iso(3,1)^*$. A pairing $\langle\cdot,\cdot\rangle$ between $\iso(3,1)$ and $\iso(3,1)^*$ is given by 
\begin{equation}\label{eq:the pairing between poincare algebra and its dual}
    \langle\mbs{X}^*,\mbs{X}\rangle=x^\mu p_\mu+\psi^{\mu\nu}j_{\mu\nu}.
\end{equation}
The adjoint action of $\iso(3,1)$ on itself is given by the brackets \eqref{eq:the poincare algebra}, from which the coadjoint action of $\msf{iso}(3,1)$ on $\msf{iso}(3,1)^*$ is determined through\footnote{We included an extra factor of $\mfk{i}$ to come up with the convenient expressions.}
\begin{equation}
    \langle \delta_{\mbs{Y}}\mbs{X}^*,\mbs{X}\rangle:=-\mfk{i}\langle \mbs{X}^*,[\mbs{X},\mbs{Y}]\rangle.
\end{equation}
The coadjoint actions are determined as follows. We expand $\mbs{X}$ and $\mbs{X}^*$ as above and $\mbs{Y}$ as $\mbs{Y}=y^\mu\mbs{P}_\mu+\frac{1}{2}\psi'^{\mu\nu}\mbs{J}_{\mu\nu}$. Substituting in \eqref{eq:the pairing between poincare algebra and its dual} and equating terms with the same coefficients gives
\begin{eqgathered}\label{eq:coadjoint action of poincare on coordinates on iso(3,1)*}
    \delta_{\mbs{P}_\mu}p_\nu=0, 
    \\
    \delta_{\mbs{P}_\mu}j_{\nu\sigma}=\eta_{\mu\nu}p_\sigma-\eta_{\mu\sigma}p_\nu, \qquad \delta_{\mbs{J}_{\nu\sigma}}p_\mu=\eta_{\mu\sigma}p_\nu-\eta_{\mu\nu}p_\sigma,
    \\
    \delta_{\mbs{J}_{\mu\nu}}j_{\rho\sigma}=\eta_{\mu\rho}j_{\nu\sigma}+\eta_{\nu\sigma}j_{\mu\rho}-\eta_{\mu\sigma}j_{\nu\rho}-\eta_{\nu\sigma}j_{\mu\sigma}.
\end{eqgathered}
As $(\mbs{P}_\mu,\mbs{J}_{\mu\nu})$ can be thought of as a basis for linear functions on $\iso(3,1)^*$, we get, by definition, the Lie--Poisson structure on $\msf{iso}(3,1)^*$ as follows
\begin{eqgathered}\label{eq:linear poisson structure on iso(3,1)*,appendix}
    \{p_\mu,p_\nu\}_{\iso(3,1)^*}:=\delta_{\mbs{P}_\mu}p_\nu=0,
    \\
    \{p_\mu,j_{\nu\sigma}\}_{\iso(3,1)^*}:=\delta_{\mbs{P}_\mu}j_{\nu\sigma}=-\delta_{\mbs{J}_{\nu\sigma}}p_\mu=\eta_{\mu\sigma}p_\nu-\eta_{\mu\nu}p_\sigma,   
    \\
    \{j_{\mu\nu},j_{\rho\sigma}\}_{\iso(3,1)^*}:=\delta_{\mbs{J}_{\mu\nu}}j_{\rho\sigma}=\eta_{\mu\rho}j_{\nu\sigma}+\eta_{\nu\sigma}j_{\mu\rho}-\eta_{\mu\sigma}j_{\nu\rho}-\eta_{\nu\sigma}j_{\mu\sigma}.
\end{eqgathered}
This familiar form is the classical analog of the Poincar\'e algebra \eqref{eq:the poincare algebra}, i.e. under quantization map $\{\cdot,\cdot\}_{\msf{iso}(3,1)^*}\mapsto -\mfk{i}[\cdot,\cdot]_{\msf{iso}(3,1)}$, \eqref{eq:linear poisson structure on iso(3,1)*,appendix} becomes \eqref{eq:the poincare algebra}. From these brackets, one can compute the brackets between any two elements of $\msf{iso}(3,1)^*$.

\smallskip Since the Poincar\'e group is connected, one can derive the finite transformations by successive application of transformations \eqref{eq:the poincare algebra} and \eqref{eq:coadjoint action of poincare on coordinates on iso(3,1)*} to get the adjoint and coadjoint action of the Poincar\'e group $\tenofo{Iso}(3,1)$, respectively. The adjoint action of an element $g=(\mbs{\Lambda},\mbs{a})\in \tenofo{Iso}(3,1)$ on the generators $(\mbs{P}_\mu,\mbs{J}_{\mu\nu})$ of $\iso(3,1)$ as \cite{Rawnsley197502} 
\begin{eqgathered}\label{eq:adjoint action of poincare group}
    \tenofo{Ad}_g\mbs{P}_\mu=(\mbs{P}\cdot\mbs{\Lambda})_\mu,
    \\
    \tenofo{Ad}_g \mbs{J}_{\mu\nu}=(\mbs{\Lambda}\cdot \mbs{J} \cdot \mbs{\Lambda})_{\mu\nu}-\left[(\mbs{P}\cdot\mbs{\Lambda})_\mu (\mbs{a}\cdot\mbs{\Lambda})_\nu-(\mbs{a}\cdot\mbs{\Lambda})_\mu(\mbs{P}\cdot\mbs{\Lambda})_\nu\right]\,,
\end{eqgathered} 
where $\mbs{\Lambda}\cdot \mbs{J} \cdot \mbs{\Lambda}=\Lambda\indices{^\mu_\rho}\Lambda\indices{^\nu_\sigma}\mbs{J}_{\mu\nu}$.\footnote{Here, we have used the notation $(\mbs{\Lambda})\indices{^\mu_\nu}=\Lambda\indices{^\mu_\nu}$.} Similarly, its coadjoint actions on the dual coordinates (the finite version of \eqref{eq:coadjoint action of poincare on coordinates on iso(3,1)*}), which is defined through
\begin{equation}
    \langle \tenofo{Ad}^*_g\mbs{X}^*,\mbs{X}\rangle=\langle\mbs{X}^*,\tenofo{Ad}_{g^{-1}}\mbs{X}\rangle,
\end{equation}
are given by\footnote{Recall that $(\mbs{\Lambda}^{-1})\indices{^\mu_\nu}=(\mbs{\Lambda})\indices{_\nu^\mu}=(\mbs{\Lambda}^T)\indices{^\mu_\nu}$, where $\mbs{\Lambda}^T$ denotes the transpose of the matrix of $\mbs{\Lambda}$.}
\begin{eqgathered}\label{eq:coadjoint action of the poincare group, appendix}
    \tenofo{Ad}^*_gp_\mu=(p\cdot\mbs{\Lambda}^{-1})_\mu,
    \\
    \tenofo{Ad}^*_g\psi_{\mu\nu}=\left(\mbs{\Lambda}^{-1}\cdot\psi\cdot\mbs{\Lambda}^{-1}\right)_{\mu\nu}-\left[(p\cdot\mbs{\Lambda}^{-1})_\mu (\mbs{a})_\nu-(\mbs{a})_\mu (p\cdot\mbs{\Lambda}^{-1})_\nu\right].
\end{eqgathered}

\paragraph{Casimir functions on coadjoint orbits.}
We have seen in \S\ref{sec:Poincare algebra} that the representation of $\iso(3,1)$ are determined by two Casimir elements $\wh{\mscr{C}}_2(\iso(3,1))$ and $\wh{\mscr{C}}_4(\iso(3,1))$, defined in \eqref{eq:quadratic casimir of poincare algebra} and \eqref{eq:quartic Casimir of the Poincare algebra}, respectively. The classical counterpart of this statement is that there are  two Casimir functions on coadjoint orbits of $\iso(3,1)$ whose values determine the coadjoint orbits. These invariants can be constructed by noting that the coadjoint actions \eqref{eq:coadjoint action of poincare on coordinates on iso(3,1)*} imply that
\begin{eqgathered}\label{eq:casimir functions on coadjoint orbits of poincare, appendix}
    \mscr{C}_2(\iso(3,1)):=-\eta^{\mu\nu}p_\mu p_\nu,\qquad \mscr{C}_4(\iso(3,1)):=\eta^{\mu\nu} w_\mu w_\nu,
\end{eqgathered}
where
\begin{equation}\label{eq:dual PL vector, appendix}
    w_\mu:=\frac{1}{2}\varepsilon_{\mu\nu\rho\sigma}p^\nu j^{\rho\sigma},
\end{equation}
is a polynomial function\footnote{Polynomial functions on a Lie co-algebra turn to elements of universal enveloping algebra of the corresponding Lie algebra.} on $\msf{iso}(3,1)^*$, the analog to the Pauli--Luba\'nski pseudo-vector \eqref{eq:definition of PL pseudo-vector}, are invariant under the coadjoint action of $\iso(3,1)$. We call $w_\mu$, the dual Pauli--Luba\'nski pseudo-vector.\footnote{This is just a terminology. The fact that it is a pseudo-vector is shown below.} Hence, \eqref{eq:casimir functions on coadjoint orbits of poincare, appendix} are the two Casimir functions on coadjoint orbits of $\iso(3,1)$ whose level sets are the coadjoint orbits of the Poincar\'e algebra. The dimension of coadjoint orbits is easy to determine. The dimension of $\msf{iso}(3,1)^*$ is $10$, just like $\msf{iso}(3,1)$, and there are two Casimir functions whose level sets are the coadjoint orbits. Hence, the dimension of all coadjoint orbits of $\iso(3,1)$ is $8$.\footnote{The dimension of coadjoint orbits of $\iso(d,1)$ is $(d+1)(d+2)/2-\lfloor (d+2)/2\rfloor$, where the second term is the number of Casimir functions on coadjoint orbits of $\iso(d,1)$. The latter is the same as the number of Casimir elements for $\iso(d,1)$ \cite[Eq. (5.7)]{ChaichianDemichevNelipa1983}.} 

\smallskip Analog to $W_\mu$, $w_\mu$ has two important properties
\begin{enumerate}
    \item [(1)] It follows from \eqref{eq:linear poisson structure on iso(3,1)*,appendix} that $w_\mu$ satisfies
    \begin{eqgathered}\label{eq:poisson bracket of dual pauli-lubanski pseudo-vector,appendix}
    \{p_\mu,w_\nu\}_{\iso(3,1)^*}=0, \qquad \{j_{\mu\nu},w_\sigma\}_{\iso(3,1)^*}=\eta\indices{_{\mu\sigma}}w_\nu-\eta\indices{_{\nu\sigma}}w_\mu,
    \\
    \{w_\mu,w_\nu\}_{\iso(3,1)^*}=\frac{1}{2}\varepsilon_{\mu\nu\rho\sigma}p^\rho w^\sigma.
\end{eqgathered}
These are the classical analogs of \eqref{eq:algebra relations for the Pauli--Lubanski vector} and \eqref{eq:Lie bracket of components of Pauli--Lubanski pseudo-vector}.
    \item [(2)] \eqref{eq:coadjoint action of the poincare group, appendix} implies that under the parity transformation, represented by the matrix $\mbs{\Lambda}=\tenofo{diag}(+1,-1,-1,-1)$, we have
\begin{eqaligned}\label{eq:transformation of poicare coalgebra coordinates under the parity}
    p_0&\mapsto +p_0, &\qquad p_i&\mapsto -p_i,
    \\
    j_{0i}&\mapsto -j_{0i}, &\qquad j_{kl}&\mapsto +j_{kl}.
\end{eqaligned} 
Therefore, under the parity transformation, we have
\begin{equation}
    w_0\mapsto -w_0, \qquad w_i\mapsto +w_i.
\end{equation}
Therefore, $w_\mu$ is also a pseudo-vector.
\end{enumerate}

\subsection{Evolution of the dual Pauli--Luba\'nski pseudo-vector}\label{sec:evolution of dual pauli-lubanski pseudo-vector}

Having determined the coadjoint orbits of $\msf{iso}(3,1)$, we would like to define dynamics on these coadjoint orbits and evaluate the time evolution of $w_\mu$. For this purpose, we need to define a set of local coordinates on a coadjoint orbits, to which we now turn.

\paragraph{Coordinate representatives in the rest frame.} To determine the parameterization of the orbits, we proceed as we did for constructing the representation theory, namely, we use \eqref{eq:coadjoint action of poincare on coordinates on iso(3,1)*} to put $\{p_\mu,j_{\mu\nu}\}$ into simple forms, which we dub ``the rest frame". Let us set the value of the first Casimir to $\mscr{C}_2(\iso(3,1))=\kappa^2$ for some real positive $\kappa$. Then, by the Lorentz boost \eqref{eq:lorentz transformation from the rest frame to a general frame} with a different set of parameters ($\mbs{P}_\mu\to p_\mu$ and $m\to\kappa$)
    \begin{eqgathered}\label{eq:lorentz transformation from the generic coordinate to the reference coordinate}
        (\mbs{\Lambda}_p)\indices{^0_0}=\frac{p_0}{\kappa}, \qquad  (\mbs{\Lambda}_p)\indices{^0_i}=\frac{p_i}{\kappa}, \qquad  (\mbs{\Lambda}_p)\indices{^i_0}=\frac{p^i}{\kappa},
        \\
        (\mbs{\Lambda}_p)\indices{^i_j}=\delta\indices{^i_j}+\frac{p^ip_j}{\kappa(\kappa+p^0)},
    \end{eqgathered}
we can take
\begin{equation}\label{eq:coordinate p in the rest frame}
    p_{\mu}^{\msf{R}}=(\kappa,0,0,0),
\end{equation}
where $\msf{R}$ denotes the rest frame.\footnote{Here, we use the coadjoint action to fix the form of $(p_\mu,j_{\mu\nu})$ while in \S\ref{sec:Poincare algebra}, we use the adjoint action to fix te form of $(\mbs{P}_\mu,\mbs{J}_{\mu\nu})$.} It is clear that the set of transformations that preserve \eqref{eq:coordinate p in the rest frame} consist of spatial rotations and translations, forming the isotropy subalgebra $\msf{so}(3)\sds\mbb{R}^{3,1}$. We can now fix the value of other coordinates $j_{\mu\nu}$ while keeping $p^{\msf{R}}_\mu$ fixed using the left-over rotations and translations. Under an infinitesimal translation with parameter $\epsilon_\mu$, \eqref{eq:coadjoint action of poincare on coordinates on iso(3,1)*} implies that
\begin{equation}
    j_{0i}\to j_{0i}+\epsilon^\mu\delta_{\mbs{P}_\mu}j_{0i}=j_{0i}-\kappa\epsilon^\mu\eta_{\mu i}.
\end{equation}
Therefore, by a judicious choice of the translation parameters $\epsilon^\mu$, we can set
\begin{equation}\label{eq:coordinate j0i in the rest frame}
    j^{\msf{R}}_{0i}=0.
\end{equation}
This will leave the rotation subalgebra $\msf{so}(3)$ to fix $j_{kl}$. Since $j_{kl}=-j_{lk}$, there are three independent coordinates $j_{kl}$, and hence they can be thought as components of a three-vector $s^i$, defined through $j_{kl}=\varepsilon_{kli}s^i$. Furthermore, using the rotation freedom, we can get an arbitrary three-vector $s^i$ from a preferred three-vector $s^{\msf{R}}_i=(0,0,s)$, by the virtue of which, we can set
\begin{equation}\label{eq:coordinate jkl in the rest frame}
    j_{kl}^{\msf{R}}=\varepsilon_{kl}s, \qquad k,l=1,2.
\end{equation}
There is no preference in choosing the form of $s^{\msf{R}}_i$; It can be taken to be $(1,0,0)$, $(0,1,0)$, or any other fixed three-vector. Notice that using $s_i=\frac{1}{2}\varepsilon_{ikl}j^{kl}$ and \eqref{eq:linear poisson structure on iso(3,1)*,appendix}, it follows that
\begin{equation}\label{eq:poisson bracket of spin 3-vector on coadjoint orbits of poincare}
    \{s_i,s_j\}_{\iso(3,1)^*}=\varepsilon\indices{_{ij}^k}s_k.
\end{equation}
This means that $s_i$ implements the action of the isotropy algebra $\so(3)$ on $\iso(3,1)^*$. This is the classical counterpart of the fact that the spin three-vector $S_i$, defined in \eqref{eq:spin 3-vector in general Lorentz frame}, satisfies the $\so(3)$ algebra.

\smallskip Using \eqref{eq:coordinate p in the rest frame}, \eqref{eq:coordinate j0i in the rest frame}, and \eqref{eq:coordinate jkl in the rest frame}, we can write the rest frame form of $w_\mu$ as
\begin{equation}\label{eq:dual pl in the rest frame}
    w_\mu^{\msf{R}}=(0,0,0,-\kappa s).
\end{equation}
This fixes the value of $\mscr{C}_2(\iso(3,1))=-\kappa^2$ through the choice of \eqref{eq:coordinate p in the rest frame} and the value of $\mscr{C}_4(\iso(3,1))=\kappa^2 s^2$ through \eqref{eq:dual pl in the rest frame}. Therefore, any pair $(\kappa,s)\in\mbb{R}_+\times\mbb{R}$ completely fixes the orbit and would be an orbit representative.\footnote{Similar to the representation theory, where there is an invariant condition $\mbs{P}^0>0$ to preserve the casual type, there is an invariant $p^0>0$ which we have ignored but has to be considered.} After the quantization, any such pair, where now $s$ becomes integer or half-integer valued, provides a label for an irreducible representation of the Poincar\'e algebra. 

\paragraph{Coordinate representatives in a general frame.}
Having found a representative $(p_\mu^{\msf{R}},j_{\mu\nu}^{\msf{R}})$ of the coordinates on a coadjoint orbit in the rest frame, we can go to an arbitrary point by the coadjoint action of Poincar\'e algebra using \eqref{eq:lorentz transformation from the generic coordinate to the reference coordinate} and an arbitrary translation. An analysis similar to what we explain here using a complicated decomposition of the Lorentz transformation \eqref{eq:lorentz transformation from the generic coordinate to the reference coordinate} is done in \cite{Andrzejewski:2020qxt}. 

\smallskip Applying the coadjoint action  \eqref{eq:coadjoint action of the poincare group, appendix} to $(p_\mu^{\msf{R}},j_{\mu\nu}^{\msf{R}})$ for the Poincar\'e-group element $g=(\mbs{\Lambda}_p,\mbs{a})$,  we find that\footnote{We use the notation $(\mbs{a})_\nu=a_\nu$.}
\begin{eqgathered}\label{eq:coordinates on coadjoint orbits, first form}
    p_\mu=\left(p^{\msf{R}}\cdot\mbs{\Lambda}_p\right)_\mu
    \\
    j_{0i}=-p_0a_i+p_ia_0-\frac{1}{\kappa}\varepsilon^{ijk}p_js_k,
    \\
    j_{kl}=-p_ka_l+p_la_k+\varepsilon_{kli}s^i+\frac{(\varepsilon\indices{_{kij}}p_l-\varepsilon\indices{_{lij}}p_k)}{\kappa(\kappa+p^0)}p^is^j.
\end{eqgathered}
One can eliminate the last term in $j_{kl}$ by defining a new set of coordinates
\begin{equation}\label{eq:the coordinate sigma}
    \Sigma_i:=-p_0a_i+p_ia_0-\frac{p^0}{\kappa(\kappa+p^0)}\varepsilon_{imn}p^ms^n,
\end{equation}
by which, we can write \eqref{eq:coordinates on coadjoint orbits, first form} as
\begin{eqaligned}
    j_{0i}&=\Sigma^i-\frac{1}{\kappa+p^0}\varepsilon_{imn}p^ms^n,
    \\
    j_{kl}&=\frac{1}{p_0}\left(-p_k\Sigma_l+p_l\Sigma_k\right)+\varepsilon_{kli}s^i.
\end{eqaligned}
Note that $p^0>0$ and the above expressions are well-defined. We can finally re-define $\Sigma_i/p_0\to\Sigma_i$ and end-up with
\begin{eqaligned}\label{eq:coordinates on coadjoint orbits of poincare}
    j_{0i}&=p_0\Sigma_i-\frac{1}{\kappa+\phi^0}\varepsilon_{imn}p^ms^n,
    \\
    j_{kl}&=-p_k\Sigma_l+p_l\Sigma_k+\varepsilon_{kli}s^i.
\end{eqaligned}
We see that the coordinates $(p_\mu,j_{\mu\nu})$ in a general frame can be written in terms of $(p_\mu,\Sigma_i,s_i)$, which in total make $10$ parameters. However, on the orbit labeled by $(\kappa,s)$, $p_0$ is uniquely determined through $-(p^0)^2+(p^i)^2=-\kappa^2$ and $p^0>0$, and $s^2=\eta^{ij}s_is_j$. These two constraints reduce the number of parameters by $2$, and we are left with $8$ parameters, which is the same as the dimension of a coadjoint orbit. This data provides a concrete way to coordinatize an orbit. For another way to provide a coordinate on coadjoint orbits see \cite[\S 2.3]{DonnellyFreidelMoosavianSperanza202012}

\paragraph{Canonical coordinates on coadjoint orbits.} To study dynamics on coadjoint orbits, we define a canonical set of ``position" $\mcal{X}^\mu$, ``momentum"  $\mcal{P}_\mu$, and spin $\mcal{S}_\mu$ coordinates whose defining relations are\footnote{In the computation of these Poisson brackets, the restriction to the corresponding coadjoint orbit, fixed by $(\kappa,s)$,
is understood.}
\begin{eqaligned}\label{eq:poisson bracket of canonical coordinates}
         \{\mcal{X}^i,\mcal{X}^j\}_{\iso(3,1)^*}&=0, &\qquad 
      \{\mcal{P}_i,\mcal{P}_j\}_{\iso(3,1)^*}&=0,
      \\
      \{\mcal{X}^i,\mcal{S}_j\}_{\iso(3,1)^*}&=0, &\qquad 
      \{\mcal{P}_i,\mcal{S}_j\}_{\iso(3,1)^*}&=0, 
      \\
      \{\mcal{X}^i,\mcal{P}_j\}_{\iso(3,1)^*}&=\delta\indices{^i_j}, &\qquad  \{\mcal{S}_i,\mcal{S}_j\}_{\iso(3,1)^*}&=\varepsilon\indices{_{ij}^k}\mcal{S}_k,
\end{eqaligned}
The ``time evolution" on coadjoint orbits, whose parameter we denote as $\tau$, is defined by the Hamiltonian $H:=\mcal{P}^0$. Notice that, from \eqref{eq:linear poisson structure on iso(3,1)*,appendix}, we have
    \be
    \{p_k,j_{0l}\}_{\iso(3,1)^*}=+\eta_{kl}p_0,
    \ee
    which, given the expression of $j_{0i}$ in \eqref{eq:coordinates on coadjoint orbits of poincare}, is satisfied if we impose
        \be
    \{p_i,\Sigma_{j}\}_{\iso(3,1)^*}=\eta_{ij}\,,\quad \{p_i,s_{j}\}_{\iso(3,1)^*}=0\,.
    \ee
These relations taken together with \eqref{eq:poisson bracket of spin 3-vector on coadjoint orbits of poincare} imply that we can identify the canonical coordinates as follows 
\begin{eqaligned}\label{eq:identification of canonical coordinates}
    \mcal{X}^i&:= -\Sigma^i, &\qquad \mcal{S}_i&:= s_i.
    \\
     \mcal{P}_0&:=p_0, &\qquad \mcal{P}_i&:= p_i.
\end{eqaligned}
The coordinates $j_{\mu\nu}$ can be written as
\begin{eqaligned}\label{eq:coorinates jmunu in canonical coordinates}
    j_{0i}&=-\mcal{X}_i\mcal{P}_0-\frac{1}{\kappa+ \mcal{P}^0}\varepsilon_{imn}\mcal{P}^m\mcal{S}^n,
    \\
    j_{kl}&=-\mcal{X}_k\mcal{P}_l+\mcal{X}_l\mcal{P}_k+\varepsilon_{kli}\mcal{S}^i.
\end{eqaligned}

\paragraph{Evolution of the dual Pauli--Luba\'nski pseudo-vector.} Finally, we can determine the time-evolution of $w_\mu$ \eqref{eq:dual PL vector, appendix}. We define the time-evolution of a quantity $\mcal{Q}$ is determined through
\begin{equation}
    \dot{\mcal{Q}}=\frac{\rd \mcal{Q}}{\rd \tau}:=\{\mcal{Q},H\}_{\iso(3,1)^*},
\end{equation}
where $H=\mcal{P}_0$ plays the role of the Hamiltonian of the system. From \eqref{eq:the coordinate sigma} (with the rescaling $\Sigma_i/p_0$ is understood) and $a^0\to \tau$, we have $\dot{\mcal{X}}^i=\mcal{P}^i/\mcal{P}^0$. Furthermore, from \eqref{eq:linear poisson structure on iso(3,1)*,appendix}, \eqref{eq:poisson bracket of canonical coordinates}, and \eqref{eq:identification of canonical coordinates}, we get $\dot{\mcal{P}}^i=\dot{\mcal{S}}^i=0$. Hence, from \eqref{eq:coorinates jmunu in canonical coordinates}, we get 
\begin{equation}
    \dot{j_{0i}}=\mcal{P}_i, \qquad \dot{j}^{kl}=0,
\end{equation}
from which we conclude that
\begin{equation}\label{eq:vanishing of time-evlution of dual pauli--lubanski pseudo-vector}
    \dot{w}_\mu=0.
\end{equation}
This shows that $w_\mu$ is a constant of motion.

\section{Details of various computations}\la{App:comp}

In this appendix, we provide the details of various computations leading to some of the results in the main body of the paper.

\subsection{Proof of results in  \S \ref{sec:gbms and poincare embeddings}}\la{sec:proofs in section 3}

\paragraph{Proof of intrinsic angular momentum algebra \eqref{SY}.}

In order to derive the algebra \eqref{SSS} of the intrinsic angular momentum \eqref{SY}, we use the coadjoint action \eqref{eq:coadjoint action on gbms*}, as well as $\d_TG=T, \d_Y G=(\cL_Y-W_Y)G$ to compute first the actions
 \be
\d_T \msf{I}_Y&=\f32 \bigintsss_S Y^Am\pa_AT\bfe +\frac{1}{2}\bigintsss_S TY^A\pa_Am\bfe
- \bigintsss_S Y^A\p_A T m\bfe
+ \f 12 \bigintsss_S m {T D_AY^A}\bfe
\cr
&=
\f12 \bigintsss_S Y^A\pa_A(Tm)\bfe
+ \f12 \bigintsss_S m {T D_AY^A}\bfe
=0
\,,
\\
\d_{Y'}\msf{I}_Y&=\underbrace{\bigintsss_S Y^A \left(\cL_{Y'}j_A + 2W_{Y'} j_A\right)\bfe}_{:=\d_{Y'}^1}
\underbrace{- \bigintsss_S \left(Y^A\p_A G-\f 12 G D_AY^A \right)\left(Y^{'B} \pa_B m+  3 W_{Y'} m\right)\bfe}_{:=\d_{Y'}^2}
\cr
&\underbrace{-\bigintsss_S\left[Y^AD_A\left(Y^{'B}\p_B-W_{Y'}\right)G -\f12D_AY^A \left(Y^{'B}\p_B-W_{Y'}\right)G\right] m\bfe}_{:=\d_{Y'}^3}\,,
\ee
from which we see immediately that the intrinsic angular momentum \eqref{SY} is supertranslation invariant. Moreover, we manipulate the second (angular momentum) action using
\be
\d_{Y'}^1&=\bigintsss_S Y^A \left(
Y^{'B} D_Bj_A +j_B  D_AY^{'B}
 + D_B Y^{'B} j_A\right)\bfe
 \cr
& =
 \bigintsss_S  \left(
-D_B (Y^{'B}Y^A)  +  Y^B D_BY^{'A}
 + Y^AD_B Y^{'B} \right)j_A\bfe
 \cr
& =
 \bigintsss_S  \left(
- Y^{'B}D_B Y^A  +  Y^B D_BY^{'A}
  \right)j_A\bfe
  \cr
& =
 \bigintsss_S  \left[
Y,Y'
  \right]_{S}^Aj_A\bfe\,,\la{JJ}
\ee
and
\be
\d_{Y'}^2&=- \bigintsss_S \left(Y^A\p_A G-\f 12 G D_AY^A \right)\left(Y^{'B} \pa_B m+  3 W_{Y'} m\right)\bfe
\cr
&=
 \bigintsss_S\left(\left[ D_B\left(Y^A\p_A G-\f 12 G D_AY^A \right) \right]Y^{'B} -  \f12\left(Y^A\p_A G-\f 12 G D_AY^A \right) D_B Y^{'B} \right)m\bfe\,.\cr
\ee
We then look at all contributions of the  action $\d_{Y'}^2+\d_{Y'}^3$ that do not contain derivatives on the Goldstone field $G$, which give
\be
 \f 12 \bigintsss_S\left(
 -Y^{'B}D_B D_AY^A 
 + Y^AD_AD_BY^{'B}\right) Gm\bfe
=\f 12 \bigintsss_S
 G D_A\left[
Y,Y'
  \right]_{S}^A m\bfe\,,\la{LJ1}
\ee
while the terms with derivatives on $G$ give
\be
&-\bigintsss_S\Big[Y^BD_BY^{'A}-\f12 Y^A D_BY^{'B}
+Y^AY^{'B}D_B
-\f12Y^{'A}D_BY^B 
\cr
&
-Y^{'B}D_BY^A-Y^AY^{'B}D_B+\f12Y^{'A}D_BY^B
+\f12 Y^AD_BY^{'B}\Big]\p_AG m\bfe
\cr
&=-\bigintsss_S\Big[Y^BD_BY^{'A}
-Y^{'B}D_BY^A
\Big]\p_AG m\bfe
\cr
&=-\bigintsss_S
\left[
Y,Y'
  \right]_{S}^A
\p_AG m\bfe\,.\la{LJ2}
\ee
Combining \eqref{LJ1} and \eqref{LJ2}
we obtain 
\be
\{\mathsf{L}_{Y}, \mathsf{J}_{Y'}\}_{\g^*}&=\d_{Y'}\mathsf{L}_{Y}=\mathsf{L}_{\left[
Y,Y'
  \right]_{S}}\,.
\ee 
Including the action \eqref{JJ}, we thus  arrive at
\be
\{\mathsf{I}_{Y}, \mathsf{J}_{Y'}\}_{\g^*}&=\d_{Y'}\mathsf{I}_{Y}=\mathsf{I}_{\left[
Y,Y'
  \right]_{S}}\,.
\ee 
In order to compute the bracket $\{\mathsf{I}_{Y}, \mathsf{I}_{Y'}\}_{\g^*}$ we need to include the action the Goldstone mode generator $\mathsf{G}_\varphi:=\bigintsss_S \varphi G \,\bfe$ on $\mathsf{P}_T$, namely
 $\d_\varphi \mathsf{P}_T=-\d_T \mathsf{G}_\varphi=-\bigintsss_S \varphi T\,\bfe$, which yields the bracket
\be
\{G(\sigma), m(\sigma')\}_{\g^*}=\d^{(2)}(\sigma, \sigma')\,.
\ee
We then have
\be
&\{\mathsf{L}_{Y}, \mathsf{L}_{Y'}\}_{\g^*}=
\bigintsss_S \bfe \bigintsss_S \bfe'
\{(Y^A\p_A G(\sigma)-\f12 G(\sigma)D_AY^A)m(\sigma),
(Y^{'B}\p_{B'} G(\sigma')-\f12 G(\sigma')D_{B'}Y^{B'})m(\sigma')\}
\cr
&=\bigintsss_S \bfe \bigintsss_S \bfe' \bigg[m(\sigma)Y^{'B}\p_{B'} G(\sigma') Y^A\p_A\d^{(2)}(\sigma, \sigma')
-\f12 m(\sigma)G(\sigma')D_{B'}Y^{B'}
Y^A\p_A\d^{(2)}(\sigma, \sigma')
\cr
&-\f12 D_AY^Am(\sigma)Y^{'B}\p_{B'} G(\sigma')\d^{(2)}(\sigma, \sigma')
+\f14 m(\sigma)D_AY^AD_{B'}Y^{B'}G(\sigma')\d^{(2)}(\sigma, \sigma')
-Y\leftrightarrow Y'\bigg]
\cr
&=\bigintsss_S 
\bigg[
Y^AD_A(Y^{'B}\p_{B} G) 
-\f12  Y^AD_A(GD_{B}Y^{'B})
-\f12m Y^{'B} D_AY^A \p_{B} G
-Y\leftrightarrow Y'
\bigg]m\,\bfe
\cr
&=\bigintsss_S 
\left[
Y^AD_AY^{'B} \p_{B} G
-\f12  G Y^AD_AD_{B}Y^{'B}
-Y\leftrightarrow Y'
\right]m\,\bfe
\cr
&=\bigintsss_S 
\left(
 [Y,Y']^B\p_{B} G
-\f12  GD_B [Y,Y']^B
\right)m\,\bfe\,,
\ee
where in the last passage we used
$Y^A[D_A,D_{B}]Y^{'B}-Y\leftrightarrow Y'=0$.
Hence, we arrive at
\be
\{\mathsf{L}_{Y}, \mathsf{L}_{Y'}\}_{\g^*}= \mathsf{L}_{\left[
Y,Y'
  \right]_{S}}\,,
\ee
as well as
\be
\{\mathsf{I}_{Y}, \mathsf{I}_{Y'}\}_{\g^*}= \mathsf{I}_{\left[
Y,Y'
  \right]_{S}}\,.
\ee

\paragraph{Proof of $\gbms$ coadjoint action \eqref{eq:coadjoint action on gbms*}.}

The derivation goes as follows. Using the natural pairing of $\gbms$ and $\gbms^*$, the right-hand side of \eqref{coadj} can be written as
\begin{equation*}\label{eq:the pairing between gbms and gbms*}
	\begin{aligned}
	&\hphantom{=}\langle \delta_{(Y_1,T_1)} (j, m )| Y_2,T_2 \rangle 
	=-\langle j ,m | Y_{12},T_{12}\rangle=-\bigintsss_{S} \big(Y_{12}^A j_A +T_{12} m\big)\bfe
	\\
		&=-\bigintsss_{S} \left[
		(\cL_{Y_1} Y_2^A) j_A 
		+\left(Y_1[T_2]-Y_2[T_1] +\frac{1}{2}T_1 \cD _AY^A_2-\frac{1}{2}T_2 \cD _AY^A_1\right) m \right]\bfe
	\\
	&=\bigintsss_{S} \left[
		Y_2 ^A (\cL_{Y_1} j_A  +j_A \cD_B Y^B)		+ 
		Y_2^A \left( \pa_A T_1 m+ \frac{1}{2} \cD_A (T_1 m ) \right) 
		+ T_2 \left(\cD _A(Y^A_1 m)  + \frac{1}{2} (\cD _AY^A_1) m \right)\right]\bfe\,,
		\end{aligned}
\end{equation*}
where the Lie derivative action on a one-form is 
\begin{equation}
    \mcal{L}_Y j_A=Y^B\cD_B j_A+ j_B\cD_AY^B\,.
\end{equation}
Collecting all terms, we have the following infinitesimal coadjoint action of $\gbms$ on $\gbms^*$
\begin{eqgathered}
    \delta_{(Y,T)}m=\cD_A(m Y^A)+\frac{m}{2} \cD_AY^A,
    \\
    \delta_{(Y,T)}j_A=\cL_{Y_1} j_A  +j_A \cD_B Y^B +\frac{3}{2}m \p_A T+\frac{1}{2} \p_A (m T).
\end{eqgathered}
 This proves \eqref{eq:coadjoint action on gbms*}.

\paragraph{Proof of $\gbms^*$ coordinate  Poisson brackets \eqref{eq:poisson brackets of coordinates on gbms*}.}

The first relation in \eqref{eq:poisson brackets of coordinates on gbms*} can be proven as follows. From the first relation of \eqref{eq:gbms charge algebra}, 
\begin{eqaligned}
    \{J_{Y_1},J_{Y_2}\}_{\g^*}&=\bigintsss_{S\times S'}Y_1^A(\sigma)Y_2^B(\sigma')\{j_A(\sigma),j_B(\sigma')\}_{\g^*}=J_{[Y_1,Y_2]_S}
    \\
    &=\bigintsss_S j_A(\sigma)\left(Y_1^B(\sigma)\pa_BY_2(\sigma)-Y_2^B(\sigma)\pa_BY_1^A(\sigma)\right).
\end{eqaligned}
where $S'$ is the sphere with local coordinates $\sigma'$. The two terms in the last relation can be written as follows. The first term becomes
\begin{eqaligned}
    \bigintsss_S j_A(\sigma)Y_1^B(\sigma)\pa_BY_2(\sigma)&=\bigintsss_{S\times S'}j_A(\sigma)Y_1^B(\sigma)\pa'_BY_2^A(\sigma')\delta^{(2)}(\sigma-\sigma')
    \\
    &=-\bigintsss_{S\times S'}j_A(\sigma)\partial'_B\delta^{(2)}(\sigma-\sigma')Y_1^B(\sigma)Y_2^A(\sigma')
    \\
    &=-\bigintsss_{S\times S'}Y_1^A(\sigma)Y_2^B(\sigma')j_B(\sigma)\partial'_A\delta^{(2)}(\sigma-\sigma').
\end{eqaligned}
Similarly, the second term becomes
\begin{eqaligned}
    -\bigintsss_S j_A(\sigma)Y_2^B(\sigma)\pa_BY_1^A(\sigma)&=-\bigintsss_{S\times S'} j_A(\sigma')\delta^{(2)}(\sigma-\sigma')Y_2^B(\sigma')\pa_BY_1^A(\sigma)
    \\
    &=+\bigintsss_{S\times S'} Y_1^A(\sigma)Y_2^B(\sigma')j_A(\sigma')\pa_B\delta^{(2)}(\sigma-\sigma').
\end{eqaligned}
We can now read-off $\{j_A(\sigma),j_B(\sigma')\}_{\g^*}$ easily 
\begin{equation}
    \{j_A(\sigma),j_B(\sigma')\}_{\g^*}=j_A(\sigma')\pa_B\delta^{(2)}(\sigma-\sigma')-j_B(\sigma)\partial'_A\delta^{(2)}(\sigma-\sigma'),
\end{equation}
which is the first relation in \eqref{eq:poisson brackets of coordinates on gbms*}. The proof of the second relation in \eqref{eq:poisson brackets of coordinates on gbms*} goes similarly. From \eqref{eq:gbms charge algebra}, we know
\begin{eqaligned}
    \{J_Y,M_T\}_{\g^*}&=\bigintsss_{S\times S'}Y^A(\sigma)T(\sigma')\{j_A(\sigma),m(\sigma')\}_{\g^*}=M_{Y[T]-TW_Y}
    \\
    &=\bigintsss_{S} m(\sigma)\left(Y^A(\sigma)\pa_AT(\sigma)-\frac{T(\sigma)}{2}\cD_AY^A(\sigma)\right).
\end{eqaligned}
The first term can be written as
\begin{eqaligned}
    \bigintsss_{S} m(\sigma)Y^A(\sigma)\pa_AT(\sigma)&=\bigintsss_{S\times S'} m(\sigma)Y^A(\sigma)\delta^{(2)}(\sigma-\sigma')\pa'_AT(\sigma')
    \\
    &=-\bigintsss_{S\times S'} Y^A(\sigma)T(\sigma')m(\sigma)\pa'_A\delta^{(2)}(\sigma-\sigma'),
\end{eqaligned}
while the second term becomes
\begin{eqaligned}
    -\bigintsss_{S} m(\sigma)\frac{T(\sigma)}{2}\cD_AY^A(\sigma)&=-\frac{1}{2}\bigintsss_{S\times S'}m(\sigma')T(\sigma')\delta^{(2)}(\sigma-\sigma')\cD_AY^A(\sigma)
    \\
    &=+\bigintsss_{S\times S'}Y^A(\sigma)T(\sigma')\left(\frac{m(\sigma')}{2}\pa_A\delta^{(2)}(\sigma-\sigma')\right).
\end{eqaligned}
We thus have
\begin{equation}
    \{j_A(\sigma),m(\sigma')\}_{\g^*}=\frac{m(\sigma')}{2}\pa_A\delta^{(2)}(\sigma-\sigma')-m(\sigma)\pa'_A\delta^{(2)}(\sigma-\sigma'),
\end{equation}
which is the desired result. Finally, $\{m(\sigma'),m(\sigma'')\}_{\g^*}=0$ is an immediate consequence of \eqref{eq:gbms charge algebra}. 

\paragraph{Proof of \eqref{cDe}.}

We want to compute
\be
D_B e^\mu_A=\p_B e^\mu_A-e^\mu_C \Gamma^C_{AB}(q).
\ee
We start computing 
\be
e^\mu_C\Gamma^C_{AB}(q)&=\f12 e^{\mu D} (\p_A q_{DB}+\p_B q_{DA}-\p_D q_{AB})
\cr
&=\f12 e^{\mu D} (\p_A (e^\nu_D e_{\nu B})+\p_B (e^\nu_D e_{\nu A})-\p_D q_{AB})
\cr
&=(\eta^{\mu\nu}+\f12 (n^\mu\bar n^\nu+\bar n^\mu n^\nu))\p_A e_{\nu B}
\cr
&+\f12 e^{\mu D}\left(e_{\nu B} \p_D e^\nu_A
+e_{\nu A} \p_D e^\nu_B
-\p_D q_{AB}\right)
\cr
&=\p_A e^\mu_{ B} +\f12 (n^\mu\bar n^\nu+\bar n^\mu n^\nu)\p_A e_{\nu B}
\ee
where we used \eqref{q-ind} and $\p_A e^\nu_D=\p_D e^\nu_A$. We thus have
\be
D_B e^\mu_A&=-\f12 (n^\mu\bar n^\nu+\bar n^\mu n^\nu)\p_A e_{\nu B}
\cr
&=
\f12 
(
n^\mu \p_A \bar n^\nu 
+
\bar n^\mu  \p_A n^\nu
)e_{\nu B}
\cr
&=
\f12 
(
\bar n^\mu  -n^\mu 
)q_{AB}\,,
\ee
where we used $
n_\mu e^\mu_A=0=\bar n_\mu e^\mu_A$ and $\p_A \bar n^\mu=-\p_A  n^\mu$  (see \eqref{pnA}).

\subsection{Proof of results in  \S \ref{sec:algebraic aspects of gbms}}\la{sec:proofs in section 4}

\paragraph{Proof of \eqref{eq:p transformation under gbms}.}

Using \eqref{eq:coadjoint action on gbms*}, we can easily derive the transformation rules \eqref{eq:p transformation under gbms}. We have
 \begin{eqaligned}
     \delta_T p_A&=\delta_T(\rho^{-1}j_A)
     \\
     &=\frac{3}{2}\rho^{-1}\left(m\partial_AT+\frac{T}{3}\partial_Am\right)
     \\
     &=\frac{3}{2}\partial_A\left(m^{\frac{1}{3}}T\right)
     \\
     &=\frac{3}{2}\partial_A(\sqrt{\rho}T),
 \end{eqaligned}
 where in going  to the second line, we have used $\delta_T\rho=0$. Similarly,
\begin{eqaligned}
    \delta_Yp_A&=\delta_Y(\rho^{-1}j_A)
    \\
    &=-\rho^{-2}\delta_Y\rho+\rho^{-1}\delta_Yj_A
    \\
    &=-\rho^{-2}\cD_A(\rho Y^A)j_A+\rho^{-1}\left(\mcal{L}_Yj_A+2W_Yj_A\right)
    \\
    &=-\rho^{-1}\cD_A(\rho Y^A)p_A+\rho^{-1}Y^A\cD_A\rho p_A+\mcal{L}_Yp_A+\cD_AY^Ap_A
    \\
    &=\mcal{L}_Yp_A.
\end{eqaligned}
This completes the proof of \eqref{eq:p transformation under gbms}.

\paragraph{Proof of \eqref{wT}.}

First notice that
\begin{eqaligned}
    \epsilon_{AB}=\sqrt{\gamma}\varepsilon_{AB}, \qquad \epsilon^{AB}=\frac{1}{\sqrt{\gamma}}\varepsilon^{AB},
\end{eqaligned}
where $\varepsilon_{AB}$ is the Levi--Civita symbol with $\varepsilon_{12}=1$. Therefore, \eqref{dq0} implies that 
\begin{equation}
    \delta_{(Y,T)}\epsilon_{AB}=0, \qquad \delta_{(Y,T)}\epsilon^{AB}=0.
\end{equation}
Hence, 
\begin{eqaligned}
    \delta_Tw&=\rho^{-1}\epsilon^{AB}\partial_A\delta_Tp_B 
    \\
    &=\frac{3}{2}\rho^{-1}\epsilon^{AB}\partial_A\partial_B(\sqrt{\rho}T)
    \\
    &=0.
\end{eqaligned}
On the other hand,
\begin{eqaligned}\label{eq:coadjoint transformation of w rho}
\delta_{Y} (w\rho) &=\epsilon^{AB} \pa_A \delta_Y {p}_B
\\
&=\epsilon^{AB}\partial_A(Y^C \cD_Cp_B+p_C\cD_BY^C)
\\
&= \epsilon^{AB} 
\pa_A ( Y^C\pa_C {p}_B + p_C\pa_B Y^C)
\\ 
&=  \epsilon^{AB} (\pa_A  Y^C) (\pa_C {p}_B - \pa_B p_C) +  Y^C \epsilon^{AB} \pa_A \pa_C  {p}_B
\\
&=\epsilon^{AB} \epsilon_{CB} (\pa_A  Y^C) w {\rho} +  Y^C \pa_C (w {\rho})
- Y^C(\pa_C \epsilon^{AB}) \pa_A\bar{p}_B
\\
&=  (\cD_C  Y^C) w {\rho}  +  Y^C \pa_C (w {\rho})
\\
&= \cD_A[w {\rho}  Y^A],
\end{eqaligned}
where we used the following results (1) in the fifth line, we have used $\pa_C {p}_B - \pa_B {p}_C = \epsilon_{CB} \rho w$; and (2) we have
\begin{eqaligned}
    \partial_C\epsilon^{AB}&=\partial_C\left(\frac{1}{\sqrt{\gamma}}\varepsilon^{AB}\right)
    =-\frac{\partial_C\sqrt{\gamma}}{\sqrt{\gamma}}\epsilon^{AB}.
\end{eqaligned}
Hence, using the definition $\cD_AY^A=1/\sqrt{\gamma}\partial_A(\sqrt{\gamma}Y^A)$, we end up with the first term in the sixth line. Therefore, 
\begin{eqaligned}
    \cD_A(w\rho Y^A)&=\delta_Y(w\rho)
    \\
    &=\delta_Yw\rho+w\delta_Y\rho 
    \\
    &=\delta_Yw\rho+w\cD_A(\rho Y^A).
\end{eqaligned}
We thus end-up with 
\begin{equation}
    \delta_Yw=Y^A\partial_Aw=Y[w],
\end{equation}
which is the desired result \eqref{wT}.

\paragraph{Proof of \eqref{eq:invariance of casimirs under gbms}.}

We trivially have $\delta_T\mscr{C}_n(\gbms)=0$, which follows from \eqref{dq0}, \eqref{eq:rho transformation under gbms}, and \eqref{wT}. Furthermore, the same equations imply that 
\begin{eqaligned}
    \delta_Y\mscr{C}_n(\gbms)&=\bigintsss_S\delta_Y(w^n\rho\bfe)
    \\
    &=\bigintsss_S\left(\delta_Yw^{n}\rho\bfe+w^n\delta_Y\rho\bfe+w^n\rho\delta_Y\bfe\right)
    \\
    &=\bigintsss_S \left(Y^A\partial_A w^n\rho\bfe+w^n \cD_A(\rho Y^A)\bfe \right)
    \\
    &=\bigintsss_S \cD_A(w^n\rho Y^A)\bfe
    \\
    &=0.
\end{eqaligned}

\paragraph{Proof of \eqref{eq:algebra of area-preserving vector fields for rescaled area form}.}

This only requires writing down the Lie bracket
\begin{eqaligned}
    [Y_\chi,Y_\psi]^A&=Y^B_\chi\partial_BY^A_\psi-Y^B_\psi\partial_BY^A_\chi
    \\
    &=m^{-\frac{2}{3}}\epsilon^{BC}\left[\partial_C\chi\partial_B\left(m^{-\frac{2}{3}}\epsilon^{AD}\partial_D\psi\right)-\partial_C\psi\partial_B\left(m^{-\frac{2}{3}}\epsilon^{AD}\partial_D\chi\right)\right]
    \\
    &=m^{-\frac{2}{3}}\epsilon^{BC}\partial_B\left(m^{-\frac{2}{3}}\epsilon^{AD}\partial_C\chi\partial_D\psi\right)-m^{-\frac{2}{3}}\epsilon^{BC}\partial_B\left(m^{-\frac{2}{3}}\epsilon^{AD}\partial_D\chi\partial_C\psi\right)
    \\
    &=m^{-\frac{2}{3}}\epsilon^{BC}\partial_B\left(m^{-\frac{2}{3}}\epsilon^{AD}\partial_C\chi\partial_D\psi\right)-m^{-\frac{2}{3}}\epsilon^{BD}\partial_B\left(m^{-\frac{2}{3}}\epsilon^{AC}\partial_C\chi\partial_D\psi\right)
    \\
    &=\left(\varepsilon^{BC}\varepsilon^{AD}-\varepsilon^{BD}\varepsilon^{AC}\right)\left[\frac{m^{-\frac{2}{3}}}{\sqrt{\gamma}}\partial_B\left(\frac{m^{-\frac{2}{3}}}{\sqrt{\gamma}}\partial_C\chi\partial_D\psi\right)\right],
\end{eqaligned}
where in the third line, we have used the anti-symmetry of $\epsilon^{BC}$, and in the last line, we have used $\epsilon^{AB}=\gamma^{-\frac{1}{2}}\varepsilon^{AB}$. We now use the following relation
\begin{eqaligned}
    \varepsilon^{BC}\varepsilon^{AD}-\varepsilon^{BD}\varepsilon^{AC}&=\delta^{BA}\delta^{CD}-\delta^{BD}\delta^{CA}-\delta^{BA}\delta^{DC}+\delta^{BC}\delta^{DA}
    \\
    &=\delta^{BC}\delta^{DA}-\delta^{BD}\delta^{CA}
    \\
    &=-\varepsilon^{AB}\varepsilon^{CD}.
\end{eqaligned}
Using this relation, we would have
\begin{eqaligned}
    [Y_\chi,Y_\psi]^A&=-\varepsilon^{AB}\varepsilon^{CD}\left[\frac{m^{-\frac{2}{3}}}{\sqrt{\gamma}}\partial_B\left(\frac{m^{-\frac{2}{3}}}{\sqrt{\gamma}}\partial_C\chi\partial_D\psi\right)\right]
    \\
    &=-m^{-\frac{2}{3}}\epsilon^{AB}\partial_B\left(m^{-\frac{2}{3}}\epsilon^{CD}\partial_C\chi\partial_D\psi\right)
    \\
    &=-m^{-\frac{2}{3}}\epsilon^{AB}\partial_B\{\chi,\psi\}_\rho.
    \\
    &=Y^A_{\{\chi,\psi\}_\rho},
\end{eqaligned}
as we wished to prove.

\paragraph{Proof of \eqref{eq:coadjoint transformation of vorticity}.}

The proof goes as follows
\begin{eqaligned}
    \delta_{Y}\mscr{S}_\chi&=\bigintsss_S\left(\delta_Y\chi \rho\omega+\chi\delta_Y(\rho w)\right)\bfe
    \\
    &=\bigintsss_S \chi \cD_A(\rho w Y^A)\bfe
    \\
    &=-\bigintsss_S Y^A\partial_A\chi \rho w\bfe
    \\
    &=-\mscr{S}_{Y[\chi]},
\end{eqaligned}
where in the third line, we have used \eqref{eq:coadjoint transformation of w rho}. Note also that in these expressions $\delta_{(Y,T)}$ denotes the coadjoint action and acts on elements of $\gbms^*$. Therefore, $\delta_{(Y,T)}\chi=0$ since it is not an element of $\gbms^*$.

\paragraph{Proof of \eqref{eq:poisson lie bracket of smeared vorticities}.}

This can be proven as follows.
From the actions \eqref{eq:coadjoint transformation of vorticity}, we see immediately that $\d_\chi \rho=0$ and, at the same time, we can compute $\d_\chi J_Y= - \d_Y \mscr{S}_\chi$ from which
\be
\d_\chi j_A =\p_A \chi \epsilon^{BC}\p_B p_C.
\ee
Therefore, we have
\begin{eqaligned}
    \{\mscr{S}_\chi,\mscr{S}_\psi\big\}_{\g^*}&=-\d_\chi \mscr{S}_\psi
=-\bigintsss_S \psi\,\epsilon^{AB} \pa_A (\rho^{-1} \d_\chi j_B) \,\bfe
\cr
&=-\bigintsss_S \psi\,\epsilon^{AB} \pa_A (\rho^{-1} \p_B \chi \epsilon^{CD}\p_C p_D) \,\bfe
\cr
&=\bigintsss_S  \rho^{-1} \epsilon^{AB} \pa_A  \psi \p_B \chi  \,\epsilon^{CD}\p_C p_D \,\bfe
\cr
&=-\bigintsss_S \{\chi,\psi\}_\rho\rho w\bfe 
\cr
    &=-\mscr{S}_{\{\chi,\psi\}_\rho}.
\end{eqaligned}

\subsection{Poincar\'e charge algebra}\la{app:Lor}

We derive first the Lorentz algebra \eqref{Lor-alg} for the generators \eqref{Lor-gen} by repeated use of the identities \eqref{nid}.
We start with the algebra of the rotation generators. By means of \eqref{eq:gbms charge algebra} we have
\begin{eqaligned}
    \{J^i, J^j\}_{\g^*}&=
\bigintsss_S(
\epsilon^{BC}\pa_C n^i  D_B ( \epsilon^{AD}\pa_D n^j)-i\leftrightarrow j)
\, j_A\,  \bfe
\cr
&=\bigintsss_S
 \epsilon^{AD} \epsilon^{BC}  D_D(\pa_C n^i \pa_B n^j)
\, j_A\,  \bfe
\cr
&=\bigintsss_S
  \epsilon^{AD}\pa_D( \epsilon^{BC}\pa_C n^i \pa_B n^j)\, j_A\,  \bfe
\cr
&=-\varepsilon^{ij}\!_k\bigintsss_S
  \epsilon^{AD}\pa_Dn^k\, j_A\,  \bfe
\cr
&=-\varepsilon^{ij}\!_k J^k.
\end{eqaligned}
where we used $\epsilon^{AB} \pa_A n^i \pa_Bn^j=\epsilon^{ij}{}_k n^k$ see \eqref{nid}.

\smallskip Next, we compute
\be
\{J_i,K_j\}_{\g^*}&=
\bigintsss_S\left[
 \epsilon^{BC}\pa_C n_i  D_B (q^{AD}\pa_D n_j)
-q^{BD}\pa_D n_j    D_B(  \epsilon^{AC}\pa_C n_i)\right]
\, j_A\,  \bfe
\cr
&=
\bigintsss_S\left[
  \epsilon^{BC}q^{AD}\pa_C n_i   D_B \pa_D n_j
-q^{BD}  \epsilon^{AC}\pa_D n_j   D_B\pa_C n_i
\right]
\, j_A\,  \bfe
\cr
&=
-\bigintsss_S\left[
  \epsilon^{BC}q^{AD} q_{BD} \pa_C n_i  n_j
-q^{BD}  \epsilon^{AC}  q_{BC}\pa_D n_j  n_i
\right]
\, j_A\,  \bfe
\cr
&=
-\bigintsss_S\left[
  \epsilon^{AC} (\pa_C n_i  n_j
-\pa_C n_j  n_i)
\right]
\, j_A\,  \bfe
\cr
&=-\varepsilon_{ij}\!^kK_k\,,
\ee
where we used $ {D}_B {D}_Cn_i= - q_{BC} n_i$ see \eqref{keydiff} and 
$n^i\pa_A n^j- n^j\pa_A n^i = \varepsilon^{ij}\!_k  \epsilon_A{}^{B}\pa_B n^k$ see \eqref{nid}.

\smallskip Next, we compute
\be
\{K_i,K_j\}_{\g^*}&=
\bigintsss_S\left(
q^{BC}\pa_C n_i   D_B (q^{AD}\pa_D n_j)
-i\leftrightarrow j\right)\, j_A\,  \bfe
\cr
&=
\bigintsss_S\left[
q^{BC}q^{AD}\pa_C n_i   D_B \pa_D n_j
-q^{BC}q^{AD}  D_B \pa_D n_i\pa_C n_j 
\right]\, j_A\,  \bfe
\cr
&=
-\bigintsss_S\left[
q^{BC}q^{AD} q_{BD} (n_j\pa_C n_i  -
 n_i\pa_C n_j)
\right]\, j_A\,  \bfe
\cr
&=
\varepsilon_{ij}{}^k \bigintsss_S\left[
q^{AC}   \epsilon_C{}^B \pa_B n_k
\right]\, j_A\,  \bfe
\cr
&= \varepsilon_{ij}{}^k J_k.
\ee

Finally, we derive the Lorentz action on the four-momentum \eqref{eq:poisson bracket of momentum and angular momentum, gbms}.
From \eqref{eq:coadjoint action on gbms*}
we have
\be
\d_{T=n_\mu} j_A=
\f32 m\pa_A n_\mu+\frac{n_\mu}{2}\pa_Am\,.
\ee
We can thus compute the brackets
\be
\{P_i,J_j\}_{\g^*}&=-
\bigintsss_SY^A_{J_j}\,\d_{T=n_i} j_A\,  \bfe
=-\bigintsss_S \,
 m \epsilon^{AB}\pa_B n_j\pa_A n_i\,  \bfe
 \cr
 &=- \varepsilon_{ij}\!^k\bigintsss_S \,
 n_k m\,  \bfe
 =-\varepsilon_{ij}\!^k P_k\,,
 \cr
\{P_0,J_j\}_{\g^*}&=-
\bigintsss_SY^A_{J_j}\,\d_{T=n_0} j_A  \bfe
=
-
\f12\bigintsss_S \epsilon^{AB}\pa_B n_j\,\p_A m\,  \bfe=0 \,,
\cr
\{P_i,K_j\}_{\g^*}&=-
\bigintsss_SY^A_{K_j}\,\d_{T=n_i} j_A\,  \bfe
=-
\bigintsss_S 
\left(
\f32 m\pa_A n_i\p^A {n_j}+\frac{1} {2} n_i\p^A {n_j}\pa_Am
\right)\,  \bfe
\cr
&=-
\bigintsss_S 
\left(
 m\pa_A n_i\p^A {n_j}-\frac{1} {2} mn_i\Delta {n_j} 
\right)\,  \bfe
=-\eta_{ij}
\bigintsss_S m \,  \bfe
=-\eta_{ij} P_0\,,
\cr
\{P_0,K_j\}_{\g^*}&=-
\bigintsss_SY^A_{K_j}\,\d_{T=n_0} j_A\,  \bfe
=-\frac{1} {2}
\bigintsss_S 
\left(
 \p^A {n_j}\pa_Am
\right)\,  \bfe
=-P_j\,.
\ee

\subsection{Proof of \eqref{YW}.}\la{app:YW}
By means of \eqref{eq:rho transformation under gbms}, \eqref{wT}, we have
\be
\d_Y S_\mu&=
\bigintsss_S  \, n_\mu\,\d_Y(\rho^\f32 w)\bfe
=\bigintsss_S  \, n_\mu\left(\rho^\f32Y^A\p_A w 
+\f32 \rho^\f12 w \cD_A(\rho Y^A)
\right)\bfe
\cr
&=\bigintsss_S  \, n_\mu \cD_A \left(\rho^\f32Y^A w\right) \bfe
+\f12 \bigintsss_S  \, n_\mu\rho^\f32 w \cD_A Y^A
\bfe
\cr
&=-\bigintsss_S  \, Y^A\p_A n_\mu  \rho^\f32 w\, \bfe
+\f12 \bigintsss_S  \, n_\mu\rho^\f32 w \cD_A Y^A
\bfe
\cr
&=
-\msf{S}[\rho^\f12 Y[n_\mu]]
+\f12 \msf{S}[\rho^\f12 n_\mu \cD_AY^A] .
\ee

\subsection{Lorentz covariance of the Pauli--Luba\'nski generator }\la{app:Wcov}

We want to use the action \eqref{YW} to verify the analog of the Lorentz action on the Pauli--Luba\'nski pseudo-vector for the quantity \eqref{eq:4-vector W for gbms}, namely 
\be
[J_{\mu\nu},S_\rho]=\mfk{i}(\eta_{\nu\rho}S_\mu-\eta_{\mu\rho}S_\nu).
\ee

We start with  the rotation generators
\be
[J_{i j},S_0]=0\,,\qquad
[J_{i j},S_k]=\mfk{i}(\eta_{j k}S_i-\eta_{i k}S_j),
\ee
whose analog is given by
\be
\{J_{ij},S_k\}_{\g^*}&=
-\delta_{Y_{ij}}S_k=\f12\msf{S}[\rho^\f12 Y_{ij}[ n_k]]
\\
&=\bigintsss_S \rho^\f12 
\varepsilon_{ij}\!^\ell\epsilon^{CD}\pa_D n_\ell \pa_C n_k\, \,\epsilon^{AB}\partial_Ap_B\bfe
\cr
&=
\bigintsss_S \rho^\f12\,
\varepsilon_{ij}\!^\ell
\varepsilon_{k\ell m}n^m
\, \epsilon^{AB}\partial_Ap_B\bfe
\cr
&=
\bigintsss_S \rho^\f12\,
(
\eta_{im}\eta_{jk}
-\eta_{ik}\eta_{jm}
)
n^m
\, \epsilon^{AB}\partial_Ap_B\bfe
\cr
&=\eta_{jk} S_i-\eta_{ik} S_j.
\ee
At the same time, the commutator $\{J_{i j},S_0\}_{\g^*}=0$ is trivially reproduced. 

\smallskip Next, we look at the boost action
\be
[J_{0i},S_0]=\mfk{i}S_i\,,\qquad [J_{0i},S_j]=\mfk{i}\eta_{ij}S_0,
\ee
whose analog  is given by
\be
\{J_{0i},S_0\}_{\g^*}&= -\delta_{ Y_{0i}}S_0=
-\f12 \msf{S}[\rho^\f12 n_0 D_AY_{0i}^A]
\cr
&=\msf{S}[\rho^\f12 n_i]
=S_i,
\ee
and 
\be
\{J_{0i},S_j\}_{\g^*}&=- \delta_{ Y_{0i}}S_j=
-\f12 \msf{S}[\rho^\f12 n_j D_AY_{0i}^A] +\msf{S}[\rho^\f12 Y_{0i}[n_j]]
\cr
&=
\bigintsss_S 
n_i n_j\,\rho ^\f12\,\epsilon^{AB}\partial_Ap_B\bfe
+\bigintsss_S 
q^{CD}\pa_D n_i \pa_C n_j\,\rho^\f12 \,\epsilon^{AB}\partial_Ap_B\bfe
\cr
&=\eta_{ij}\bigintsss_S 
\rho ^\f12\,\epsilon^{AB}\partial_Ap_B\bfe
\cr
&=\eta_{ij}S_0,
\ee
where we used the first identity in \eqref{nid}.

\subsection{Pauli--Luba\'nski generator algebra }\la{app:PLalg}

We give the proof of  \eqref{PLalg}.
Let us start with $\mu=i,\nu=j$. In this case we have
\be
- \varepsilon_{ij \rho \sigma} P^\rho S^\sigma  &=  \varepsilon_{ijk} (P^0S^k - P^k S^0)
\cr
&=\varepsilon_{ijk}M^\f43 \bigintsss_S
\f{(n^k-v^k)}{(1-v\cdot \widehat{n})}\,
\epsilon^{AB} \pa_A {p}_B\,
\bfe\,.
\ee
We write the LHS of \eqref{PLalg}
 as
\be\la{bella}
&\msf{S}[\{\rho_v^{\f12}n_\mu,\rho_v^{\f12}n_\nu\}_\rho]
=
\bigintsss_S
\epsilon^{AB}\,
\partial_A( \rho^{\f12}n_\mu)
\,
\partial_B(\rho^{\f12} n_\nu)\,
w\,
\bfe
\cr
&=\varepsilon_{\mu\nu}^{\hphantom{\mu\nu}\rho 0}
\bigintsss_S 
n_\rho
 \,\epsilon^{CD}\partial_Cp_D\,
\bfe
+\f{1}2 \bigintsss_S 
\epsilon^{AB} (n_\mu\p_B n_\nu- n_\nu\p_B n_\mu) \rho^{-1}\, \partial_A  \rho
 \,\epsilon^{CD}\partial_Cp_D\,
\bfe,
\cr
\ee
and use the relations
\be
&\f12 \rho_{v}^{-1}\, \partial_A  \rho_{v}=\f{v^\ell\p_A n_\ell}{(1-v\cdot \widehat{n})}\\
&\epsilon^{AB} (n_i\p_B n_j- n_j\p_B n_i)
=-\varepsilon_{ij}\!^k\p^A n_k,
\ee
from which, using \eqref{nid}, we have
\be
\msf{S}[\{\rho_v^{\f12}n_i,\rho_v^{\f12}n_j\}_\rho]
&=\varepsilon_{ij}\!^k
\bigintsss_S 
\left(n_k-\f{v^\ell\p_A n_\ell \p^A n_k}{\gamma_{v}(1-v\cdot \widehat{n})} \right)
\,
\epsilon^{AB} \pa_A {p}_B\,
\bfe
\cr
&=
\varepsilon_{ij}\!^k
\bigintsss_S 
\left(n_k-\f{v^\ell(\eta_{\ell k}-n_\ell n_k)}{(1-v\cdot \widehat{n})} \right)
\,
\epsilon^{AB} \pa_A {p}_B\,
\bfe
\cr
&=
\varepsilon_{ij}\!^k
\bigintsss_S 
\f{\left(n_k-v_k\right)}{(1-v\cdot \widehat{n})} 
\,
\epsilon^{AB} \pa_A {p}_B\,
\bfe\,,
\ee
as desired.
Next we consider $\mu=i, \nu=0$. In this case we have
\be
- \varepsilon_{\mu\nu \rho \sigma} P^\rho S^\sigma  &=  \varepsilon_{ijk}  P^k S^j 
\cr
&=
\varepsilon_{ijk}M^\f43 \bigintsss_S
\f{n^j v^k }{(1-v\cdot \widehat{n})}
\,
\epsilon^{AB} \pa_A {p}_B\,
\bfe\,.
\ee
We use \eqref{bella} and the relation
\be
\epsilon^{AB} (n_i\p_B n_0- n_0\p_B n_i)
=-\epsilon^{AB}\p_B n_i,
\ee
to compute
\be
\msf{S}[\{\rho_v^{\f12}n_i,\rho_v^{\f12}n_0\}_\rho]
&=
\bigintsss_S 
\f{v^j\epsilon^{AB}\p_A n_i\p_B n_j}{(1-v\cdot \widehat{n})}
\,
\epsilon^{AB} \pa_A {p}_B\,
\bfe
\cr
&=\varepsilon_{ijk}
\bigintsss_S 
\f{v^j n^k }{(1-v\cdot \widehat{n})}
\,
\epsilon^{AB} \pa_A {p}_B\,
\bfe\,,
\ee
where we used \eqref{nid} again. This concludes the derivation of  \eqref{PLalg}.

\section{Conformal transformation}\la{App:conf}

We look at a transformation 
$q_{AB} \to \gamma_{AB}= \omega^{-2} q_{AB}$ and we consider a function in $\mathbb{R}_{\Delta}$ such that $\widetilde{\phi}= \omega^\Delta \phi$. Under this transformation we have that 
\be
\cD_{\langle A}\cD_{B\rangle}\widetilde\phi&=
{D}_{\langle A}{D}_{B\rangle}\widetilde\phi +  \frac{2}{\omega} D_{\langle A} \omega D_{B\rangle}\widetilde\phi\,,
\cr
{D}_{\langle A}{D}_{B\rangle}\widetilde\phi &=
{D}_{\langle A}(\omega^{\Delta} {D}_{B\rangle}\phi +\Delta  \phi \omega^{\Delta-1 } D_{B\rangle}\omega )
\cr
&= \omega^{\Delta} \left[
{D}_{\langle A} {D}_{B\rangle}\phi + 
2 \Delta \frac{{D}_{\langle A} \phi D_{B\rangle}\omega }{\omega}
+\Delta \phi \left( \frac{{D}_{\langle A}{D}_{B\rangle}\omega }{\omega}
+  (\Delta-1) \frac{{D}_{\langle A}\omega {D}_{B\rangle}\omega }{\omega^2}.
\right) \right]\,,
\cr
\frac{1}{\omega} D_{\langle A} \omega D_{B\rangle}\widetilde\phi &=
\omega^{\Delta}   \left[\frac{ D_{\langle A} \omega {D}_{B\rangle}\phi}{\omega} +\Delta  \phi  \frac{D_{\langle A} \omega D_{B\rangle}\omega}{\omega^2} \right].
\ee
Summing both gives 
\be
\cD_{\langle A}\cD_{B\rangle}\widetilde\phi
&=
\omega^{\Delta} \left[
{D}_{\langle A} {D}_{B\rangle}\phi + 
2( \Delta+1) \frac{{D}_{\langle A} \phi D_{B\rangle}\omega }{\omega}
+\Delta \phi \left( \frac{{D}_{\langle A}{D}_{B\rangle}\omega }{\omega}
+ (\Delta +1) \frac{{D}_{\langle A}\omega {D}_{B\rangle}\omega }{\omega^2}
\right) \right]\,.
\ee
Therefore, if one chooses $\Delta= -1$, we get that 
\be \la{DDphi}
\cD_{\langle A}\cD_{B\rangle}\widetilde\phi
=
\omega^{-1} \left[
{D}_{\langle A} {D}_{B\rangle}\phi -   \phi \left( \frac{{D}_{\langle A}{D}_{B\rangle}\omega }{\omega}
\right) \right].
\ee 
This means that we need to introduce the symmetric traceless tensor (Liouville stress tensor) such that 
\be 
{T}_{AB}(\gamma) = T_{AB} (q)
-   \frac{{D}_{\langle A}{D}_{B\rangle}\omega }{\omega}\,,
\ee 
and we have that $\left({D}_{\langle A}{D}_{B\rangle} +  {T}_{AB}(q)\right)$ is a conformal operator of weight\footnote{It maps $V_{(-1,0)}\to V_{(1,2)}$} $(2,2)$ acting on weight $-1$ scalars. From \eqref{DDphi}, we have
\be 
\left(\cD_{\langle A}\cD_{B\rangle} + {T}_{AB}(\gamma)\right)[\omega^{-1} \phi ]
= \omega^{-1} 
\left({D}_{\langle A}{D}_{B\rangle} + {T}_{AB}(q)\right)\phi\,.
\ee 
Note that we also have 
\be 
\frac{{R}(\gamma)}{2} = \omega^{-2} \left(\frac{R(q)}{2}  + \frac{\Delta \omega}{\omega} + \frac{{D}_{C}\omega {D}^C\omega }{\omega^2}\right)\,.
\ee

\section{Condensate field decomposition}\la{App:decom}

\subsection{Supertranslation harmonic decomposition}\la{App:Gold}
In this section we are showing that there exists a mode $C_{\va P}(n)$  such that 
\bea 
\frac{P^4}{(P\cdot n)^3} &=& 
 {\tau}\cdot P - 3 \widehat{n} \cdot P 
 + \f12 \Delta (\Delta +2 ) C_{\va P}(n)\cr
& = & \tau \cdot P - 3 \widehat{n} \cdot P
 {-}D_A D_B C_{\va P}^{AB}(n)\,,
\eea 
where $\tau^\mu=(1, \vec{0})$ and $\widehat{n}^\mu = (0, \vec{n})$.
The proposal is that $C_{\va P}(n)$ is given by 
\be 
C_{\va P}(n)=(n\cdot P) \ln  \left(\f{-n\cdot P}{M}\right)\,,
\ee 
such that \be 
C_{\va P}^{AB}(n) =-2D^{\langle A} D^{B \rangle } C_{\va P}(n)= 4\frac{ (D^{\langle A} n \cdot P)( D^{B \rangle } n \cdot P)}{ (n\cdot P)}
\ee 
corresponds to the leading soft factor for a (hard) particle of 4-momentum $P^\mu$. 
We use that 
\bea
n:=(1,\vec{n}),\qquad 
\bar{n} =(1,-\vec{n})
\eea 
satisfy $n\cdot \bar{n}=-2$ and
\be 
\Delta n_\mu = \bar{n}_\mu-n_\mu, \qquad D_A n_\mu D^An_\nu =  \eta_{\mu\nu} + \frac12( n_\mu\bar{n}_\nu +
\bar{n}_\mu n_\nu ).
\ee 
Then, noticing  that 
\be
\Delta (\Delta +2 )(n\cdot P) \ln M=
\Delta((n+\bar n)\cdot P) \ln M=0,
\ee
due to  $D_A (n+\bar{n})=0$,
we compute first
\bea 
(\Delta +2)
[(n\cdot P) \ln (-n\cdot P)]
&=& D^A [ (D_An\cdot P)   \ln(-n\cdot P) +  D_An \cdot P ] + 2 (n\cdot P) \ln (-n\cdot P)\cr
&=& ((\Delta+2) n\cdot P)   \ln(-n\cdot P) +   \frac{(D^An\cdot P)( D_An \cdot P)}{(n\cdot P) } + \Delta n\cdot P \cr
&=& (\bar{n}+n)\cdot P \ln(-n\cdot P)
+ \frac{P^2 + (\bar{n}\cdot P)(n\cdot P)}{(n\cdot P) } + (\bar{n}-n)\cdot P \cr
&=&(\bar{n}+n)\cdot P \ln(-n\cdot P)
+ \frac{P^2}{(n\cdot P) } +(2\bar{n}-n)\cdot P \,.
\eea 
Next, applying $\Delta $ to this result, we get that 
\be
\Delta 
&(\Delta +2)
[(n\cdot P) \ln (-n\cdot P)]\cr
&=  (\bar{n}+n)\cdot P D^A \left[\frac{D_An\cdot P}{(n\cdot P)}\right]
- D^A \left[\frac{P^2 D_An \cdot P}{(n\cdot P)^2 } \right] + \Delta (2\bar{n}-n)\cdot P \cr
&=  (\bar{n}+n)\cdot P  \left[\frac{\Delta n\cdot P}{(n\cdot P)} -  \frac{(D_A n\cdot P)(D^An\cdot P)}{(n\cdot P)^2}\right]
\cr
&- \left[\frac{P^2 \Delta n \cdot P }{(n\cdot P)^2 } - 2\frac{P^2 (D_A n\cdot P)(D^An\cdot P) }{(n\cdot P)^3 } \right] + 3(n-\bar{n})\cdot P \cr
&=
(\bar{n}+n)\cdot P  \left[\frac{(\bar{n}-n)\cdot P}{(n\cdot P)} -  \frac{P^2 + (\bar{n}\cdot P)(n\cdot P)}{(n\cdot P)^2}\right]
\cr
&- P^2 \left[\frac{ (\bar{n}-n)\cdot P }{(n\cdot P)^2 } - 2\frac{P^2 + (\bar{n}\cdot P)(n\cdot P)}{(n\cdot P)^3 } \right] + 3(n-\bar{n})\cdot P \cr
&=
(\bar{n}+n)\cdot P  \left[-\frac{n\cdot P}{(n\cdot P)} -  \frac{P^2}{(n\cdot P)^2}\right]
- P^2 \left[-\frac{ (\bar{n}+n)\cdot P }{(n\cdot P)^2 } - 2\frac{P^2 }{(n\cdot P)^3 } \right] + 3(n-\bar{n})\cdot P \cr
&=
\left[ 2\frac{P^4 }{(n\cdot P)^3}  + P^2  \frac{ (\bar{n}+n)\cdot P}{(n\cdot P)^2}   -  P^2 \frac{(\bar{n}+n)\cdot P }{(n\cdot P)^2}\right] - (\bar{n}+n)\cdot P +3(n-\bar{n})\cdot P\cr
&= 2\frac{P^4 }{(n\cdot P)^3}
- (n+\bar{n})\cdot P+ 3(n-\bar{n})\cdot P\,,
\ee
where we used that $D_A (n+\bar{n})=0$, and $n+\bar{n}=2\tau, n-\bar{n}=2\widehat{n}$.

\subsection{Angular momentum harmonic decomposition}\la{App:AMGold}
For a particle of momentum $P$ and  angular momentum $J= P\wedge x $ we have that the angular momentum aspect is 
\be
j^{\va (P,J)}_A(n)= P^4 \frac{(n^\mu D_An^\nu)}{(-n\cdot P)^4} J_{\mu\nu}\,.
\ee 
We want to verify that this aspect can be decomposed in spherical harmonic components as
\be
j^{\va (P,J)}_A(n) = (n^\mu D_A n^\nu)  J_{\mu\nu} +\f23 D_{\langle A } D_B D_{C \rangle } C^{BC}_{(\va P,J)}(n)\,.
\ee
The first term $J_A:= (n^\mu D_A n^\nu)  J_{\mu\nu}$ is the $\ell=1$ components of $j_A(n)$ while the second term involving $C^{BC}_{(\va P,J)}(n)$ includes the $\ell\geq 2$ components. This factor is  the subleading soft factor, given by 
\be
C^{BC}_{(\va P,J)}(n)=\f{(D^{\langle B} n \cdot P)( n^\mu D^{C \rangle }  n^\nu J_{\mu\nu})}{(n\cdot P)}\,.
\ee

Let us verify that this decomposition is correct. 
We work in complex coordinates for simplicity and denote $D:=D_z, \bar D:=D^z$---a similar but more cluttered  derivation can be repeated for general coordinates---. We can use
\be
&D^2n_\mu=0= {\bar D}^2n_\mu\,, \quad D\bar D n_\mu= \f12(\bar{n}_\mu-n_\mu)
\,,
\\
&
D n_{(\mu} \bar Dn_{\nu)} =  \f12\eta_{\mu\nu} + \frac14( n_\mu\bar{n}_\nu +
\bar{n}_\mu n_\nu )
\,,
\ee
to compute first
\be
&-D^2 C^{zz}_{\va (P,J)}(n)=D^2\f{(\bar D n \cdot P)( n^\mu \bar D  n^\nu J_{\mu\nu})}{(-n\cdot P)}
\cr
&=
-\f{D^2[(\bar D n \cdot P)( n^\mu \bar D  n^\nu J_{\mu\nu})]}{(n\cdot P)}
+2\f{D[(\bar D n \cdot P)( n^\mu \bar D  n^\nu J_{\mu\nu})]D n\cdot P}{(n\cdot P)^2}
\cr
&
-2\f{(\bar D n \cdot P)( n^\mu \bar D  n^\nu J_{\mu\nu})(D n\cdot P)^2}{(n\cdot P)^3}
\cr
&=
-\f12\f{D[(\bar{n}\cdot P-n \cdot P )( n^\mu \bar D  n^\nu J_{\mu\nu})]}{(n\cdot P)}
-\f{D[(\bar D n\cdot P)( D n^\mu \bar D  n^\nu J_{\mu\nu})]}{(n\cdot P)}
-\f12\f{D[(\bar D n\cdot P)(  n^\mu \bar   n^\nu J_{\mu\nu})]}{(n\cdot P)}
\cr
&+\f{(\bar{n}\cdot P-n \cdot P) ( D n\cdot P)( n^\mu \bar D  n^\nu J_{\mu\nu})}{(n\cdot P)^2}
+2\f{(\bar D n \cdot P)( D n^\mu \bar D  n^\nu J_{\mu\nu})D n\cdot P}{(n\cdot P)^2}
\cr
&+\f{(\bar D n \cdot P)(  n^\mu \bar   n^\nu J_{\mu\nu})D n\cdot P}{(n\cdot P)^2}
\cr
&
-\f{(P^2 + (\bar{n}\cdot P)(n\cdot P))( n^\mu   \bar D n^\nu  J_{\mu\nu}) (D n\cdot P)}{(n\cdot P)^3}\,.
\ee
If we take $J_{0i}=0$, then all the terms proportional to $n^\mu \bar   n^\nu J_{\mu\nu}$ vanish
 and
\be
&-\f12\f{D[(\bar{n}\cdot P-n \cdot P )( n^\mu \bar D  n^\nu J_{\mu\nu})]}{(n\cdot P)}
-\f{D[(\bar D n\cdot P)( D n^\mu \bar D  n^\nu J_{\mu\nu})]}{(n\cdot P)}
\cr
&=-\f{(\bar{n}\cdot P-n \cdot P )(D n^\mu \bar D  n^\nu J_{\mu\nu})}{(n\cdot P)}
-\f12\f{(D \bar{n}\cdot P-D n \cdot P )( n^\mu \bar D  n^\nu J_{\mu\nu})}{(n\cdot P)}
\cr
&-\f12\f{(\bar D n\cdot P)(( \bar  n^\nu - n^\nu)D n^\mu  J_{\mu\nu})}{(n\cdot P)}
\cr
&=-\f{(\bar{n}\cdot P-n \cdot P )(D n^\mu \bar D  n^\nu J_{\mu\nu})}{(n\cdot P)}
-\f12\f{(D \bar{n}\cdot P-D n \cdot P )( n^\mu \bar D  n^\nu J_{\mu\nu})}{(n\cdot P)}
\cr
&+\f{(\bar D n\cdot P)( n^\nu D n^\mu  J_{\mu\nu})}{(n\cdot P)}
\,,
\ee
where we used 
\be
(\bar n^\nu-n^\nu) J_{\mu\nu}
=-2n^\nu J_{\mu\nu}
\ee
due to $J_{0i}=0$.
We can thus write
\be
-D^2 C^{zz}_{\va (P,J)}(n)&=D^2\f{(\bar D n \cdot P)( n^\mu \bar D  n^\nu J_{\mu\nu})}{(-n\cdot P)}
\cr
&=\f12\f{ ( n^\mu \bar D  n^\nu J_{\mu\nu})}{(n\cdot P)^2}
\left[2( D n\cdot P)(\bar{n}\cdot P-n \cdot P)
-(D \bar{n}\cdot P-D n \cdot P )(n\cdot P)
\right]
\cr
&
+\f{
D n^\mu \bar D  n^\nu J_{\mu\nu}
}{(n\cdot P)^2}
\left[
(P^2 + (\bar{n}\cdot P)(n\cdot P))
-(\bar{n}\cdot P-n \cdot P )(n\cdot P)
\right]
\cr
&+\f{(\bar D n\cdot P)( n^\nu D n^\mu  J_{\mu\nu})}{(n\cdot P)}
\cr
&-\f{(P^2 + (\bar{n}\cdot P)(n\cdot P))( n^\mu   \bar D n^\nu  J_{\mu\nu}) (D n\cdot P)}{(n\cdot P)^3}
\cr
&=\f{ ( n^\mu \bar D  n^\nu J_{\mu\nu})}{(n\cdot P)^2}
( D n\cdot P)(\bar{n}\cdot P)
\cr
&
+
D n^\mu \bar D  n^\nu J_{\mu\nu}
\left(1+
\f{P^2}{(n\cdot P)^2} 
\right)
\cr
&+\f{(\bar D n\cdot P)( n^\nu D n^\mu  J_{\mu\nu})}{(n\cdot P)}
\cr
&-\f{(P^2 + (\bar{n}\cdot P)(n\cdot P))( n^\mu   \bar D n^\nu  J_{\mu\nu}) (D n\cdot P)}{(n\cdot P)^3}\,,
\ee
where in the first line of the last passage we used
\be
D (n+\bar n)=0\,.
\ee
We now compute the $D$-derivative of these four terms separately. We have
\be
&D\left[
\f{ ( n^\mu \bar D  n^\nu J_{\mu\nu})}{(n\cdot P)^2}
( D n\cdot P)(\bar{n}\cdot P)
\right]
\cr
&=
\f{ ( Dn^\mu \bar D  n^\nu J_{\mu\nu})}{(n\cdot P)^2}
( D n\cdot P)(\bar{n}\cdot P)
-
{\f{ ( n^\mu \bar D  n^\nu J_{\mu\nu})}{(n\cdot P)^2}
( D n\cdot P)^2
}
\cr
&
{
-2\f{ ( n^\mu \bar D  n^\nu J_{\mu\nu})}{(n\cdot P)^3}
( D n\cdot P)^2(\bar{n}\cdot P)
}
\cr
&=
\underbrace{\f{ ( Dn^\mu \bar D  n^\nu J_{\mu\nu})}{(n\cdot P)^2}
( D n\cdot P)(\bar{n}\cdot P)}_{\square}
-\underbrace{\f{ ( n^\mu \bar D  n^\nu J_{\mu\nu})}{(n\cdot P)^3}
( D n\cdot P)^2 \left( n+2 \bar{n} \right)\cdot P}_{\bigtriangleup}
\,,
\ee
and
\be
&D\left[
D n^\mu \bar D  n^\nu J_{\mu\nu}
\left(1+
\f{P^2}{(n\cdot P)^2} 
\right)
\right]
\cr
&=
{
-
 n^\nu D n^\mu  J_{\mu\nu}
\left(1+
\f{P^2}{(n\cdot P)^2} 
\right)
}
- 2\f{D n^\mu \bar D  n^\nu J_{\mu\nu}}{(n\cdot P)^3}P^2(Dn\cdot P)\,,
\ee
and 
\be
&D\left[
\f{(\bar D n\cdot P)( n^\nu D n^\mu  J_{\mu\nu})}{(n\cdot P)}
\right]
\cr
&=
\f12 \f{((\bar n-n)\cdot P)( n^\nu D n^\mu  J_{\mu\nu})}{(n\cdot P)}
-\f12 \f{( n^\nu D n^\mu  J_{\mu\nu})
(P^2+(\bar  n\cdot P)(  n\cdot P))
}{(n\cdot P)^2}
\cr
&=
-
\f12 n^\nu D n^\mu  J_{\mu\nu}
\left(1+
\f{P^2}{(n\cdot P)^2} 
\right)\,,
\ee
and 
\be
&D\left[
-\f{(P^2 + (\bar{n}\cdot P)(n\cdot P))( n^\mu   \bar D n^\nu  J_{\mu\nu}) (D n\cdot P)}{(n\cdot P)^3}
\right]
\cr
&=
{
-\f{ ((\bar{n}-n)\cdot P)( n^\mu   \bar D n^\nu  J_{\mu\nu}) (D n\cdot P)^2}{(n\cdot P)^3}
}
\cr
&
-\f{(P^2 + (\bar{n}\cdot P)(n\cdot P))( Dn^\mu   \bar D n^\nu  J_{\mu\nu}) (D n\cdot P)}{(n\cdot P)^3}
\cr
&
+3\f{(P^2 + (\bar{n}\cdot P)(n\cdot P))( n^\mu   \bar D n^\nu  J_{\mu\nu}) (D n\cdot P)^2}{(n\cdot P)^4}
\cr
&=
\underbrace{\f{ ( n^\mu   \bar D n^\nu  J_{\mu\nu}) (D n\cdot P)^2}{(n\cdot P)^3}(n+2\bar{n})\cdot P
}_{\bigtriangleup}
\cr
&
-\f{P^2( Dn^\mu   \bar D n^\nu  J_{\mu\nu}) (D n\cdot P)}{(n\cdot P)^3}
-\underbrace{\f{( Dn^\mu   \bar D n^\nu  J_{\mu\nu}) }{(n\cdot P)^2}(D n\cdot P)(\bar{n}\cdot P)}_{\square}
\cr
&
+3\f{P^2 ( n^\mu   \bar D n^\nu  J_{\mu\nu}) (D n\cdot P)^2}{(n\cdot P)^4}
\,.
\ee
The contributions marked with the same symbol cancel each other and, combining the remaining ones, we arrive at
\be
-D^3 C^{zz}_{\va (P,J)}(n)
&=
- \f32n^\nu D n^\mu  J_{\mu\nu}
\cr
&- \f32\f{P^2 
}{(n\cdot P)^2} n^\nu D n^\mu  J_{\mu\nu}
\cr
&+3\f{P^2}{(n\cdot P)^4}
\left (
 n^\mu   \bar D n^\nu  J_{\mu\nu}(D n\cdot P)^2
 - Dn^\mu \bar D  n^\nu J_{\mu\nu}(D n\cdot P)( n\cdot P)
 \right).
 \la{almost}
\ee

The third line can be expanded with 
\be
Dn^{(\sigma} \bar D n^{\nu)}
&=
\f12 q^{z\bar z} (D_z n^{\sigma}  D_{\bar z} n^{\nu}
+D_z n^{\nu}  D_{\bar z} n^{\sigma}
)
\cr
&=\f12 Dn^\sigma \bar D n^\nu+
\f12 \bar Dn^\sigma  D n^\nu
\cr
&=\f12\eta^{\sigma\nu} + \frac14( n^\sigma\bar{n}^\nu +
\bar{n}^\sigma n^\nu )\,,
\ee
from which
\be
Dn^\sigma \bar D n^\nu
=-\bar Dn^\sigma  D n^\nu
+\eta^{\sigma\nu} + \frac12( n^\sigma\bar{n}^\nu +
\bar{n}^\sigma n^\nu )\,.
\la{Awesome}
\ee
We then have
\be
\bar D n^\nu (Dn\cdot P)&=
- D n^\nu (\bar Dn\cdot P)+P^\nu
+\frac12( (P\cdot n) \bar{n}^\nu +
(P\cdot \bar{n})n^\nu )\,,
\ee
and
\be
n^\mu   \bar D n^\nu  J_{\mu\nu}(D n\cdot P)^2&=-n^\mu D n^\nu J_{\mu\nu} (D n\cdot P) (\bar Dn\cdot P)
+n^\mu P^\nu J_{\mu\nu}(D n\cdot P)
\cr
&=-\f12 n^\mu D n^\nu J_{\mu\nu}(P^2+ (\bar{n}\cdot P)(n\cdot P))
+n^\mu P^\nu J_{\mu\nu}(D n\cdot P)\,,
\la{Awesome1}
\ee
where we used $J_{0i}=0$ again, and
\be
 Dn^\mu \bar D  n^\nu J_{\mu\nu}(D n\cdot P)( n\cdot P)&=
P^\nu Dn^\mu J_{\mu\nu} ( n\cdot P)
\cr
&
+\f12   Dn^\mu J_{\mu\nu} ( n\cdot P)
( (P\cdot n) \bar{n}^\nu +
(P\cdot \bar{n})n^\nu )\,.\la{Awesome2}
\ee
If we now plug \eqref{Awesome1} and \eqref{Awesome2} into \eqref{almost}, we arrive at
\be
-D^3 C^{zz}_{\va (P,J)}(n)&=
- \f32n^\nu D n^\mu  J_{\mu\nu}
\cr
&-\f32 \f{ P^4 }{(n\cdot P)^4} n^\mu D n^\nu J_{\mu\nu}
\cr
&+3 \f{P^\nu J_{\mu\nu}}{(n\cdot P)^4} \left(
n^\mu (Dn\cdot P)- Dn^\mu (n\cdot P)
\right)\,.
\ee

It is now immediate to see that for the orbital part of $J_{\mu\nu}$, namely $L_{\mu\nu}= P_\mu x_\nu - P_\nu x_\mu$, we have
\be
P^\nu L_{\mu\nu} \left(
n^\mu (Dn\cdot P)- Dn^\mu (n\cdot P)
\right)=0\,.
\ee
As for the intrinsic spin we have $P^\nu S_{\mu\nu}=0$, we finally arrive at the desired result
\be
 \f{P^4 }{(n\cdot P)^4}  n^\mu D n^\nu J_{\mu\nu}=
 n^\mu D n^\nu  J_{\mu\nu}
+\f23 D^3 C^{zz}_{\va (P,J)}(n)\,.
\ee

\section{Details on the Kerr metric and Casimirs}\label{sec:details on the kerr metric and casimirs}

In the main body of the paper, we have constructed the Casimir functionals and the spin charge for the Kerr metric. In this appendix, we provide the details of this construction. We start with the large-distance expansion of the Kerr metric in the Bondi-Sachs coordinates, from which we read off the phase space mass $\mcal{M}$ and angular momentum $\mcal{J}_A$ aspects. We then construct Casimir functionals and the spin charge for the Kerr spacetime. 

\subsection{Writing the Kerr metric in the Bondi--Sachs coordinates.}\label{sec:writing the kerr metric in the bondi-sachs coordinates}

We start with writing the Kerr metric in the Bondi-Sachs coordinates, which we denote as $\{u,r,\theta,\varphi\}$, at large values of the luminosity radius $r$. We follow the procedure explained in \cite{FletcherLun200309} as follows: (1) We first consider the Kerr metric written in the generalized Bondi--Sachs (GBS) coordinates derived in \cite[Eq. (48)]{FletcherLun200309}. Denoting the coordinates by $\{u,\bar{r},\theta,\varphi\}$, the GBS coordinate system is defined by $g_{\bar{r}\,\bar{r}}=g_{\bar{r}\theta}=g_{\bar{r}\varphi}=0$ \cite[Eq. (4)]{FletcherLun200309} (2) We then use \cite[Eq. (6)]{FletcherLun200309}
\begin{equation}\label{eq:the bondi-sachs gauge}
    g_{\theta\theta}g_{\varphi\varphi}-g_{\theta\varphi}^2=r^4\sin^2\theta,
\end{equation}
the so-called Bondi--Sachs gauge, to define the luminosity radius $r$ in the Bondi--Sachs coordinates \cite{BondivanderBurgMetzner196208,Sachs196212} in terms of $\br$. $r$ in principle can be written exactly in terms of $\br$; one just needs to solve the Bondi--Sachs gauge condition for arbitrary values of $r$. However, we only need $r\to\infty$ limit and only consider this case. (3) We use the coordinate transformation $\{u,\br,\theta,\varphi\}\mapsto\{u,r,\theta,\varphi\}$ to write the metric in the Bondi--Sachs coordinates.

\paragraph{The Kerr solution in the generalized Bondi-Sachs coordinates.} The Kerr solution of mass $M$ and spin angular momentum $J$ \cite{Kerr196309} is given by\footnote{Notice that we have written the metric (and its expansion \eqref{eq:kerr metric in the gbs coordinates} in the generalized Bondi--Sachs coordinates) in the signature $(-,+,+,+)$, as is used in this paper, rather than the signature $(+,-,-,-)$ used in \cite{FletcherLun200309}.}
\begin{eqaligned}
    \rd s^2_{\tenofo{BL}}=&-\left(1-\frac{2M\wt{r}}{\wt{\rho}^2}\right)\rd\wt{t}^2-\frac{4a\wt{r}\sin^2(\wt{\theta})}{\wt{\rho}^2}\rd\wt{t}\rd\wt{\varphi}+\frac{\wt{\rho}^2}{\wt{\Delta}}\rd\wt{r}^2
    \\
    &+\wt{\rho}^2\rd\wt{\theta}^2+\sin^2(\wt{\theta})\left(\wt{A}^2+\frac{2aM\wt{r}\sin^2(\wt{\theta})}{\wt{\rho}^2}\right)\rd\wt{\varphi}^2,
\end{eqaligned}
where $\{\wt{t},\wt{r},\wt{\theta},\wt{\varphi}\}$ are Boyer--Lindquist coordinates \cite{BoyerLindquist196702}, and
\begin{eqgathered}
    \wt{\rho}^2:= \wt{r}^2+a^2\cos^2(\wt{\theta}), \qquad \wt{\Delta}^2:=\wt{r}^2-2M\wt{r}+a^2,\qquad \wt{A}^2:= \wt{r}^2+a^2,
\end{eqgathered}
and the reduced angular momentum is
\begin{equation}\label{eq:reduced angular momentum}
    a:= \frac{J}{Mc}.
\end{equation}
The generalized Bondi--Sachs coordinate system, which we denote as $\{u,\bar{r},\theta,\varphi\}$, is defined by \cite[Eq. (4)]{FletcherLun200309}
\begin{equation}\label{eq:definition of gbs coordinates}
    g_{\bar{r}\,\bar{r}}=g_{\bar{r}\theta}=g_{\bar{r}\varphi}=0.
\end{equation}
To write the Kerr metric in this coordinate system, certain conditions have to be imposed \cite[\S 4]{FletcherLun200309}. The following relation between $\{\wt{t},\wt{r},\wt{\theta},\wt{\varphi}\}$ and $\{u,\bar{r},\theta,\varphi\}$ is necessary and sufficient to fulfill these conditions \cite[Eqs. (19-22)]{FletcherLun200309}
\begin{eqaligned}
    \wt{t}&=u+f(\br,\theta), &\qquad \wt{r}&=\br,
    \\
    \wt{\theta}&=\wt{\theta}(\br,\theta), &\qquad \wt{\varphi}&=\varphi+g(\br,\theta),
\end{eqaligned}
for some functions $f(\br,\theta)$ and $g(\br,\theta)$. Then, the Kerr metric in the GBS coordinates takes the following form \cite[Eq. (48)]{FletcherLun200309}\footnote{We have used the same notation as \cite{FletcherLun200309}. Some of these notations such as $\rho$ conflict with the notation we used in the paper. This hopefully will not make any trouble for the careful reader.}
\begin{eqaligned}\label{eq:kerr metric in the gbs coordinates}
    \rd s^2_{\tenofo{GBS}}=&-\left(1-\frac{2M\br}{\rho^2}\right)\rd u^2-\left(\frac{2\rho^2}{B}\right)\rd u\rd\br-2\left(\left[1-\frac{2M\br}{\rho^2}\right]\cdot\left[\frac{a\cos\theta}{C^2\cosh^2\alpha}\right]\right)\rd u\rd\theta
    \\
    &-\left(\frac{4aM\br D^2}{\rho^2C^2}\right)\rd u\rd\varphi+\left(\frac{\br\left[\br\rho^2C^2\cosh^2\alpha+2a^2M\cos^2\theta\right]}{\rho^2C^4\cosh^4\alpha}\right)\rd\theta^2
    \\
    &+\left(\frac{4a^2M\br D^2\cos\theta}{\rho^2 C^4\cosh^2\alpha}\right)\rd\theta\rd\varphi+\left(\frac{D^2\left[B^2C^2\cosh^2\alpha+a^2\Delta\cos^2\theta\right]}{\rho^2C^4\cosh^2\alpha}\right)\rd\varphi^2,
\end{eqaligned}
where the parameters are given by 
\begin{eqaligned}
    \rho^2&:= A^2-a^2\frac{D^2}{C^2},&\qquad \Delta&:= \bar{r}^2-2M\bar{r}+a^2,
    \\
        A^2&:= \bar{r}^2+a^2, &\qquad B^2&:= A^4-a^2\Delta,
    \\
    C&:=1+\tanh\alpha\sin\theta,&\qquad D&:=\tanh\alpha+\sin\theta\,,
\end{eqaligned}
and $\alpha=\alpha(r,a,M)$ is given by \cite[Eq. (38)]{FletcherLun200309}
\begin{equation}\label{eq:definition of function alpha}
    \alpha(\br)=-\bigintsss_{\br}^\infty\frac{a\rd s}{\sqrt{s^4+a^2s^2+2a^2Ms}}.
\end{equation}

\paragraph{Large-$\br$ expansion in the generalized Bondi--Sachs coordinates.} We can now take the large-$\br$ behavior of the metric \eqref{eq:kerr metric in the gbs coordinates}.

\smallskip In the following, we would need the large-$\br$ expansion of $\alpha$. We have ($\br\ge 0$)
\begin{eqaligned}\label{eq:large-r value of alpha}
    \alpha&=-\bigintsss_{\br}^\infty\frac{a\rd s}{\sqrt{s^4+a^2s^2+2a^2Ms}}
    \\
    &=-\bigintsss_\br^\infty \frac{a\rd s}{s^2}\left(1-\frac{a^2}{2s^2}-\frac{a^2M}{s^3}+\mcal{O}(s^{-3})\right)
    \\
    &=-\bigintsss_\br^\infty \frac{a\rd s}{s^2}+\mcal{O}(s^{-4})
    \\
    &=-\frac{a}{\br}+\mcal{O}(\br^{-3}).
\end{eqaligned}
The components of the Kerr metric in the large-$\bar{r}$ expansion are given as follows. First, we have
\begin{eqaligned}
    \bg_{uu}&=-1+\frac{2M\bar{r}}{\rho^2}=-1+\frac{2M}{\br}+\mcal{O}(\br^{-2}),
    \\
    \bg_{u\br}&=-\frac{\rho^2}{B}
    \\
    &=-1-\frac{a^2}{\br^2}\left(\frac{1}{2}-\frac{D^2}{C^2}\right)+\mcal{O}(\br^{-3})
    \\
    &=-1+\frac{a^2}{\br^2}\left(\frac{1}{2}-\cos^2\theta\right)+\mcal{O}(\br^{-3}),
\end{eqaligned}
where in the case of $g_{u\br}$, we have used the fact that as $\br\to\infty$, then $\alpha(\br)\to 0$, which implies $\tanh\alpha\sim\alpha\to 0$. Hence, 
\begin{eqaligned}\label{eq:large-r expansion of D2/C2}
    \frac{D^2}{C^2}&=\left(\frac{\tanh\alpha+\sin\theta}{1+\tanh\alpha\sin\theta}\right)^2
    \\
    &\sim (\tanh\alpha+\sin\theta)^2(1-\tanh\alpha\sin\theta)^2
    \\
    &\sim \sin^2\theta+2\tanh\alpha\sin\theta\cos^2\theta,
    \\
    &\sim \sin^2\theta-\frac{2a}{\br}\sin\theta\cos^2\theta,
\end{eqaligned}
where we have used \eqref{eq:large-r value of alpha} in the last line. In the expansion of $g_{u\br}$, we only use the first term. Next, we have
\begin{eqaligned}
    \bg_{u\theta}&=-\left(1-\frac{2M\br}{\rho^2}\right)\frac{a\cos\theta}{C^2\cosh^2\alpha}
    \\
    &=-\left(1-\frac{2M}{\br}\right)(a\cos\theta)(1-2\tanh\alpha\sin\theta)+\mcal{O}(\br^{-2})
    \\
    &=-\left(1-\frac{2M}{\br}\right)(a\cos\theta)\left(1+2\frac{a}{\br}\sin\theta\right)+\mcal{O}(\br^{-2})
    \\
    &=-a\cos\theta+\frac{2a\cos\theta}{\br}(M-a\sin\theta)+\mcal{O}(\br^{-2}),
\end{eqaligned}
where in the third line, we have used \eqref{eq:large-r value of alpha}. Next, we have
\begin{eqaligned}
    \bg_{u\varphi}&=-\frac{2aM\br}{\rho^2}\frac{D^2}{C^2}
    \\
    &=-\frac{2aM}{\br}\left(1-\frac{a^2}{\br^2}\left(1-\frac{D^2}{C^2}\right)\right)\frac{D^2}{C^2}
    \\
    &=-\frac{2aM\sin^2\theta}{\br}+\mcal{O}(\br^{-2}),
\end{eqaligned}
where in the third line, we have use \eqref{eq:large-r expansion of D2/C2}. Next, we have written the expansion for $g_{\theta\theta}$
\begin{eqaligned}
    \bg_{\theta\theta}&=\frac{\br(\br\rho^2C^2\cosh^2\alpha+2a^2M\cos^2\theta)}{\rho^2C^4\cosh^4\alpha}.
\end{eqaligned}
Since there are terms of order $\br^2$, we ignore all terms of order $\br^{-n}$ with $n\ge 1$. The second term can be expanded as
\begin{eqaligned}
    \frac{2a^2\br M\cos^2\theta}{\rho^2C^4\cosh^4\alpha}&=\frac{2a^2M\cos^2\theta}{\br}\left(1-\frac{a^2}{\br^2}\left(1-\frac{D^2}{C^2}\right)\right)\left(1+\frac{4a}{\br}\sin\theta\right)\left(1-\frac{2a^2}{\br^2}\right)
    \\
    &\sim \mcal{O}(\br^{-1}),
\end{eqaligned}
and can be safely ignored. The first term gives
\begin{eqaligned}
    \frac{\br^2}{C^2\cosh^2\alpha}&=\br^2\left(1-2\tanh\alpha\sin\theta+3\tanh^2\alpha\sin^2\theta\right)(1-\alpha^2)+\mcal{O}(\br^{-1})
    \\
    &=\br^2\left(1+\frac{2a}{\br}\sin\theta+\frac{3a^2}{\br^2}\sin^2\theta\right)\left(1-\frac{a^2}{\br^2}\right)+\mcal{O}(\br^{-1})
    \\
    &=\br^2+2a\br\sin\theta+a^2(3\sin^2\theta-1)+\mcal{O}(\br^{-1}).
\end{eqaligned}
Putting together, we thus have
\begin{equation}
    \bg_{\theta\theta}=\br^2+2a\br\sin\theta+a^2(3\sin^2\theta-1)+\mcal{O}(\br^{-1}).
\end{equation}
Next, 
\begin{eqaligned}
    \bg_{\theta\varphi}&=\frac{2a^2M\br\cos\theta D^2}{\rho^2C^4\cosh^2\alpha}
    \\
    &=\frac{2a^2M\cos\theta}{\br}\left(1-\frac{a^2}{\br^2}\left(1-\frac{D^2}{C^2}\right)\right)\left(1+\frac{4a}{\br}\sin\theta\right)\left(1-\frac{a^2}{\br^2}\right)+\mcal{O}(\br^{-2})
    \\
    &=\frac{2a^2M\cos\theta}{\br}+\mcal{O}(\br^{-2}).
\end{eqaligned}
Finally, we have
\begin{eqaligned}
    \bg_{\varphi\varphi}&=\frac{D^2(C^2B^2\cosh^2\alpha+a^2\Delta\cos^2\theta)}{\rho^2C^4\cosh^2\alpha}.
\end{eqaligned}
The first term can be expanded as
\begin{eqaligned*}
    \frac{D^2B^2}{\rho^2C^2}&=\left(\sin\theta-\frac{a}{\br}\right)^2\left(\br^2+a^2-\frac{2Ma^2}{\br}\right)\left(1-\frac{a^2}{\br^2}\cos^2\theta\right)\left(1+\frac{2a}{\br}\sin\theta+\frac{3a^2}{\br^2}\sin^2\theta\right)+\mcal{O}(\br^{-1})
    \\
    &=\left(\br^2\sin^2\theta-2a\br\sin\theta+a^2+a^2\sin^2\theta\right)\left(1+\frac{2a}{\br}\sin\theta+\frac{3a^2}{\br^2}\sin^2\theta\right)+\mcal{O}(\br^{-1})
    \\
    &=\br^2\sin^2\theta-2a\br\sin\theta\cos^2\theta+a^2(1-4\sin^2\theta\cos^2\theta)+\mcal{O}(\br^{-1}).
\end{eqaligned*}
Similarly, the second term gives
\begin{eqaligned*}
    \frac{a^2\cos^2\theta D^2\Delta}{\rho^2C^4\cosh^2\alpha}=a^2\sin^2\theta\cos^2\theta+\mcal{O}(\br^{-1}).
\end{eqaligned*}
We thus get
\begin{equation}
    \bg_{\varphi\varphi}=\br^2\sin^2\theta-2a\br\sin\theta\cos^2\theta+a^2(1-3\sin^2\theta\cos^2\theta)+\mcal{O}(\br^{-1}).
\end{equation}

\paragraph{Large-$r$ expansion in the Bondi--Sachs coordinates.}

We can now find the luminosity radius in the Bondi--Sachs coordinates through the relation $g_{\theta\theta}g_{\varphi\varphi}-g_{\theta\varphi}^2=r^4\sin^2\theta$ to write down $\br$ in terms of $r$. By substituting the components of the metric in \eqref{eq:the bondi-sachs gauge} and solving for $r$, we find that\footnote{Note that we have solved this equation in the large-$r$ limit. In principle, one can find the exact solution and write $\rd^2s_{\tenofo{GBS}}$ in the Bondi--Sachs coordinates. However, the large-$r$ form of the solution is enough for our purpose, and hence we restrict ourselves to that situation.}
\begin{equation}
    \br=r+aF(\theta)+\frac{a^2}{8r}G(\theta)+\mcal{O}(r^{-2}),
\end{equation}
where
\begin{eqgathered}
    F(\theta):= \cot(2\theta)\cos\theta,
    \\
    G(\theta):=\frac{1+4\sin^2\theta-8\sin^4\theta}{\sin^2\theta}.
\end{eqgathered}
We now do a coordinate transformation from the GBS $\bar{X}:= \{u,\br,\theta,\varphi\}$ to the Bondi--Sachs coordinates $X:= \{u,r,\theta,\varphi\}$ using
\begin{equation}
    g_{\mu\nu}(X)=\left.\frac{\partial \bar{X}^{\bar\mu}}{\partial X^\mu}\frac{\partial\bar{X}^{\bar\nu}}{\partial X^\nu}\bg_{\bar\mu\bar\nu}(\bar{X})\right|_{\bar{X}=\bar{X}(X).}
\end{equation}
First, notice that
\begin{eqaligned}
  \frac{\partial\br}{\partial r}&=1-\frac{a^2}{8r^2}G(\theta), 
  \\
     \frac{\partial\br}{\partial\theta}&=aF'(\theta)+\frac{a^2}{8r}G'(\theta)
     \\
     &=-\frac{a}{2\sin^2\theta}\left(2\sin\theta\sin(2\theta)+\cos\theta\cos(2\theta)\right)-\frac{a^2}{4r}\left(4\sin(2\theta)+\frac{\cos\theta}{\sin^3\theta}\right).
\end{eqaligned}
We get
\begin{equation}
    g_{uu}(X)=\bg_{uu}(\bar{X})\Big|_{\bar{X}=\bar{X}(X)}=-1+\frac{2M}{r}+\mcal{O}(r^{-2}).
\end{equation}
Next, we have\footnote{This expansion has been done in \cite[Appendix D]{Barnich:2011mi}. However, there is a discrepancy between our $g_{ur}$ component in \eqref{eq:large-r limit of g-ur in Bondi-Sachs coordinates, appendix} and the expression in the latter reference. This discrepancy is important in the construction of shear $C_{AB}$ as we do in \S\ref{sec:gravitational casimir functionals, appendix}.}
\begin{eqaligned}\label{eq:large-r limit of g-ur in Bondi-Sachs coordinates, appendix}
    g_{ur}(X)&=\left.\frac{\partial\br}{\partial r}\bg_{ur}(\br)\right|_{\bar{X}=\bar{X}(X)}
    \\
    &=\left.\left(1-\frac{a^2}{8r^2}G(\theta)\right)\left(-1+\frac{a^2}{\br^2}\left(\frac{1}{2}-\cos^2\theta\right)\right)\right|_{\bar{X}=\bar{X}(X)}
    \\
    &=-1+\frac{a^2}{r^2}\left(\frac{1}{2}-\cos^2\theta+\frac{1}{8}G(\theta)\right)+\mcal{O}(r^{-4}).
\end{eqaligned}
Next component is
\begin{eqaligned}\label{eq:large-r limit of g-utheta in Bondi-Sachs coordinates}
    g_{u\theta}(X)&=\bg_{u\theta}(X)+\frac{\partial\br}{\partial\theta}\bg_{u\br}(X)
    \\
    &=\frac{a\cos\theta}{2\sin^2\theta}+\frac{a\cos\theta}{4r}\left(8M+\frac{a}{\sin^3\theta}\right)+\mcal{O}(r^{-2}).
\end{eqaligned}
The other components are given by
\begin{eqaligned}\label{eq:large-r limit of g-uphi in Bondi-Sachs coordinates}
    g_{u\varphi}(X)=\bg_{u\varphi}(X)=-\frac{2aM\sin^2\theta}{r}+\mcal{O}(r^{-2}),
\end{eqaligned}
and
\begin{eqaligned}\label{eq:large-r expansion of thetatheta component of the kerr metric in the bs coordinates}
    g_{\theta\theta}(X)&=\bg_{\theta\theta}(X)=r^2+\frac{a}{\sin\theta}r+\frac{a^2}{2\sin^2\theta}+\mcal{O}(r^{-1}),
\end{eqaligned}
and
\begin{equation}\label{eq:large-r expansion of thetaphi component of the kerr metric in the bs coordinates}
    g_{\theta\varphi}(X)=\bg_{\theta\varphi}(X)=\frac{2a^2M\cos\theta}{r}+\mcal{O}(\br^{-2}),
\end{equation}
and finally
\begin{eqaligned}\label{eq:large-r expansion of phiphi component of the kerr metric in the bs coordinates}
    g_{\varphi\varphi}(X)&=\bg_{\varphi\varphi}(X)=r^2\sin^2\theta-ar\sin\theta+\frac{a^2}{2}+\mcal{O}(r^{-1}).
\end{eqaligned}

\subsection{Gravitational Casimir functionals}\label{sec:gravitational casimir functionals, appendix}

Having the metric in the Bondi--Sachs coordinates, we can now compute the Casimir functionals for the Kerr metric. We do this in a few steps.

\paragraph{Covariant mass and covariant momentum.}
We next compute covariant mass and momentum. Consider the metric in the Bondi gauge given in \eqref{eq:BondiMetric} with coefficients for the large-$r$ expansion as in \eqref{eq:FallOff}.  From \eqref{eq:BondiMetric}, we see that
\begin{eqaligned*}
    g_{uA}&=-\Gamma_{AB}\Upsilon^B+\mcal{O}(r^{-3})
    \\
    &=-\left(q_{AB}+\frac{1}{r}C_{AB}\right)\left(\frac{1}{2}D_CC^{CB}+\frac{1}{r}\left(\frac{2}{3}\cJ^B-\frac{1}{2}C^{BC}D^DC_{DC}-\frac{1}{16}\partial^B(C_{CD}C^{CD}\right)\right)+\mcal{O}(r^{-3})
    \\
    &=-\frac{1}{2}q_{AB}D_CC^{CB}+\frac{1}{r}\left(\frac{2}{3}\cJ_A-\frac{1}{16}\partial_A(C_{CD}C^{CD})\right)+\mcal{O}(r^{-2}).
\end{eqaligned*}
If we denote the coefficient of $r^{(n)}$ in the large-$r$ expansion of $g_{\mu\nu}$ as $g_{\mu\nu}^{(n)}$, we see that 
\begin{eqaligned}\label{eq:momentum in terms of metric components I}
    \cJ_A&=\frac{3}{2}g_{uA}^{(-1)}+\frac{3}{32}\partial_A(C_{BC}C^{BC}).
\end{eqaligned}
On the other hand, from \eqref{eq:BondiMetric} and \eqref{eq:FallOff}, we have
\begin{eqaligned}
    g_{ur}&=-e^{2\beta}=-1-2\beta+\mcal{O}(r^{-4})
    \\
    &=-1+\frac{1}{16}\frac{C_{AB}C^{AB}}{r^2}+\mcal{O}(r^{-4}),
\end{eqaligned}
and hence, we have
\begin{equation}\label{eq:expression for C.C}
    C_{AB}C^{AB}=16g_{ur}^{(-2)}.
\end{equation}
Together with \eqref{eq:momentum in terms of metric components I}, we thus conclude that
\begin{equation}\label{eq:general formula for covariant momentum}
    \cJ_A=\frac{3}{2}\left(g_{uA}^{(-1)}+\partial_Ag_{ur}^{(-2)}\right).
\end{equation}
On the other hand, we can compute the covariant mass as follows
\begin{eqaligned}\label{eq:general formula for covariant mass}
    \mcal{M}&=M+\frac{1}{8}N^{AB}C_{AB}
    \\
    &=M+\frac{1}{16}\partial_u(C_{AB}C^{AB})
    \\
    &=M+\partial_ug_{ur}^{(-2)},
\end{eqaligned}
where in the third line, we have used \eqref{eq:expression for C.C}. Using these formulas, we can write the explicit form of covariant mass $\mcal{M}$ and covariant momentum $\cJ=\cJ_A\rd\sigma^A$; the former is easy: Note from \eqref{eq:large-r limit of g-ur in Bondi-Sachs coordinates, appendix} that $g_{ur}$ is $u$-independent and hence \eqref{eq:general formula for covariant mass} implies
\begin{equation}\label{eq:explicit form of the covariant mass for the Kerr metric}
    \mcal{M}=M.
\end{equation}
On the other hand, we see from \eqref{eq:large-r limit of g-ur in Bondi-Sachs coordinates, appendix}, \eqref{eq:large-r limit of g-utheta in Bondi-Sachs coordinates}, \eqref{eq:large-r limit of g-uphi in Bondi-Sachs coordinates}, and \eqref{eq:general formula for covariant momentum} that
\begin{eqaligned}\label{eq:explicit form of the components of the covariant momentum for the Kerr metric}
   \cJ_\theta&=+3aM\cos\theta,
    \\
    \cJ_\varphi&=-3aM\sin^2\theta.
\end{eqaligned}
Hence, the covariant momentum one-form is given by
\begin{equation}
   \cJ=3aM(\cos\theta\,\rd\theta-\sin^2\theta\,\rd\varphi).
\end{equation}
Having the expression for the covariant mass and covariant momentum, we are now ready to construct the Casimir functional for the Kerr metric.

\paragraph{Asymptotic shear for the Kerr metric.} Before proceeding further and for completeness, let us record the components of the asymptotic shear tensor $C_{AB}$ for the Kerr metric, and confirm \eqref{eq:expression for C.C}. From \eqref{eq:BondiMetric} and \eqref{gamma}, we look into the components
\begin{equation}
    r^2\Gamma_{AB}\rd\sigma^A\rd\sigma^B=(r^2q_{AB}+rC_{AB}+\mcal{O}(r^0))\rd\sigma^A\rd\sigma^B, \qquad A,B=\theta,\varphi,
\end{equation}
from which it follows that
\begin{equation}
    C_{AB}=g^{(1)}_{AB}, \qquad A,B=\theta,\varphi,
\end{equation}
From \eqref{eq:large-r expansion of thetatheta component of the kerr metric in the bs coordinates}, \eqref{eq:large-r expansion of thetaphi component of the kerr metric in the bs coordinates}, and \eqref{eq:large-r expansion of phiphi component of the kerr metric in the bs coordinates}, we find that
\begin{eqaligned}
    C_{\theta\theta}&=\frac{a}{\sin\theta}, 
    \\
    C_{\theta\varphi}&=C_{\varphi\theta}=0,
    \\
    C_{\varphi\varphi}&=-a\sin\theta.
\end{eqaligned}
Taking into account that\footnote{Recall that we raise and lower indices by the leading order round metric $q_{AB}$ on the sphere.}
\begin{eqaligned}
    C^{\theta\theta}&=q^{\theta\theta}q^{\theta\theta}C_{\theta\theta}=\frac{a}{\sin\theta},
    \\
    C^{\theta\varphi}&=C^{\varphi\theta}=0,
    \\
    C^{\varphi\varphi}&=q^{\varphi\varphi}q^{\varphi\varphi}C_{\varphi\varphi}=-\frac{a}{\sin^3\theta},
\end{eqaligned}
we can compute 
\begin{equation}\label{eq:expression for C.C, 2nd version}
    C_{AB}C^{AB}=\frac{2a^2}{\sin^2\theta}.
\end{equation}
It is easy to see that
\begin{eqaligned}
    \eqref{eq:expression for C.C}-\eqref{eq:expression for C.C, 2nd version}&=
    16g_{ur}^{(-2)}-\frac{2a^2}{\sin^2\theta}
    \\
    &=16a^2\left(\frac{1}{2}-\cos^2\theta+\frac{1}{8}G(\theta)\right)-\frac{2a^2}{\sin^2\theta}
    \\
    &=0,
\end{eqaligned}
which confirms the expression \eqref{eq:expression for C.C}.

\paragraph{Moment map for the Kerr metric.} The next piece of information is the moment map $\mu_{\gbms}^{\va \text{Kerr}}:\phasespace_{\va \text{ENR}}^{\va \text{Kerr}}\to\gbms^*$ for the $\gbms$ action on $\phasespace_{\va \text{ENR}}^{\va \text{Kerr}}$, the non-radiative strongly electric phase space for the Kerr spacetime. From \eqref{eq:moment map for gbms} and \eqref{eq:explicit form of the covariant mass for the Kerr metric} (which implies $D_A\mcal{M}=0$), we conclude that
\begin{equation}
    \mu_{\gbms}^{\va \text{Kerr}}(\mcal{M})=m, \qquad 
    \mu_{\gbms}^{\va \text{Kerr}}(2^{-1}\mcal{J})=j.
\end{equation}
Recall that $j=j_A\rd\sigma^A$. We are now in a position to construct phase space quantities, i.e. gravitational vorticity, gravitational Casimirs, and gravitational spin charge, for the Kerr metric.

\paragraph{Gravitational vorticity.} The gravitational vorticity for the Kerr metric can be computed easily. From \eqref{eq:pulled-back vorticity}, we see that
\begin{eqaligned}
    \mbf{w}^{\va \text{Kerr}}&:={\mu^{\va \text{Kerr}}_{\gbms}}^*(w(m,j))
    \\
    &=\frac{1}{2}\mcal{M}^{-\frac{2}{3}}\epsilon^{AB}\partial_A\left(\mcal{M}^{-\frac{2}{3}}\mcal{J}_B\right),
\end{eqaligned}
which using \eqref{eq:explicit form of the covariant mass for the Kerr metric} and \eqref{eq:explicit form of the components of the covariant momentum for the Kerr metric} can be explicitly written as
\begin{eqaligned}\label{eq:gravitational corticity for the kerr metric}
    \mbf{w}^{\va \text{Kerr}}=-3a M^{-\frac{1}{3}}\cos\theta,
\end{eqaligned}
where we have used $\epsilon^{\theta\varphi}=\frac{1}{\sin\theta}$. 

\paragraph{Gravitational Casimir functionals.} Next, we compute the gravitational Casimir functionals using \eqref{eq:pulled-back casimirs}
\begin{eqaligned}
    \mscr{C}_n(\phasespace^{\va \text{Kerr}}_{\va \text{ENR}})&:={\mu^{\va \text{Kerr}}_{\gbms}}^*(\msf{C}_n(\gbms))
    \\
    &=\bigintsss_S \mcal{M}^{\frac{2}{3}}\mbf{w}^n \bfe.
\end{eqaligned}
This can be explicitly computed using \eqref{eq:gravitational corticity for the kerr metric} and \eqref{eq:explicit form of the covariant mass for the Kerr metric} as
\begin{eqaligned}
    \mscr{C}_n(\phasespace^{\va \text{Kerr}}_{\va \text{ENR}})&=\bigintsss_S \mcal{M}^{\frac{2}{3}}\mbf{w}^n \bfe,
   \\
   &=\frac{1}{4\pi}\cdot(-3a)^n\cdot M^{\frac{2-n}{3}}\cdot\bigintsss_S\rd\theta\rd\varphi\,\cos^n\theta\sin\theta,
\end{eqaligned}
where the factor of $1/4\pi$ in the second line comes from our normalization of integrals over the sphere (see \eqref{eq:area-form preserved by gbms}). We thus end up with the explicit form of the gravitational Casimirs for the Kerr metric
\begin{equation}
    \mscr{C}_n(\phasespace^{\va \text{Kerr}}_{\va \text{ENR}})=
    \left\{
    \begin{aligned}
    \frac{(-3a)^n}{n+1}&M^{\frac{2-n}{3}}, &\qquad n&=0,2,4,\ldots,
    \\
    &0, &\qquad n&=1,3,5,\ldots.
    \end{aligned}
    \right.
\end{equation}
In deriving this expression, we used
\begin{eqaligned}
    \bigintsss_0^\pi\rd\theta\,\cos^n\theta\sin\theta=
    \left\{
    \begin{aligned}
        &\frac{2}{n+1}, &\qquad n&=0,2,4,\ldots,
        \\
        &\hphantom{nn}0, &\qquad n&=1,3,4,\ldots.
    \end{aligned}
    \right.
\end{eqaligned}

\bibliography{References}
\bibliographystyle{JHEP}

\end{document}